%% file: tese.tex
\documentclass[defaultstyle,10pt]{01.thesis}

\input{packages.tex}
\input{pagesetup.tex}

\makeatletter
\let\origps@plain\ps@plain
\newcommand\MakePlainPagestyleEmpty{\let\ps@plain\ps@empty}
\newcommand\MakePlainPagestylePlain{\let\ps@plain\origps@plain}
\makeatother

\begin{document}
\pagestyle{begin}
\input{1.cover.tex}
\input{3.Acknowledgments.tex}
\input{4.Abstract.tex}
\input{5.Keywords.tex}
\input{4.Resumo.tex}

\input{5.PalavrasChave.tex}
\input{6.Tables.tex}
\acresetall
\input{7.Acronim.tex}

\input{8.symbols.tex}
\pagestyle{documentsimple}
\tableofcontents
\input{2.1.preface.tex}
\input{1.Introduction/main.tex}

\MakePlainPagestyleEmpty
\part{Classical Thermodynamic Systems in Curved Spacetimes}
\thispagestyle{empty}
\cleardoublepage
\MakePlainPagestylePlain
\input{I.1ShellIntro/main.tex}

\cleardoublepage
\input{I.3ShellCharged/main.tex}

\cleardoublepage
\input{I.4ShellExtremal/main.tex}

\cleardoublepage
\input{I.5ExtremalLimits/main.tex}

\cleardoublepage
\input{I.2ShellD/main.tex}

\cleardoublepage
\MakePlainPagestyleEmpty
\part{Quantum Thermodynamic Systems in Curved Spacetimes}
\thispagestyle{empty}
\cleardoublepage
\MakePlainPagestylePlain
\input{II.1GFinCS/main.tex}

\input{II.2QFTinCS/main.tex}

\input{II.3BTZ/main.tex}
\input{II.4Lifshitz/main.tex}

\cleardoublepage
\input{II.55Dcharged/main.tex}

\cleardoublepage
\input{II.65D6D/main.tex}

\cleardoublepage
\input{II.7SymRest/main.tex}

\cleardoublepage
\input{98.Conclusions/conclusions.tex}
\cleardoublepage
\begin{appendices}
	\begin{appendix}
		\input{AppendiceA/appendixA.tex}

		\input{AppendiceB/appendixB.tex}

	\end{appendix}
\end{appendices}
\cleardoublepage
\addcontentsline{toc}{chapter}{Bibliography}
\nocite{apsrev41Control}
\bibliographystyle{apsrev4-1}
\bibliography{99.Bibliography/biblio.bib}

\end{document}

%% file: packages.tex
\usepackage[english]{babel}

\usepackage[latin1]{inputenc}

\usepackage{ulem} 
\usepackage{datetime}
\newdateformat{todaythesis}{%
\monthname[\THEMONTH]  \THEYEAR}

\usepackage{latexsym}

\usepackage{amsmath, amsthm, amssymb, amsfonts, amsbsy}

\usepackage{multirow}
\usepackage{colortbl}
\usepackage{supertabular}
\usepackage{pdflscape}

\usepackage{graphics}
\usepackage{graphicx}
\usepackage{epsfig}
\usepackage[hang,small,bf]{subfigure}
\usepackage{dcolumn}
\usepackage{bm}
\usepackage{booktabs}
\usepackage{rotating}
\usepackage{multirow}
\usepackage{array}

\usepackage[font=small,labelfont=bf,textfont=normalfont]{caption}

\usepackage[square,numbers,sort&compress]{natbib}

\usepackage{acronym}
\usepackage[titletoc,title,header]{appendix}
\usepackage[noauto]{chappg}

\usepackage{00.extra_functions}

\usepackage{float}
\usepackage{00.symlist}

\newcommand{\equationref}[1]{Equation (\ref{#1})}

\newcommand{\g}[0]{\gamma}
\newcommand{\dd}[0]{\partial}
\newcommand{\al}[0]{\alpha}
\newcommand{\be}[0]{\beta}
\newcommand{\de}[0]{\delta}
\newcommand{\e}[0]{\varepsilon}
\newcommand{\ze}[0]{\zeta}
\newcommand{\si}[0]{\sigma}
\newcommand{\la}[0]{\lambda}

\newcommand{\bs}[1]{\textbf{#1}}
\newcommand{\ma}[1]{\mathcal{#1}}
\newcommand{\ea}[1]{\begin{align}#1\end{align}}
\newcommand{\eq}[1]{\begin{equation}#1\end{equation}}

\usepackage{slashed}
\usepackage{braket}
\usepackage{mathrsfs}

%% file: pagesetup.tex
\usepackage[top=2.5cm, bottom=2.5cm, inner=2.5cm, outer=2.5cm]{geometry}

\usepackage{fancyhdr}
\fancypagestyle{plain}{
   \fancyhead{} 
   
   \fancyfoot[C]{\bfseries\small\thepage}
}
\fancypagestyle{begin}{%
   \fancyhead{}

   \fancyfoot[C]{\bfseries\small\thepage}
}
\fancypagestyle{document}{%
	\fancyhf{}
	\fancyhead[LE]{\bfseries\nouppercase{\leftmark}}
	\fancyhead[RO]{\bfseries\nouppercase{\rightmark}}
	\fancyfoot[C]{\bfseries\small\thepage}
	\addtolength{\headheight}{2pt} 
}
\fancypagestyle{documentsimple}{%
	\fancyhf{}
	\fancyfoot[C]{\bfseries\small\thepage}
	\addtolength{\headheight}{2pt} 
}
\graphicspath{{Figures/}}

%% file: 1.cover.tex
\setcounter{page}{1} \pagenumbering{Alph}

\pdfbookmark[0]{Title}{Title}

\thispagestyle{empty}
\begin{flushleft} ~\\ \vspace{-18mm} \hspace{-9mm}  \includegraphics[width=40mm]{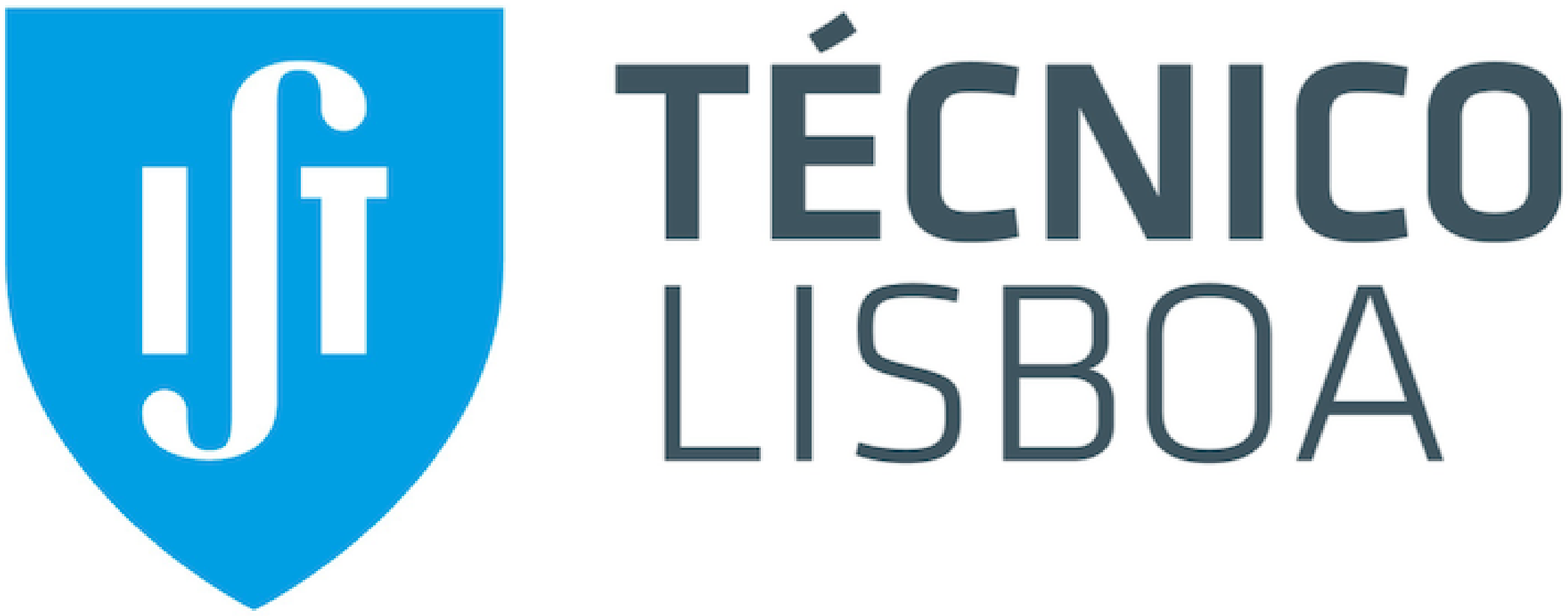}
\\ \vspace{5mm}
\begin{centering}
\LARGE \textbf{UNIVERSIDADE DE LISBOA}\\
\LARGE \textbf{INSTITUTO SUPERIOR T\'{E}CNICO}\\
\vspace{5cm}
\LARGE \textbf{Classical and Quantum Thermodynamic}\\
\LARGE \textbf{Systems in Curved Spacetime}
\\ \vspace{30mm}
\Large \textbf{Gonçalo Martins Quinta} \\
\vspace{15mm}
\flushleft
\large \textbf{Supervisor}: Doctor Jos\'{e} Pizarro de Sande e Lemos   \large \\
\large \textbf{Co-Supervisor}: Doctor Antonino Flachi   \large \\

\centering
\vspace{15mm}
\textbf{Thesis approved in public session to obtain the PhD Degree in Physics}\\
\vspace{3mm}
\textbf{Jury final classification: Pass with Distinction and Honour}
\\ \vspace{30mm}

\vspace{13mm}

\Large \textbf{2018} \\
\end{centering}
\let\thepage\relax
\end{flushleft}
\pagebreak

\clearpage

\thispagestyle{empty}
\cleardoublepage

\setcounter{page}{1} \pagenumbering{roman}

\setcounter{page}{10} \pagenumbering{arabic}

\baselineskip 18pt 

\thispagestyle{empty}
\begin{flushleft} ~\\ \vspace{-18mm} \hspace{-9mm}  \includegraphics[width=40mm]{Figures/Cover/istlogo}
\\ \vspace{5mm}
\begin{centering}
\LARGE \textbf{UNIVERSIDADE DE LISBOA}\\
\LARGE \textbf{INSTITUTO SUPERIOR T\'{E}CNICO}\\
\vspace{10mm}
\LARGE \textbf{Classical and Quantum Thermodynamic}\\
\LARGE\textbf{Systems in Curved Spacetime}
\\ \vspace{10mm}
\Large \textbf{Gonçalo Martins Quinta} \\
\vspace{5mm}
\flushleft
\large \textbf{Supervisor}: Doctor Jos\'{e} Pizarro de Sande e Lemos   \large \\
\large \textbf{Co-Supervisor}: Doctor Antonino Flachi   \large \\

\centering
\vspace{10mm}
\textbf{Thesis approved in public session to obtain the PhD Degree in Physics}\\
\vspace{1mm}
\textbf{Jury final classification: Pass with Distinction and Honour}
\\ \vspace{10mm}
\textbf{Jury}

{\flushleft
\textbf{Chairperson:} Doctor Jorge Manuel Rodrigues Crispim Rom\~{a}o, Instituto Superior T\'{e}cnico, Universidade de Lisboa \\
\vspace{5mm}
\textbf{Members of the Committee:}\\
\begin{itemize}
\item[] Doctor J\'{u}lio C\'{e}sar Fabris, Universidade Federal do Esp\'{i}rito Santo, Brasil \\
\item[] Doctor Jos\'{e} Pizarro de Sande e Lemos, Instituto Superior T\'{e}cnico, Universidade de Lisboa \\
\item[] Doctor Carlos Alberto Ruivo Herdeiro, Instituto Superior T\'{e}cnico, Universidade de Lisboa \\
\item[] Doctor Oleg Borisovich Zaslavskii, V. N. Karazin Kharkiv National University, Ukraine
\end{itemize}
}

\vspace{5mm}

\textbf{Funding Institution:}\\
Grant SFRH/BD/92583/2013 from Funda\c{c}\~{a}o para a Ci\^{e}ncia e Tecnologia (FCT).

\vspace{13mm}

\Large \textbf{2018} \\
\end{centering}
\let\thepage\relax
\end{flushleft}
\pagebreak

\clearpage

\thispagestyle{empty}
\cleardoublepage

\setcounter{page}{1} \pagenumbering{roman}

\setcounter{page}{10} \pagenumbering{arabic}

\baselineskip 18pt 

%% file: 3.Acknowledgments.tex
\pdfbookmark{Acknowledgments}{Acknowledgments}
\begin{acknowledgments}

A PhD is a long journey and a lot of life happens in four years. Thankfully, they were spent in wonderful company, so let me give credit where credit is due.

First of all, I am extremely grateful to my supervisor Prof. Jos\'{e} Lemos, for all things taught both in life and in physics, and for giving me the opportunity to pursue my own scientific projects. I am also indebted to my co-supervisor Prof. Antonino Flachi, for his top-grade mentoring, hospitality and good humor, and for being able to do it even from the other side of the world. You will be an excellent father!

I would also like to express my gratitude to all members of CENTRA, in particular the gravity group, for the excellent and stimulating working environment (and for the memorable dinners). Particular mention goes to all my office mates, who made our office the best one to work in all of IST. Special thanks also go to Dulce Concei\c{c}\~{a}o, for turning the indecipherable bureaucratic world into a breeze.

I gratefully acknowledge the hospitality of the International College of Liberal Arts and the physics group at Keio University, where parts of this work have been done. I am also very grateful to the friends that made my stay in Japan a wonderful experience, with special thanks to all the Philips and Julias, Erika, Max, Nate, Hillary and Yuuki. I miss you all!

For making me temporally forget about the housing prices in Lisbon, I thank all my house mates. You have made it wonderful to come back home after work.

And for last, I save my most warming thanks to all my family and loved ones. Mother, I know it took a lot to even get me to the starting point. No good research can be done without happiness, peace of mind and memes. For having given me all of that, I save my special thanks to my friends Miguel Orcinha, Rui Andr\'{e}, M\'{a}rio Aires, Jo\~{a}o Lu\'{i}s and Gon\c{c}alo Guiomar. You guys set my standards for human beings real high.

\end{acknowledgments}
\clearpage
\thispagestyle{empty}
\cleardoublepage

%% file: 4.Abstract.tex
\begin{abstract}

Systems at finite temperature make up the vast majority of realistic physical scenarios. Indeed, although zero temperature is often accompanied by simpler mathematics, the richness in physical results is evident when one considers the system to have temperature. This is even more so if the background geometry where the physical system resides has a general curvature. This thesis will be dedicated to the study of this type of physical systems, where thermodynamics and general relativity equally contribute to the dynamics.

The first part of the thesis will be devoted to the study of classical thermodynamic systems in curved spacetime, namely thin matter shells at finite temperature. These objects partition spacetime into separate pieces, and their very existence is conditioned by the so-called junctions conditions. The latter conditions allow us to carefully study both the mechanical and thermodynamics of the shell and, in particular, they give rise to a well-defined notion of entropy. The shell can then be taken to its black hole limit, providing an alternative way to study black hole thermodynamics. We will do this for different geometries, obtaining as byproduct a plausible answer for the debated value of the entropy of an extremal black hole.

In the second part of this thesis, we shall review the standard formalisms to study quantum field theory in curved spacetimes, in order to explore quantum properties of thermodynamic systems in the presence of gravity. Massive quantum scalar fields at finite temperature will be the systems of choice, whereby various instances of vacuum polarisation will be calculated in a variety of black hole geometries. Both numerical and analytic results will be obtained, and new addition formulas for a certain class of transcendental functions will be derived. This part will culminate with a careful numerical study of symmetry restoration of a self-interacting scalar field around a charged black hole, where we verify insights present in the literature.

\end{abstract}

%% file: 5.Keywords.tex
\clearpage
\thispagestyle{empty}
\cleardoublepage

%% file: 4.Resumo.tex
\begin{resumo}

Sistemas a uma temperatura finita constituem a vasta maioria de cen\'{a}rios fisicamente realistas. De facto, embora a aus\^{e}ncia de temperatura seja frequentemente acompanhada de simplicidade matem\'{a}tica, a riqueza de resultados f\'{i}sicos \'{e} evidente quando se considera que o sistema tem temperatura. Isto \'{e} ainda mais verdade ainda se a geometria de fundo onde o sistema reside tiver uma curvature geral. Esta tese ser\'{a} dedicada a estudar este tipo de sistema f\'{i}sico, onde termodin\^{a}mica e relatividade geral contribuem igualmente para a din\^{a}mica.

A primeira parte da tese ser\'{a} dedicada ao estudo de sistemas termodin\^{a}micos cl\'{a}ssicos em espa\c{c}o-tempo curvos, nomeadamente camadas finas de mat\'{e}ria com temperatura finite. Este tipo de objectos divide o espa\c{c}o-tempo em peda\c{c}os separados, e a sua própria exist\^{e}ncia \'{e} condicionada pelas chamadas condi\c{c}ões de jun\c{c}\~{a}o. Estas últimas permitem-nos estudar cuidadosamente tanto a mec\^{a}nica como termodin\^{a}mica da camada e, em particular, d\~{a}o origem a uma no\c{c}\~{a}o de entropia bem definida. A camada pode assim ser levada para o seu limite de buraco negro, fornecendo um m\'{e}todo alternativo para o estudo da termodin\^{a}mica de buracos negros. Iremos faz\^{e}-lo para geometrias diferentes, obtendo como subproduto uma resposta plaus\'{i}vel para o debatido valor da entropia de um buraco negro extremal.

Na segunda parte desta tese, iremos rever os formalismos usuais para estudar teoria do campo qu\^{a}ntico em espa\c{c}os-tempo curvos, de forma a explorar as propriedades qu\^{a}nticas de sistemas termodin\^{a}micos na presença de gravidade. Campos qu\^{a}nticos escalares massivos com temperatura finita ser\~{a}o a escolha de sistema, para os quais se ir\~{a}o calcular v\'{a}rias inst\^{a}ncias de polariza\c{c}\~{a}o de v\'{a}cuo em diferentes geometrias de buracos negros.
Ir\~{a}o ser obtidos resultados tanto num\'{e}ricos como anal\'{i}ticos, e novas f\'{o}rmulas de adi\c{c}\~{a}o de fun\c{c}\~{o}es transcendentais ser\~{a}o derivadas. Esta parte ir\'{a} culminar no estudo num\'{e}rico detalhado da quebra de simetria de um campo escalar auto-interactivo em torno de um buraco negro carregado, onde iremos verificar intui\c{c}\~{o}es j\'{a} estabelecidas na literatura.

\end{resumo}

%% file: 5.PalavrasChave.tex
\begin{palavraschave}
Buracos negros, teoria de campo quântico, termodinâmica, camadas finas, polarisação de vácuo
\end{palavraschave}
\clearpage
\thispagestyle{empty}
\cleardoublepage

%% file: 6.Tables.tex
\dominitoc
\dominilof
\dominilot

\pdfbookmark[1]{List of Figures}{lof}
\listoffigures
\clearpage
\thispagestyle{empty}
\cleardoublepage



%% file: 7.Acronim.tex
\pdfbookmark[1]{List of Acronyms}{loac}

\chapter*{Acronyms}


\begin{acronym}
\acro{ADM}{Arnowitt-Deser-Misner}
\acro{GR}{General Relativity}
\acro{QFT}{Quantum Field Theory}
\acro{QFTCS}{Quantum Field Theory in Curved Spacetime}
\end{acronym}

\clearpage
\thispagestyle{empty}
\cleardoublepage

%% file: 8.symbols.tex
\chapter*{Notation and conventions}

Whenever unspecified, the metric is assumed to have a $(-1,1,1,1)$ signature, a Rienman tensor defined by $R^{\al}{}_{\be\g\de} = \Gamma^{\al}{}_{\be\de,\g} +\cdots$, and a Ricci tensor defined by $R_{\al \be} = R^{\mu}{}_{\al \mu \be}$. Greek indices $(\al,\be,\cdots)$ run from 0 to 3, while latin ones $(a,b,\cdots)$ run from 1 to 3. Unless explicitely stated otherwise, we will use natural units $\varepsilon_0=c=k_{B} = \hbar = 1$. Operators will be denoted with a hat $(\hat{\mathcal{O}}, \cdots)$.

\clearpage
\thispagestyle{empty}

\cleardoublepage
\baselineskip 18pt

%% file: 2.1.preface.tex
\begin{preface}
Official research presented in this thesis has been carried out at Centro de Astrof\'{i}sica e Gravita\c{c}\~{a}o (CENTRA) in the Physics Department of Instituto Superior T\'{e}cnico, and was supported by Funda\c{c}\~{a}o para a Ci\^{e}ncia e Tecnologia (FCT), through Grant No.~SFRH/BD/92583/2013.

I declare that this thesis is not substantially the same as any that I have submitted for a degree, diploma or other qualification at any other university and that no part of it has already been or is concurrently submitted for any such degree, diploma or other qualification.

The majority of the work presented in Part I was done in collaboration with Professor Jos\'{e} Sande Lemos and Professor Oleg Zaslavskii.
Part II was entirely made in collaboration with Professor Professor Jos\'{e} Sande Lemos and Professor Antonino Flachi. Parallel research was also done in collaboration with my colleague Rui Andr\'{e}. Chapters 3, 4, 5, 10, 11 and 12 have been published. Chapters 6 and 13 have been submitted and Chapter 9 is being prepared for submission. The publications presented in this thesis are the following:
\begin{itemize}
   \item[6.] Gon\c{c}alo M. Quinta, Antonino Fachi, Jos\'{e} P. S. Lemos, ``Quantum vacuum polarization around a Reissner-Nordstr\"{o}m black hole in five dimensions'', Phys. Rev. D \textbf{97}, 025023 (2018); arXiv:1712.08171. (Chapter 11)
   \item[5.] Gon\c{c}alo M. Quinta, Antonino Fachi, Jos\'{e} P. S. Lemos, ``Black Hole Quantum Vacuum Polarization in Higher Dimensions'', Phys. Rev. D \textbf{94}, 105001 (2016);   arXiv:1609.06794. (Chapter 12)
   \item[4.] Gon\c{c}alo M. Quinta, Antonino Fachi, Jos\'{e} P. S. Lemos, ``Vacuum polarization in asymptotically Lifshitz black holes'', Phys. Rev. D \textbf{93}, 124073 (2016); arXiv:1604.00495. (Chapter 10)
   \item[3.] Jos\'{e} P. S. Lemos, Gon\c{c}alo M. Quinta, Oleg B. Zaslavski, ``Entropy of extremal black holes: Horizon limits through charged thin shells in a unified approach'', Phys. Rev. D \textbf{93}, 084008 (2016); arXiv:1603.01628. (Chapter 3)
   \item[2.] Jos\'{e} P. S. Lemos, Gon\c{c}alo M. Quinta, Oleg B. Zaslavski, ``Entropy of a self-gravitating electrically charged thin shell and the black hole limit'', Phys. Rev. D \textbf{91}, 104027 (2015); arXiv:1503.00018. (Chapter 4)
   \item[1.] Jos\'{e} P. S. Lemos, Gon\c{c}alo M. Quinta, Oleg B. Zaslavski, ``Entropy of an extremal electrically charged thin shell and the extremal black hole'', Phys. Lett. B \textbf{750}, 306-311 (2015); arXiv:1505.05875. (Chapter 5)
\end{itemize}
Publications published in proceedings are:
\begin{itemize}
    \item[2.] Gon\c{c}alo M. Quinta, Jos\'{e} P. S. Lemos,
        ``Lifshitz black holes and vacuum polarization'' In:
        \emph{Proceedings, 14th Marcel Grossmann Meeting on Recent
                        Developments in Theoretical and Experimental General
                        Relativity, Astrophysics, and Relativistic Field Theories
                        (MG14) (In 4 Volumes): Rome, Italy, July 12-18, 2015},
        (2017).
    \item[1.] Gon\c{c}alo M. Quinta and Jos\'{e} P. S. Lemos,
        ``Spherical thin shells in $d$-dimensional general
                        relativity: Thermodynamics and entropy''. In:
        \emph{Proceedings, 14th Marcel Grossmann Meeting on Recent
                        Developments in Theoretical and Experimental General
                        Relativity, Astrophysics, and Relativistic Field Theories
                        (MG14) (In 4 Volumes): Rome, Italy, July 12-18, 2015},
        (2017).
\end{itemize}
The publications which have recently submitted are the following:
\begin{itemize}
\item[2.] Gon\c{c}alo M. Quinta, Antonino Flachi, Jos\'{e} P. S. Lemos, ``Symmetry restoration of interacting quantum fields around charged black holes''. (Chapter 13)
\item[1.] Jos\'{e} P. S. Lemos, Rui Andr\'{e}, Gon\c{c}alo M. Quinta, ``Thermodynamics of thin matter shells in $d$ dimensions''. (Chapter 6)
\end{itemize}
Further work published by the author during the duration of the thesis but not discussed in this work is:
\begin{itemize}
   \item[2.] Gon\c{c}alo M. Quinta, Rui Andr\'{e}, ``Classifying quantum entanglement
   through topological links'', Phys. Rev. A \textbf{97}, 042307 (2018); arXiv:1803.08935.
   \item[1.] Jos\'{e} P. S. Lemos, Francisco J. Lopes, Gon\c{c}alo M. Quinta, Vilson T. Zanchin, ``Compact stars with a small electric charge: the limiting radius to mass relation and the maximum mass for incompressible matter'', Eur.Phys.J. C \textbf{75}, 76 (2015); arXiv:1408.1400.
\end{itemize}

\end{preface}
\clearpage
\thispagestyle{empty}
\cleardoublepage

%% file: 1.Introduction/main.tex
\pagenumbering{arabic} \setcounter{page}{1}
\chapter{General introduction}
\label{cap:int}

\section{Quantum gravity and the semiclassical limit}

Mankind's understanding of gravity, described by Einstein's theory of General Relativity (GR), is indisputable. The recently centenary theory characterizes the movement of bodies in a gravitational field with an accuracy that keeps surprising physicists even today. Equally astonishing is the empirical success of quantum field theory in describing the dynamics of the microscopic world with record breaking precision. However, despite these achievements, we know that both theories must suffer modifications at very high energies, converging into a single quantum gravity theory of nature. Many attempts have been made at finding such a unification, with string theory and loop quantum gravity being the most known ones. The program of string theory is to describe elementary particles and force carriers as strings in higher-dimensional spacetimes, while loop quantum gravity attempts to use quantized loops in order to quantize spacetime itself. All of these theories, however, possess shortcomings of various sorts. As we stand, it is safe to say that a full understanding of how gravity incorporates the quantum attribute of the fields responsible for all fundamental forces has not been achieved yet.

The lack of a full quantum theory of gravity is, nevertheless, only relevant at energies of the order of the Planck scale $E_{P} \sim 10^{19} \, \textrm{GeV}$, which is absurdly higher than the energy scales $E\sim 100 \, \textrm{GeV}$ usually involved in particle physics or gravity. In the monumental scale difference between these two regimes lies the semiclassical approximation of quantum gravity. Quantum field theory in curved spacetime (QFTCS) is the framework which describes this approximation (see \cite{Birrell:1982ix,Parker:2009uva,Toms:2012bra} for standard introductory literature), dictating how elementary particles behave in gravitational fields at energies below the Planck scale. It does so by considering the fields to lie in a curved background, whose geometry is still ruled by the Einstein equations. This approximation is inevitable, due to the lack of knowledge of how the dynamics of spacetime is intertwined with that of quantized fields. As an approximation, it stands in the same place as studying the Schr\"{o}edinger equation in the presence of a classical electric field, rather than using its quantized version. Nevertheless, just as the quantized energies of the atom arise therein, equally fascinating consequences emerge from considering quantized fields in curved spacetimes.

\section{Physics in $d$ dimensions}

Physicists have turned to the possibility of extra dimensions beyond the four known ones for a long time. One of the most serious attempts dates back to the twenties, where spacetime was assumed to have one more spatial dimension. This theory, nowadays know as Kaluza-Klein theory (see \cite{Nordstrom:1914,Kaluza:1921} for the original articles and \cite{Appelquist:1987nr} for an English translation), encoded the Einstein equations, Maxwell equations and predicted an extra scalar field known as dilaton; all of this possible by just assuming spacetime to have five dimensions rather than four. The idea was that the extra dimension could somehow be compacted in a very small scale, and the hope was to construct a unified theory of gravity and electromagnetism, but as more fundamental forces were discovered, it became clear it could not provide the correct picture. Despite this, it served as inspiration for other more robust theories such as string theory, which needs a minimum of 26 dimensions to be physically consistent. Another famous example where higher dimensions make an appearance is in the AdS/CFT correspondence \cite{Maldacena:1997re,Gubser:1998bc,Witten:1998qj}, which conjectures that weakly-coupled AdS gravity geometries in $d+1$ dimensions can be mapped one-to-one into strongly coupled conformal quantum field theories in $d$ dimensions. Although the total number of dimensions is of course relevant, the most important point is that the existence of a bulk in an extra dimension is what allows the correspondence in the first place. Even though this conjecture has not been formally proven yet, it has already been used to study a variety of strongly coupled quantum systems by using methods familiar to numerical relativists. Reviews on different aspects of the duality can be consulted in \cite{Aharony:1999ti,DHoker:2002nbb,Hartnoll:2009sz,McGreevy:2009xe,CasalderreySolana:2011us,Kim:2012ey,Adams:2012th}.

More recently, a flurry of interest has risen in another application of $d$ dimensional spacetimes, namely in the large $d$ limit \cite{Soda:1993xc,Grumiller:2002nm,Yoshino:2002br,Berti:2009kk,Hod:2010zza,Hod:2011zza,Hod:2011zzb,Hod:2011dwa,Coelho:2012sya,Coelho:2012sy,Coelho:2013zs,Caldarelli:2012hy,Emparan:2013moa,Giribet:2013wia,Emparan:2013xia,Prester:2013gxa,Emparan:2013oza,Emparan:2014cia,Emparan:2014jca,Emparan:2013wqa,Emparan:2014aba,Emparan:2015rva,Bhattacharyya:2015dva,Emparan:2015hwa,Emparan:2016sjk,Emparan:2016ylg,Emparan:2016ipc,Dandekar:2016jrp,Andrade:2018nsz,Andrade:2018rcx}. If one interprets the number of dimensions as a free parameter, it becomes plausible to expand for $d \to \infty$ and express quantities as a series in $1/d$. The main advantage of this expansion is that black hole geometries become non-flat in only a small region around the black hole horizon, simplifying a great deal both numerical and analytic problems in black hole physics.

A great deal of this thesis will be devoted to studying the nature of both classical and quantum systems in curved spacetime. In particular, we will see later on that extra dimensions add a lot of content to physical results, although the large $d$ approximation is particularly difficult to implement in a quantum field theoretic context.

\section{Finite temperature classical systems in curved spacetime}

It is just as reasonable to consider a thermodynamic system in flat spacetime as it is in a curved one. However, nature has hinted from time to time that General Relativity and Thermodynamics actually shared some much more deep connections. The first glimpse of this was obtained by Bekenstein \cite{Bekenstein:1972tm,Bekenstein:1973ur,Bekenstein:1974ax,Bekenstein:1980jp}, who found that differential quantities associated to them were related in a similar fashion to that of the first law of thermodynamics, which inspired him to conjecture that black hole should have an associated entropy and temperature. Even though this hypothesis was only confirmed later by Hawking using a quantum treatment, classical systems were enough to infer that such properties should hold. In fact, classical arguments were enough to recover the first law of thermodynamics from the Einstein equations later on, showing beyond doubt that the geometry of spacetime somehow has encoded in it the statistical laws of thermodynamics. Nonetheless, the microscopic origin of such degrees of freedom, namely in the entropy associated to horizons, remains an unsolved problem (see \cite{Gerlach:1976kk,Jacobson:1993vj,Visser:1993nu,Banados:1993qp,Susskind:1993ws} for some attempts at a solution).

Despite all advances already made regarding the thermodynamics of the gravitational field \cite{Jacobson:1995ab,Padmanabhan:2002ma,Padmanabhan:2002jr,Padmanabhan:2002sha,Padmanabhan:2009vy}, much more clarifications are still in order. In fact, even though a quantum theory of gravity should not create any distinction between gravitational and material degrees of freedom, it is still a subject of study at the phenomenological level nonetheless. Indeed, a phenomenological classical treatment could shed some light on the features of a definite unified treatment of quantum interactions, as was shown to happen in the past. We would thus be interested in a system which contains both gravitational and material degrees of freedom but which does not introduce too many complexities due to the matter constitution. A particularly simple system which satisfies these requirements is a spherically symmetric self-gravitating thin matter shell at a finite temperature.

A thin shell is an infinitesimally thin surface which partitions spacetime into an interior region and an exterior region. Since it corresponds to a singularity in the metric of the spacetime, the thin shell must satisfy some conditions in order for the entire spacetime to be a valid solution of the Einstein equations. Such conditions are called junctions conditions, and relate the stress-energy tensor of the shell to the extrinsic curvature of the spacetime through Israel's massive thin shell formalism \cite{Israel:1966rt, Lanczos:1922, Chase:1970, Poisson:2004}. Thus, the material degrees of freedom of the shell are related to the gravitational degrees of freedom through the gravitational field equations, and so the thermodynamics of the shell is deeply connected to the structure of spacetime. Indeed, Davies, Ford and Page \cite{Davies:1986wx} and Hiscock \cite{Hiscock:1989uj}, have shown the usefulness of studying thin shells in (3+1)-dimensional general relativistic spacetimes from a thermodynamic viewpoint.

Another reason which motivates the use of thin shells is the fact that they can be taken to their gravitational radius, i.e., the black hole limit. One can, for example, calculate the entropy of a shell for given spacetimes and see what value it assumes in the black hole limit. Thus, the black hole thermodynamic properties can be studied by a much more direct computation than the usual black hole mechanics if thin shell are used, an idea which was developed by Brown and York \cite{Braden:1990hw} and Martinez \cite{Martinez:1996ni} and which is going to be used throughout this work. A similar approach used for the study of black holes through quasi-black holes has also been proposed by Lemos and Zaslavskii \cite{Lemos:2010,Lemos:2009uk,Lemos:2010kw}.

\section{Finite temperature quantum systems in curved spacetime}

Although classical arguments can have reasonable predictive power regarding quantum properties of systems in curved spacetime, a quantum treatment is necessary if one wishes to probe deeper into physical manifestations of quantum effects. To consider this is to use the semiclassical Einstein equations, whose difference from the classical version is the inclusion of the quantum-average of the stress-energy tensor operator rather than the classical one. Although this constitutes an approximation, since a full quantum theory is not available, these semiclassical Einstein equations represent a middle-ground between the classical version and a full-fledged quantum form based on the firm foundations of Quantum Field Theory (QFT) and GR. This proves enough to obtain already a plethora of new predictions, like the Hawking effect \cite{Hawking:1974sw,Fredenhagen:1989kr}, the Fulling-Davies-Unruh effect \cite{Davies:1974th,Fulling:1972md,Unruh:1976db}, cosmological particle creation and the production of curvature fluctuations during inflation \cite{Guth:1982ec,Hawking:1982cz,Mukhanov:1981xt,Starobinsky:1982ee}. These kind of predictions are of utmost importance, since a full quantum theory of gravity should also predict these phenomena. Indeed, candidate theories are selected based on whether or not they fullfil this criterion.

Ideally, the most complete problem to be adressed in QFTCS would be to fully solve the semiclassical Einstein equations. This problem, however, has proven to be particularly complex, even numerically, due to the sheer complexitity of the average of the stress-energy tensor operator. The latter quantity, however, turns out to be a function of a simpler object, called vacuum polarisation, which is by definition the coincidence limit of the Green function associated to the quantum field permeating spacetime. This quantity indicates the rate of spontaneous particle creation, and is strongly influenced by the curvature of spacetime. By introducing a temperature in the environment, the vacuum polarisation will thermalize with it, resulting in one of the most physically rich setups in physics, where gravity, quantum mechanics and thermodynamics all come into play.

A black hole radiates with a Hawking temperature, so it serves as an ideal thermal bath for the quantum field to be immersed in. This was the setup consired by the pioneering work of Candelas \cite{Candelas:1980zt}, where he calculated the vacuum polarisation of a massless minimally coupled quantum scalar field at the horizon of a Schwarschild black hole. This worked was extended by Candelas and Howard \cite{Candelas:1984pg} and Fawcett and Whitting \cite{Fawcett:1981fw} to the entire exterior region of the black hole. The interior region was studied by Candelas and Jensen \cite{Candelas:1985ip}, while Howard and Candelas \cite{Howard:1984qp,Howard:1985yg} and Fawcett \cite{Fawcett:1983dk} calculated the average value of the stress-energy tensor operator for the entire Schwarschild geometry. Many other geometries were investigated, some of which included massive and non-minimally coupled fields. This thesis will contribute to the program of understanding quantum field activity around black holes through the calculation of vacuum polarisation in geometries not previously addressed.

\section{Outline of the thesis}

This thesis structure is intended to be self-contained, in the sense that any physics graduate student should be able to get a good grasp of the mathematical and physical concepts being treated in it, without having to consult any other literature in order to replicate the vast majority of the calculations.

The thesis is organized as follows. It is divided into two parts: part I will be concerned with classical thermodynamic systems in curved space and part II will be devoted to quantum systems in curved space at finite temperature. In terms of chapters, Chapter 2 will give a concise introduction of the thin shll formalism in GR, playing a foundational role in the development of all the chapters in part I. Chapter 3 will deal with thin shells in $d$ dimensions and their thermodynamic properties both generally and in the black hole limit, were they are taken to their gravitational radius. Chapter 4 will repeat the same study, this time regarding charged shells in four dimensions. Chapter 5 will be concerned with the delicate limit of extremaly charged shells, which will allow us to draw conclusions regarding the entropy of extremal black holes, a topic still in debate. Chapter 6 will re-evaluate the results of Chapter 5 in an alternative approach that deepens out understading of the extremal limit. Part II of the thesis then starts will Chapter 7, which introduces the main mathematical tools used in QFT in a curved background at finite temperature. Chapter 8 will make use of the concepts of Chapter 7, applying them to QFTCS at finite temperature using the proper time formalism introduced by Schwinger. Chapter 9 will be the first application of the results of Chapters 7 and 8, where we will calculate the thermal Green functions of a scalar field in the background geometry of a BTZ black hole. Chapter 10 will be dedicated to the calculation of the vacuum polarisation of a free scalar field around a lifshitz black hole in four dimensions. In Chapter 11 we will consider one extra spacetime dimension and derive numerical results for the vacuum polarisation of a scalar field around a charged black hole. In Chapter 12, we shall generalise the methods used in previous chapters to an arbitrary number of dimensions, using them explicitely to the five and six dimensional cases. Chapter 13 will consider the more complicated and physically realistic case of a self-interacting massive scalar field around a charged black hole in four dimensions, where phase transitions of the field will be seen to occur at certain distances from the black hole horizon.

\cleardoublepage

%% file: I.1ShellIntro/main.tex
\chapter{The Thin Shell Formalism}
\label{cap:chapterI1}
\input{I.1ShellIntro/outline.tex}
\input{I.1ShellIntro/1.section.tex}
\input{I.1ShellIntro/2.section.tex}

\input{I.1ShellIntro/3.section.tex}

%% file: I.1ShellIntro/outline.tex
\section{Introduction}

As in flat spacetime, thermodynamics in curved spaces remains exactly of the same form, i.e. the main laws are still obeyed by any physical systems with well defined statistical properties. However, the simplest the physical system chosen, the easier it will be to extract meaningful information from it. In this chapter, we will review a mathematical formalism, developed in \cite{Israel:1966rt}, which allows us to study a very simple system, given by a two dimensional hypersurface containing matter, also called a thin matter shell. These results presented here will be pillar to the development of all remaining chapters in the first part of this thesis.

%% file: I.1ShellIntro/1.section.tex
\section{Introductory definitions}

Consider a $d-1$ dimensional hypersurface $\Sigma$ that partitions a $d$ dimensional spacetime into two regions $\ma{V}^{+}$ and $\ma{V}^{-}$. Each region is covered by a coordinate patch $x^{\al}_{\pm}$, where the plus or minus signs correspond to the regions $\ma{V}^{+}$ or $\ma{V}^{-}$, respectively. The problem we are interested in is the following: what conditions must the metric satisfy in order for both regions to be smoothly joined at $\Sigma$?

\begin{figure}[h!]
  \centering
    \includegraphics[width=0.6\textwidth]{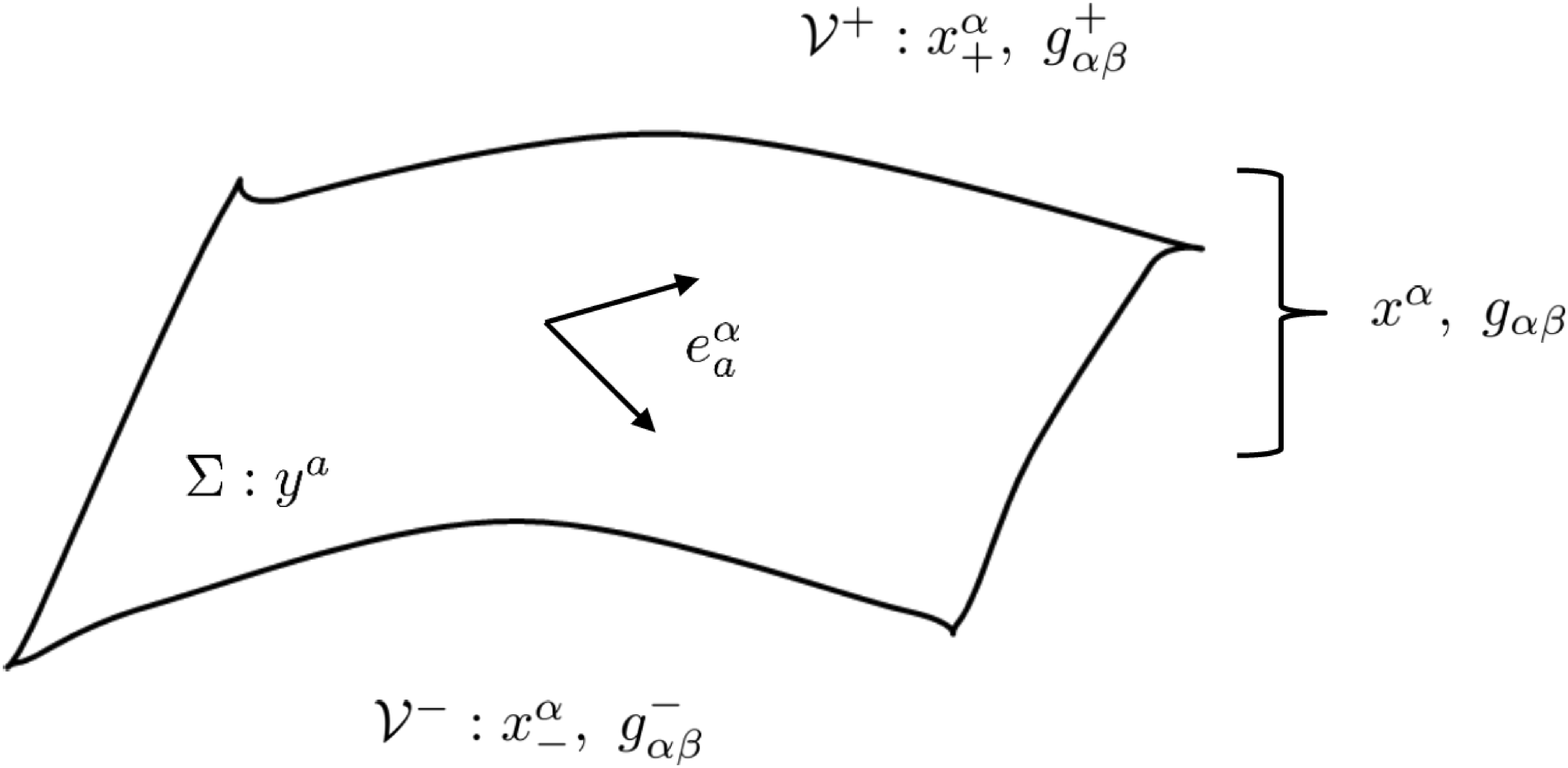}
    \caption{A hypersurface parametrized by coordinates divides spacetime into two sections, each with each own coordinate system. A third coordinate system is introduced in a region containing both regions of spacetime, as is continuous through the hypersurface.}
    \label{Shell1}
\end{figure}

\begin{figure}[h!]
  \centering
    \includegraphics[width=0.4\textwidth]{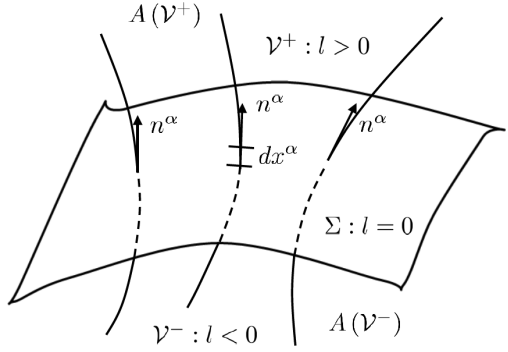}
    \caption{Geodesics cut the hypersurface in an orthogonal fashion, allowing the definition of a normal vector field to the hypersurface.}
    \label{Shell2}
\end{figure}

To address this question, first assume that the hypersurface is parametrized by a coordinate system $y^a$ which is the same on both sides of the hypersurface\footnote{Henceforth, Greek letters will be used for indexes of the spacetime and roman letters will represent the indexes on the hypersurface.}. Suppose as well that a third continuous coordinate system $x^{\al}$ overlaps with $x_+^{\al}$ and $x_-^{\al}$ in open regions of $\ma{V}^{+}$ and $\ma{V}^{-}$. We will make all the calculations in the $x^{\al}$ coordinates but they are merely temporary since the final results will not depend on them. The setup is depicted in Fig.~(\ref{Shell1}).
Now, if
\begin{equation}\label{Par}
x^{\al} = x^{\al}(y^a)
\end{equation}
are the parametric equations that describe the hypersurface, then differentiating (\ref{Par}) with respect to $y^a$ yields the vectors
\begin{equation}\label{tv}
e^{\al}_a = \frac{\dd x^{\al}}{\dd y^{a}},
\end{equation}
which are tangent to the lines of constant $y^{a}$ on $\Sigma$. Perpendicular to (\ref{tv}) are the normal vectors $n^{\al}$, which we choose to point from $\ma{V}^{-}$ to $\ma{V}^{+}$. To find such normal field, we start by piercing $\Sigma$ orthogonally  with a congruence of geodesics and parametrize their proper distance $l$ such that $l>0$ in $\ma{V}^{+}$, $l<0$ in $\ma{V}^{-}$ and $l=0$ at $\Sigma$. This implies that a displacement away from the hypersurface along any geodesic will be of the form $dx^{\al} = n^{\al} dl$ where $dl$ is the infinitesimal proper distance from $\Sigma$ to a point $P$ along a geodesic. We also have that
\begin{equation}\label{normal}
n_{\al} = \e \frac{\dd l}{\dd x^{\al}}
\end{equation}
where $\e = n^{\al}n_{\al}$. The only values that $\e$ can have are $+1$ or $-1$, in which case the hypersurface is said to be respectively timelike or spacelike. The situation is schematically shown in Fig.~(\ref{Shell2}).
It will also prove useful to introduce the notation for the jump of a quantity across $\Sigma$, that is
\begin{equation}
[A] \equiv A\left(\ma{V}^{+}\right)\big|_{\Sigma} - A\left(\ma{V}^{-}\right)\big|_{\Sigma}
\end{equation}
where $A$ is some function of spacetime. Since both $x^{\al}$, $y^a$ and $l$ are continuous across $\Sigma$, we arrive at the result
\begin{equation}\label{ne}
\left[n^{\al}\right] = \left[e^{\al}_a\right] = 0.
\end{equation}

It will be essential as well to use the concept of induced metric which arises when one wants to know the metric on an hypersurface alone instead of on the whole spacetime. Its definition comes out naturally using a path on the hypersurface. Let $x^{\al}(\lambda)$ be a path in $\Sigma$, where $\lambda$ is the curve parameter. Then an infinitesimal line segment along this path is given by
\begin{equation}
dx^{\al} = \frac{dx^{\al}}{d\lambda} d\lambda = e^{\al}_a dy^a
\end{equation}
which can be substituted in the line element $ds^2$ of the space time metric, giving
\begin{equation}
ds^2=g_{\al\be}dx^{\al}dx^{\be} = h_{ab} dy^a dy^b
\end{equation}
where $h_{ab}$ are the components of the so-called \textit{induced metric}, given explicitly by
\begin{equation} \label{hab}
h_{ab} = g_{\al\be} e^{\al}_a e^{\be}_b.
\end{equation}
The quantity (\ref{hab}) is invariant under a $x^{\al} \to x^{\al'}$ transformation and transforms like a tensor for $y^a \to y^{a'}$ transformations. Such a quantity is called a three-tensor. These tensors will play an important part in defining a formalism independent of the coordinates $x^{\al}$. The induced metric, in particular, is used to raise or lower indexes of three-tensors on the hypersurface which will be done frequently in the calculations throughout this work.

%% file: I.1ShellIntro/2.section.tex
\section{First junction condition}

The entire spacetime metric can be written using the language of distributions. More precisely, it can be expressed as
\begin{equation}\label{gab}
g_{\al\be} = \Theta(l)g_{\al\be}^+ + \Theta(-l)g_{\al\be}^-
\end{equation}
where $g_{\al\be}^{\pm}$ are the metrics in the regions $\ma{V}^+$ and $\ma{V}^-$ expressed\footnote{From now on, all quantities with a + or - sign are to be interpreted as seen from $\ma{V}^+$ or $\ma{V}^-$, respectively.} in the coordinate system $x^{\al}$ and $\Theta(l)$ is the Heaviside distribution, defined as
\begin{equation}\label{heaviside}
\Theta(l) = \left\{ \begin{array}{lll}
       +1, \hspace{3mm} \textrm{if} \hspace{2mm} l>0 \\
       \\
       -1, \hspace{3mm} \textrm{if} \hspace{2mm} l<0 \\
			 \\
			 \textrm{indeterminate}, \hspace{3mm} \textrm{if} \hspace{2mm} l=0
\end{array}
\right. .
\end{equation}
The distribution (\ref{heaviside}) has some important properties that will often be used in the calculations to follow, namely
\begin{equation}\label{propH}
\Theta^2(l) = \Theta(l), \quad \Theta(l)\Theta(-l) = 0, \quad \frac{d}{dl}\Theta(l) = \delta(l)
\end{equation}
where $\delta(l)$ is the Dirac distribution. The question of whether or not the spacetimes in each side of $\Sigma$ join smoothly is then equivalent to asking if the metric of the whole spacetime is a valid solution of the Einstein equations
\begin{equation}\label{Einst}
R_{\al\be} - \frac{1}{2}g_{\al\be}R = 8\pi G_d T_{\al\be}
\end{equation}
where $R_{\al\be}$ is the Ricci curvature tensor, $R \equiv g_{\al\be} R^{\al\be}$ is the Ricci scalar curvature and $G_d$ is Newton's gravitational constant in $d$ dimensions. The Ricci tensor is obtained from the Riemann tensor through $R_{\al\be} = R^{\mu}{}_{\al\mu\be}$, where the Riemann tensor has the explicit form
\begin{equation}\label{Riemann}
R^{\al}{}_{\be\g\de} = \Gamma^{\al}{}_{\de\be,\g} - \Gamma^{\al}{}_{\g\be,\de} + \Gamma^{\al}{}_{\g\lambda}\Gamma^{\lambda}{}_{\de\be} - \Gamma^{\al}{}_{\de\lambda}\Gamma^{\lambda}{}_{\g\be}
\end{equation}
where the notation $f_{,\mu} = \frac{\dd f}{\dd x^{\mu}}$ was used and $\Gamma$ are the Christoffel symbols given by
\begin{equation}\label{Christ}
\Gamma^{\al}{}_{\be\de} = \frac{1}{2}g^{\al\lambda}\left(g_{\lambda\be,\de} + g_{\lambda\de,\be} - g_{\be\de,\lambda}\right).
\end{equation}
If (\ref{gab}) is to be a valid solution, then the geometrical quantities of which (\ref{Einst}) is made of must be correctly defined as distributions. The first concerning quantity that arises is the derivative of the metric when calculating (\ref{Christ}). Using equations (\ref{normal}) and (\ref{propH}), a simple calculation shows that
\begin{equation}\label{dgab}
g_{\al\be,\g} = \Theta(l)g_{\al\be,\g}^+ + \Theta(-l)g_{\al\be,\g}^- + \e \delta(l)\left[g_{\al\be}\right]n_{\g}.
\end{equation}
The first two terms are well behaved but the last one will induce terms of the form $\Theta(l)\delta(l)$ when the Christoffel symbols are calculated and terms like those are not defined as distributions. The only way to avoid this is if
\begin{equation}\label{C0}
\left[g_{\al\be}\right] = 0
\end{equation}
but this holds only in the $x^{\al}$ coordinates since from the beginning they are assumed to be continuous across $\Sigma$, i.e. the condition $[x^{\al}]=0$ is satisfied a priori and therefore so is (\ref{C0}). However, this doesn't need to be true in other coordinate systems, thus revealing the need for a relation which is independent of the coordinates $x^{\al}$. This can be achieved by doing
\begin{equation}
\left[g_{\al\be}\right]e^{\al}_a e^{\be}_b = \left[g_{\al\be}e^{\al}_a e^{\be}_b\right] = 0
\end{equation}
where equation (\ref{ne}) was used. Since the quantity inside the square brackets is the induced metric (\ref{hab}), which is a three-tensor, we arrive at the coordinate $x^{\al}$ independent relation
\begin{equation}\label{1JC}
\left[h_{ab}\right] = 0\,,
\end{equation}
also called the first junction condition.  In other words, (\ref{1JC}) states that the induced metric in $\Sigma$ must be the same viewed from either side of it. This condition must always be satisfied if the hypersurface is to have a well defined geometry. It also implies the relation (\ref{C0}) while keeping the coordinate independence which can also be seen by the fact that (\ref{1JC}) produces only six conditions while (\ref{C0}) produces ten: the difference corresponds to the four equations arising from the necessary additional condition $[x^{\al}]=0$ when there is no coordinate independence.

%% file: I.1ShellIntro/3.section.tex
\section{Second junction condition}

The result (\ref{1JC}) from last section guarantees that the Christoffel symbols will not contain any problematic terms. Indeed, calculating (\ref{Christ}) using (\ref{gab}) and the first junction condition (\ref{1JC}), one is lead to the result
\begin{equation}
\Gamma^{\al}{}_{\be\g} = \Theta(l)\Gamma^{+}{}^{\al}_{\be\g} + \Theta(-l)\Gamma^{-}{}^{\al}_{\be\g}.
\end{equation}
Only the derivatives of this last expression are need to construct (\ref{Riemann}), which are straightforwardly shown to be
\begin{equation}
\Gamma^{\al}{}_{\be\g,\delta} = \Theta(l)\Gamma^{+}{}^{\al}_{\be\g,\delta} + \Theta(-l)\Gamma^{-}{}^{\al}_{\be\g,\delta} + \e \delta(l)\left[\Gamma^{\al}{}_{\be\g}\right]n_{\delta}
\end{equation}
where the Dirac distribution shows up again. Using this, the Riemann tensor is readily calculated, giving
\begin{equation}\label{RIE}
R^{\al}{}_{\be\g\delta} = \Theta(l)R^{+}{}^{\al}_{\be\g\delta} + \Theta(-l)R^{-}{}^{\al}_{\be\g\delta} + \delta(l)A^{\al}{}_{\be\g\delta}
\end{equation}
where
\begin{equation}\label{A}
A^{\al}{}_{\be\g\delta} = \e\left(\left[\Gamma^{\al}{}_{\be\delta}\right]n_{\g} - \left[\Gamma^{\al}{}_{\be\g}\right]n_{\delta}\right).
\end{equation}
Looking at (\ref{RIE}), we see that there's still a $\delta(l)$ term. Again, this term is not problematic per se since it is well defined as a distribution but it does represent a curvature singularity at $\Sigma$. Thus it poses no mathematical issues this time, since it will not be multiplied by any other distribution. The term $A^{\al}{}_{\be\g\delta}$ is even a tensor since the difference between two Christoffel symbols transforms like one. Hence, one must study this term by finding out the specific form for the $\delta(l)$ part of the Einstein equations.

We begin by noting that $g_{\al\be}$ is continuous across $\Sigma$, thus its tangential derivatives must also be continuous and so $g_{\al\be,\g}$ can only have a discontinuity along the normal vector $n^{\al}$. Mathematically, this means that there must exist a tensor $\kappa_{\al\be}$ such that
\begin{equation}\label{gaby}
\left[g_{\al\be,\g}\right] = \kappa_{\al\be}n_{\g}
\end{equation}
which solved for $\kappa_{\al\be}$ gives
\begin{equation}
\kappa_{\al\be} = \e \left[g_{\al\be,\g}\right] n^{\g}.
\end{equation}
\equationref{gaby} can be inserted in the calculation of $\left[\Gamma^{\al}{}_{\be\g}\right]$, resulting in
\begin{equation}
\left[\Gamma^{\al}{}_{\be\g}\right] = \frac{1}{2}\left(\kappa^{\al}{}_{\be}n_{\g} + \kappa^{\al}{}_{\g}n_{\be} - \kappa_{\be\g}n^{\al}\right).
\end{equation}
By making use of this last expression, we arrive at the explicit form for the $\delta(l)$ part of the Riemann tensor
\begin{equation}
A^{\al}{}_{\be\g\delta} = \frac{\e}{2}\left(\kappa^{\al}{}_{\delta}n_{\be}n_{\g}-\kappa^{\al}{}_{\g}n_{\be}n_{\delta} - \kappa_{\be\delta}n^{\al}n_{\g}+\kappa_{\be\g}n^{\al}n_{\delta}\right).
\end{equation}
Contracting the first and third indices, one obtains the singular part of the Ricci tensor
\begin{equation}
A_{\al\be} \equiv A^{\nu}{}_{\be\nu\delta} = \frac{\e}{2}\left(\kappa_{\nu\al}n^{\nu}n_{\be} + \kappa_{\nu\be}n^{\nu}n_{\al} - \kappa n_{\al}n_{\be} - \e \kappa_{\al\be}\right)
\end{equation}
where $\kappa \equiv \kappa^{\al}{}_{\al}$. Contracting the remaining indexes results in the singular part of the Ricci scalar curvature
\begin{equation}
A \equiv A^{\al}{}_{\al} = \e \left(\kappa_{\al\be}n^{\al}n^{\be} - \e \kappa\right).
\end{equation}
We can now construct the $\delta$ part of the Einstein equations, which is simply
\begin{equation}
A_{\al\be}-\frac{1}{2}g_{\al\be}A \equiv 8 \pi G_d S_{\al\be}
\end{equation}
where $S_{\al\be}$ is the associated stress-energy tensor. This expression appears in the total stress-energy tensor among two others terms:
\begin{equation}
T_{\al\be} = \Theta(l)T_{\al\be}^+ + \Theta(-l)T^-_{\al\be} + \delta(l)S_{\al\be}.
\end{equation}
The first two terms are associated to the regions $\ma{V}^{\pm}$, so the $\delta(l)$ term must be associated to $\Sigma$, that is, it is the surface stress-energy tensor of the hypersurface. This implies that when such stress-energy tensor is non-null there must exist a distribution of energy where the hypersurface is located, also called a thin shell. Writing out the terms explicitly, we get
\begin{equation}\label{S}
16\pi\e G_d S_{\al\be}  =  \kappa_{\nu\be}n^{\nu}n_{\al} + \kappa_{\nu\al}n^{\nu}n_{\be} - \kappa n_{\al}n_{\be}
 - \e \kappa_{\al\be} - \left(\kappa_{\mu\nu}n^{\mu}n^{\nu} - \e \kappa\right)g_{\al\be}.
\end{equation}
Notice, however, that $S_{\al\be}n^{\be}=0$, or in other words, this stress-energy tensor is tangent to $\Sigma$ and therefore can be completely written in terms of tangent vectors to that hypersurface, like $e^{\al}_a$. This means that $S^{\al\be}=S^{ab}e^{\al}_a e^{\be}_b$, or conversely $S_{ab}=S_{\al\be}e^{\al}_a e^{\be}_b$, where $S_{ab}$ is a symmetric three-tensor. Decomposing the metric in its normal and tangential components with respect to $\Sigma$
\begin{equation}
g^{\al\be}  =  h^{ab} e^{\al}_a e^{\be}_b + n^{\al}n^{\be}\,,
\end{equation}
and multiplying both sides of (\ref{S}) by $e^{\al}_a e^{\be}_b$, one arrives at an expression for $S_{ab}$, namely
\begin{equation}\label{Snew}
16\pi G_d S_{ab} = -\kappa_{\al\be} e^{\al}_a e^{\be}_b + h^{mn}\kappa_{\mu\nu}e^{\mu}_m e^{\nu}_n h_{ab}.
\end{equation}
Note that even though $\kappa_{\al\be}$ allows the calculation of explicit formulas, it would be desirable to express those formulas as functions of more usual geometric quantities that characterize hypersurfaces. One such important quantity is the extrinsic curvature, whose tensor components $K_{ab}$ are defined as
\begin{equation} \label{Kab}
K_{ab} = n_{\al;\be}e^{\al}_a e^{\be}_b.
\end{equation}
Indeed, this quantity appears quite naturally in (\ref{Snew}). To see this, we start by calculating the jump in the covariant derivative of the normal vector components
\begin{equation}
\left[n_{\al;\be}\right] = - \left[\Gamma^{\g}{}_{\al\be}\right] n_{\g} = \frac{1}{2}\left(\e \kappa_{\al\be}-\kappa_{\g\al}n_{\be}n^{\g} - \kappa_{\g\be}n_{\al}n^{\g}\right)
\end{equation}
where it was used the fact that $\left[n_{\al,\be}\right] = 0$ which in turn follows from (\ref{ne}). From this it is possible to calculate the jump in extrinsic curvature, which is
\begin{equation}\label{Kabl}
\left[K_{ab}\right] = \left[n_{\al;\be}\right]e^{\al}_a e^{\be}_b = \frac{\e}{2}\kappa_{\al\be}e^{\al}_a e^{\be}_b.
\end{equation}
Defining $K \equiv h^{ab}K_{ab}$ and inserting (\ref{Kabl}) in (\ref{Snew}), we obtain
\begin{equation}\label{Sab}
S_{ab} = -\frac{\e}{8\pi G_d}\left(\left[K_{ab}\right]-\left[K\right]h_{ab}\right),
\end{equation}
which gives a relation between the surface stress energy-tensor and the jump in extrinsic curvature. Now, a smooth transition across $\Sigma$ is the same as saying that there can be no $\de(l)$ part in the Einstein equations which can only happen if $S_{ab}=0$. From (\ref{Sab}), it is immediately seen that this can only happen if
\begin{equation}\label{2JC}
\left[K_{ab}\right] = 0
\end{equation}
The above relation is called the second junction condition and since $K_{ab}$ is a three-tensor, this condition is also independent of the $x^{\al}$ coordinates. Together with (\ref{1JC}), they form the set of necessary conditions that must be satisfied in order for the spacetimes in each side of $\Sigma$ to connect smoothly. However, although (\ref{1JC}) must always be satisfied, equation (\ref{2JC}) needs not. If that's the case, there is also a physical interpretation to it: the smoothness in the transition across $\Sigma$ no longer exists because there is a thin matter shell present at the hypersurface, with stress-energy tensor given by (\ref{Sab}).

Finally, notice that because of (\ref{ne}) the indexes in both junction conditions can be raised or lowered freely without changing anything. In fact, the most direct and clean way of obtaining the surface stress-energy tensor of the shell is by using one contravariant and one covariant index, i.e.
\begin{equation}\label{Sac}
S^{a}{}_{b} = -\frac{\e}{8\pi G_d}\left(\left[K^{a}{}_{b}\right]-\left[K\right]h^{a}{}_{b}\right).
\end{equation}
Ample use of the above equation will be made throughout the first part of this thesis.

%% file: I.3ShellCharged/main.tex
\chapter{Thermodynamics of an electrically charged thin shell}
\label{cap:chapterI3}
\input{I.3ShellCharged/outline.tex}
\input{I.3ShellCharged/1.section.tex}

\input{I.3ShellCharged/2.section.tex}

\input{I.3ShellCharged/3.section.tex}

\input{I.3ShellCharged/4.section.tex}

\input{I.3ShellCharged/6.section.tex}
\input{I.3ShellCharged/7.section.tex}
\input{I.3ShellCharged/conclusions.tex}

%% file: I.3ShellCharged/outline.tex
\section{Introduction}

Each part of this chapter is dedicated to the first application of the thin shell formalism of this thesis and will serve as template of the main results one may expect to retrieve from the use of the formalism for a system at finite temperature. We will investigate electrically charged shells, as in \cite{Kuchar:1968}. Using the first law of thermodynamics together with the junctions conditions of Chapter 2, we shall obtain a general expression for the entropy of the shell and study a variety of properties associated to its main thermodynamic quantities. This will reveal a considerably complicated thermodynamic analysis of the shell, in particular the computation of the stability regions for the parameters contained in the thermal equation of state. It will, however, complement existing work on the entropy of charged black holes \cite{Braden:1990hw,Peca:1998cs} and lay the groundwork on top of which we will construct the next two chapters.

%% file: I.3ShellCharged/1.section.tex
\section{The thin-shell spacetime}
\label{thinsh}

We start with the Einstein-Maxwell equations in 3+1 dimensions
\eq{
G_{\al\be}=8\pi G\,
T_{\al\be}\,,
\label{ein4}
}
\eq{
\nabla_{\beta}F^{\alpha\beta}=4\pi J^{\alpha}\,,
\label{max2}
}
where $F_{\alpha\beta}$ is the
Faraday-Maxwell tensor and
$J_\alpha$ is the electromagnetic four-current.
The other Maxwell equation $\nabla_{[\gamma}F_{\alpha\beta]}=0$,
where $[...]$ means anti-symmetrization,
is automatically satisfied for a properly defined
$F_{\alpha\beta}$.

We consider now a two-dimensional timelike
massive electrically charged shell with radius $R$,
which we will call $\Sigma$.  The shell partitions spacetime into two
parts, an inner region $\mathcal{V}_-$ and an outer region
$\mathcal{V}_+$.
In order to find a global spacetime solution for the
Einstein equation, Eq.~(\ref{ein4}), we will use the thin-shell
formalism developed in the last chapter. First, we specify the metrics on each side of the shell.
In
the inner region $\mathcal{V}_-$ ($r< R$) we assume the spacetime
is flat, i.e.
\ea{\label{LEI}
ds_-^2  = g_{\al\be}^- dx^\alpha dx^\beta = -dt_-^2 + dr^2 + r^2\, d\Omega^2\,,\quad r< R\,,
}
where $t_-$ is the inner time coordinate,
polar coordinates
$(r,\theta,\phi)$ are used, and
$d\Omega^2 = d\theta^2 + \sin^2\theta\,d\phi^2$.
In the outer region $\mathcal{V}_+$ ($r> R$),
the spacetime is described by the
Reissner-Nordstr\"{o}m line element
\ea{\label{LEO2}
ds_+^2 & = g_{\al\be}^+ dx^\alpha dx^\beta =  -\left(1 - \frac{2Gm}{r} + \frac{G Q^2}{r^2}\right)
dt_+^2 + \frac{dr^2}{1 - \dfrac{2Gm}{r} + \dfrac{G Q^2}{r^2}}  + r^2 d\Omega^2 \,, \quad r> R\,,
}
where $t_+$ is the outer time coordinate, and again
$(r,\theta,\phi)$ are polar coordinates, and
$d\Omega^2 = d\theta^2 + \sin^2\theta\,d\phi^2$.
The constant $m$ is to be  interpreted as
the ADM mass, or energy, and
$Q$ as the electric charge.
Finally,
on the hypersurface itself, $r=R$,
the metric
$h_{ab}$ is that of a 2-sphere with an additional time dimension,
such that,
\eq{
ds_{\Sigma}^2 = h_{ab} dy^a dy^b =
-d\tau^2 + R^2(\tau) d\Omega^2\,, \quad r= R\,,
\label{intrinsmetr2}
}
where we have chosen $y^a=(\tau,\theta,\phi)$ as
the time and spatial coordinates on the
shell. We have adopted the convention to use Latin indices for the
components on the hypersurface. The time coordinate $\tau$ is the
proper time for an observer located at the shell. The shell radius is
given by the parametric equation $R= R(\tau)$ for an observer on the
shell. On each side of the hypersurface, the parametric equations for
the time and radial
coordinates are denoted by
$t_{-}=T_{-}(\tau)$, $r_{-}=R_{-}(\tau)$,
and $t_{+}=T_{+}(\tau)$, $r_{+}=R_{+}(\tau)$.
Viewed from each side of
the shell, the induced metric is given by
\eq{
h^{-}_{ab} = g^{-}_{\al\be} \, e^{\al}_{-}{}_a \,
e^{\be}_{-}{}_b\,,
\quad
h^{+}_{ab} = g^{+}_{\al\be} \, e^{\al}_{+}{}_a \,
e^{\be}_{+}{}_b\,,
}
where $e^{\al}_{-}{}_a$
and $e^{\al}_{+}{}_a$ are tangent vectors to the hypersurface
viewed from the inner and outer regions, respectively.
With these last expressions, we have all
the necessary information to employ the formalism developed in
Chapter 1. For electrically charged
systems, this was first displayed in
\cite{Kuchar:1968}.

Regarding the first junction condition (\ref{1JC}), it
immediately implies that
\eq{
h^{-}_{ab}=h^{+}_{ab}=h_{ab}\,,
} or explicitly
\eq{\label{J1}
-\left(1 - \frac{2Gm}{r} + \frac{G Q^2}{r^2}\right) \dot{T}_+^2 + \frac{\dot{R}_+^2}{\left(1 -
\dfrac{2Gm}{r} + \dfrac{G Q^2}{r^2}\right)} = -\dot{T}_-^2 +
\dot{R}_-^2 = -1\,,
}
where a dot denotes differentiation with respect to $\tau$.
Proceeding to the second junction condition, given by Eq.~(\ref{2JC}),
we must first calculate the extrinsic curvature components using Eq.~(\ref{Kab}),
for which one may derive the general expressions
\ea{
K^{\tau}_-{}_{\tau} &= \frac{\ddot{R}}{\sqrt{1+\dot{R}^2}}\,,
\label{a1}\\
K^{\tau}_+{}_{\tau} &= \frac{-\frac{G \dot{m}}{R\,\dot{R}}-\frac{G
Q^2}{R^3}+\frac{G m}{R^2}+\ddot{R}}{\sqrt{1-\frac{2Gm}{R} + \frac{G
Q^2}{R^2}+\dot{R}^2}}\,, \label{a2}\\
K^{\phi}_-{}_{\phi} = K^{\theta}_-{}_{\theta} &=
\frac{1}{R}\sqrt{1+\dot{R}^2}\,,\label{a3}\\
K^{\phi}_+{}_{\phi} = K^{\theta}_+{}_{\theta} &=
\frac{1}{R}\sqrt{1-\frac{2Gm}{R} + \frac{G Q^2}{R^2}+\dot{R}^2}\,.
\label{a4}
}
Using Eqs.~(\ref{a1})-(\ref{a4}) in Eq.~(\ref{Sac}), one can
calculate the non-null components of the stress-energy tensor $S_{ab}$
of the shell. In particular, we will assume a static shell as in Chapter 2, such that
$\dot{R}=0$, $\ddot{R}=0$, and $\dot{m}=0$. In that case, we are led to
\ea{
S^{\tau}{}_{\tau} & = \frac{\sqrt{1-\frac{2Gm}{R} + \frac{G Q^2}{R^2}}
- 1}{4 \pi G R}\,, \label{S14} \\
S^{\phi}{}_{\phi} = S^{\theta}{}_{\theta} &  =
\frac{\sqrt{1-\frac{2Gm}{R}+\frac{G Q^2}{R^2}}-1}{8 \pi G R} +
\frac{\frac{m G}{R}-\frac{G Q^2}{R^2}}{8 \pi G R
\sqrt{1-\frac{2Gm}{R}+\frac{G Q^2}{R^2}}} \label{S24}\,.
}
To
further advance, one needs to specify what kind of matter the shell
is made of, which we will consider to be a perfect fluid with surface
energy density $\sigma$ and pressure $p$. This implies that the
stress-energy tensor will be of the form
\eq{
S^{a}{}_{b} = (\sigma +p) u^a u_b + p
h^{a}{}_{b}\,,
\label{perffluid}
}
where
$u^a$ is the three-velocity of a shell element. We thus find that
\eq{
S^{\tau}{}_{\tau} = -\sigma\,,
\label{lam}
}
\eq{
S^{\theta}{}_{\theta} = S^{\phi}{}_{\phi} = p
\label{press}\,.
}
Combining Eqs.~(\ref{lam})-(\ref{press})
with Eqs.~(\ref{S14})-(\ref{S24}) results in the equations
\ea{
\sigma = & \frac{1-\sqrt{1-\frac{2Gm}{R} + \frac{G Q^2}{R^2}}}{4 \pi G
R}\,, \label{SS12} \\
p = & \frac{\sqrt{1-\frac{2Gm}{R}+\frac{G
Q^2}{R^2}}-1}{8 \pi G R} +
\frac{\frac{m G}{R}-\frac{G Q^2}{R^2}}{8
\pi G R \sqrt{1-\frac{2Gm}{R}+\frac{G Q^2}{R^2}}}
\label{SS22}\,.
}
It is now useful to define the shell's redshift
function $k$ as
\eq{
k=\sqrt{
1 - \frac{2Gm}{R} + \frac{G Q^2}{R^2}}
\label{red0}\,,
}
which allows Eqs.~(\ref{SS12})-(\ref{SS22}) to be written as
\ea{
\sigma = & \frac{1-k}{4\pi G R} \label{Sig1}\,, \\
p = & \frac{R^2(1-k)^2 - GQ^2}{16 \pi G R^3 k}\,.
\label{pQk1}
}
From the energy density $\sigma$ of the shell we can define
the rest mass $M$ through the equation
\eq{
\sigma = \frac{M}{4\pi R^2} \,.
\label{sigmarestmasssigma}
}
Note that from Eqs.~(\ref{Sig1}) and (\ref{sigmarestmasssigma})
one has
\eq{\label{M1}
M = \frac{R}{G}(1-k).
}
Using Eqs.~(\ref{red0}) and (\ref{M1}), we are led to
an equation for the ADM mass $m$,
\eq{\label{m0}
m = M - \frac{G M^2}{2R} + \frac{Q^2}{2R}\,.
}
This equation is intuitive on physical grounds as it states that the
total energy $m$
of the shell is given by its mass $M$ minus the energy
required to built it against the action of gravitational and
electrostatic forces,
i.e., $- \frac{G M^2}{2R} + \frac{Q^2}{2R}$.
For $Q=0$, we recover the result derived in
\cite{Martinez:1996ni}.
Note that Eq.~(\ref{m0}) is also purely a
consequence of the Einstein equation encoded in the junction
conditions, i.e., although no information about the matter fields of
the shell has been given, we know that they must have an ADM mass
given by Eq.~(\ref{m0}).

There are other variables one may use to obtain more symmetric results, namely the gravitational radius $r_+$ and
the Cauchy horizon $r_-$ of the shell
spacetime, which are given by the zeros
of the $g_{00}^o$ component in Eq.~(\ref{LEO2}), of the form
\eq{
r_{+} = G\,m +\sqrt{G^2m^2 - G Q^2}\,,
\label{horradi}
}
\eq{
r_{-} = G\,m - \sqrt{G^2m^2 - G Q^2}\,,
\label{horradicauch}
}
respectively.
The gravitational radius $r_+$ is also the horizon radius when the
shell radius $R$ is inside $r_+$, i.e., the spacetime contains a black
hole. Although they have the same expression, conceptually, the
gravitational and horizon radii are distinct.  Indeed, the
gravitational radius is a property of the spacetime and matter,
independently of whether there is a black hole or not. On the other
hand, the horizon radius exists only when there is a black hole. The gravitational radius $r_+$
and the Cauchy horizon $r_-$
in Eqs.~(\ref{horradi})-(\ref{horradicauch})
can be inverted to give
\eq{
m=\frac{1}{2G}\left( r_{+} +  r_{-}\right)\,,
\label{invhorradi}
}
\eq{
Q=\sqrt{
\frac{
r_{+}r_{-}}{G}
}\,.
\label{invhorradicauch}
}
From Eq.~(\ref{horradi}) one can define
the gravitational area $A_+$ as
\eq{\label{areaplus}
A_+=4\pi\,r_+^2\,,
}
which is also the event horizon area when there is
a black hole. The area of the shell itself, from Eq.~(\ref{intrinsmetr2}), is simply given by
\eq{
A=4 \pi R^2\,.
\label{area1}
}
Using Eqs.~(\ref{horradi})-(\ref{horradicauch})
implies that $k$ in Eq.~(\ref{red0}) can be written as
\eq{\label{red2}
k=\sqrt{\Big(1-\frac{r_+}{R}\Big)\Big(1-\frac{r_-}{R}\Big)}\,.
}
Having taken care of the shell's intrinsic details, we must turn to Eq.~(\ref{max2}). The
Faraday-Maxwell tensor $F_{\alpha\beta}$ is usually defined in terms
of an electromagnetic four-potential $A_\alpha$ by
\eq{
F_{\al\be} = \partial_{\al}A_{\be}-\partial_{\be}A_{\al}\,.
\label{empot}
}
Although we did not develop it in Chapter 2, there exists as well a number of junction conditions that the quantities in electromagnetism must satisfy when two different mediums are separated by a hypersurface. These are the conditions that the electromagnetic field obeys in classical electromagnetism, written in a covariant way. We will not derive them, instead citing them as we progress.

To use the thin-shell formalism related to the electric part
we must first specify the vector potential $A_\alpha$
on each side of the shell.
We assume an electric ansatz for the
electromagnetic four-potential $A_\alpha$, i.e.,
\eq{
A_{\alpha} = (-\phi,
0,0,0)\,,
\label{empotspec}
}
where $\phi$ is the electric potential.
In
the inner region $\mathcal{V}_i$ ($r < R$) the spacetime
is flat,
so the Maxwell equation
$\nabla_{\beta}F^{\alpha\beta}=\frac{1}{\sqrt{-g}}
\partial_{\beta}\left(\sqrt{-g}F^{\alpha\beta}\right)
=0$ has as
a constant solution for the inner
electric potential $\phi_i$
which, for convenience, can be written as
\eq{\label{phiin}
\phi_i= \frac{Q}{R}+{\rm constant}\,,\quad r< R\,,
}
where $Q$ is a constant,
to be interpreted as the conserved electric charge.
In the outer region $\mathcal{V}_o$ ($r> R$),
the spacetime is Reissner-Nordstr\"om
and the
Maxwell equation
$\nabla_{\beta}F^{\alpha\beta}=\frac{1}{\sqrt{-g}}
\partial_{\beta}\left(\sqrt{-g}F^{\alpha\beta}\right)
=0$ now yields
\eq{\label{phiout}
\phi_o= \frac{Q}{r}+{\rm constant}\,,\quad r>R\,.
}

Due to the existence of electricity in the shell, another important
set of restrictions must also be considered.  These restrictions are
related to the discontinuity present in the electric field across the
charged shell. We are interested in the projection
\eq{
A_a = A_{\al} \,e^{\al}_a
}
of the four-potential in the shell's
hypersurface, since it will contain
quantities which are intrinsic to the shell.
Indeed, following \cite{Kuchar:1968},
\eq{\label{potjunct1}
\left[A_{a}\right]=0\,,
}
with
$A_{i\,a} = (-\phi_{i}, 0,0)$, and
$A_{o\,a} = (-\phi_{o}, 0,0)$
being the vector potential at $R$,
on the shell, seen from each side of it.
Thus, the constants in Eqs.~(\ref{phiin}) and
(\ref{phiout}) are indeed the same and so
at $R$
\eq{
\phi_o =
\phi_i \,,\quad r=R\,.
}
Following \cite{Kuchar:1968} further, the tangential
components $F_{ab}$ of the electromagnetic tensor $F_{\al\be}$ must
change smoothly across $\Sigma$, i.e.
\eq{\label{JC1}
\left[F_{ab}\right]=0\,,
}
with
\eq{
F^{i}_{ab} = F^{i}_{\al\be} e^{\al}_{i}{}_a \,
e^{\be}_{i}{}_b\,,\quad
F^{o}_{ab} = F^{o}_{\al\be} e^{\al}_{o}{}_a \,
e^{\be}_{o}{}_b\,,
}
while the normal components $F_{a\perp}$ must change by a jump
as,
\eq{\label{JC2}
\left[F_{a\perp}\right] = 4\pi \sigma_e u_{a}\,,
}
where
\eq{
F^{i}_{a\perp} = F^{i}_{\al\be}  e^{\al}_{i}{}_a \,
n^{\be}_{i}\,,\quad
F^{o}_{a\perp} = F^{o}_{\al\be}  e^{\al}_{o}{}_a \,
n^{\be}_{o}\,,
}
and $\sigma_eu_{a}$ is the surface electric current,
with  $\sigma_e$ being the density of charge and
$u_{a}$ its 3-velocity, defined on the shell.
One can then show that Eq.~(\ref{JC1}) is
trivially satisfied, while Eq.~(\ref{JC2}) leads to the single
nontrivial equation at $R$, on the shell,
\eq{\label{JCn1}
\frac{\partial\phi_o}{\partial r}
-\frac{\partial\phi_i}{\partial r}
= - 4\pi \sigma_e\,,\quad r=R\,.
}
Then, from Eqs.~(\ref{phiin}), (\ref{phiout}),
and (\ref{JCn1}) one obtains
\eq{\label{PhiJC22}
\frac{Q}{R^2}=4\pi \sigma_e \,,
}
relating the total charge $Q$, the charge density $\sigma_e$,
and the shell's radius $R$ in the expected manner.

%% file: I.3ShellCharged/2.section.tex
\section{Thermodynamics
and stability conditions for the thin shell}
\label{thermo}

As in Chapter 3, we start
with the assumption that the shell
in static equilibrium
possesses a well-defined
temperature $T$ and an entropy $S$ which is a function of
three variables $M$, $A$, $Q$, i.e.,
\begin{equation}\label{entropy0}
S=S(M,A,Q)\,,
\end{equation}
where the arguments $(M,A,Q)$ can be considered as three generic parameters.
In this case, they are
the shell's
rest mass $M$, area $A$, and charge $Q$.
The first law of thermodynamics can thus be written as
\begin{equation}\label{TQ22}
T dS = dM + pdA - \Phi dQ
\end{equation}
where $dS$
is the differential of the entropy of the shell,
$dM$ is the
differential of the rest mass,
$dA$ is the differential of the area of the shell,
$dQ$ is the differential of the charge,
and $T$, $p$ and
$\Phi$ are the temperature, the pressure, and the thermodynamic
electric potential of the shell, respectively.
In order to find the entropy $S$,
one thus needs three equations of state, namely,
\ea{
p & =p(M,A,Q)\,, \label{press02} \\
\beta & =\beta(M,A,Q)\,, \label{temper} \\
\Phi & =\Phi(M,A,Q)\,, \label{electr}
}
where
\begin{equation}\label{beta2}
\be \equiv \frac1T
\end{equation}
represents the inverse temperature.

It is important to note that the temperature
and the thermodynamic
electric potential
play the role of
integration factors, which implies that there will be integrability
conditions that must be specified in order to guarantee the existence
of an expression for the entropy, i.e. that the differential $dS$ is
exact. These integrability conditions essentially assert that the cross derivatives of each term of the differential $dS$ must be equal, implying that
\begin{align}
\left(\frac{\dd \be}{\dd A}\right)_{M,Q} & = \hspace{3mm}
\left(\frac{\dd \be p}{\dd M}\right)_{A,Q}\,, \label{Ione} \\
\left(\frac{\dd \be}{\dd Q}\right)_{M,A} & = - \left(\frac{\dd \be
\Phi}{\dd M}\right)_{A,Q}\,, \label{Itwo} \\
\left(\frac{\dd \be p}{\dd Q}\right)_{M,A} & = - \left(\frac{\dd \be
\Phi}{\dd A}\right)_{M,Q} \,.
\label{Ithree}
\end{align}
These equations enable one to determine the
relations between the three equations of state of
the system.

With the first law of thermodynamics given in Eq.~(\ref{TQ22}),
one is able
to perform a thermodynamic study of the local
intrinsic stability of the shell.
To have thermodynamic stability the
following inequalities should hold
\begin{equation}\label{B1}
\left(\frac{\dd^2 S}{\dd M^2}\right)_{A,Q} \leq 0\,,
\end{equation}
\begin{equation}\label{B2}
\left(\frac{\dd^2 S}{\dd A^2}\right)_{M,Q} \leq 0\,,
\end{equation}
\begin{equation}\label{B3}
\left(\frac{\dd^2 S}{\dd Q^2}\right)_{M,A} \leq 0\,,
\end{equation}
\begin{equation}\label{B4}
\left(\frac{\dd^2 S}{\dd M^2}\right)\left(\frac{\dd^2 S}{\dd
A^2}\right) - \left(\frac{\dd^2 S}{\dd M \dd A}\right)^2 \geq 0\,,
\end{equation}
\vspace{2mm}
\begin{equation}\label{B5}
\left(\frac{\dd^2 S}{\dd A^2}\right)\left(\frac{\dd^2 S}{\dd
Q^2}\right) - \left(\frac{\dd^2 S}{\dd A \dd Q}\right)^2 \geq 0\,,
\end{equation}
\vspace{2mm}
\begin{equation}\label{B6}
\left(\frac{\dd^2 S}{\dd M^2}\right)\left(\frac{\dd^2 S}{\dd
Q^2}\right) - \left(\frac{\dd^2 S}{\dd M \dd Q}\right)^2 \geq 0\,,
\end{equation}
\vspace{2mm}
\begin{equation}\label{B7}
\left(\frac{\dd^2 S}{\dd M^2}\right) \left(\frac{\dd^2 S}{\dd Q \dd
A}\right) - \left(\frac{\dd^2 S}{\dd M \dd A}\right) \left(\frac{\dd^2
S}{\dd M \dd Q}\right) \geq 0\,.
\end{equation}
The derivation of these expressions
follows the rationale presented in
\cite{Callen:1985}, and a detailed presentation is available in Appendix
\ref{ap:a}.

%% file: I.3ShellCharged/3.section.tex
\section{The thermodynamic independent variables
and the three equations of state}
\label{eqsos}

We will work from now onwards with the more useful three
independent variables $(M,R,Q)$ instead of
$(M,A,Q)$.
The rest mass $M$ of the shell is from
Eq.~(\ref{sigmarestmasssigma}) given by
\begin{equation}
M = 4\pi R^2 \, \sigma \,,
\label{restmasssigma}
\end{equation}
where $\sigma$ is calculated from Eq.~(\ref{Sig1})
and $R$ is the radius of the shell.
The first law of thermodynamics
written in generic terms
is simpler when expressed
using the area $A$ of the shell, but here
it is
handier to use the
radius $R$ in this specific study.
The radius $R$ is related to the area $A$
through Eq.~(\ref{intrinsmetr2}), i.e.,
\begin{equation}
R= \sqrt
{
\dfrac{A}{4\,\pi}}.
\label{radiusarea}
\end{equation}
As for the charge $Q$,
using Eq.~(\ref{PhiJC22}), it is given by
\begin{equation}
Q =  4\pi R^2 \, \sigma_e .
\end{equation}
We should now look at
Eq.~(\ref{m0}) and
Eqs.~(\ref{horradi})-(\ref{horradicauch})
as functions of $(M,R,Q)$, i.e.
\begin{equation}\label{m}
m(M,R,Q) = M - \frac{G M^2}{2R} + \frac{Q^2}{2R}\,,
\end{equation}
and
\begin{equation}
r_{+}(M,R,Q) = G\,m(M,R,Q) +\sqrt{G^2m(M,R,Q)^2 - G Q^2}\,,
\label{horradi2}
\end{equation}
\begin{equation}
r_{-}(M,R,Q) = G\,m(M,R,Q) - \sqrt{G^2m(M,R,Q)^2 - G Q^2}\,,
\label{horradicauch2}
\end{equation}
respectively.
The function $k$ in Eq.~(\ref{red2}) is also a function of
$(M,R,Q)$,
\begin{align}\label{red2}
k(r_{+}(M,R,Q),r_{-}(M,R,Q),R)=
\sqrt{\Big(1-\frac{r_+(M,R,Q)}{R}\Big)\Big(1-
\frac{r_-(M,R,Q)}{R}\Big)}\,.
\end{align}

Expressing the pressure equation of state in the form of
Eq.~(\ref{press02}), we obtain from Eqs.~(\ref{SS22})
and (\ref{m0}) [or Eq.~(\ref{m})],
\begin{equation}\label{pinMRQ}
p(M,R,Q) = \frac{GM^2-Q^2}{16 \pi R^2 (R-GM)}\,,
\end{equation}
or changing from the variables $(M,R,Q)$ to $(r_+,r_-,R)$
which is more useful, we find
[see Eqs.~(\ref{pQk1}) and (\ref{invhorradicauch})],
\begin{align} \label{pQk}
&p(r_+,r_-,R) = \frac{R^2(1-k)^2
 - r_+ r_-}{16 \pi G R^3 \,k}\,,
\end{align}
where  $k$ can be seen as $k=k(r_+,r_-,R)$
as given in Eq.~(\ref{red2})
and $r_+$ and $r_-$ are functions of $\,(M,R,Q)$.
This
reduces to the expression obtained in \cite{Martinez:1996ni} in the limit
$Q=0$ or $r_-=0$.
This equation, Eq.~(\ref{pQk}), is a pure
consequence of the Einstein equation, encoded in the junction
conditions.

Turning now to the temperature equation of state (\ref{temper}), we
will need to focus on the integrability condition
(\ref{Ione}). Changing
from the variables $(M,R,Q)$ to $(r_+,r_-,R)$, Eq.~(\ref{Ione})
becomes
\begin{equation}
\left(\frac{\dd \be}{\dd R}\right)_{r_+,r_-} =
\be \frac{R(r_+ + r_-)-2r_+ r_-}{2 R^3 k^2}
\end{equation}
which has the analytic solution
\begin{equation}\label{BS5}
\be(r_+,r_-,R) = b(r_+,r_-) \,k
\end{equation}
where
$k$ is
a function of $r_+$, $r_-$, and $R$, as given in
Eq.~(\ref{red2}), and
$b(r_+,r_-)\equiv \be(r_+,r_-,\infty)$ is an arbitrary function,
representing the inverse of the temperature of the shell if its radius
were infinite.
Hence, from Eq.~(\ref{BS5}),
we recover Tolman's formula.

The remaining equation of state to be studied is the electric
potential.  Using Eqs.~(\ref{M1}) and (\ref{red2}), one can deduce
$\left(\frac{\partial M}{\partial A}\right)_{r_{+},r_{-}}=-p$, i.e.,
\begin{equation}
\left(\frac{\partial M}{\partial R}\right)_{r_{+},r_{-}}=-8\pi\,R\,p\,.
\label{mp}
\end{equation}
Then, it follows from Eqs.~(\ref{Ione})-(\ref{Ithree})
and Eq.~(\ref{mp}) that the differential
equation
\begin{equation}
\left(\frac{\dd p}{\dd Q}\right)_{M,R} + \frac{1}{8 \pi
R}\left(\frac{\dd \Phi}{\dd R}\right)_{r_+,r_-} + \Phi \left(\frac{\dd
p}{\dd M}\right)_{R,Q} = 0\,,
\label{phieqdiff}
\end{equation}
holds, where the second term has been expressed in the variables
$(r_+,r_-,R)$ and the other terms in the variables $(M,R,Q)$ for the
sake of computational simplicity.
Then, after using Eq.~(\ref{pinMRQ})
in Eq.~(\ref{phieqdiff}),
we obtain that Eq.~(\ref{phieqdiff}) takes the form
\begin{equation}
R^{2}\,
\left(\frac{\partial \Phi k}{\partial R}\right)_{r_+,r_-}
- \frac{\sqrt{r_+ r_-}}
{\sqrt{G}}=0\,.
\label{above}
\end{equation}
The solution of Eq.~(\ref{above}) is then
\begin{equation}\label{Phi02}
\Phi(r_+,r_-,R) = \frac{\phi(r_+,r_-) - \frac{\sqrt{r_+ r_-}}
{\sqrt{G}R}}{k}
\end{equation}
where $\phi(r_+,r_-)\equiv \Phi(r_+,r_-,\infty)$ is an arbitrary
function that corresponds physically to the electric potential of the
shell if it were located at infinity.  This thermodynamic electric
potential $\Phi$ is the difference in the
electric potential $\phi$ between
infinity and $R$, blueshifted from infinity to $R$ (see a similar
result in \cite{Braden:1990hw,Peca:1998cs} for an
electrically charged black hole in a grand canonical
ensemble).  We also see that by changing to the variables
$(r_+,r_-,R)$ we are able somehow
to reduce the number of arguments of the
arbitrary function from three to two. This is a feature of the Einstein
equations in conjunction with the first law of thermodynamics.

It is convenient to define a function $c(r_+,r_-)$
through $c(r_+,r_-) \equiv \frac{\phi(r_+,r_-)}{Q}$, i.e.,
\begin{equation}\label{cr+r-}
c(r_+,r_-) \equiv \sqrt{G} \frac{\phi(r_+,r_-)}{\sqrt{r_+ r_-}}\,,
\end{equation}
where we have used $Q=\sqrt{r_+r_-/G}$
as given in Eq.~(\ref{invhorradicauch}).
Then, Eq.~(\ref{Phi02}) is written as
\begin{equation}\label{Phi}
\Phi(r_+,r_-,R) = \dfrac{ {c(r_+,r_-)}
- \dfrac{1}{R}  }
{k}\,\sqrt{\frac{r_+r_-}{G}}\,.
\end{equation}

%% file: I.3ShellCharged/4.section.tex
\section{Entropy of the thin shell and the black hole limit}
\label{entro}

At this point we have all the necessary information to calculate the
entropy $S$. By inserting the
equations of state for the pressure,
Eq.~(\ref{pQk}),
for the temperature,
Eq.~(\ref{BS5}), and for the
electric potential,
Eq.~(\ref{Phi}), as well as
the differential of $M$ given in
Eq.~(\ref{M1}) and the differential of
the area $A$ or of the radius $R$,
into the first law,
Eq.~(\ref{TQ22}),
we
arrive at the entropy differential
\begin{align}\label{dSQ}
dS = b(r_+,r_-) & \frac{1-c(r_+,r_-) r_-}{2G} dr_+ + b(r_+,r_-)\frac{1-c(r_+,r_-) r_+}{2G} dr_-\,,
\end{align}
Now, Eq.~(\ref{dSQ}) has its own integrability condition if $dS$ is to
be an exact differential. Indeed, it must satisfy the equation
\begin{equation}\label{bc5}
\frac{\dd b}{\dd r_-} (1-r_- c) - \frac{\dd b}{\dd r_+}(1-r_+ c) =
\frac{\dd c}{\dd r_-} b r_- - \frac{\dd c}{\dd r_+} b r_+.
\end{equation}
This shows that in order to obtain a specific expression for the
entropy one can choose either $b$ or $c$,
and the other remaining function can
be obtained by solving the differential equation (\ref{bc5}) with
respect to that function. Since Eq.~(\ref{bc5}) is
a differential equation there is still some freedom
in choosing the other remaining function.
In the first examples
we will choose to specify the
function $b$ first and from it obtain an expression for $c$.
We also give examples where
the function $c$ is specified first.

From Eq.~(\ref{dSQ}) we obtain
\begin{equation}\label{entr5}
S=S(r_+,r_-)\,,
\end{equation}
so that the entropy is a function of $r_+$ and $r_-$ alone.
In fact $S$ is a function of $(M,R,Q)$,
but the functional dependence has to be
through $r_+(M,R,Q)$ and $r_-(M,R,Q)$,
i.e., in full form
\begin{equation}\label{entr6}
S(M,R,Q)=S(r_+(M,R,Q),r_-(M,R,Q))\,.
\end{equation}
This result shows that the entropy of the thin
charged shell
depends on the $(M,R,Q)$ through
$r_+$ and $r_-$ which themselves
are specific functions of $(M,R,Q)$.

It is also worth noting the following feature.
From Eq.~(\ref{entr6}) we see that
shells with the same $r_+$ and $r_-$, i.e., the
same ADM mass $m$ and charge $Q$,
but different radii $R$, have the
same entropy.
Let then
an observer sit at infinity and measure $m$
and $Q$ (and thus $r_+$ and $r_-$).
Then, the observer
cannot distinguish the entropy of shells with
different radii.  This is a kind of thermodynamic
mimicker, as a shell near its own
gravitational radius and another one far from
it have the same entropy.

Let us now consider a charged thin shell, for which the differential of
the entropy has been deduced to be Eq.~(\ref{dSQ}).
We are free
to choose an equation of state for the inverse temperature.
Let us pick for convenience the following inverse temperature
dependence,
\begin{equation}\label{bBH}
b(r_+,r_-) = \gamma\,
\frac{r_+^2}{r_+-r_-}\,,
\end{equation}
where $\gamma$ is some constant with
units of inverse mass times inverse radius,
i.e., units of angular momentum.

For a charged shell we must also specify the function $c(r_+,r_-)$,
whose form can be taken from
the differential equation (\ref{bc5}) upon
substitution of the function (\ref{bBH}).
There is a family of solutions
for $c(r_+,r_-)$
but for
our purposes here we choose
the following specific solution,
\begin{equation}\label{cBH}
c(r_+,r_-) = \frac{1}{r_+}\,.
\end{equation}
The reason for the choices above becomes clear when
we discuss the shell's gravitational radius, i.e., black hole
limit.
Inserting the choice for $b(r_+,r_-)$, Eq.~(\ref{bBH}),
along with the  choice for the function
$c(r_+,r_-)$,  Eq.~(\ref{cBH}), in the
differential (\ref{dSQ}) and integrating, we obtain the entropy
differential for the
shell
\begin{equation}\label{diffSS}
dS = \frac{\gamma}{2\, G} \,r_+\, dr_+\,.
\end{equation}
Thus, the entropy of the shell is
\eq{
S=\frac{\gamma}{4\, G} \,r_+^2+ S_0\,,
}
where $S_0$ is an integration constant.
Imposing that when the shell vanishes
(i.e., $M=0$ and $Q=0$, and so
$r_+=0$) the entropy vanishes,
we have that $S_0$ is zero,
and so
$S=\frac{\gamma}{4\, G} \,r_+^2$.
Thus, we can write
the entropy $S(M,R,Q)$ as
\begin{equation}\label{SSa}
S = \frac{\gamma}{16\pi G}\, A_+\,,
\end{equation}
where $A_+$ is the gravitational
area of the shell, as given in
Eq.~(\ref{areaplus}).
This result shows that the entropy of this thin
charged shell
depends on $(M,R,Q)$ through
$r_+^2$ only, which itself is
a specific function of $(M,R,Q)$.

Now, the constant $\g$ should be determined
by the properties of the matter in the shell,
and cannot be decided a priori. The thermodynamic stability of the uncharged case ($Q=0$, i.e.,
$r_-=0$) can be worked out
\cite{Martinez:1996ni} and elucidates the issue.  In the
uncharged case the nontrivial stability conditions are given by
Eqs.~(\ref{B1}) and (\ref{B4}).
Equation~(\ref{B1}) gives immediately
$R\leq\frac32 r_+$, i.e., $R\leq3Gm$. On the other hand,
Eq.~(\ref{B4}) yields $R\geq r_+$, i.e., $R\geq2Gm$. Thus, the
stability conditions yield the following range for $R$, $r_+\leq
R\leq\frac32 r_+$, or in terms of $m$, $2Gm\leq R\leq 3Gm$. This is
precisely the range for stability found by York \cite{York:1986it} for a
black hole in a canonical ensemble in which a spherical massless thin
wall at radius $R$ is maintained at fixed temperature $T$. In
\cite{York:1986it} the criterion used for stability is that the heat
capacity of the system should be positive, and physically such a tight
range for $R$ means that only when the shell, at a given temperature
$T$, is sufficiently close to the horizon can it smother the
black hole enough
to make it thermodynamically stable. The positivity of the heat
capacity is equivalent to our stability conditions, Eqs.~(\ref{B1})
and (\ref{B4}) in the uncharged case.

The stability conditions, Eqs.~(\ref{B1})-(\ref{B7}), for the general
charged case cannot be solved analytically in this instance, they
require numerical work, which will shadow what we want to
determine. Nevertheless, the approach followed in
\cite{Braden:1990hw,Peca:1998cs} for the heat capacity of a charged black
hole in a grand canonical ensemble gives a hint of the procedure that
should be followed.

Although $\gamma$ should be determined
by the properties of the matter in the shell,
there is a case in which the properties
of the shell have to adjust to the
environmental properties of the spacetime, which is
the case when
$R
\to r_+$.
This case is special because the free parameters of the shell have to adjust to the environmental properties of the spacetime, independently of the matter fields which make up the shell. To understand this, one must recall that the thermal stress energy tensor on the shell,
corresponding to a temperature $T_{0}$, can be represented in the form \cite{Anderson:1995fw,Loranz:1995gc}
\begin{equation}\label{backReact}
T^{a}{}_{b}=\frac{(T_{0})^{4}-(T_{H})^{4}}{(g_{00})^{2}} f^{a}{}_{b}\,,
\end{equation}
where, $f^{a}{}_{b}$ is some tensor finite on the horizon, and $T_{H}$ is the Hawking tempeature of the black hole associated to the metric in question. The Hawking temperature is the temperature that a black hole possesses, which is in general given by
\eq{
T_H = {\kappa \over 2\pi}\,,
}
where $\kappa$ is the surface gravity of the black hole. For a black hole associated to the metric of Eq.~(\ref{LEO2}), we have
\eq{
T_{\rm bh} = \frac{\hbar}{4 \pi} \frac{r_+-r_-}{r_+^2}\,,
}
where $\hbar$ is Planck's constant. Now, in the horizon limit we have $g_{00} \to 0$, so the
requirement of the finiteness of $T^{a}{}_{b}$ entails $T_{0}=T_{H}$, i.e. if one takes the shell to its gravitational radius, the integrity of the shell will remain only if it is at the Hawking temperature. Physically, this corresponds to containing the backreaction of the shell's quantum fields, such that the shell is not destroyed. Thefore, we need the shell to be at the Hawking temperature when taking it to its gravitational radius,
so we must choose
\begin{equation}
\gamma= \frac{4\pi}{\hbar} \,,
\label{ga}
\end{equation}
i.e., $\gamma$ depends on fundamental constants.
Then,
\begin{equation}\label{thawk}
b(r_+,r_-) = \frac{1}{T_{\rm bh}}
=\frac{4 \pi} {\hbar}
\frac{r_+^2}{r_+-r_-}\,.
\end{equation}
In this case the entropy of the shell is
$S= \frac14\frac{A_+}{G\hbar}$, i.e.,
\begin{equation}\label{SS}
S=
\frac{1}{4}\frac{A_+}{A_p}\,,
\end{equation}
where $l_p=\sqrt{G\hbar}$ is the Planck length, and $A_p=l_p^2$ the
Planck area.
Note now that the entropy
given in Eq.~(\ref{SS})
is the black hole Bekenstein-Hawking entropy
$S_{\rm bh}$ of a charged black hole since
\begin{equation}\label{SSBH}
S_{\rm bh}= \frac{1}{4}\frac{A_+}{A_p}\,,
\end{equation}
where $A_+$ is here the horizon area.  Thus, when we
take the shell to its own gravitational radius the entropy is the
Bekenstein-Hawking entropy.  The limit also implies that the pressure
and the thermodynamic electric potential go to infinity as $1/k$,
according to Eqs.~(\ref{pQk}) and (\ref{Phi}), respectively.  Note,
however, that the local inverse temperature goes to zero as $k$, see
Eq.~(\ref{BS5}), and so the local temperature of the shell also goes
to infinity as $1/k$. We thus see that the well-controlled
infinities cancel out precisely to give the Bekenstein-Hawking entropy (\ref{SS}).

Note that the shell at its own gravitational radius, at least in
the uncharged case, is thermodynamically stable, since in this case
stability requires $r_+\leq R\leq\frac32 r_+$, as mentioned above. In
addition, our approach and the approach followed in
\cite{Pretorius:1997wr} to find the black hole entropy have some
similarities. The two approaches use matter fields, i.e., shells,
to find the black hole entropy. Here we use a static shell
that decreases its own radius $R$ by steps, maintaining its staticity
at each step. In \cite{Pretorius:1997wr} a reversible contraction of a thin
spherical shell down to its own gravitational radius was examined, and
it was found that the black hole entropy can be defined as the
thermodynamic entropy stored in the matter in the situation that the
matter is compressed into a thin layer at its own gravitational
radius.

Finally we note that the extremal limit $\sqrt{G}m=Q$ or $r_+=r_-$
is well defined from above.  Indeed, when one takes the limit $r_+\to
r_-$ one finds that $1/b(r_+,r_-)=0$ (i.e., the Hawking temperature is
zero) and the entropy of the extremal black hole is still given by
$S_{\rm extremal\,bh}= \frac{1}{4}\frac{A_+}{A_p}$.  It is well known
that extremal black holes and in particular their entropy have to be
dealt with care.  If, ab initio, one starts with the analysis for an
extremal black hole one finds that the entropy of the extremal black
hole has a more general expression than simply being equal to one
quarter of the area \cite{Pretorius:1997wr,Lemos:2010kw}.  This
extremal shell is an example of a Majumdar-Papapetrou matter system.
Its pressure $p$ is zero, and it remains zero, and thus finite, even
when $R\to r_+$.  This limit of $R\to r_+$ is called a quasiblack
hole, which in the extremal case is a well-behaved one.

It is important to stress that the requirement $b=T_{\mathrm{bh}}^{-1}$
is compulsory
only for shells that approach their own gravitational
radius. Otherwise, if we consider the radius of the shell within some
constrained region outside the gravitational radius, the shell
temperature can be arbitrary since away from the horizon, quantum
backreaction remains modest and does not destroy the thermodynamic
state.
One can discuss whole classes of functions $b(r_{+},r_{-})\neq
T_{\mathrm{bh}}^{-1}$. The choice (\ref{cBH}) for $c$ is also only necessary
for shells at the gravitational radius limit.
According to Eq.~(\ref{cr+r-}), this gives us
$\phi =\frac{ \sqrt{r_-} }{ \sqrt{G\, r_{+} }}$, i.e.,
\begin{equation}
\phi =\frac{Q}{r_{+}}  \label{rn}
\end{equation}
that coincides with the standard expression for the electric potential
for the Reissner-Nordstr\"{o}m black hole. In addition,
Eq.~(\ref{Phi})
acquires the form
\begin{equation}
\Phi =\frac{Q}{k}\left(\frac{1}{r_{+}}-\frac{1}{R}\right)
\label{frn}
\end{equation}
that coincides entirely with the corresponding formula for the
Reissner-Nordstr\"{o}m black hole in
a grand canonical ensemble \cite{Braden:1990hw}.  Meanwhile, in this
case there is no black hole.  Moreover, considering the uncharged
case, $Q\rightarrow0$ or
$r_{-}\rightarrow 0$, it is seen from Eq.~(\ref{dSQ}) that the
quantity $c$ drops out from the entropy, so the choice of $c$ is
relevant for the charged case only.

There are similarities between the thin-shell approach and
the black hole mechanics approach \cite{Bardeen:1973gs}.
These are evident if we express the
differential of the entropy of the charged shell (\ref{dSQ}) in terms
of the black hole ADM
mass $m$ and charge $Q$, given in terms of the
variables $(r_+,r_-)$ by
Eqs.~(\ref{invhorradi})-(\ref{invhorradicauch}).
The differential for the entropy of the shell reads,
in these variables,
\begin{equation}
T_0 dS = dm - c\, Q\, dQ
\end{equation}
where we have defined $T_0 \equiv 1/b(r_+,r_-)$ which is the
temperature the shell would possess if located at infinity.
Here, $T_0 =1/b(r_+,r_-)$ and $c=c(r_+,r_-)$ should be seen as
$T_0 (m,Q)=1/b(m,Q)$ and $c(m,Q)$, respectively,
since $r_+$ and $r_-$ are functions of
$m$ and $Q$.
As we have
seen, if we take the shell to its gravitational radius, we must fix
$T_0 = T_{\rm bh}$ and $c = 1/r_+$.
This suggests that $Q/r_+$ should play the role of
the black hole electric potential $\Phi_{\rm bh}$, which in fact is
true, as shown in Eq.~(\ref{rn}).
This implies that the
conservation of energy of the shell is expressed as
\begin{equation}\label{1T7}
T_{\rm bh} dS_{\rm bh} = dm - \Phi_{\rm bh}\, dQ\,.
\end{equation}
We thus
see that the first law of thermodynamics for the shell
at its own gravitational radius is
equal to the energy conservation for the
Reissner-Nordstr\"om
black hole.

%% file: I.3ShellCharged/6.section.tex
\section{The thin shell
with another specific equation of state for the
temperature}
\label{eqstate}

The previous equation of state is not prone to a simple stability
analysis. Here we give another equation of state
that permits finding both  an expression for the shell's entropy
and performing a simple stability analysis.

We must first
specify an adequate thermal equation of state for $b(r_+,r_-)$.
A possible
simple choice is a power law in the ADM mass $m$, i.e.,
$b(r_+,r_-)$ has the form
\begin{equation}\label{bF6}
b(r_+,r_-) = 2G \, a(r_++r_-)^{\al}
\end{equation}
where $a$ and $\al$ are free coefficients
related to the properties of the shell.
Power laws occur frequently in thermodynamic systems,
and so this is a natural choice as well.
The simple choice above
allows one to find the form of the function $c$.
Indeed, the integrability equation
(\ref{bc5}) gives that the function $c$
can be put in the form
$c(r_+,r_-) = 2G \frac{f(r_+ r_-)}{(r_+ + r_-)^{\al}}$, where
$f(r_+ r_-)$ is an arbitrary function
of the product $r_+ r_-$ and supposedly also depends on the
intrinsic constants of the matter that makes up the shell.
For convenience we choose $f(r_+ r_-)=d\,(r_+ r_-)^{\delta}$,
where $d$ and $\delta$ are parameters
that reflect the shell's properties, so that
\begin{equation}\label{cpower}
c(r_+,r_-) = 2G \,d\,  \frac{(r_+ r_-)^{\delta}}
{(r_+ + r_-)^{\al}}\,.
\end{equation}
The gravitational constant $G$ was introduced in
Eqs.~(\ref{bF6}) and (\ref{cpower}) for convenience.
Inserting Eqs.~(\ref{bF6})-(\ref{cpower})
into Eq.~(\ref{dSQ}) and integrating,
gives the entropy
\begin{equation}\label{SF6}
S(r_+,r_-) = a \, \left[\frac{(r_++r_-)^{\al+1}}{\al+1} - d \,
\frac{(r_+ r_-)^{\de+1}}{\de+1}\right] \,,
\end{equation}
where the constant of integration $S_0$
has been put to zero, as expected in the limit $r_+
\to 0$ and $r_- \to 0$.
Again, the entropy of this thin
charged shell
depends on $(M,R,Q)$ through
$r_+$ and $r_-$ only, which in turn are
specific functions of $(M,R,Q)$.

We consider positive temperatures and
positive electric potentials, so
\begin{equation}\label{adgeq0}
a>0\,,\quad d>0\,.
\end{equation}
We consider only
\begin{equation}\label{ageq0}
\al>0 \,,
\end{equation}
for the simplicity of the upcoming stability analysis. Although this
choice somewhat narrows down the range of cases to which the analysis
is applicable, it only rules out the cases where $-1<\al<0$, since for
values $\al\leq-1$ it would give a diverging entropy in the limit $r_+
\to 0$ and $r_- \to 0$, something which is not physically
acceptable. Indeed, in such a limit we would expect the entropy to be
zero which requires $\al > -1$.

Proceeding to the thermodynamic stability treatment, we start with
Eq.~(\ref{B1}), which can be shown to be equivalent to
\begin{equation}\label{B1s}
r_+ r_- - 2 R^2 k^2 \al + (1-k^2)R^2 \geq 0.
\end{equation}
Solving for $k$, this leads to the restriction
\begin{equation}\label{k1s}
k \leq \sqrt{\frac{1}{2\al +1}\left(1+\frac{r_+ r_-}{R^2}\right)}.
\end{equation}
Going now to Eq.~(\ref{B2}), it gives
\begin{align}\label{B2s}
[r_+ r_- - (1-k)^2 R^2]&[\al (r_+ r_- - (1-k)^2 R^2) + 3(r_+ r_- + (1-k^2)R^2)] \leq 0.
\end{align}
Since the second multiplicative term on the left must be positive,
one can solve for $k$ and obtain the set of values which satisfy
the inequality,
\begin{align}\label{k2s}
\frac{\al}{\al+3} - \sqrt{\frac{9}{(\al+3)^2} + \frac{r_+ r_-}{R^2}}
\leq k
\leq\frac{\al}{\al+3} + \sqrt{\frac{9}{(\al+3)^2} +
\frac{r_+ r_-}{R^2}}\,.
\end{align}
As for Eq.~(\ref{B3}), it reduces to
\begin{equation}
\frac{d R(2\de +1)(r_+ r_-)^{\de}}{\left(\frac{r_+ r_-}{R} +
(1-k^2)R\right)^{\al}} \geq \frac{R^2(1-k^2)+(2 \al + 1)r_
+ r_-}{R^2(1-k^2)+r_+ r_-}.
\end{equation}
Although one cannot conclude anything directly from the above
inequality, it is nonetheless worth noting that the right-hand
side is greater than zero, and so $\de$ must obey the condition
\begin{equation}\label{intDe}
\de \geq -\frac{1}{2}.
\end{equation}
Regarding Eq.~(\ref{B4}), it is possible to show that it implies
the condition
\begin{align}\label{B4s}
r_+^2 r_-^2 (\al+3) - 2r_+ r_-R^2(2k^2\al + 2k^2 -
k + \al -1)
& + (1-k)^2 R^4 (3k^2 \al + k^2 + 2\al k + \al -1) \leq 0,
\end{align}
which does not provide any information on its own since it is
a polynomial of order four in the variable $k$. Nonetheless,
it does need to be satisfied once a region of allowed
values for $k$ is known, which will be ascertained shortly.

Concerning Eq.~(\ref{B5}), we are led to
\begin{equation}
\frac{d R(2\de +1)(r_+ r_-)^{\de}}{\left(\frac{r_+ r_-}{R} +
(1-k^2)R\right)^{\al}} \leq
\frac{r_+^2 r_-^2(3\al + 1) + 2(1-k)r_+ r_-R^2(2\al (k-1) +
2k -1)-(1-k)^3 R^4(k(\al+3) - \al + 3)}{\left[(1-k)^2 R^2 -
r_+ r_-\right]\left[(k-1) R^2 (k(\al+3) - \al + 3)\right]},
\end{equation}
which does not contain any new information.
On the other hand,
when Eq.~(\ref{B6}) is simplified to
\begin{align}
\frac{d R(2\de +1)(r_+ r_-)^{\de}}{\left(\frac{r_+ r_-}{R} +
(1-k^2)R\right)^{\al}}
 \geq \frac{R^2(1-k^2)+(2 \al + 1)r_+ r_-
- 2R^2 k^2 \al }{R^2(1-k^2)+r_+ r_- -2R^2k^2\al}\,,
\end{align}
and one notices that the numerator on the right side
must be positive, another constraint on $k$ naturally appears,
namely
\begin{equation}\label{k6s}
k \leq \sqrt{\frac{1}{2\al +1} + \frac{r_+ r_-}{R^2}}.
\end{equation}

Finally, the last condition (\ref{B7}) gives the inequality
\begin{equation}
r_+ r_-(\al+1) - R^2 \left[(\al+1)k^2 + \al -1\right] \geq 0
\end{equation}
which constricts the values of $k$ to be within the interval
\begin{equation}\label{k7s}
k \leq \sqrt{-\frac{\al -1}{\al +1} + \frac{r_+ r_-}{R^2}}.
\end{equation}

The definitive region of permitted values for $k$ is
the intersection of the conditions (\ref{k1s}), (\ref{k2s}),
(\ref{k6s}) and (\ref{k7s}). It is possible to show that
such an intersection gives the range
\begin{equation}\label{intK}
\frac{\al}{\al+3} - \sqrt{\frac{9}{(\al+3)^2} + \frac{r_+ r_-}{R^2}}
\leq k \leq \sqrt{\frac{r_+ r_-}{R^2}-\frac{\al -1}{\al +1}}
\end{equation}
where $\al$ must be restricted to
\begin{equation}\label{intAl}
\al \geq \frac{1+\frac{r_+ r_-}{R^2}}{1-\frac{r_+ r_-}{R^2}}.
\end{equation}
Returning to Eq.~(\ref{B4s}), it is now possible to verify if
the interval (\ref{intK}) satisfies said condition,
which indeed it does.

If one takes the shell to its own gravitational radius,
the chosen temperature equation of state (\ref{bF6})
is wiped out,
and a new
equation of state
sets in  to adapt to
the quantum spacetime properties.
The new equation of state
is then given by Eq.~(\ref{thawk})
and the black hole entropy
(\ref{SS}) follows.

%% file: I.3ShellCharged/7.section.tex
\section{Other equations of state}
\label{other}

Naturally, other equations of state can be sough.
We give four examples, one fixing  $b(r_+,r_-)$
and three others
fixing $c(r_+,r_-)$.

If we fix
the inverse temperature
\begin{equation}
b(r_+,r_-) = \gamma\,
\frac{r_+^2}{r_+-r_-}\,,
\end{equation}
for some $\gamma$, as we did before,
then generically, from Eq.~(\ref{bc5}), we find
\begin{equation}
c(r_+,r_-) = \frac{a(r_+r_-)(r_+-r_-)+r_-}{r_+^2} \,,
\end{equation}
where $a(r_+r_-)$ is an arbitrary function
of integration
of the product $r_+r_-$
and presumably
also depends on the intrinsic constants
of the matter that makes up the shell.
Then, from Eq.~(\ref{dSQ}), the entropy is
\begin{equation}
S(r_+,r_-)=\frac{\gamma}{4G}\left(r_+^2 +
\int_{0}^{r_{+}r_{-}}
\left(1-a(x)\right)\,dx\right)\,,
\end{equation}
where we are assuming zero entropy
when $r_+=0$.
In the example we gave previously we have
put $a(r_+r_-)=1$, so that
$c(r_+,r_-) = \frac{1}{r_+}$. This case
$a(r_+r_-)=1$ gives precisely that the
entropy of the shell
is proportional to the area
of its gravitational radius
and for $\gamma= \frac{4\pi}{\hbar}$
gives  that the
entropy of the shell is equal to
the corresponding black hole entropy
as we have discussed previously.
Of course, many
other choices can be given
for $a(r_+r_-)$ and quite
generally the entropy
will be a function of $r_+$
and $r_-$.

Inversely, instead of $b(r_+,r_-)$ one can give $c(r_+,r_-)$, which could be
\begin{equation}
c(r_+,r_-) =\frac{1}{r_{+}}\,,
\end{equation}
as for the black hole case. The
integrability condition, Eq.~(\ref{bc5}),
for the temperature
then gives
\begin{equation}
b(r_+,r_-)=\frac{h(r_{+})}{r_{+}-r_{-}}\,,
\label{neweos1}
\end{equation}
where $h(r_+)$ is a function that can be
fixed according to the matter
properties of the shell.
Then, from Eq.~(\ref{dSQ}), the entropy is
\begin{equation}
S(r_+)=\frac{1}{2G}\int_{0}^{r_{+}}\,
\frac{h(x)}{x}\,dx\,,
\end{equation}
where it is implied that the function $h(x)$ vanishes at $x=0$ rapidly
enough so that the entropy goes to zero when $r_+=0$.
If we choose $h(r_+)=\frac{4\pi}{\hbar}
r_+^2$, then one recovers
the black hole temperature and the black hole
entropy for the shell.

Another equation of state one can choose
for $c(r_+,r_-)$ is
\begin{equation}
c(r_+,r_-) =\frac{1}{r_{-}}\,,
\end{equation}
for which the integrability condition
gives
\begin{equation}
b(r_+,r_-) =\frac{h(r_{-})}{r_{+}-r_{-}}\,,
\end{equation}
where $h(r_-)$ is a function that can be
fixed in accord with the matter
properties of the shell.
In this case, from Eq.~(\ref{dSQ}),  the entropy
of the shell depends on $r_{-}$ only,
and is given by
\begin{equation}
S(r_-)=\frac{1}{2G}\int_{0}^{r_{-}}\frac{h(x)}{x}\,dx\,,
\end{equation}
where we are assuming zero entropy
when $r_-=0$. Yet another example can be obtained if one puts
\begin{equation}
c(r_+,r_-) =c(r_{+}r_{-})\,,
\end{equation}
i.e., $c$ is a function of the product
$r_{+}r_{-}$
and may also depend on the
intrinsic constants of the matter that makes up the shell.
The integrability condition
then gives
\begin{equation}
b=b_{0}\,,
\end{equation}
where $b_0$ is a constant, and so
in this case, the temperature measured at infinity
does not depend on
$r_{+}$ or $r_{-}$.  The entropy is, in that case,
\begin{equation}
S(r_+,r_-)
=\frac{b_{0}}{2G}
\left(r_{+}+r_{-}-\int_{0}^{r_{+}r_{-}} c(x)\,dx\right)\,,
\end{equation}
where we are assuming zero entropy
when $r_+=0$ and $r_-=0$. One could study in detail these four cases
for the thermodynamics of a shell
performing
in addition a stability analysis for each one.
We refrain here to do so. Certainly other interesting
cases can be thought of.

%% file: I.3ShellCharged/conclusions.tex
\section{Conclusions}

We have considered the thermodynamics of a self-gravitating
electrically charged thin shell, generalizing previous works on
the
thermodynamics of self-gravitating thin-shell systems.  Relatively to
the simplest shell where there are two independent thermodynamic state
variables, namely, the rest mass $M$ and the size $R$ of the shell, we
have now a new independent state variable in the thermodynamic system,
the electric charge $Q$, out of which, using the first law of
thermodynamics and the equations of state, one can construct the
entropy of the shell.  Due to the additional variable, the
charge $Q$, the calculations are somewhat more complex, although
the richness in physical results increases in the same proportion.

The equations of state one has to give are the
pressure $p(M,R,Q)$, the
temperature $T(M,R,Q)$,
and the electric potential $\Phi(M,R,Q)$.
The pressure can be obtained from dynamics alone, using the thin-shell
formalism and the junction conditions for a flat interior and a
Reissner-Nordstr\"om exterior.  The form of the temperature and of the
thermodynamic electric potential are obtained using the integrability
conditions that follow from the first law of thermodynamics.

The differential for the entropy in its final form shows
remarkably
that the entropy must be a function of $r_+$ and $r_-$ alone,
i.e., a function of the
intrinsic properties of the shell spacetime.
Thus, shells with the same $r_+$ and $r_-$
(i.e., the same ADM mass $m$ and charge $Q$)
but different radii $R$, have the
same entropy. From the thermodynamics
properties alone of the shell one cannot
distinguish a shell near its own gravitational
radius from a shell far from it.
In a sense, the shell can mimic a black
hole.

The differential for the entropy in its final form
gives that $T$ and $\Phi$
are related through an integrability condition.  One has
then to
specify either $T$ or $\Phi$ and the form of the other
function is somewhat constrained.
We gave two example
cases and mentioned other possibilities.

Many interesting
equations of state can be chosen, and some were given where a full thermodynamic stability analysis was possible.  However at the
gravitational radius all turn into the Hawking
equation of state, i.e., the Hawking temperature.
Since the area of the shell $A$
is equal to the gravitational radius area $A_+$ when the shell is at its own gravitational
radius, and $S=\frac{1}{4}\frac{A_+}{A_p}$
in this limit,
we conclude that
the entropy of the shell is proportional to its
own area $A$. This indicates in a sense
that all its fundamental degrees of freedom
have been excited, hinting that indeed one may infer thermodynamic properties from black holes
using thin matter shells.

%% file: I.4ShellExtremal/main.tex
\chapter{Thermodynamics of an extremal electrically charged thin shell}
\label{cap:chapterI1}
\input{I.4ShellExtremal/outline.tex}

\input{I.4ShellExtremal/1.section.tex}

\input{I.4ShellExtremal/2.section.tex}

\input{I.4ShellExtremal/3.section.tex}

\input{I.4ShellExtremal/4.section.tex}

\input{I.4ShellExtremal/conclusions.tex}

%% file: I.4ShellExtremal/outline.tex
\section{Introduction}

Wide debate is centered around the entropy of an extremal black hole. On one hand, such a black hole has zero temperature, according to the Hawking temperature formula, and so it should have zero entropy according to one of the formulations of the third law of thermodynamics \cite{Callen:1985}. Hawking \cite{Hawking:1974sw,Hawking:1994ii} and Teitelboim \cite{Teitelboim:1994az} have also given some topological arguments which point to the same conclusion. On the other hand, there is no convincing reason why the Bekenstein-Hawking entropy formula should not be valid in the extremal case. After all, working out the entropy of non-extremal black
holes and taking the extremal limit $m=Q$ yields $S=A_+/4$, see, e.g.,
\cite {Bardeen:1973gs,Hawking:1974sw,Braden:1990hw,Lemos:2015gna}. In this case, the
thermodynamic argument would not hold, the extremal black hole could
be a system of minimum energy and degenerate ground state and such
systems can have entropy even at zero temperature. String theory also claims that the entropy should be a quarter of the black hole area (see \cite{Strominger:1996sh,Sen:2014aja,Ghosh:1994mm,Zaslavsky:1997ha,Zaslavsky:1997kp,Mann:1997hm,Zaslavsky:1998cb,Mitra:1998tv,Kiefer:1998rr,Wang:1999gj,Wang:1999sk,Hod:2000pa,Carroll:2009maa,Edery:2010cx} for discussions on this topic).

This chapter is dedicated to the study of the limit of an extremal black hole formed by taking an extremal charged shell to its gravitational radius. This will allow a careful analysis of the entropy form the onset, similar to how it was done in the last chapter.

%% file: I.4ShellExtremal/1.section.tex
\section{The extremal charged thin shell spacetime}

\label{characteristics}

As in Chapter 4, we will be considering the case of a four-dimensional
spherically symmetric spacetime and a spherical thin shell at some
radius $R$ separating an inner region $\mathcal{V}_{i}$ with flat
metric and an outer region $\mathcal{V}_{o}$ with an extremal
Reissner-Nordstr\"{o}m line element. Thus, for the inner region the
metric is
\eq{
ds_{i}^{2} =g_{\alpha \beta }^{i}dx^{\alpha }dx^{\beta }=
 -dt_{i}^{2}+dr^{2}+r^{2}\,d\Omega ^{2}\,,\quad r\leq R\,, \label{LEI}
}
where $x^{\alpha }=(t_{i},r,\theta ,\phi )$ are the inner coordinates,
with $ t_{i}$ being the inner time, and $(r,\theta ,\phi )$ polar
coordinates, and $ d\Omega ^{2}=d\theta ^{2}+\sin ^{2}\theta \,d\phi
^{2}$. For the outer region the metric is
\begin{align}
ds_{o}^{2}& =g_{\alpha \beta }^{o}dx^{\alpha }dx^{\beta }=
 -\left( 1-\frac{m}{r}\right) ^{2}dt_{o}^{2}+\frac{dr^{2}}{
\left( 1-\dfrac{m}{r}\right) ^{2}} +r^{2}d\Omega ^{2}\,,\quad r\geq R\,, \label{LEO}
\end{align}
where $x^{\alpha }=(t_{o},r,\theta ,\phi )$ are the outer coordinates,
with $ t_{o}$ being the outer time, and $(r,\theta ,\phi )$ polar
coordinates. In the extremal case, the ADM mass and charge are related by
\begin{equation}
m=Q\,.  \label{extremalrel}
\end{equation}
On the hypersurface itself, $r=R$, the
metric is that of a 2-sphere with an additional time dimension,
such that the line element is
\begin{equation}
ds_{\Sigma }^{2}=h_{ab}dy^{a}dy^{b}=-d\tau ^{2}+R^{2}(\tau )d\Omega
^{2}\,,\quad r=R\,,  \label{intrinsmetr}
\end{equation}
where we have chosen $y^{a}=(\tau ,\theta ,\phi )$ as the time and
spatial coordinates on the shell. The time coordinate $\tau $ is again the
proper time for an observer located at the shell and the shell radius is
given by the parametric equation $R=R(\tau )$ for an observer on the
shell.  We consider once again a static shell so that $R(\tau)={\rm constant}$.
On each side of the hypersurface, the parametric equations for the
time and radial coordinates are denoted by $ t_{i}=T_{i}(\tau )$,
$r_{i}=R_{i}(\tau )$, and $t_{o}=T_{o}(\tau )$, $ r_{o}=R_{o}(\tau )$.

As usual, the shell will be composed of a perfect fluid, with a stress-energy
tensor $S^{a}{}_{b}$ given by Eq.~(\ref{perffluid}). One then finds
through the junction conditions
\begin{align}
\sigma & = \frac{m}{4\pi R^2} \,,  \label{SS1} \\
p & = 0 \,.  \label{SS20}
\end{align}
Matter for which $p=0$ and
totally supported by electric forces
against gravitational collapse
is called extremal matter or, sometimes, electrically
counterpoised dust.
The rest mass of the shell $M$ is defined as
\begin{equation}
\sigma = \frac{M}{4\pi R^2}\,,  \label{SS1restmass}
\end{equation}
and so in the extremal case
\begin{equation}
M=m\,.  \label{admmassrestmass}
\end{equation}

The gravitational radius $r_+$ of the shell
is given by the zero of the  $g_{00}^o$ in Eq.~(\ref{LEO}).
It is actually a double zero: one gives the
gravitational radius $r_+$, the other the
Cauchy horizon $r_-$ of the shell. The double zero
means that for the
extremal
spacetime
the two radii
coincide,
\begin{equation}
r_+= r_-\,,  \label{zeror+=r-}
\end{equation}
and we call it $r_+$
from now on. The zero of the $g_{00}^o$ in Eq.~(\ref{LEO})
then gives
\begin{equation}
r_+= m\,,  \label{zeror+}
\end{equation}
and so
\begin{equation}
r_+= r_-=m=Q=M\,.  \label{zeror+general}
\end{equation}
Following the example of previous chapters, we define the shell's redshift function $k$ as
\begin{equation}  \label{red}
k=1-\frac{r_+}{R}
\end{equation}
and the area $A$ of the shell
\begin{equation}
A=4 \pi R^2.  \label{area1}
\end{equation}
Repeating the same treatment of Chapter 4 for the junction conditions of the electromagnetic field, we obtain
\begin{equation}  \label{PhiJC}
\frac{Q}{R^2}=4\pi \sigma_e \,.
\end{equation}
In addition, the shell should always be
outside its own gravitational radius, so
\begin{equation}
R \geq r_+\,.  \label{notrapped2}
\end{equation}
Then the physical allowed values for $k$ in Eq.~(\ref{red}) are in the
interval $[0,1]$. Since the pressure of the matter in the shell is
zero and the energy density is considered positive the energy conditions, weak, strong, and dominant,
are always obeyed for $R \geq r_+$.

It is worth noting that in the limit $R\to r_+$ there are subtleties
connected with the behavior of the boundary's geometry.
Indeed, there is a discontinuity because of the timelike
character of the boundary from the inside and the light-like character
of the boundary from the outside (see \cite{Lemos:2009uk} for
details). However, here they are essentially irrelevant since in what
follows we consider the external region only.

%% file: I.4ShellExtremal/2.section.tex
\section{Entropy of an extremal charged thin shell}

\label{entropyreview}

\subsection{Entropy and the first law of thermodynamics for an extremal
charged thin shell}

\label{E}

As before, we assume that the shell
in static equilibrium at radius $R$ has a well defined local
temperature $T$ and an entropy $S$. The entropy $S$ is a function of
the shell's rest mass $M$, area $A$, and charge $Q$, i.e.,
\begin{equation}  \label{entropy0}
S\equiv S(M,A,Q)\,.
\end{equation}
The first law of thermodynamics can be then written as
\begin{equation}  \label{TQ0}
T dS = dM + pdA - \Phi dQ\,,
\end{equation}
or, defining the inverse temperature $\beta$,
\begin{equation}  \label{beta}
\beta \equiv \frac1T
\end{equation}
one has
\begin{equation}  \label{TQ}
dS = \beta\left( dM + pdA - \Phi dQ\right)\,.
\end{equation}
Unlike the non-extremal case of Chapter 4, the extremal case is a special one, since the the extremality condition will
constraint the possible configurations. From Eq.~(\ref{zeror+general}), we
have for an extremal shell
\begin{equation}  \label{equalQM}
dQ=dM\,.
\end{equation}
Thus, the number of independent variables reduces to two, namely, $M$
and $A$, and so, $p=p(M,A)$, $\beta=\beta(M,A)$, and
$\Phi=\Phi(M,A)$. It is more convenient to work out with the shell's
radius $R$ than its area $A$, which can be done from
Eq.~(\ref{area1}), so the equations of state are of the form
\begin{equation}  \label{eqsstate}
p=p(M,R)\,,\; \beta=\beta(M,R)\,,\; \Phi=\Phi(M,R)\,.
\end{equation}
Now, from Eq.~(\ref{SS20}), one has that the equation
of state for the pressure is
\begin{equation}  \label{presseqsstate}
p(M,R)=0\,.
\end{equation}
Thus, the the first law (\ref{TQ}) is now
\begin{equation}  \label{secondlawextremal0}
dS = \beta\left(1-\Phi\right)dM\,,
\end{equation}
and, since from
Eq.~(\ref{zeror+general}) we have $M=r_+$ and $dM=dr_+$, one can write
the first law as
\begin{equation}  \label{secondlawextremal}
dS = \beta\left(1-\Phi\right)dr_+\,,
\end{equation}
where now
\begin{equation}  \label{eqsstate2}
\beta=\beta(r_+,R)\,,\; \Phi=\Phi(r_+,R)\,.
\end{equation}
The integrability condition for Eq.~(\ref{secondlawextremal})
reduces to a simple equation, namely,
\begin{equation}  \label{integextr}
\beta\left(1-\Phi\right)=s(r_+)\,,
\end{equation}
where $s$ is a function of $r_+$ alone and is arbitrary as long as it
gives a positive meaningful entropy.
Since $\beta\geq0$ and $s\geq0$, we have the
following constraint on $\Phi$,
\begin{equation}  \label{constraintPhi}
\Phi\leq1\,.
\end{equation}
The result given in Eq.~(\ref{integextr}), that the most
general function of the product of two functions (namely,
$\beta$ and
$1-\Phi$) of $r_+$ and $R$ is a function of $r_+$ alone, is new and
interesting. Using now Eq.~(\ref{secondlawextremal})
together with Eq.~(\ref{integextr}) yields
\begin{equation}  \label{dSQ2}
dS = s(r_+) \, dr_+ \,.
\end{equation}
The function $s(r_+)$ is thus a kind of entropy density.
Integrating Eq.~(\ref{dSQ2}), we conclude that the
entropy of the extremal
shell is given by
\begin{equation}  \label{S7}
S= S(r_+)\,,\quad R\geq r_+\,,
\end{equation}
where we have assumed that the constant of integration is zero for the same reason as in the previous chapters. Thus
the entropy of an extremal charged thin shell is a function of $r_+$
alone.  Depending on the choice of $s(r_+)$ we can obtain a wide range
of values for the entropy $S(r_+)$ of the shell. Since $\beta(r_+,R)$
and $\Phi(r_+,R)$ are arbitrary as long as they obey the constraint
(\ref{integextr}), this shows that the extremal case is indeed quite
special. Such a result does not appear in the non-extremal case of Chapter 2.

\subsection{Choices for the matter equations of state of an extremal
charged thin shell}
\label{E3}

If the shell was non-extremal, we would have obtained the equation (\ref{BS5}) of state for the temperature and so, taking the limit to the extremal shell, i.e., $r_+=r_-$, we would find
$
\beta(r_{+},R)=b(r_{+})\,k(r_{+},R)\,,
$
where $k(r_{+},R)$ is given in
Eq.~(\ref{red2}). Now, in the extremal case, the only integrability condition is
Eq.~(\ref{integextr}), and it has nothing to do with Tolman's formula.
However, among all other possible choices, Tolman's
formula  $\beta(r_{+},r_-,R)=b(r_{+},r_-)\,k(r_{+},r_-,R)$,
allows for a nontrivial generalization.
For nonextremal shells, one finds from the integrability conditions
that
$b=b(r_{+},r_-)$, i.e.,
$b$ cannot depend on $R$.
For extremal shells, on the other hand,
nothing prevents us from including in $b$
a dependence not only on $r_{+}=r_-$, but
also on $R$. As a result, the generic Tolman
formula in the extremal case must be
\begin{equation}
\beta(r_{+},R)=b(r_{+},R)\, k\,.
\label{tolmanextremalpostulated}
\end{equation}
As usual, the function $b(r_+,R\to\infty)$
represents the inverse of the temperature of the shell if it
were located at infinity. With the choice for $\beta$
given in Eq.~(\ref{tolmanextremalpostulated}), one finds that
Eq.~(\ref{integextr}) yields
\begin{equation}  \label{Phi0}
\Phi(r_+,R) = \frac{\phi(r_+,R) - \frac{r_+} {R}}{k}\,,
\end{equation}
where we have defined
$\phi(r_+,R)\equiv 1-\frac{s(r_+)}{b(r_+,R)}$, i.e.,
$\phi$ is such that
\begin{equation}  \label{integextrbphi}
b\left(1-\phi\right)=s(r_+)\,.
\end{equation}
From Eq.~(\ref{Phi0}) one sees that
$\phi(r_+,R\to\infty)$ represents the electric
potential of the shell if it were located at infinity.

We could proceed and give specific equations
for $b(r_+,R)$ and $\phi(r_+,R)$, and determine the thermodynamic
properties of the shell including its thermodynamic stability.  We
refrain from doing it here, and study instead some particular instances
that allow us to take
the black hole limit.

%% file: I.4ShellExtremal/3.section.tex
\section{Entropy of an extremal black hole}
\label{bh}

We are now interested in taking the extremal black hole limit, which consists in taking the shell to its gravitational radius $R=r_+$,
which is a somewhat delicate process for an extreme shell.
Firstly, we need to fix the shell at some radius $R>r_{+}$ and
choose appropriately the functions $\beta$ and $\Phi$,
or $b$ and $\phi$,
and only afterwards send the shell to $R=r_{+}$.
However, we know that the Hawking temperature measured at infinity for
an extremal black hole is $T_H=0$, so we must choose the temperature at
infinity $T_\infty$ as $T_\infty=0$, i.e., $b=\infty$. Thus from
Eq.~(\ref{tolmanextremalpostulated})
we also find $\beta=\infty$ and so the temperature on
the shell is zero when it is infinitesimally close to the horizon.
Now we have to find $\phi$ and $\Phi$ such that the products
$b(1-\phi)=s$ and $\beta(1-\Phi)=s$ remain finite, and equal to
some function $s(r_+)$. It becomes clear that we must have $\phi=1$
and $\Phi=1$ when the shell approaches the gravitational radius, but
any function which obeys this limit will be valid, so long as the product
$b(1-\phi)$ remains finite. If these conditions are satisfied for the shell
at any radius $R>r_+$, then it can be safely taken to its gravitational radius.

Taking now the shell to its own gravitational radius $R=r_+$,
i.e., take the black hole limit, the entropy differential for the shell
will depend solely on $r_+$ through the
function $s(r_+)$, which is arbitrary. Thus, we conclude that the entropy of
the extremal shell in the extremal black hole limit is given by
\begin{equation}  \label{S7777}
S=S(r_+) \,,\quad R= r_+,
\end{equation}
constituting the extremal black hole limit
of an extremal shell. Such a configuration is also called a quasiblack hole.

Our approach implies that
the entropy of an extremal black hole can assume the form of any well-behaved
function of $r_+$. The precise function of the entropy depends on the
constitution of the matter that collapsed to form the black
hole. Depending on the choice of $s$ that, in turn, depends on
the choices for $\beta$ and
$\Phi$, we can obtain any function of $r_+$ for the entropy $S$ of the
extremal black hole.  The fact that the entropy in the extremal case
is model-dependent agrees with previous work
\cite{Lemos:2010kw} and more early studies \cite{Pretorius:1997wr}.
Of course, a particular class of entropies for the extremal
black hole would be the Bekenstein-Hawking entropy $S(r_+)=\frac14
A_+=\pi r_+^2$.
In summary,
our result is quite different from the non-extremal case, where the
entropy can only have the Bekenstein-Hawking functional dependence
$S(r_{+})=\frac{1}{4}A_{+}$ \cite{Lemos:2015gna}.

Note, that, although the importance of the product $\beta \left( 1-\Phi
\right)$ has been raised in the
extremal black hole context in \cite{Hod:2000pa}
(see also \cite{Ghosh:1996dq}), the result that the most general function
of the product $\beta \left( 1-\Phi \right) $ is a well-behaved,
but otherwise arbitrary, function of $r_{+}$ is
new. There are additional differences between
\cite{Hod:2000pa,Ghosh:1996dq} and our work. For example, in
\cite{Hod:2000pa,Ghosh:1996dq}, the product $ \beta
(1-\Phi)$ enters the path integral over fluctuating geometries, so it
appears in a quantum context. In doing so, finite nonzero $\beta$
are not forbidden.
However, for such $\beta$ quantum backreaction
destroys the extremal horizon \cite{Anderson:1995fw}. In our approach,
we consider a shell, not a black hole, and thus we can adjust $\beta $
and $\Phi $
at the shell radius in such a way that, for any $R$ close to $r_{+}$, the
backreaction remains finite \cite{Lemos:2010kw}.

One may now speculate on constraints that the entropy of the extremal black hole should have. For
instance, the
initial Bekenstein arguments for non-extremal black holes \cite{Bekenstein:1973ur}
proved that an entropy proportional to $A_{+}^{1/2}$ should be
discarded on the basis of the second law of
thermodynamics. However, since extremal black holes have a
different character
from non-extremal ones, these arguments do not hold here. Another
possible
constraint is the following. For the usual, non-extremal,
black holes the entropy is
$S(r_{+})=\frac{1}{4}A_{+}$. In this case, when one takes the shell to its
own gravitational radius the pressure at the shell blows up, $p\rightarrow
\infty $ \cite{Lemos:2015gna}, and the spacetime
is assumed to take the Hawking
temperature. In a sense, this means that all possible degrees of freedom are
excited and the black hole takes the Bekenstein-Hawking entropy which is
the maximum possible entropy. Taking the extremal limit
from a non-extremal black hole, one finds
that in this particular limit the extremal black hole
entropy is the Bekenstein-Hawking entropy, suggesting that
the maximum entropy that an extremal black hole
can take is the Bekenstein-Hawking entropy.
Therefore, in this regard,
the range of values for the entropy of an
extremal black hole is
\begin{equation}  \label{Srange}
0\leq S(r_{+})\leq \frac{1}{4}A_{+}\,,
\end{equation}
or $0\leq S(r_{+})\leq \pi r_+^2$.
The case studied by Ghosh and Mitra
\cite{Hod:2000pa,Ghosh:1996dq} has $S\propto r_{+}$ and hence it is
within our limits. Table 1 below summarizes the comparison between
an extremal shell  at its own
gravitational radius with $T_\infty=0$, which we
have called a special shell,
and an extremal black hole with $T_\infty=0$.

\vskip 0.2cm
{\hskip -0.5cm
\begin{tabular}
[c]{|l|l|l|l|l|l|}\hline
Case & $T$ at $\infty$ & Local $T$ on $r_+$ &
Backreaction on $r_+$ & $\Phi$  &
Entropy\\\hline
Special shell & 0 & $0$ & Finite &
$1$ on
$R$ &
Well-defined
$S(r_+)$\\\hline
Black hole & 0 & $0$  & Finite & $1$ on $r_+$ &
In debate,
$S=0$, $\frac{A_+}{4}$, $S(r_+)$\\\hline
\end{tabular}
}
\vskip -0.1cm
\noindent
Table 1. Comparison between
a special extremal shell at its own
gravitational radius
and an extremal black hole both with $T_\infty=0$.

%% file: I.4ShellExtremal/4.section.tex
\section{A generic shell at the gravitational
radius limit}
\label{bh2}

Another interesting shell configuration can also be considered, where
the back reaction remains finite even with the shell at the gravitational radius. Suppose that the shell
has a small nonzero local temperature (i.e., finite large
$\beta$), rather than zero, keeping in mind
Eq.~(\ref{tolmanextremalpostulated})
and Eq.~(\ref{Phi0}), as well as the
constraints (\ref{constraintPhi}) and
(\ref{integextrbphi}).
From Eq.~(\ref{tolmanextremalpostulated}) we see that
the product $bk$ is the most relevant quantity.
Let us now prepare the shell at any $R$ such that
$b={\bar b}/k$, i.e.,
$b={\bar b}/(1-r_+/R)$, for some ${\bar b}$ finite.
It follows that $\beta={\bar b}$ is finite, and holds true
for any $R>r_+$, as well as for the temperature
measured at infinity $T_\infty=1/b$.
Regarding the potential of the shell, we fix it such that
$(1-\phi)=(1-\bar\phi)k$ and
$(1-\Phi)=(1-\bar\phi)$
for any $R>r_+$. Note that
the potential measured at infinity
$\phi$ is less than one.

We are now in a position to take the gravitational radius limit $R=r_+$. In this case, the shell
has been prepared such that $k$ goes to zero but is compensated by a large $b$, such that the local
shell temperature $T=1/\be$ remains bounded. The temperature measured at infinity $T_\infty=1/b$ is zero
and thus
coincides with the Hawking temperature,
$T_{H}=0$.
As a consequence, the
quantum backreaction in this case remains finite and controllable,
even for $R=r_{+}$.
Since $\beta$ is finite and $\beta(1-\Phi)=s(r_+)$, this also means that $\Phi<1$.
Therefore, the entropy of the shell at $R=r_+$ is again an arbitrary function $S=S(r_+)$.
This thought process was also used in \cite{Lemos:2009uk} in
a general discussion of the entropy for the extremal case. There,
any $\beta$ and $\Phi<1$ obeying Eq.~(\ref{integextr}) were concluded to be suitable,
and the entropy also came out as an arbitrary
function of
$r_{+}$.
This case sharply contrasts with the extremal black hole case where
any $T_\infty=b^{-1}\neq0$ leads to $\beta\rightarrow0$ and infinite
local temperature $\beta^{-1}$ on the horizon with divergent
backreaction that destroys the horizon. For the latter extremal black hole scenario, see \cite{Hawking:1994ii}, where nothing is said about quantum backreaction
and it is argued that the entropy is zero. We note, however, the results of
\cite{Anderson:1995fw}, which show that the backreaction grows unbound if
$T_H$ is not zero. Table 2 below summarizes the comparison between
an extremal shell  at its own
gravitational radius with $T_\infty=0$ and
$\beta$ finite nonzero, which we
have called a generic shell,
and an extremal black hole with $T_\infty$
not zero.

\vskip 0.5cm
{\hskip -0.5cm
\begin{tabular}
[c]{|l|l|l|l|l|l|}\hline
Case & $T$ at $\infty$ & Local $T$ on $r_+$ &
Backreaction on $r_+$ & $\Phi$  & Entropy\\\hline
Generic shell & 0 & Finite, not 0 & Finite & $<1$
on $R$ & Well-defined $S(r_+)$\\\hline
Black hole & $T_\infty\neq0$ & $\infty$ & Infinite &
$1$ & Not known or undefined\\\hline
\end{tabular}
}
\vskip -0.1cm
\noindent
Table 2. Comparison between
a generic  extremal shell at its own
gravitational radius
and an extremal black hole with infinite temperature at
the horizon.
\vskip 0.2cm

%% file: I.4ShellExtremal/conclusions.tex
\section{Conclusions}
Upon consideration of spherically symmetric
systems and through the formalism
of thin matter shells and their thermodynamics properties,
we have shown a possible solution for the ongoing debate concerning the
entropy of an extremal black hole.
Although a full quantum
theory of gravity would be necessary to fully understand the result obtained, it is
nonetheless interesting to see that the use of the junction conditions
through the Einstein equation
leads inevitably to the suggestion that extremal black holes are a
different class of objects than non-extremal black holes, due to the
fact that their entropy depends on the particularities of the matter
distribution which originated the black hole.

%% file: I.5ExtremalLimits/main.tex
\chapter{The different limits of the thermodynamics of extremal shells}
\label{cap:chapterI5}
\input{I.5ExtremalLimits/outline.tex}
\input{I.5ExtremalLimits/1.section.tex}

\input{I.5ExtremalLimits/2.section.tex}

\input{I.5ExtremalLimits/3.section.tex}

\input{I.5ExtremalLimits/conclusions.tex}

%% file: I.5ExtremalLimits/outline.tex
\section{Introduction}

As was seen in Chapter 5, a number of preparations must be made before taking an extremal shell to its gravitational radius and that, depending on the shell considered, the backreaction on the shell could be bounded or not. In this chapter we will re-evaluate the problem of extremaly charged shells from a different point of view, using new variables that introduce various different ways in which the extremality of the shell may be studied.

%% file: I.5ExtremalLimits/1.section.tex
\section{The three extremal horizon limits}
\label{geo}

We will take as starting point the results of Chapter 4, namely the main themodynamic quantities of a non-extremal charged shell. To study independently the limit of an extremal shell and the limit of a
shell being taken to its gravitational radius, it will prove fruitful to
define the variables $\varepsilon$ and $\delta$ through the equations
\begin{equation}
1-\frac{r_{+}}{R}=\varepsilon ^{2}\,,  \label{e}
\end{equation}
\begin{equation}
1-\frac{r_{-}}{R}=\delta ^{2}\,.  \label{d}
\end{equation}
Since the extremal horizon limit involves taking $R = r_+ = r_-$, it is clearly seen from Eqs.~(\ref{e}) and (\ref{d}) that the variables $
\varepsilon$ and $\delta$ are the most natural ones to take the extremal limit.
There are however different limits depending on which and how $\varepsilon$
and $\delta$ are taken to zero. There are three physically relevant cases which follow.

\vskip 0.3cm \noindent \textbf{Case 1.} In this case we do $r_+\neq r_-$ and
$R\to r_+$, i.e.,
\begin{equation}
\delta ={O}(1)\,,\quad\varepsilon \to 0\,.  \label{de1}
\end{equation}
After all the calculations are done and finished and we have an expression
for the entropy, we may then take the $\delta\to0$ limit to get at the
gravitational radius an extremal shell. According to Eq.~(\ref{e}), this
means bringing the shell to its gravitational radius. It follows from (\ref
{d}) that $r_{+}\neq r_{-}$, so the horizon limit is taken but not the extremal one,
and thus the shell remains nonextremal throughout the whole process.

\vskip 0.3cm \noindent \textbf{Case 2.} In this case we do $R\rightarrow
r_{+}$ and $r_{+}\rightarrow r_{-}$, i.e., \noindent
\begin{equation}
\delta =\frac{\varepsilon }{\lambda}\,,\quad\varepsilon \to 0\,,  \label{de}
\end{equation}
where it is assumed that the new parameter $\lambda$ remains constant in the
limiting process and that it must satisfy $\lambda \leq 1$ due to $r_{+}\geq
r_{-}$. The limit $\varepsilon \rightarrow 0$ means that
$R\rightarrow r_{+}$ and $r_{+}\rightarrow r_{-}$ simultaneously, in such a
way that $\delta \sim \varepsilon $. In other words, the horizon limit is
accompanied with the extremal one.

\vskip 0.3cm \noindent \textbf{Case 3.} In this case we do $r_+=r_-$ and $
R\to r_+$, i.e.,
\begin{equation}
\delta=\varepsilon \,,\quad\varepsilon \to 0\,.  \label{de2}
\end{equation}
As a consequence, we have $r_{+}=r_{-}$ from the very beginning., which corresponds to the
extremal shell. This case was analyzed in Chapter 5, so we will simply state
the results and use them for comparison.

\section{The three extremal horizon limits for the mass and electric charge}

\label{mq}

We start by using Eqs.~(\ref{e}) and (\ref{d}) in Eq.~(\ref{red2}), we immediately get
that the redshift function is
\begin{equation}
k(R,\varepsilon,\delta)=\varepsilon \delta\,.  \label{ked}
\end{equation}
In these variables it explicitely depends on $\varepsilon$ and $\delta$ but not on $R$. Moreover, from Eqs.~(\ref{M1}) and (\ref{invhorradicauch}), we immediately see that
\begin{equation}
M(R,\varepsilon,\delta)=R(1-\varepsilon \delta )\,,  \label{med}
\end{equation}
\begin{equation}
Q(R,\varepsilon,\delta)=R\sqrt{(1-\varepsilon ^{2})(1-\delta ^{2})}\,.
\label{qed}
\end{equation}
We then have the following limits for the charge and rest mass of the shell.

\vskip 0.3cm \noindent \textbf{Case 1.} For $r_+\neq r_-$ and as $
R\rightarrow r_{+}$, i.e., for $\delta ={O}(1)$ and as $\varepsilon \to 0$,
we get from Eqs.~(\ref{med})-(\ref{ked})
\begin{equation}
M(r_+,\varepsilon,\delta)=r_+\,,\quad Q(r_+,\varepsilon,\delta)=r_+ \,.
\label{mqk1}
\end{equation}

\vskip 0.3cm \noindent \textbf{Case 2.} For $R\rightarrow r_{+}$ and $
r_{+}\rightarrow r_{-}$, i.e., for $\delta =\frac{\varepsilon }{\lambda} $,
with $\lambda$ kept fixed according to Eq.~(\ref{de}), and $\varepsilon \to
0 $ we get from Eqs.~(\ref{ked})-(\ref{med})
\begin{equation}
M(r_+,\varepsilon,\delta)=r_+\,,\quad Q(r_+,\varepsilon,\delta)=r_+ \,.
\label{mqk2}
\end{equation}

\vskip 0.3cm \noindent \textbf{Case 3.} For $r_+=r_-$ and as $R\to r_+$,
i.e., for $\delta=\varepsilon$ and $\varepsilon \to 0$ it is seen from Eq.~(
\ref{pd}) that
\begin{equation}
M(r_+,\varepsilon,\delta)=r_+\,,\quad Q(r_+,\varepsilon,\delta)=r_+ \,.
\label{mqk3}
\end{equation}
Not surprisingly, the three limits here yield the same result, identified as the
mass-charge-radius extremal condition.

\section{The three horizon limits for the pressure, electric potential and temperature}

\label{pphit}

\subsection{Pressure limits}

In order for the non-extremal electric charged shell to remain static, its
surface pressure must have a specific functional form, given by
Eq.~(\ref{pQk1}), which in terms of the variables $\varepsilon $ and $\delta$ can be readily written as
\begin{equation}
p(R,\varepsilon,\delta) =\frac{1}{16\pi R}\frac{(\delta -\varepsilon )^{2}}{
\delta \varepsilon }\,.  \label{pd}
\end{equation}
From this we deduce the following three different limits for the pressure.

\vskip 0.3cm \noindent \textbf{Case 1.} For $r_+\neq r_-$ and as $
R\rightarrow r_{+}$, i.e., for $\delta ={O}(1)$ and as $\varepsilon \to 0$,
we get from Eq.~(\ref{pd})
\begin{equation}
p(r_+,\varepsilon,\delta) = \frac{\delta }{16\pi r_+\varepsilon }\sim \frac{1
}{ \varepsilon }\,,  \label{pdiv2}
\end{equation}
so the pressure is divergent in this case as $1/\varepsilon$.

\vskip 0.3cm \noindent \textbf{Case 2.} For $R\rightarrow r_{+}$ and $
r_{+}\rightarrow r_{-}$, i.e., for $\delta =\frac{\varepsilon }{\lambda} $,
with $\lambda$ kept fixed according to Eq.~(\ref{de}), and $\varepsilon \to
0 $ we get from Eq.~(\ref{pd})
\begin{equation}
p(r_+,\varepsilon,\delta) =\frac{1}{ 16\pi r_{+}}\frac{(1-\lambda )^{2}}{
\lambda }\,.  \label{p3}
\end{equation}
The above result asserts that the pressure will remain finite but nonzero in this horizon
limit for the extremal shell.

\vskip 0.3cm \noindent \textbf{Case 3.} For $r_+=r_-$ and as $R\to r_+$,
i.e., for $\delta=\varepsilon$ and $\varepsilon \to 0$ it is seen from Eq.~(\ref{pd}) that
\begin{equation}
p(r_+,\varepsilon,\delta)=0\,.  \label{pe}
\end{equation}
The result $p=0$ holds in fact at any radius, including the horizon limit.

\subsection{Electric potential limits}

The electric potential $\Phi $ of the shell must also assume a specific form
if the shell is to remain static, which we derived in Eq.~(\ref{Phi02}). In this case, we shall make use of the result that mimicks the black hole, given by Eq.~(\ref{frn}). In terms of $\varepsilon $ and $\delta$ we have then
\begin{equation}
\Phi (R,\varepsilon ,\delta )=\sqrt{\frac{1-\delta ^{2}}{1-\varepsilon ^{2}}}
\,\,\frac{\varepsilon }{\delta }\,.  \label{cd}
\end{equation}
It is now straightforward to analyze the three limiting cases under
discussion.

\vskip 0.3cm \noindent \textbf{Case 1.} For $r_+\neq r_-$ and as $
R\rightarrow r_{+}$, i.e., for $\delta ={O}(1)$ and as $\varepsilon \to 0$,
we get from Eq.~(\ref{cd}),
\begin{equation}
\Phi(r_+,\varepsilon,\delta)=0\,.
\end{equation}

\vskip 0.3cm \noindent \textbf{Case 2.} For $R\rightarrow r_{+}$ and $
r_{+}\rightarrow r_{-}$, i.e., for $\delta =\frac{\varepsilon }{\lambda}$,
with $\lambda$ kept fixed according to Eq.~(\ref{de}), and $\varepsilon \to
0 $ we get
\begin{equation}
\Phi(r_+,\varepsilon,\delta)=\lambda\,,
\end{equation}
with $0\leq\lambda \leq 1$.

\vskip0.3cm \noindent \textbf{Case 3.} For $r_{+}=r_{-}$ and as $
R\rightarrow r_{+}$, i.e., for $\delta =\varepsilon $ and $\varepsilon
\rightarrow 0$ it would seem from Eq.~(\ref{cd}) that $\Phi
(r_{+},\varepsilon ,\delta )=1$. However, this case is special since from
the very beginning we should proceed in a different way, so the forms (\ref{Phi02}) and (\ref{cd})
resulting from the
integrability condition are no longer
valid here.
As it is shown in Chapter 5, the calculations for this case lead to
the inequality (\ref{constraintPhi}), i.e.
\begin{equation}
\Phi (r_{+},\varepsilon ,\delta )\leq 1\,.  \label{f2}
\end{equation}
Thus, if we take an extremal shell from the very beginning, the electric
potential in general differs from what is obtained by the extremal limit
from the nonextremal state.

\subsection{Temperature limits}

Assuming that the shell has a well defined temperature, the integrability
conditions imposed from the first law of thermodynamics result in Eq.~(\ref{BS5}). Taking the shell
to be at the black black hole temperature (\ref{thawk}), we will have then have the local temperature at the shell in terms of
$\varepsilon$ and $\delta$ as
\begin{equation}
T(R,\varepsilon,\delta)= \frac{T_{H}}{k}=\frac{\delta ^{2}-\varepsilon ^{2}}{
4\pi R\delta \varepsilon (1-\varepsilon ^{2})^{2}}\,.  \label{tloc}
\end{equation}
The limits will thus be the following.

\vskip 0.3cm \noindent \textbf{Case 1.} For $r_+\neq r_-$ and as $
R\rightarrow r_{+}$, i.e., for $\delta ={O}(1)$ and as $\varepsilon \to 0$,
we get from Eq.~(\ref{tloc}),
\begin{equation}  \label{tdiv}
T(r_+,\varepsilon,\delta)= \frac{\delta }{4\pi r_{+}\varepsilon }\sim
\frac1\varepsilon \,,
\end{equation}
which is divergent.

\vskip 0.3cm \noindent \textbf{Case 2.} For $R\rightarrow r_{+}$ and $
r_{+}\rightarrow r_{-}$, i.e., for $\delta =\frac{\varepsilon }{\lambda}$ and $\varepsilon \to
0 $, we get from Eq.~(\ref{tloc})
\begin{equation}
T(r_+,\varepsilon,\delta) =\frac{1-\lambda ^{2}}{4\pi r_{+}\lambda }\,.
\label{t3}
\end{equation}
We see that it remains finite and nonzero. It is also worth noting a simple formula that
follows from (\ref{t3}) and relates the pressure and temperature in this
horizon limit, namely
\eq{
\frac{p}{T}=\frac{1}{4}\frac{1-\lambda }{1+\lambda}\,.
}

\vskip 0.3cm \noindent \textbf{Case 3.} For $r_+=r_-$ and as $R\to r_+$,
i.e., for $\delta=\varepsilon$ and $\varepsilon \to 0$, one may choose the temperature at infinity to be zero while the local temperature remains finite, as was done in Chapter 5.

%% file: I.5ExtremalLimits/2.section.tex
\section{The three extremal horizon limits of the entropy}
\label{s}

To obtain the distinct limits for the entropy, one needs the first law of thermodynamics
expressed in terms of the variables $(R,\varepsilon,\delta)$. This can be done straightforwardly
by using Eq.~(\ref{dSQ}) together with Eqs.~(\ref{e}), (\ref{d}), arriving at
\eq{
TdS=a_{1}dR+a_{2}d\varepsilon +a_{3}d\delta \label{1LTd}\,,
}
where
\ea{
a_{1}&=1-\delta \varepsilon +\frac{(\delta -\varepsilon )^{2}}{2\delta
\varepsilon }+\frac{(1-\delta ^{2})(1-\varepsilon ^{2})}{\delta \epsilon }
(1-Rc)\,, \\
a_{2}&=-\delta R\left[ 1+\frac{1-\delta ^{2}}{\delta ^{2}}(1-Rc)
\right]\,, \\
a_{3}&=-\varepsilon R\left[ 1+\frac{1-\varepsilon ^{2}}{
\varepsilon ^{2}} (1-Rc)\right]\,.
}
Imposing in addition that the electric
potential assumes the form of Eq.~(\ref{cBH}), enables us to simplify
the coefficients $a_1$, $a_2$, and $a_3$, into
\ea{
a_{1}&=\frac{\delta
^{2}-\varepsilon ^{2}}{2\delta \varepsilon }\,, \\
a_{2}&=-\delta R\left[ 1-
\frac{\varepsilon ^{2}}{\delta ^{2}}\left( \frac{ 1-\delta ^{2}}{
1-\varepsilon ^{2}}\right) \right]\,, \\
a_{3}&=0\,.
}
We may now use Eq.~(\ref{tloc}) and write the differential for the entropy as
\begin{equation}
dS=2\pi \,R \left(1-\varepsilon^2\right)^2\,dR
-4\pi\,R^2\varepsilon\left(1-\varepsilon^2\right)d\varepsilon\,,  \label{ds}
\end{equation}
can be integrated to give
\begin{equation}
S(r_+,\epsilon,\delta)=\pi \,R^2 \left(1-\varepsilon^2\right)^2\,,
\label{ds2}
\end{equation}
where we have put the integration constant to zero. Using Eq.~(\ref{e}), we finally obtain
\begin{equation}
S(r_+)=\frac{A_+}{4}\,  \label{sbh00}
\end{equation}
where $A_+$ is the gravitational radius area, or the horizon area when the
shell is pushed into the gravitational radius.

\vskip 0.3cm \noindent \textbf{Case 1.} For $r_+\neq r_-$ and as $
R\rightarrow r_{+}$, i.e., for $\delta ={O}(1)$ and as $\varepsilon \to 0$,
we get the same result as Eq.~(\ref{sbh00}). This is general for
any nonextremal black hole. We can now take the extremal limit $\delta\to0$
and obtain that the entropy of an extremal charged black hole is by
continuity $S(r_+)=\frac{A_+}{4}$, the Bekenstein-Hawking entropy.

\vskip 0.3cm \noindent \textbf{Case 2.} For $R\rightarrow r_{+}$ and $
r_{+}\rightarrow r_{-}$, i.e., for $\delta =\frac{\varepsilon }{\lambda}$ and $\varepsilon \to
0 $, we obtain from Eq.~(\ref{sbh00}), $S(r_+)=\frac{A_+}{4}$. This means that in the case
where the shell goes to the gravitational radius simultaneously with the
extremal limit, one also gets the Bekenstein-Hawking entropy.

\vskip 0.3cm \noindent \textbf{Case 3.} For $r_+=r_-$ and as $R\to r_+$,
i.e., for $\delta=\varepsilon$ and $\varepsilon \to 0$, the entropy cannot
be handled in this manner and should be considered separately, as was done in Chapter 5.
There we obtained that the entropy is not fixed unambiguously
for a given $r_{+}$, and so it is any physical well behaved function of $r_+$, or
if one prefers, of $A_+$.

%% file: I.5ExtremalLimits/3.section.tex
\section{The physical origin of the entropy in each horizon limit}

\label{sdisc}

It is instructive to trace in more detail how the entropy arises from the first law,
i.e. which quantities are responsible for the degrees of freedom contained within
the entropy.

\vskip 0.3cm \noindent \textbf{Case 1.} For $r_+\neq r_-$ and as $
R\rightarrow r_{+}$, i.e., for $\delta ={O}(1)$ and as $\varepsilon \to 0$,
let us, for simplicity, take $\varepsilon =\mathrm{constant}\ll 1$. Then, in
the first law Eq.~(\ref{TQ22}), and from Eq.~(\ref{pdiv2}), we can retain solely the
term due to the pressure and, taking into account Eq.~(\ref{tdiv}), we
obtain the result (\ref{sbh00}). Thus, the pressure term gives the whole
contribution to the entropy.

\vskip 0.3cm \noindent \textbf{Case 2.} For $R\rightarrow r_{+}$ and $
r_{+}\rightarrow r_{-}$, i.e., for $\delta =\frac{\varepsilon }{\lambda}$ and $\varepsilon \to
0 $, all three terms in the first law give a contribution to the entropy.
Hence, the mass, pressure and electric potential terms are all responsible
for the shells entropy.

\vskip 0.3cm \noindent \textbf{Case 3.} For $r_+=r_-$ and as $R\to r_+$,
i.e., for $\delta=\varepsilon$ and $\varepsilon \to 0$, and according to
Eq.~(\ref{pe}), the first and third terms in Eq.~(\ref{TQ22}) contribute to
the entropy. As a consequence, cases 1 and 3 are complementary to each other regarding
the origin of the entropy.
\vskip 0.3cm We summarize all these results in the following table.

\vskip 0.5cm
\begin{tabular}
[c]{|c|c|c|c|c| m{2.3cm} |}\hline
Case & Pressure $p$ & Potential $\Phi$ & Local temperature $T$ & Entropy &
Contribution from \\ \hline
1 & divergent like $\varepsilon^{-1}$ & 1 & infinite & $A_+/4$ &
pressure\\\hline
2 & finite nonzero & any$<1$ & finite nonzero & $A_+/4$ & mass, pressure and
potential\\\hline
3 & 0 & any$\leq1$ & finite nonzero & a function of $A_+$&
mass and potential\\\hline
\end{tabular}
\vskip 0.4cm
\noindent
Table 1. The contributions of the pressure
$p$, electric potential $\Phi$, and temperature $T$,
to the
extremal black hole entropy $S$, according to the first law.
\label{tabent}

It is worth stressing that the results presented in the above table refer in
general not to black holes but to shells. Only in the horizon limit do these
results apply to black holes. Usually, if one considers the extremal limit
of a nonextremal black hole, it remains in the same topological class during
the limiting transition, so it is not surprising that in the extremal limit
one obtains the Bekenstein-Hawking value. However, in our case, we obtained
a somewhat more general statement, since the exact value of the shell's entropy coincides
with that of a black hole for a given $r_{+}$ independently of $R$. The only case where
this does not happen is for an extremal shell taken from the onset, where it is
seen to possess an entropy which is an arbitrary function of $r_+$.

\section{Role of the backreaction}

\label{back}

As was detailed in Sec.~\ref{sec:state} of Chapter 3, the finiteness of the backreaction is responsible for the integrity of the shell.
As it was seen through Eq.~(\ref{backReact}), in the horizon limit, the
requirement of finite backreaction implied $T_{0}=T_{H}$, even for a black hole.
However, while for the nonextremal black hole one has $T_{H}\neq 0$, in the extremal case $T_{H}=0$, whence
\begin{equation}
T^{a}{}_{b}=\frac{(T_{0})^{4}}{(g_{00})^{2}} f^{a}{}_{b}\,.
\end{equation}
Thus, the attempt to put $T_{0}\neq 0$ for a black hole according to the prescriptions given
in \cite{Hawking:1994ii,Teitelboim:1994az}, leads to infinite stresses, since $\frac{T_{0}^{4}}{
(g_{00})^{2}}$ diverges as one approaches the horizon. As a consequence, this would imply that the horizon would be destroyed \cite{Anderson:1995fw,Loranz:1995gc}. However, when we deal with a shell instead of a black hole, an intermediate
case can be accomplished, in particular simultaneously $T_{0}\rightarrow 0 $ and $
g_{00}\rightarrow 0$ in such a way that $T$ is kept bounded. This is
realized in Case 2, according to Eq.~(\ref{t3}), and in Case 3.

%% file: I.5ExtremalLimits/conclusions.tex
\section{Conclusions}

We found what happens
when calculating the entropy and other thermodynamic
quantities
when different limiting transitions for a shell are taken,
as well as how they are related to each other when the
radius of the shell approaches the gravitational radius,
i.e., when it turns into a black hole.

We saw that the limits in cases
1 and 2 agree with respect to the entropy but
disagree in the behavior of all other quantities. Cases 2 and 3
disagree in what concerns the entropy but agree in the behavior of the
local temperature and electric potential. We also observed that Case 2 is
intermediate between 1 and 3.

The results obtained showed how careful one should be in the
calculations when a system close to extremality approaches the horizon.
It would also be interesting to trace whether and
how these subtleties can affect calculations in quantum field theory.

%% file: I.2ShellD/main.tex
\chapter{Thermodynamics of a $d$-dimensional thin matter shell}
\label{cap:chapterI2}
\input{I.2ShellD/outline.tex}
\input{I.2ShellD/1.section.tex}
\input{I.2ShellD/2.section.tex}

\input{I.2ShellD/3.section.tex}

\input{I.2ShellD/4.section.tex}

\input{I.2ShellD/5.section.tex}

\input{I.2ShellD/6.section.tex}

\input{I.2ShellD/conclusions.tex}

%% file: I.2ShellD/outline.tex
\section{Introduction}

Marking the ending of Part I of this thesis, this chapter will generalize the calculations for spacetimes with any dimension, in this case with neutral shells, both for simplicity sake and to concentrate on the effect of the extra dimensions in the thermodynamics of the shell. Using the junction conditions of Chapter 2, we shall again obtain conditions for the equations of state of the shell, which will allow the calculation of its entropy. Due to the higher simplicity of the problem compared to the charged case, we will be able to perform a detailed study of the shell's thermodynamically stability.

%% file: I.2ShellD/1.section.tex
\section{The thin shell spacetime}

Consider a spherically symmetric timelike $(d-1)$ hypersurface $\Sigma$ with radius $R$ that partitions a $d$-dimensional spacetime in an inner region $\ma{V}_{-}$ and outer region $\ma{V}_{+}$. The full spacetime geometry will have to be a solution of the Einstein equations (for $d>3$)
\begin{equation}
	G_{ab} = 8 \pi G_d T_{ab}\,,
\end{equation}
where $G_d$ is Newton's gravitational constant in $d$ dimensions. In the inner region we will assume a flat spacetime, i.e.
\begin{equation} \label{LE1}
	ds^2_{-} = g^{-}_{\al\be} dx^{\al} dx^{\be} = - dt_{+}^2 + dr^2 + r^2 d\Omega^2_{d-2}\,, \quad (r<R)
\end{equation}
where $t_+$ is the inner time coordinate, $(r,\theta_1,\ldots,\theta_{d-2})$ are the polar coordinates generalized to $(d-1)$ dimensions and
\begin{equation}
	d\Omega^2_{d-2} = d\theta^2_1 + \sum_{i=2}^{d-2}\left( \prod_{j=1}^{i-1} \sin^2\theta_j \right)d\theta_i^2,
\end{equation}
is the corresponding differential of the solid angle
\begin{equation}\label{O}
	\Omega_{d-2}=\frac{2\pi^{(d-1)/2}}{\Gamma\left(\frac{d-1}{2} \right)}\,.
\end{equation}
As for the outer region, we will consider the spacetime to be described by the Schwarzschild-Tangherlini metric
\begin{equation} \label{LE2}
	ds^2_{+} = g^{+}_{\al\be} dx^{\al} dx^{\be} = - \left(1-\frac{2m\mu}{r^{d-3}}\right)dt_{-}^2 + {dr^2 \over \left(1-\frac{2m\mu}{r^{d-3}}\right)} + r^2 d\Omega^2_{d-2}\,, \quad (r>R)
\end{equation}
where we defined the outer time coordinate $t_+$ and the quantity
\begin{equation}\label{mudef}
	\mu \equiv \frac{8\pi G_d}{(d-2)\Omega_{d-2}}\,,
\end{equation}
and where $m$ is the so called Arnowitt-Deser-Misner (ADM) mass. The metric (\ref{LE2}) possesses a gravitational radius $r_+$ given by
\begin{equation}\label{rplus}
	r_+=\left( 2\mu m \right)^{1/(d-3)}\,.
\end{equation}
with a multiplicity of $d-3$. Regarding the hypersurface itself, let $\tau$ be the proper time of an observer comoving with it, and suppose the evolution of the shell is parametrized by the equations $R(\tau)=r|_\Sigma$ and $T(\tau)=t_\pm|_\Sigma$. Then the shell will be characterized by the geometry
\begin{equation}
ds_{\Sigma}^2 = h_{ab} dx^a dx^b = -d\tau^2 + R^2(\tau) d\Omega^2_{d-2}\,, \quad r=R\,.
\end{equation}
where the coordinates on the shell have been chosen as $y^{a} = (\tau,\theta_1,\ldots,\theta_{d-2})$. Viewed from each side of the shell, we then have
\eq{
h^{-}_{ab} = g^{-}_{\al\be} \, e^{\al}_{i}{}_a \,
e^{\be}_{i}{}_b\,,
\quad
h^{+}_{ab} = g^{+}_{\al\be} \, e^{\al}_{o}{}_a \,
e^{\be}_{o}{}_b\,.
}
We are now in a position to apply the junction conditions of Chapter 2. The first junction condition, given by Eq.~(\ref{1JC}) states that
\eq{
h^{-}_{ab} = h^{+}_{ab} = h_{ab}\,,
}
or, applied to the case in question,
\eq{
-\left(1-\frac{2m\mu}{r^{d-3}}\right) \dot{T}_{+} + {\dot{R}_{+} \over \left(1-\frac{2m\mu}{r^{d-3}}\right)} = -\dot{T}_{+} + \dot{R}_{+} = -1\,,
}
where a dot denotes differentiation with respect to $\tau$. The second junction condition, given by Eq.~(\ref{2JC}), requires us to first evaluate the extrinsic curvature components on both sides of the shell, whereby one obtains
\ea{
K^{\tau}_-{}_{\tau} &= \frac{\ddot{R}}{\sqrt{1+\dot{R}^2}}\,,
\label{a1N}\\
K^{\tau}_+{}_{\tau} &= \frac{-\frac{\mu \dot{m}}{R^{d-3}\,\dot{R}}
-(d-3)\frac{\mu m}{R^{d-2}}+\ddot{R}}{\sqrt{1-\frac{2\mu m}{R^{d-3}}+\dot{R}^2}}\,, \label{a2N}\\
K^{\phi}_-{}_{\phi} = K^{\theta}_-{}_{\theta} &=
\frac{1}{R}\sqrt{1+\dot{R}^2}\,,\label{a3N}\\
K^{\phi}_+{}_{\phi} = K^{\theta}_+{}_{\theta} &=
\frac{1}{R}\sqrt{1-\frac{2\mu m}{R^{d-3}}+\dot{R}^2}\,.
\label{a4N}
}
We can now make use of Eqs.~(\ref{a1N})-(\ref{a4N}) in Eq.~(\ref{Sac}) and
calculate the non-null components of the stress-energy tensor $S_{ab}$
of the shell. We shall be interested only in the static case, for which
$\dot{R}=0$, $\ddot{R}=0$, and $\dot{m}=0$, giving the components
\ea{
S^{\tau}{}_{\tau} & = \frac{(d-2)}{8 \pi G_d
R} \left(\sqrt{1-\frac{2\mu m}{R^{d-3}}}-1\right)\,, \label{S14N} \\
S^{\phi}{}_{\phi} = S^{\theta}{}_{\theta} &  =
\frac{(d-3)}{8 \pi G_d R} \left(\sqrt{1-\frac{2\mu m}{R^{d-3}}}-1 \right)-
\frac{(d-3)\frac{\mu m}{R^{d-2}}}{8
\pi G_d \sqrt{1-\frac{2\mu m}{R^{d-3}}}} \label{S24N}\,.
}
At this point, one needs to specify to some level the type of matter that the shell is made of. We will consider it to be a perfect fluid with surface
energy density $\sigma$ and pressure $p$, implying that the
stress-energy tensor will be of the form
\eq{
S^{a}{}_{b} = (\sigma +p) u^a u_b + p
h^{a}{}_{b}\,,
\label{perffluidD}
}
where $u^a$ is the three-velocity of a shell element. The non-zero entries of this tensor are
\eq{
S^{\tau}{}_{\tau} = -\sigma\,,
\label{lamD}
}
\eq{
S^{\theta}{}_{\theta} = S^{\phi}{}_{\phi} = p
\label{pressD}\,,
}
which, combined with Eqs.~(\ref{S14N}) and (\ref{S24N}), returns the relations
\ea{
\sigma = & \frac{(d-2)}{8 \pi G_d
R} \left(1-\sqrt{1-\frac{2\mu m}{R^{d-3}}}\right)\,, \label{SS1D} \\
p = &
\frac{(d-3)}{8 \pi G_d R} \left(\sqrt{1-\frac{2\mu m}{R^{d-3}}}-1 \right)-
\frac{(d-3)\frac{\mu m}{R^{d-2}}}{8
\pi G_d \sqrt{1-\frac{2\mu m}{R^{d-3}}}}
\label{SS2D}\,.
}
Note that Eqs.~(\ref{SS2D}) and (\ref{SS2D}) are purely a
consequence of the Einstein equation which is encoded in the junction
conditions. Thus, although no information about the matter fields of
the shell has been given, we know that they must have an energy density
and pressure equation of the form
(\ref{SS1D}) and (\ref{SS1D}), otherwise no mechanical equilibrium can be achieved.

One quantity that we have begun to see and shall often see appearing in the calculations is the gravitational redshift, defined as
\begin{equation}\label{kD}
	k = \sqrt{1-\frac{2\mu m}{R^{d-3}}} = \sqrt{ 1- \left( \frac{r_+}{R} \right)^{d-3}}\,.
\end{equation}
Using the variable $k$, Eqs.~(\ref{SS1D}) and (\ref{SS2D}) become considerably simpler, begin given by
\begin{eqnarray}
	&\si = \dfrac{(d-2)}{8 \pi G_d
R}(1-k), \label{M1N} \\
	&p = \dfrac{(d-3)(1-k)^2}{16\pi G_d R k}\,. \label{E22}
\end{eqnarray}
The surface mass density $\si$ of the shell can also be straightforwardly defined as the rest mass divided by the total area of the shell, i.e.
\eq{\label{siD}
\sigma = {M \over \Omega_{d-2} R^{d-2}}\,,
}
so, using Eq.~(\ref{M1N}) and the definition (\ref{mudef}), one arrives at
\eq{
M = \dfrac{R^{d-3}}{\mu}(1-k)\,. \label{M1S}
}
The above relation can be inverted to give the ADM mass of the shell
\eq{
m = M - \frac{\mu M^2}{2 R^{d-3}}\,, \label{E1}
}
which corresponds physically to the energy required to build the shell against the gravitational force.

%% file: I.2ShellD/2.section.tex
\section{Thermodynamics
and stability conditions for the thin shell}
\label{thermoD}

We now turn to the thermodynamic
side and to the
calculation of the entropy of the shell and use units in which the Boltzmann constant is one.
We start
with the assumption that the shell
in static equilibrium
possesses a well-defined
temperature $T$ and an entropy $S$, where the latter is a function of
two variables $M$, $A$, i.e.,
\begin{equation}\label{entropy0D}
S=S(M,A)\,.
\end{equation}
The first law of thermodynamics for the shell then reads
\begin{equation}\label{1TD}
T dS = dM + pdA\,,
\end{equation}
where we see that the volume of the system has been changed to an area, which has one dimension less due to the fact that it is a hypersurface. We also have that $dS$
is the differential of the entropy of the shell,
$dM$ is the
differential of the rest mass,
$dA$ is the differential of the area of the shell,
and $T$ and $p$ are the temperature and the pressure, respectively.
In order to find the entropy $S$,
one thus needs two equations of state, namely,
\ea{
p & =p(M,A,Q)\,, \label{press02D} \\
\beta & =\beta(M,A,Q)\,, \label{temperD}
}
where
\begin{equation}\label{betaD}
\be \equiv \frac1T
\end{equation}
is the inverse temperature. The most important remark now is that Eq.~(\ref{1TD}) can only be an exact differential for the entropy if the integrability condition
\begin{equation}\label{ICD}
	\left(\frac{\partial \beta}{\partial A}\right)_{M} = \left(\frac{\partial (\beta p)}{\partial M}\right)_A
\end{equation}
is satisfied, i.e. if the cross derivatives are equal.
Following~\cite{Callen:1985}, we can analyze the local stability of the system in relation to the entropy fundamental equation $S=S(M,A)$. As in Chapter 3, we will have a set of inequalities that need to be satisfied, which are obtained through the limit $Q \to 0$, giving
\begin{equation}\label{Stab1}
	\left( \dfrac{\partial^2 S}{\partial M^2} \right)_A \leq 0.
\end{equation}
\begin{equation}\label{Stab2}
	\left( \dfrac{\partial^2 S}{\partial A^2} \right)_M \leq 0 ,
\end{equation}
\begin{equation}\label{Stab3}
	\left( \frac{\partial^2 S}{\partial M^2} \right) \left( \frac{\partial^2 S}{\partial A^2} \right) - \left( \frac{\partial^2 S}{\partial M \partial A} \right)^2 \geq 0.
\end{equation}

%% file: I.2ShellD/3.section.tex
\section{The thermodynamic independent variables and the two equations of state}

We will switch from the thermodynamic variables $(M,A)$ to the more useful variables $(M,R)$ which, as well shall see, will overall simplify the calculation to follow. Thus, we want to use
\eq{
R = \left({A \over \Omega_{d-2}}\right)^{1/(d-2)}
}
and the rest mass, from Eq.~(\ref{siD}),
\eq{
M = \si \, \Omega_{d-2} R^{d-2}\,.
}
One should now consider Eqs.~(\ref{E1}), (\ref{rplus}) and (\ref{kD}) as functions of $(M,r)$, i.e.
\ea{
m(M,R) & = M - \frac{\mu M^2}{2 R^{d-3}}\,, \label{mMR} \\
r_+(M,R) & = \left[ 2\mu \, m(M,R) \right]^{1/(d-3)} \label{rplusMR}\,, \\
k(r_+(M,R),R) & = \sqrt{ 1- \left( \frac{r_+(M,R)}{R} \right)^{d-3}} \label{kMR}\,.
}
We may now express the pressure equation of state (\ref{E2}) in the form (\ref{press02D}), where we obtain
\eq{
p = {(d-3) \over 16 \pi G_d R^{d-2}} {\mu^2 M^2 \over R^{d-3} - \mu M}\,.
}
Turning now to the temperature equation of state, another variable change will prove fructuous, namely $(M,R)$ to $(r_+, R)$, which allows us to write the integrability condition (\ref{ICD}) as
\eq{
\left({\dd \be \over \dd R}\right)_{r_+} = \be {(d-3)(1-k^2) \over 2k^2 R}
}
which as for solution
\eq{\label{beD}
\beta(r_+,R) = b(r_+) k(r_+,R)\,,
}
where $b(r_+)$ is an arbitrary function. Since $k\rightarrow 1$ as $R\rightarrow \infty$, $b$ provides the inverse temperature if the shell was placed at infinity, i.e. $b(r_+) \equiv \be(r_+,R\to \infty)$. This is also know as the Tolman formula, which dictates that the temperature in a curved spacetime suffers a gravitational redshift. The fact that the function $b$ is arbitrary is a consequence of our ignorance with respect to the details of the matter content.

%% file: I.2ShellD/4.section.tex
\section{Entropy of the thin shell and the black hole limit}
\label{sec:state}

With the results of the previous section, we have every ingredient necessary to calculate the entropy of the shell. By inserting the
equations of state for the pressure,
Eq.~(\ref{E22}),
for the temperature,
Eq.~(\ref{beD}) as well as
the differential of $M$ given in
Eq.~(\ref{M1S}) and the differential of
the area $A$ or of the radius $R$,
into the first law,
Eq.~(\ref{1TD}),
we
arrive at the entropy differential
\eq{
dS =\beta(r_+,R) \frac{(d-3)}{2\mu k}r_+^{d-4} dr_+\,,
}
which when integrated over the shell's mass (or equivalently over the gravitational radius) gives
\begin{equation}\label{S}
	S(r_+) = \frac{(d-3)}{2\mu} \int^{r_+}_{0} b(r')(r')^{d-4}dr',
\end{equation}
where the integration constant is fixed under the condition that $S(0)=0$, i.e. such that the entropy is zero when the rest mass of the shell is taken to zero. This provides the equation of state for the shell's entropy for any acceptable equation of state for $b(r_+)$.

To proceed to a more detailed study of the thermodynamic properties of
the shell, let us assume a power-law function as an equation of state for $b(r_+)$, of the form
\begin{equation}\label{C1}
	b(r_+) = \frac{4\pi\eta(a+1)}{\hbar(d-3)} \frac{r_+^{a(d-2)+1}}{l_p^{a(d-2)}},
\end{equation}
where $\eta$ and $a$ are free parameters,
$ \frac{a+1}{d-3}$ appears for future convenience,
and
${l_p=(G_d\hbar)^{1/(d-2)}}$ is the Planck length for a
$d$-dimensional spacetime. This choice is the same as
in \cite{Martinez:1996ni}, and leads to a similar expression for the entropy
after substitution in (\ref{S}), namely
\begin{equation}\label{S1}
	S = \frac{\eta \, \Omega_{d-2}}{4} \left( \frac{r_+}{l_p} \right)^{(a+1)(d-2)}.
\end{equation}
Note that
\begin{equation}\label{a>-1}
	a>-1\,,
\end{equation}
otherwise the entropy would diverge in the limit ${r_+ \rightarrow 0}$.

A specially important particular case of the power law equation of state (\ref{C1}) is the black hole limit which occurs when the shell reaches its gravitational radius, i.e. when $R\to r_+$. In this case, as was first noted in Sec.~\ref{sec:state} of Chapter 3, one must note that as the shell approaches its own
gravitational radius, quantum fields are inevitably present and their
backreaction will diverge unless we choose the
black hole Hawking temperature
$T_{\rm bh}$
for the temperature of the shell.
For a black hole associated to the metric (\ref{LE2}), we have
\eq{
T_H= \frac{\hbar}{4\pi} \frac{(d-3)}{r_+}\,,
}
which is equivalent to setting the free parameters of Eq.~(\ref{C1}) to ${a=0}$ and ${\eta_0= 4\pi (d-3)}$.
The entropy Eq.~(\ref{S}) then becomes
\begin{equation}\label{SBH}
	S=\frac{\Omega_{d-2}r_+^{d-2}}{4G_d \hbar} \, = \, \frac{A_\text{shell}}{4 A_p},
\end{equation}
where $A_p = l_p^{d-2}$ is the Planck area in $d$ dimensions. We thus recover the area law for the entropy of a black hole of the same size. Unlike the actual black hole case, however, this is a thermodynamically stable configuration with a positive heat capacity, since Eq.~(\ref{Stab1}) is satisfied, i.e.
\begin{equation}
	\left( \frac{\partial T}{\partial M} \right)_A \geq 0
\end{equation}
also holds. There is however a price to pay for this regarding energy conditions, since the shell's pressure diverges as $1/k$ in this limit. Nonetheless, the local tempeature goes to zero as $k$, which is exactly the necessary way in order for the shell to possess a finite entropy when it is taken to its gravitational radius. In this case, we find
\eq{
S = {A_+ \over 4 A_p}\,,
}
and thus recover the Bekenstein-Hawking entropy of the corresponding black hole.

%% file: I.2ShellD/5.section.tex
\section{Local thermodynamic stability}

Rather then considering a particular case of the power law equation of state (\ref{C1}), one may study the region of parameters where
the shell is thermodynamically stable, which amounts to solving Eqs.~(\ref{Stab1})-(\ref{Stab3}). We will do so in this section.

Starting from Eq.~(\ref{Stab1}), one finds that
\eq{
\left(\frac{\partial^2S}{\partial M^2} \right)_A =
\frac{2(1+a)(d-2)S(M,A)}{(d-3)^2M^2(1-k)^2}
\left[ k^2(d-1+2a(d-2)) - (d-3) \right]\,,
}
thus, Eq.~(\ref{Stab1}) yields
\begin{equation}\label{quad1}
	k^2(d-1+2a(d-2))-(d-3)\leq 0.
\end{equation}
If ${a \leq -(d-1)/(2(d-2))}$, Eq.~(\ref{quad1}) is always satisfied.
Since we have imposed $a>-1$, Eq.~(\ref{a>-1}), we have
for $-1<a \leq -(d-1)/(2(d-2))$, Eq.~(\ref{quad1}) is satisfied.
On the other hand, for ${a > -(d-1)/(2(d-2))}$,
Eq.~(\ref{quad1}) is satisfied when
$-k_0<k<k_0$, with
\eq{
k_0= \sqrt{\frac{d-3}{d-1 + 2a(d-2) }}\,.
}
Since we have $0\leq k$, this condition can be rewritten as
$0<k<k_0$.
Now, note that
the expression inside the square root in $k_0$,
i.e., $\frac{d-3}{d-1 + 2a(d-2) }$,
is always greater than one if ${a<-1/(d-2)}$.
Then, since $k\leq1$ from equation above,
we have that, if ${a\leq-1/(d-2)}$, then
Eq.~(\ref{quad1}) is always satisfied.
Since we had also found
that for $-1<a \leq -(d-1)/(2(d-2))$ we would have Eq.~(\ref{quad1})
satisfied, then we can extend this range to
$-1<a \leq-1/(d-2)$.
Noticing now that for
$a >-1/(d-2)$ the expression inside the square root in $k_0$
is always smaller than one, then this imposes a requirement
on $k$ of the form $k\leq k_0$, with $k_0\leq1$.
So in brief, Eq.~(\ref{quad1}) is always satisfied for
\eq{\label{sta00}
-1<a \leq-{1 \over (d-2)}\,,
}
or when
\begin{equation}\label{stabk0}
	0<k<k_0,\,\quad {\rm for}\; -1/(d-2)<a\,,
\end{equation}
with
\begin{equation}\label{k0}
	 k_0= \sqrt{\frac{d-3}{d-1 + 2a(d-2) }}\,
\end{equation}
and $k_0\leq1$ in this case. Moving on to the condition of Eq.~(\ref{Stab2}), one gets
\eq{
\left( \frac{\partial^2S}{\partial A^2} \right)_M =
\frac{(1+a)S(M,A)}{A^2(d-2)(1-k)}
\left[ (1-k)(2+a)(d-2)-2(2d-5)  \right]\,.
}
Thus, Eq.~(\ref{Stab2}) yields
\begin{equation}\label{quad2}
	-k(d-2)(a+2)-2(2d-5)+(d-2)(a+2)\leq0\,.
\end{equation}
So we have that
$k \geq k_1$
where
\eq{\label{k11}
k_1 = \frac{a-2\frac{d-3}{d-2}}{a+2}\,.
}
Reminding that $a>-1$ from Eq.~(\ref{a>-1}), we note that $k_1\leq0$ for $-1<a\leq 2\frac{d-3}{d-2}$ and, since $0\leq k < 1$, Eq.~(\ref{quad2}) is always satisfied. For $a>2\frac{d-3}{d-2}$ we have $0<k_1<1$, so Eq.~(\ref{quad2}) is satisfied if $k_1 \leq k < 1$.
So in brief, Eq.~(\ref{quad1}) is always satisfied for
\eq{\label{sta01}
-1<a\leq 2\frac{d-3}{d-2}\,,
}
or when
\begin{equation}\label{stabk01}
	k_1 \leq k < 1\,\quad {\rm for}\; a>2\frac{d-3}{d-2}\,,
\end{equation}
with $k_1$ given by Eq.~(\ref{k11}). Finally, for the condition of
Eq.~(\ref{Stab3}), we obtain
\eq{
\frac{\partial^2 S}{\partial M \partial A}  = \frac{2(1+a)S(M,A)}{M(d-3)(1-k)} \left[ -(1-k)(1+a)(d-2) + 2d-5+a(d-2) \right]\,.
}
As a consequence, one can find that Eq.~(\ref{Stab3}) implies
\begin{align}
	&k^2(2+a(d-1))+\frac{2k(d-3)(1+a(d-2))}{d-2} +a(d-3) \leq 0 \label{quadratic}\,.
\end{align}
For ${-1<a\leq-2/(d-1)}$, the inequality is always satisfied by any ${0\leq k <1}$.
For ${a>-2/(d-1)}$, the inequality is satisfied by ${k_{2-}\leq k \leq k_2}$, where $k_2$
and $k_{2-}$
are the roots in Eq.~(\ref{quadratic}). Since ${k_{2-}<0}$, it can be discarded, and the inequality is satisfied by any ${0\leq k \leq k_2}$, where
\begin{align}
	k_2 = &-\frac{(d-3)(1+a(d-2))}{(2+a(d-1))(d-2)} + \frac{\sqrt{(d-3)(d-3-2a(1+a(d-2))(d-2)) }}{(2+a(d-1))(d-2)}.
\end{align}
However, for ${-2/(d-1)<a \leq -1/(d-2)}$ we have ${k_2 \geq 1}$, so the inequality is satisfied by any ${0\leq k < 1}$. For ${-1/(d-2)<a \leq 0}$ we know that ${0\leq k_2 < 1}$, so the inequality is satisfied by ${0\leq k \leq k_2}$. For $a>0$, note that ${k_2<0}$, so the inequality cannot be satisfied. Summarizing, the solutions satisfying Eq.~(\ref{sta00})
are thermodynamically stable. However, solutions with
\begin{equation}\label{sta222}
	 -\frac{1}{d-2} < a \leq 0
\end{equation}
are only thermodynamically stable if
\begin{equation}\label{k222}
	0 \leq k < k_2\,.
\end{equation}
In other words, for a given value of its rest mass $M$, the thin shell's
radius is bounded from above as
\eq{
\frac{R^{d-3}}{\mu M} \leq \frac{1}{1-k_2}\,.
}
Notably, for $a=0$, the only thermodynamically stable solution is at $k=0$ with ${R^{d-3}=\mu M}$, which amounts to placing the shell at its gravitational radius. Solutions with
\begin{equation}\label{sta2222}
a>0\,,
\end{equation}
are always unstable, i.e., for $a>0$, there are no thermodynamically stable configurations.

\begin{figure}[h!]
\centering
\begin{minipage}{.4\textwidth}
  \centering
  \includegraphics[width=1.1\linewidth]{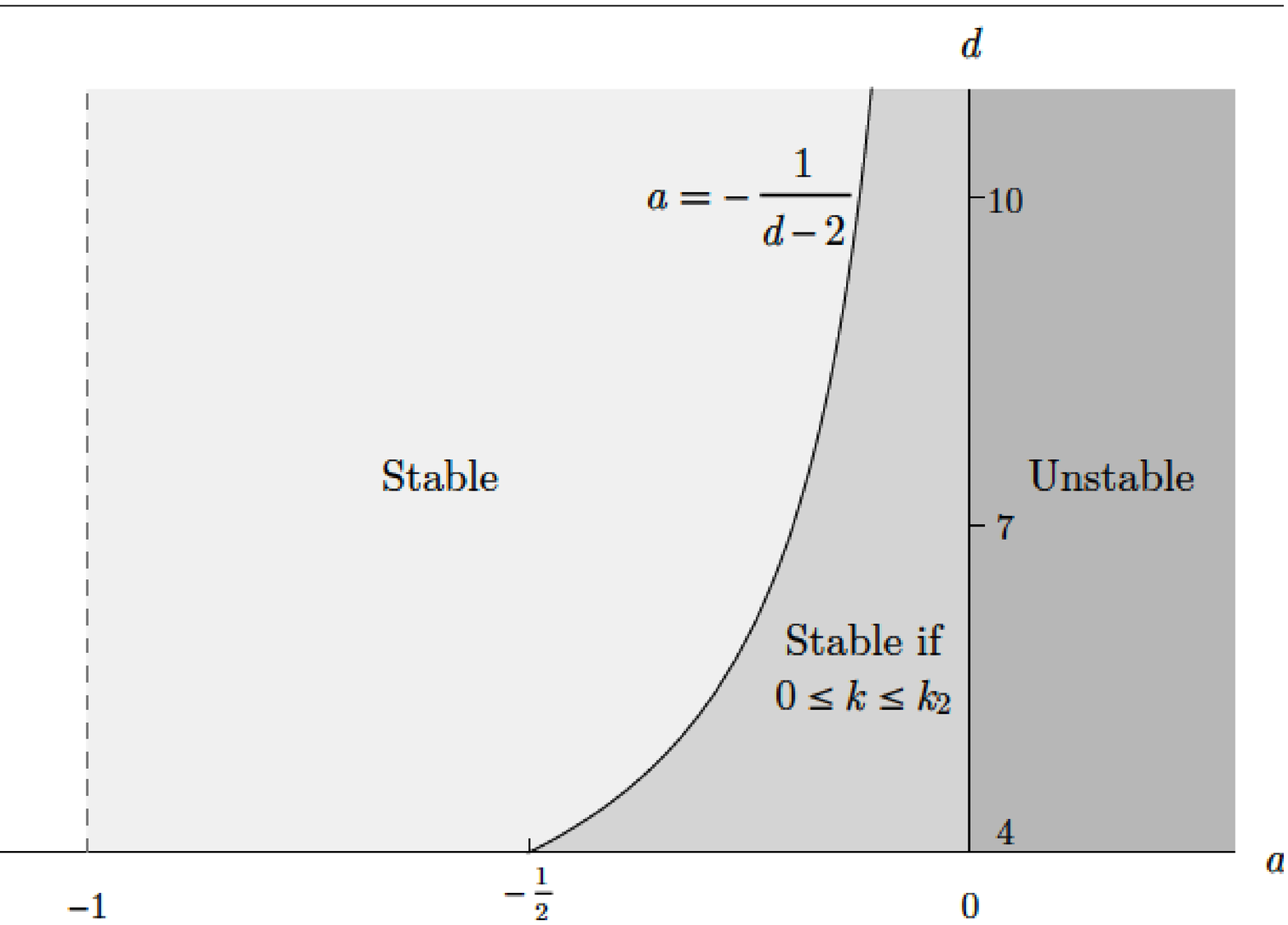}
  \captionof{figure}{Thermodynamical stability of the shell in the parameter space of $a$ and $d$.}
  \label{Pspace3}
\end{minipage}%
\hspace{10mm}
\begin{minipage}{.5\textwidth}
  \centering
  \vspace{5mm}
  \includegraphics[width=1.0\linewidth]{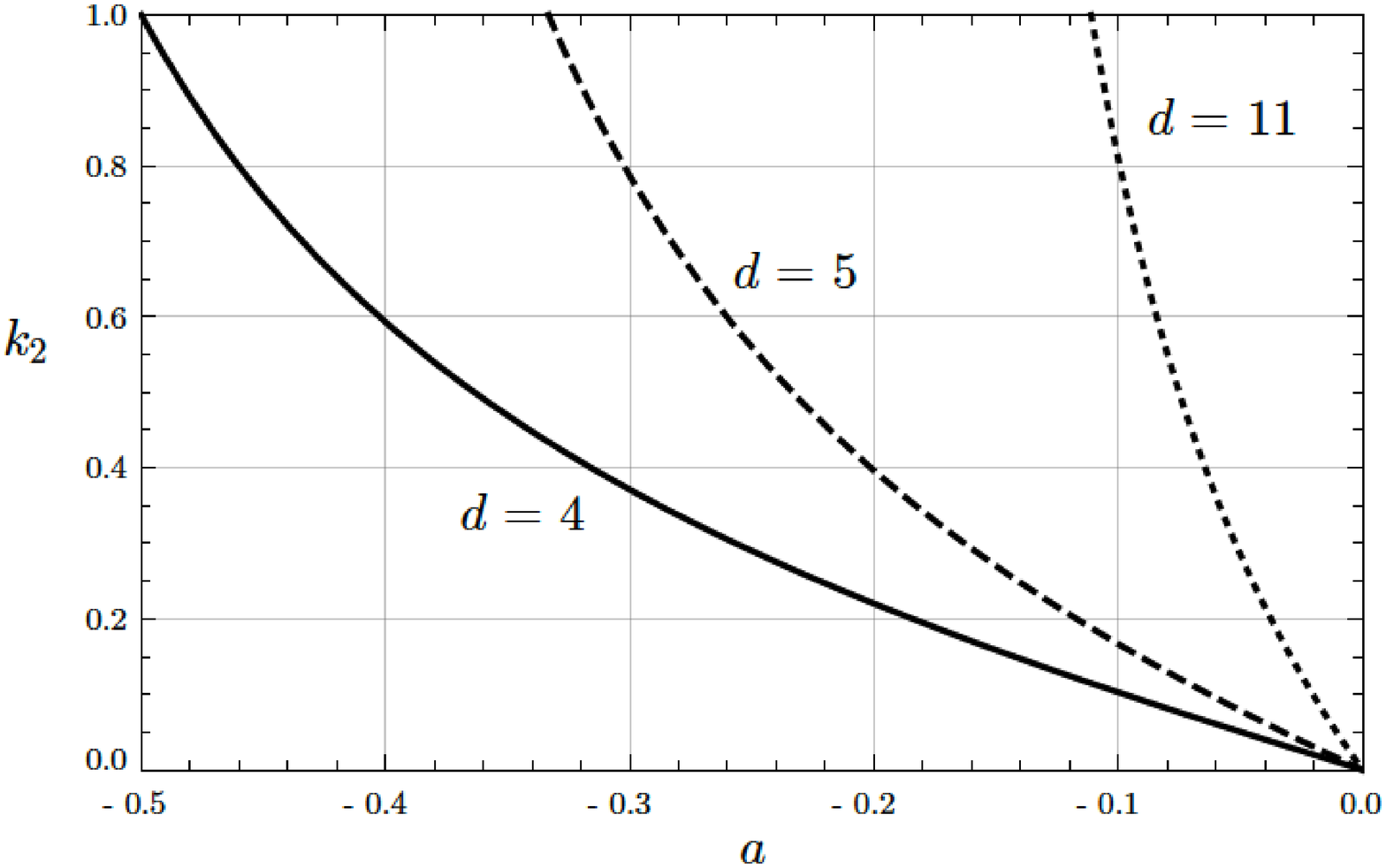}
  \captionof{figure}{Thermodynamical stability of the shell in terms of $k$. Any $k$ to the left of each plot is a stable configuration.}
  \label{k2Plot}
\end{minipage}
\end{figure}

The intersection of all the results of this section is summarized in Figs.~(\ref{Pspace3}) and (\ref{k2Plot}).

%% file: I.2ShellD/6.section.tex
\section{The holography entropy bound and the large $d$ limit}

With a number of restrictions for a thermodynamically stable static shell for the equation of state (\ref{C1}), it is relevant to inquire whether the holographic entropy bound is automatically satisfied, or if both the junction and stability conditions still allow for configurations whose entropy exceeds the bound. Properly stated in the notation we are using (one time dimension and $d-1$ spatial dimensions), at any instant, the entropy inside a $(d-1)$-sphere, should be less than or equal to one quarter of the $(d-2)$-surface area enclosing the volume, in Planckian units, i.e.
\eq{\label{Sbound}
S \leq S_{max} = {A \over l^2_p}\,.
}
Thus, the bound is set by the shell's size through
\begin{equation}\label{Bbound1}
	\frac{R}{r_+} \geq \eta^{1/(d-2)} \left( \frac{r_+}{l_p} \right)^{a}.
\end{equation}
Clearly, Eq.~(\ref{Bbound1}) will be dependent on the shell's energy with respect to the Planck scale, setting a lower limit on how small the shell can be. So let us describe the shell's gravitational radius in terms of $l_p$, through ${r_+=n\,l_p}$, with $n>1$. Then, Eq.~(\ref{Bbound1}) can be rewritten as
\begin{equation}\label{Bbound2}
	\frac{R}{r_+}\geq n^{a} \eta^{1/(d-2)},
\end{equation}
where it is relevant to stress that we have established that $-1 < a \leq 0$ for all cases. The most straightforward solution to satisfying Eq.~(\ref{Bbound2}) is demanding $\eta \leq 1$.
For the particular case $a=0$, the inequality becomes a simple ratio of the shell's size to its energy, and we can see that the bound adds no additional information to ${R \geq r_+}$.
Since $\eta$ is a constant to be set by the matter properties of the shell, should there be configurations with $\eta > 1$, any shell placed at
\begin{equation}
	r_+<R< \eta^{1/(d-2)}\left( \frac{r_+}{l_p}\right)^a l_p
\end{equation}
would violate the bound. This result is, of course, contingent to our choice for an equation of state for the temperature, in Eq.~(\ref{C1}), but it is nonetheless interesting to see how holographic bound would translate into the properties of the shell, and shows that it introduces possibly useful restraints to the free parameters in the equation of state.

When generalizing a physical system to higher dimensions, one might expect its entropy to increase in response.
This might have some implications on whether or not the system stays within some entropic bound.
Here, we will take the limit $d\rightarrow \infty$, and see how the thin shell acts in response.
To do so, it proves useful to write the solid angle of Eq.~(\ref{O}) in the following way, using the Stirling approximation,
\begin{equation}\label{stirling}
	\Omega_{d-2} = \sqrt{\frac{2}{e}} \left( \frac{2\pi e}{d-3} \right)^{\frac{d-2}{2}}\left(1+\mathcal{O}\left( 1/d \right) \right).
\end{equation}
Although the approximation works better as $d \rightarrow \infty$, it is also a great fit for any $d>3$.
With the power law equation of state~(\ref{C1}), the shell's entropy will go as
\begin{equation}\label{Sd}
	S \sim \frac{\eta}{d^{d/2}} \left( \frac{r_+}{l_p} \right)^{d(a+1)}.
\end{equation}
Clearly we have that $S\rightarrow 0$, and this is because the solid angle of Eq.~(\ref{stirling}) converges very quickly to zero, with $d^{-d/2}$.
Instead of setting $\eta$ as a constant, we could include the $d^{-d/2}$ factor into ${\eta_d \equiv \eta \,d^{-d/2}}$ and set it as our problem's constant. However, instead of analyzing the shell's entropy by itself, it makes more sense to look at how the large $d$ limit affects the distance to the holographic bound. Computing the ratio between Eqs.~(\ref{Sd}) and (\ref{Sbound}), the solid angle terms cancel out, and we find
\begin{equation}\label{ratio1}
	\frac{S}{S_{max}} \sim  \eta \left[ \frac{\left(r_+/l_p\right)^{a+1}}{R/l_p} \right]^{d},
\end{equation}
which, so long as $R\geq r_+$ and $\eta \leq 1$, means that the bound is satisfied more easily in the large $d$ limit for a given set of points $(r_+,R)$.

Since the ADM mass $m$ only has units of length for $d=4$, the $d\rightarrow \infty$ might lead to a different response if we fix the shell's energy rather than its gravitational radius, i.e., instead of computing the ratio of Eq.~(\ref{ratio1}) in terms of $(r_+,R)$, let us compute them for the energy. From Eq.~(\ref{rplus}), we get for the large $d$ limit
\begin{equation}
	\frac{S}{S_{max}} \sim \eta \left[  \frac{\left(  m^{1/d}\sqrt{d} \,/l_p \right)^{a+1}}{R/l_p} \right]^{d},
\end{equation}
which will clearly converge even faster due to Eq.~(\ref{rplus}).
This is because the holographic bound focuses mainly on the shell's radius, rather than on an entropy-to-energy relation, but one might get a different behavior on the large $d$ limit if we were to compute the ratio using different entropic bounds.

%% file: I.2ShellD/conclusions.tex
\section{Conclusions}

The work done in this chapter extends the direct integration of a static spherically symmetric thin matter shell's entropy from the first law of thermodynamics to any dimension $d\geq4$, where the $d=3$ case can also be found in~\cite{Lemos:2013dsa} for flat spacetimes, or in~\cite{Lemos:2014eva} for BTZ spacetimes where the black hole limit can also be taken.

Closely following the formulation in~\cite{Martinez:1996ni}, we retrieve the first law in its entropy representation.
The pressure equation of state $p=p(M,A)$ is fixed by the spacetime junction conditions, and the temperature equation of state $\beta=\beta(M,A)$ must have the form $\beta = k(r_+,R)\,b(r_+)$ in order to satisfy the integrability condition for the entropy, after writing the entropy differential with respect to the shell's radius $R$ and gravitational radius $r_+$.
Integrating the first law, we find that the entropy is given as a function of the gravitational radius alone, $S=S(r_+)$ in Eq.~(\ref{S}).

With the temperature equation of state now controlled completely by $b(r_+)$, we explicitly derive the thermodynamic stability conditions for a power law equation of state~(\ref{C1}), and find that the most particular case corresponds to $a=0$, where stability is only satisfied at the gravitational radius.
Additional constraints on the temperature equation of state should be set by the properties of the matter constituting the thin shell, however, instead of making any additional statements on its nature, we note that the holographic bound might also be used to retrieve further constraints for a specific equation of state.

For the black hole limit, placing the shell at its gravitational radius demands that it be at the Hawking temperature, which turns out to be a thermodynamically stable configuration, leading to a positive heat capacity, unlike a black hole. In this limit, the shell's entropy reproduces the Bekenstein-Hawking area law, as a generalization of previous results in~\cite{Martinez:1996ni,Lemos:2014eva}.

For the equation of state considered in Sec.~\ref{sec:state}, we find that the holographic bound is not harder to satisfy in the large $d$ limit, since the ratio in Eq.~(\ref{ratio1}) will converge to zero.

%% file: II.1GFinCS/main.tex
\chapter{Mathematical Preliminaries}
\label{cap:chapterII2}
\input{II.1GFinCS/outline.tex}

\input{II.1GFinCS/1.section.tex}
\input{II.1GFinCS/2.section.tex}

\input{II.1GFinCS/3.section.tex}

\input{II.1GFinCS/4.section.tex}

\input{II.1GFinCS/5.section.tex}

%% file: II.1GFinCS/outline.tex
\section{Outline}

Over the course of this chapter, we will introduce a variety of mathematical results indispensable to understand both physically and mathematically the intricacies of quantum mechanical calculations in curved spacetimes. In particular, we will review the notion of Green functions, how to calculate them using a very general method introduced by Schwinger, and how one can generalize them to finite temperature scenarios.

%% file: II.1GFinCS/1.section.tex
\section{Green functions in curved spaces}

Consider the general differential equation
\eq{\label{OpeFi}
\ma{F}^{ij}\left[u_j(x)\right] = f^i(x)
}
where the indexes represent some degrees of freedom, $\ma{F}^{ij}$ is a differential operator and $u_j$ and $f^i$ are functions of a point $x$, belonging to some $d$-dimensional manifold $M$. The indexes are raised and lowered according to the mathematical context. If we use the coordinates $x^{\mu}$ on the manifold, an infinitesimal volume element $dv_x$ at a point $x$ can be written as
\eq{
dv_x = \sqrt{|g(x)|} \, dx^0 \, dx^1 \cdots dx^{d} \equiv \sqrt{|g(x)|} \, d^{d}x^{\mu}\,,
}
where $g$ is the determinant of the manifold's metric. Using the above relation, the generalized Dirac delta function $\de(x,x')$ on a curved manifold is naturally derived from
\eq{\label{DefDelta}
\int_{M} \de(x,x') \, dv_x = 1\,.
}
This can be compared to the Dirac delta $\de(x-x')$ defined through
\eq{
\int_{M} \de^d (x-x') \, d^d x^{\mu} = 1
}
from where one obtains the relation between the two definitions
\eq{
\de(x,x') = {\de^d (x-x') \over \sqrt{|g(x)|}}\,.
}
The Green function $G(x,x')$ associated to the operator $\ma{F}$ is then defined as the function which satisfies the relation
\eq{\label{DefGreen}
\ma{F}^{ij}\left[G_{jk}(x,x')\right] =  \de^{i}{}_{k}\de^{(d)}(x-x')\,.
}
The usefulness of the Green function is demonstrated by performing the integration
\eq{\label{IntF}
\int_{M} \ma{F}^{ij}\left[G_{jk}(x,x')\right] f^k(x') d^d{x'} = \int_{M} \de^{i}{}_{k} \de^{(d)}(x-x') f^k(x') d^d{x'} = f^i(x) = \ma{F}^{ij}\left[u_j(x)\right]\,.
}
Due to (\ref{OpeFi}), $\ma{F}$ is a differential operator which acts on the point $x$ only, so we may write
\eq{
\int_{M} \ma{F}^{ij}\left[G_{jk}(x,x')\right] f^k(x') d^d{x'} = \ma{F}^{ij}\left[\int_{M} G_{jk}(x,x') f^k(x') d^d{x'}\right]\,,
}
which can be compared to the last equality of (\ref{IntF}) to obtain
\eq{
u^i(x) = \int_{M} G^{i}{}_{j}(x,x') f^j(x') \, d^d{x'}\,,
}
that is, any solution of a differential equation can be expressed as an integral of the associated Green function.

%% file: II.1GFinCS/2.section.tex
\section{Schwinger proper time formalism}

A particularly elegant and powerful formalism to calculate Green functions was developed by Schwinger \cite{Schwinger:1951}, which not only changes the mathematical problem at hand but also re-expresses it in a more notation friendly way for quantum mechanic calculations. The technique is called Proper Time Formalism and was first developed for flat manifolds, and further generalized by De Witt \cite{DeWitt:1965} for curved manifolds. We will summarize here the curved space case.

To begin the derivation of this formalism, we begin by defining the states $\ket{x} \equiv \ket{x^0,\cdots,x^d}$ by the relation
\eq{
\hat{x}^{\mu} \ket{x} = x^{\mu} \ket{x}\,,
}
i.e. such that they are the eigenstates of the operator\footnote{We will always use the hat notation for operators.} $\hat{x}^{\mu}$  which measures the coordinate $x^{\mu}$. Since we are dealing with a continuous eigenvalue spectrum, the normalization of these states is such that the integral of $\braket{x | x'}$ over the entire domain of eigenvalues is one, i.e.
\eq{\label{NormBraket}
\int_M \braket{x | x'} \, d^d x^{\mu} = 1\,,
}
implying that the completeness relation is
\eq{\label{CR}
\int_M \ket{x} \bra{x} \, d^d x^{\mu} = 1\,.
}
Using the property $\de(x,x') g(x) = \de(x,x') g(x')$ of the Dirac delta function, one can express the relation (\ref{DefDelta}) as
\eq{
\int_{M} \de(x,x') \, \sqrt{|g(x)|}  \, d^d x^{\mu} = \int_{M} \de(x,x') \, |g(x)|^{1/4} |g(x')|^{1/4}  \, d^d x^{\mu} = 1
}
which, upon comparing with (\ref{NormBraket}), leads to
\eq{\label{braketdelta}
\braket{x | x'} =  |g(x)|^{1/4} \, \de(x,x') \, |g(x')|^{1/4}\,.
}
The idea is to consider the Green function as the matrix element of an operator, such that
\eq{\label{OpeGreen}
G(x,x') \equiv \bra{x} \hat{G} \ket{x'}\,.
}
That way, we can also define the operator $\hat{\ma{F}}$ by the rule
\eq{\label{OpeFG}
\ma{F}\left[\bra{x} \hat{G} \ket{x'}\right] \equiv \bra{x} \hat{\ma{F}} \hat{G} \ket{x'}
}
as well as the operator $|\hat{g}|$ by
\eq{
|\hat{g}| \ket{x} = |g(x)| \ket{x}\,.
}
In particular, we will only consider physically situations where $\hat{\ma{F}}$ and $|\hat{g}|$ commute, or equivalently
\eq{
\ma{F}\left[|g(x)| \, u(x)\right] = |g(x)| \ma{F}\left[u(x)\right]
}
for any function $u(x)$. This will also be true for any power of $|g(x)|$ of course. Since $\braket{x | x'} = \bra{x} 1 \ket{x'}$, we can use (\ref{braketdelta}), (\ref{OpeGreen}) and (\ref{OpeFG}) in the defining relation of the Green function (\ref{DefGreen}), obtaining the operator equation
\eq{\label{OpeEqGreen}
\hat{G} = i |\hat{g}|^{-1/4} \, {1 \over i\hat{\ma{F}}} \, |\hat{g}|^{-1/4}\,,
}
where we have inserted a factor of $i$ for reasons which will become clear shortly. One may now use the identity
\eq{\label{InvF}
{1 \over i\hat{\ma{F}}} = \int^{\infty}_0 e^{-i \tau \hat{\ma{F}}} d\tau
}
in (\ref{OpeEqGreen}) and apply $\bra{x}$ to the left and $\ket{x'}$ to recover the Green function, now written as
\eq{
G(x,x') = |g(x)|^{-1/4} \left(i \int^{\infty}_0  \braket{x; \tau|x'; 0}  d\tau \right) |g(x')|^{-1/4} \,.
}
where we have defined
\eq{\label{TransProb}
\braket{x; \tau|x'; 0} \equiv \bra{x} e^{-i \tau \hat{\ma{F}}}  \ket{x'}\,.
}
The fact that we have the determinant evaluated at different points is a matter of convention due to the way (\ref{OpeEqGreen}) is written, but is also a direct consequence of assumption that $\hat{\ma{F}}$ and $|\hat{g}|$ commute, so we opt for the symmetric choice here. The definition (\ref{TransProb}) is motivated by the introduction of the notation
\eq{
\ket{x; \tau} \equiv e^{i \tau \hat{\ma{F}}} \ket{x; 0} \equiv e^{i \tau \hat{\ma{F}}} \ket{x}
}
which greatly emphasizes that the problem at hand is mathematically identical to that of a system governed by the ``Hamiltonian'' $\hat{\ma{F}}$, an operator function of the ``proper time'' parameter $\tau$, evolving from an initial state labeled by the value of spacetime coordinates $x'$ at time $\tau=0$ to a final state characterized by the value $x$ at time $\tau$. This interpretation is what gives the name ``proper time formalism'' to this procedure. In this picture, a state $\ket{x; 0} \equiv \ket{x}$ characterized by the eigenvalue $x$ at time $\tau=0$ will evolve into a state
\eq{\label{evo}
\ket{x; \tau} = e^{i \tau \hat{\ma{F}}} \ket{x; 0} \equiv e^{i \tau \hat{\ma{F}}} \ket{x}
}
at a later time $\tau$. The analogy with quantum mechanics calculations can be taken even further. As with any other operator, $\hat{x}^{\mu}$ has an associated operator, denoted here $\hat{p}^{\mu}$, which generates translations in the space of the $\ket{x}$ vectors. In quantum mechanics, such a generator is conventionally characterized by the relation
\eq{\label{DefGen}
-{i \over \hbar}\de x^{\mu} \hat{p}_{\mu} \ket{x; \tau} = \ket{x+\de x; \tau}
}
where $\de x^{\mu} \equiv \de x$ is an infinitesimal translation in the eigenvalues of $x^{\mu}$. Since the spectrum of those eigenvalues is continuous, one can also express $\hat{p}_{\mu}$ in the differential form
\eq{
\hat{p}_{\mu} \ket{x; \tau} = -{\hbar \over i}{\ket{x+\de x; \tau} - \ket{x; \tau} \over \de x^{\mu}} = -{\hbar \over i}{\de \ket{x; \tau} \over \de x^{\mu}} \equiv -{\hbar \over i} {\dd \over \dd x^{\mu}} \ket{x; \tau}
}
The operator $\hat{p}^{\mu}$ is itself characterized by the eigenvalues and eigenvectors
\eq{
\hat{p}^{\mu} \ket{p} = p^{\mu} \ket{p}\,,
}
where we call $p^{\mu}$ the momentum eigenvector. Since products of $\hat{p}_{\mu}$ represent successive derivatives, it will always be possible to represent the differential operator $\ma{F}$ of (\ref{OpeF}) as a function $\hat{\ma{F}}(\hat{p}^{\mu},\hat{x}^{\mu})$. The commutation relations
\eq{
[\hat{x}^{\mu},\hat{p}_{\nu}] = i\de^{\mu}{}_{\nu}
}
can also be proven from the definition of the generator (\ref{DefGen}), as well as the dynamical equations
\ea{
{d\hat{x}^{\mu} \over d\tau} & = i[\hat{\ma{F}},\hat{x}^{\mu}]\,, \label{OEM1} \\
{d\hat{p}^{\mu} \over d\tau} & = i[\hat{\ma{F}},\hat{p}^{\mu}]\,. \label{OEM2}
}
Finally, deriving (\ref{TransProb}) with respect to the proper time $\tau$, using the more common notation
\eq{\label{HK}
K(\tau; x,x') \equiv \braket{x; \tau|x'; 0}
}
and considering indexes for more generality, we arrive at the differential equation
\eq{\label{SchroEq}
i {\dd \over \dd \tau} K^{i}{}_{k}(\tau; x,x') = \ma{F}^{i}{}_{j} K^{j}{}_{k}(\tau; x,x')\,.
}
Consequently, the problem of evaluating the Green function of a differential equation has been converted to the purely analogous quantum mechanical problem of finding the ``transition probability'' $\braket{x; \tau|x'; 0}$. There are two ways to do this: either by solving the operator equations of motion (\ref{OEM1}) and (\ref{OEM2}); or by solving the Schr\"{o}dinger type equation (\ref{SchroEq}). The quantity $K^{i}{}_{j}(\tau; x,x')$ is also often called heat kernel since Eq.~(\ref{SchroEq}) becomes the heat equation if we analytically continue $\tau$ to imaginary values.

%% file: II.1GFinCS/3.section.tex
\section{Physical interpretation of Green functions in QFT}

Although Green functions are mathematical objects with a wide range of applications in physics, their interpretation depends on the physical situation at hand. To derive their meaning in the case of Quantum Field Theory, we will consider the case of a scalar field $\hat{\phi}(x)$ in a Minkowski spacetime, obeying the Klein-Gordon equation
\eq{\label{KG}
(\square - m^2)\hat{\phi}(x) = 0\,.
}
The field is quantized i.e. the commutation relations
\ea{
[\hat{\phi}(\bm{x},t),\hat{\phi}(\bm{x}',t)] & = 0\,, \label{CR1} \\
[\hat{\phi}(\bm{x},t),\dd_t \hat{\phi}(\bm{x}',t)] & = i\delta^3(\bm{x}-\bm{x}')\,, \label{CR2}
}
hold for equal times. In flat spacetime, one can always define a global time with respect to which all observers can agree that a certain particle has positive (or negative) energy, so that the particle concept is well defined. This means the field is allowed the Fourier expansion
\eq{\label{solKG1}
\hat{\phi}(x) = \int {d^3 \textbf{p} \over (2\pi)^3} {1\over \sqrt{2 \omega_{\textbf{p}}}} \left(\hat{a}_{\textbf{p}} \, e^{i p \cdot x} + \hat{a}^{\dagger}_{\textbf{p}} \, e^{-i p \cdot x}\right)
}
where $\hat{a}^{\dagger}_{\textbf{p}}$ and $\hat{a}_{\textbf{p}}$ are the creation and annihilation operators which create and destroy excitations in the modes with four-momentum $p$. Since the vacuum is defined as the state $\ket{0}$ which satisfies $\hat{a}_{\textbf{p}} \ket{0}$, we have that
\eq{
\hat{\phi}(x) \ket{0} = \int {d^3 \textbf{p} \over (2\pi)^3}  \, e^{-i p \cdot x} \ket{\textbf{p}}
}
where $\ket{\textbf{p}} \equiv (1/\sqrt{2 \omega_{\textbf{p}}}) \hat{a}^{\dagger}_{\textbf{p}} \ket{0}$ denotes a relativistic state with defined momentum $\textbf{p}$. The state $\hat{\phi}(x) \ket{0}$ is exactly the Fourier expansion of a state with definite position $\textbf{x}$, created at time $t$, i.e. we have
\eq{
\hat{\phi}(x) \ket{0} = \ket{\textbf{x},t} \equiv \ket{x}
}
where $x = (\textbf{x},t)$ is the spacetime point where the particle was created. Since we are dealing with a scalar field, this particle has spin 0. We can now ask what is the probability amplitude, for a (relativistic) particle, that a measurement of the position at the instant $t$ will yield $\bm{x}$ after having been observed to have the spatial coordinates $\bm{x}'$ at the instant $t'$, i.e. the probability amplitude to go from a spacetime point $x'$ to a different spacetime point $x$. The answer will be
\eq{\label{PA}
\braket{\bm{x},t|\bm{x}',t'} \equiv \braket{x|x'} = \bra{0}\hat{\phi}(x)\hat{\phi}(x')\ket{0} = \bra{0}\hat{T}\{\hat{\phi}(x)\hat{\phi}(x')\}\ket{0}
}
where the time ordering operator $\hat{T}$, defined by
\eq{
\hat{T}\{\hat{\phi}(x)\hat{\phi}(x')\} =
\hat{\phi}(x)\hat{\phi}(x')~\theta(t-t') + \hat{\phi}(x')\hat{\phi}(x)~\theta(t'-t) =
\left\{ \begin{array}{ll}
       \hat{\phi}(x)\hat{\phi}(x')\,, \quad \textrm{if} ~ (t-t') > 0 \\
       \\
      \hat{\phi}(x')\hat{\phi}(x)\,, \quad \textrm{if} ~ (t-t') < 0
\end{array}
\right.
}
can be naturally included, with $\theta$ being the Heaviside function, in order to obtain a result which is valid independently of the casual order of the instants $t'$ and $t$. Indeed, if $t'$ was an instant at a latter time than $t$, then the probability amplitude (\ref{PA}) would be $\braket{\bm{x}',t'|\bm{x},t}$ instead but even in that case that quantity would still be represented by $\bra{0}\hat{T}\{\hat{\phi}(x)\hat{\phi}(x')\}\ket{0}$. Thus, in all generality, the probability amplitude of a free scalar particle to propagate from spacetime points $x'$ to $x$ is given by $\bra{0}\hat{T}\{\hat{\phi}(x)\hat{\phi}(x')\}\ket{0}$. In order to calculate it, it is convenient to know the differential equation it obeys. If we differentiate it with respect to $t$, we find that
\eq{\label{dt}
\dd_t\bra{0}\hat{T}\{\hat{\phi}(x)\hat{\phi}(x')\}\ket{0} = \bra{0}\hat{T}\{\dd_t\hat{\phi}(x)\hat{\phi}(x')\}\ket{0}+\de(t-t')\bra{0}[\hat{\phi}(x),\hat{\phi}(x')]\ket{0}
}
where the result $\dd_x \theta(x) = \de(x)$ was used. Since $\de(t-t')$ forces $t=t'$ and at equal times we have $\left[\hat{\phi}(x),\hat{\phi}(x')\right]=0$ due to the commutation relation (\ref{CR1}), then the second term in Eq.~(\ref{dt}) is zero. Using this information and differentiating again, we find
\eq{
\dd^2_t\bra{0}\hat{T}\{\hat{\phi}(x)\hat{\phi}(x')\}\ket{0} = \bra{0}\hat{T}\{\dd^2_t\hat{\phi}(x)\hat{\phi}(x')\}\ket{0}+\de(t-t')\bra{0}[\dd_t\hat{\phi}(x),\hat{\phi}(x')]\ket{0}\,,
}
where again $\de(t-t')$ forces equal times, allowing the use of the commutation relation (\ref{CR2}) to reduce the above equation to
\eq{\label{ddt}
\dd^2_t\bra{0}\hat{T}\{\hat{\phi}(x)\hat{\phi}(x')\}\ket{0} = \bra{0}\hat{T}\{\dd^2_t\hat{\phi}(x)\hat{\phi}(x')\}\ket{0}-i\delta^4(x-x')\,.
}
As for the derivatives with respect to $\bm{x}$, it is immediately seen that
\eq{\label{ddx}
\dd^2_{\bm{x}}\bra{0}\hat{T}\{\hat{\phi}(x)\hat{\phi}(x')\}\ket{0} = \bra{0}\hat{T}\{\dd^2_{\bm{x}}\hat{\phi}(x)\hat{\phi}(x')\}\ket{0}\,.
}
Using Eqs.(\ref{ddt})-(\ref{ddx}), it is readily shown that
\eq{
(\square-m^2)\bra{0}\hat{T}\{\hat{\phi}(x)\hat{\phi}(x')\}\ket{0} = \bra{0}\hat{T}\{(\square-m^2)\hat{\phi}(x)\hat{\phi}(x')\}\ket{0}+i\de^4(x-x')\,,
}
and since $\hat{\phi}(x)$ satisfies the Klein-Gordon equation (\ref{KG}), the first term on the right is zero, leading to
\eq{\label{diffG}
(\square-m^2)\left[i \bra{0}\hat{T}\{\hat{\phi}(x)\hat{\phi}(x')\}\ket{0} \right] = -\de^4(x-x')\,,
}
which is exactly the equation obeyed by the Green function of the Klein-Gordon equation (\ref{KG}), leading to the conclusion that
\eq{\label{QFGF}
G(x,x') = i \bra{0}\hat{T}\{\hat{\phi}(x)\hat{\phi}(x')\}\ket{0}\,.
}
In other words, we have shown that Green functions in QFT are interpreted as the probability amplitudes to propagate a particle from one spacetime point to another. A similar procedure to the one used in this section can also be applied for particles with different spins. Furthermore, although we performed the calculations in flat spacetime, the result is covariant, implying that the interpretation for curved spacetimes remains valid.

%% file: II.1GFinCS/4.section.tex
\section{Thermal Green functions and finite temperature QFT}

All results involving Green functions until now have been concerned only with systems described by a pure state, namely the vacuum state, which are suitable to describe zero temperature scenarios. The lack of a statistical distribution of states throughout different energy eigenvalues makes it impossible to define a concept of temperature. As a consequence, to tackle finite temperature systems, we must consider cases where the system is described by a mixed state instead of a pure state.

We begin by considering standard assumptions from statistical quantum mechanics, namely the existence of a set of pure states $\ket{\psi_i}$ which are eigenstates of the Hamiltonian with energy eigenvalues $E_i$. These states will also be eigenstates of the number operator $N$, with eigenvalues $n_i$, indicating the number of quanta in each state. Having access to both the number of particles and their energy means we can introduce a temperate $T$ and study the system using a grand canonical ensemble of states. As a consequence, each possible value of an operator will be weighted by a factor of
\eq{
\rho_i = e^{-\be (E_i - \mu n_i)}\,,
}
so the probability of a system being in a state $\ket{\psi_i}$, i.e. of having the energy $E_i$, will be $\rho_i/Z$, where $Z$ is the partition function, defined as
\eq{
Z = \sum_{j} e^{-\be(E_j - \mu n_j)} \equiv e^{-\be \Omega}\,,
}
where
\eq{
\be = {1 \over k_{\textrm{B}} T}\,,
}
with $k_{\textrm{B}}$ begin the Boltzmann's constant, $\mu$ the chemical potential and $\Omega$ is the so-called thermodynamic potential. For a system in a pure state $\ket{\psi_i}$, the average value of certain operator $\hat{A}$ is given by
\eq{
\braket{\hat{A}} = {\braket{\psi_i | \hat{A} | \psi_i} \over \braket{\psi_i | \psi_i}}\,.
}
For a system at temperature $T = (k_{\textrm{B}} \be)^{-1}$, the ensemble average of the operator $\hat{A}$ will be given by the average of all $\braket{\psi_i | \hat{A} | \psi_i}$, weighted by their respective probabilities, i.e.
\eq{\label{Tave}
\braket{\hat{A}}_{\be} = {1 \over Z} \sum_{i} \rho_i {\braket{\psi_i | \hat{A} | \psi_i}}\,,
}
where we use $\braket{\ldots}_{\be}$ to denote the ensemble average. This invites the definition of the operator
\eq{
\hat{\rho} = \exp\left[-\be(\hat{H}-\mu \hat{N})\right]\,,
}
called density operator, such that
\eq{
\rho_i = \braket{\psi_i | \hat{\rho} | \psi_i}
}
and
\eq{
Z = \textrm{Tr} \hat{\rho}\,.
}
Since $\ket{\psi_i}$ is an eigenstate of $\hat{\rho}$, we have $\sum_{i} \rho_i {\braket{\psi_i | \hat{A} | \psi_i}} = \textrm{Tr} \hat{\rho} \hat{A}$, which means that
\eq{
\braket{\hat{A}}_{\be} = {\textrm{Tr}\hat{\rho} \hat{A} \over \textrm{Tr} \hat{\rho} }\,.
}
The above result is enough to determine the form of Green functions in a system at finite temperature. We first note that, from Eq.~(\ref{QFGF}), we have
\eq{
G(x,x') = i\braket{\hat{\phi}(\textbf{x},t)\hat{\phi}(\textbf{x}',t')} \theta(t-t') + i\braket{\hat{\phi}(\textbf{x}',t')\hat{\phi}(\textbf{x},t)} \theta(t'-t)\,,
}
so the concept of Green functions at finite temperature, or thermal Green functions, can be obtained by simply using thermal averages in the form of Eq.~(\ref{Tave}) instead of pure state averages. Since the quantum field is an operator like any other, we have the Heisenberg equations of motion
\eq{\label{preGFbe}
\hat{\phi}(\textbf{x},t) = e^{i\hat{H}(t-t_0)}\hat{\phi}(\textbf{x},t_0)e^{-i\hat{H}(t-t_0)}\,,
}
so, considering a zero chemical potential $\mu$ for simplicity, we can write
\ea{
\braket{\hat{\phi}(\textbf{x},t)\hat{\phi}(\textbf{x}',t')}_{\be} & = \textrm{Tr}\left[ e^{-\be\hat{H}}\hat{\phi}(\textbf{x},t)\hat{\phi}(\textbf{x}',t')\right] / \textrm{Tr} \hat{\rho} \nonumber \\
& = \textrm{Tr}\left[ e^{-\be\hat{H}}\hat{\phi}(\textbf{x},t)e^{\be\hat{H}}e^{-\be\hat{H}}\hat{\phi}(\textbf{x}',t')\right] / \textrm{Tr} \hat{\rho} \nonumber \\
& = \textrm{Tr}\left[ e^{-\be\hat{H}}\hat{\phi}(\textbf{x}',t') \hat{\phi}(\textbf{x},t+i\be)\right] / \textrm{Tr} \hat{\rho} \nonumber \\
& = \braket{\hat{\phi}(\textbf{x}',t')\hat{\phi}(\textbf{x},t+i\be)}_{\be} \label{PartGF}
}
where we made use of the trace property $\textrm{Tr}AB=\textrm{Tr}BA$ and the fact that the number operator commutes with the quantum field. Switching primes in the variables leads us to a similar relation, and substituting it along with Eq.~(\ref{PartGF}) in Eq.~(\ref{preGFbe}), we reach the conclusion that a thermal Green function $G_{\be}(x,x')$ has the following property
\eq{\label{PerGF}
G_{\be}(\textbf{x},t;\textbf{x}',t') = G_{\be}(\textbf{x},t+i\be;\textbf{x}',t')\,,
}
i.e. a thermal Green function is periodic in imaginary time with period $\be$. This remark strongly motivates a change of variables to imaginary time, i.e. to perform a Wick rotation $t = i\tau$. If we switch to imaginary time coordinates, by dimensional analysis frequencies will also shift as $\omega \to -i\omega$. When this fact is taken into account in Eqs.~(\ref{solKG1}) and (\ref{QFGF}), one arrives at the result
\eq{
G(i \tau,\textbf{x};i \tau',\textbf{x}') = i G(t,\textbf{x};t',\textbf{x}')\,.
}
In the literature this is commonly denoted as
\eq{
G_{\textrm{E}}(\tau,\textbf{x};\tau',\textbf{x}') \equiv G(i \tau,\textbf{x};i \tau',\textbf{x}')
}
also called Euclidean Green function. If we deal with stationary spacetimes (which will be the case throughout this work) the dependence of the Green function in the time coordinates will always be of the form $G(x,x') = G(t-t';\textbf{x},\textbf{x}')$, since all integrals will be from $t$ to $t'$ and there are no quantities depending on the time coordinates. This fact, together with Eq.~(\ref{PerGF}), leads to following relation for the thermal Euclidean Green function
\eq{\label{PerGE}
G_{\textrm{E}}(\tau-\tau';\textbf{x},\textbf{x}') = G_{\textrm{E}}(\tau-\tau'+\be;\textbf{x},\textbf{x}')
}
where we will stop using the subscript $\be$ to denote thermal quantities, as it will be implicit in the quantities themselves every time a $\be$ appears. Eq.~(\ref{PerGE}) implies that the Euclidean Green function is periodic in the real variable $\Delta\tau=\tau-\tau'$ with period $\be$, so if we impose the restriction $\tau\in [0,2\pi]$, we will also have $\Delta\tau \in [0,2\pi]$. This allows us to expand the function $G_{\textrm{E}}(\Delta\tau;\textbf{x},\textbf{x}')$ in a Fourier series as
\eq{\label{GExpN}
G_{\textrm{E}}(\tau,\textbf{x};\tau',\textbf{x}') = {1\over \be} \sum^{\infty}_{n=-\infty} e^{i \omega_n \Delta\tau} G_n(\textbf{x},\textbf{x}')\,,
}
where $G_n(\textbf{x},\textbf{x}')$ are the Fourier coefficients of the Green function and
\eq{
\omega_n = {2\pi n\over \be}
}
are called the Matsubara frequencies. It is often the case, as will be throughout this work, that the Green function is invariant with respect to rotations of the spatial vectors $\textbf{x}$ and $\textbf{x'}$. In such cases, we know that the mode Green functions $G_n(\textbf{x},\textbf{x}')$ will possess an expansion in hyperspherical harmonics, of the form \cite{Avery:1934}
\ea{\label{GExpL}
G_n(\textbf{x},\textbf{x}') = {1 \over 2\pi} {\Gamma\left({N\over 2}\right) \over \pi^{(N/ 2)}} \sum_{l=0}^{\infty} \left(l+{N \over 2}\right) C^{(N/2)}_l\left(\cos \g\right) G_{nl}(r,r')\,.
}
In this expression, $N=d-3$ with $d$ the number of spacetime dimensions, $C^{(N/2)}_l$ are the Gegenbauer polynomials, $\g$ is the geodesic distance on the $(d-2)$-sphere and $G_{nl}(r,r')$ are called the radial mode Green functions. A particularly important result regarding Gegenbauer polynomials is
\eq{\label{GenConst}
C^{(N/2)}_l(1) = {\Gamma(N+l) \over \Gamma(N)\Gamma(l+1)}\,,
}
which will be frequently used. As a final remark, since the differential equations Green functions obey involve Dirac deltas, it is necessary to use a suitable representations for them. In light of Eqs.~(\ref{GExpN}) and (\ref{GExpL}), we shall thoroughly use the choices
\eq{\label{deTau}
\delta(\tau-\tau') = {1\over \beta} \sum_n e^{i\omega_n (\tau - \tau')}
}
and
\eq{\label{deO}
\delta(\Omega,\Omega') = {1 \over 2\pi} {\Gamma\left({N\over 2}\right) \over \pi^{(N/ 2)}} \sum_{l=0}^{\infty} \left(l+{N \over 2}\right) C^{(N/2)}_l\left(\cos \g\right)
}
where $\Omega$ is respective to the angular coordinates appearing in the hyperspherical decomposition of the spatial vector $\textbf{x}$.

%% file: II.1GFinCS/5.section.tex
\section{Synge's world function}

In this section we will briefly introduce two quantities that will repeatedly appear through the majority of calculations involving renormalization in curved spacetimes. The first of them is most naturally derived form the concept of geodesics, which is a path in curved space that obeys the differential equation
\eq{\label{GeoCond}
{d^2 x^{\mu} \over d\tau^2} + \Gamma^{\mu}{}_{\al\nu} {d x^{\al} \over d\tau} {d x^{\nu} \over d\tau} = 0\,,
}
It is also straightforward to see that this is equivalent to
\eq{
\dot{x}^{\nu} \nabla_{\nu}\dot{x}^{\mu} = 0
}
from where one may deduce that
\eq{\label{DefC}
\dot{x}^{\mu} \dot{x}_{\mu} = g_{\mu\nu} \dot{x}^{\mu} \dot{x}^{\nu} = \left({ds \over d\tau}\right)^2 = C
}
where $C$ is some constant. If we now calculate the geodesic distance, we see that
\eq{
s(x,x') = \int^{x}_{x'} ds = \int^{\tau}_{\tau'} {ds \over d\tau} d\tau = \pm \sqrt{C} (\tau -\tau')\,.
}
Solving for the constant $C$ and comparing with (\ref{DefC}), we conclude that, for a geodesic curve,
\eq{
\left({ds \over d\tau}\right)^2 = 2\, {\si(x,x') \over (\tau-\tau')^2}\,,
}
where the definition
\eq{
\si(x,x') = {s^2(x,x') \over 2}
}
is called ``Synge's world function'' \cite{Synge:1960}. It is an example of a bi-scalar quantity, which is defined as a quantity which transforms as a scalar under coordinate transformations taken at the points $x$ and $x'$. One of its most useful identities is
\eq{\label{RelSi}
{1 \over 2}\si_{,\mu}\,\si_{,}{}^{\mu} = \si
}
which tells us that the vector $\si_{,\mu}$ has a length equal to the geodesic distance between the points $x$ and $x'$. We have the same for differentiation with respect to $x'$, usually denoted with a prime in the corresponding index, i.e.
\eq{
{1 \over 2}\si_{,\nu'}\,\si_{,}{}^{\nu'} = \si\,.
}
Furthermore, since
\eq{
\si_{,\mu} = s(x,x'){\dd s \over \dd x^{\mu}}\,,
}
one can also infer that $\si_{,\mu}$ is tangent to the geodesic at $x$ and oriented in the direction $x'\to x$. The quantity $\si(x,x')$ is central for the study of geodesics, as it single handedly contains all knowledge about the geodesic content of a curved space. One particular information one can derive from it is the surface where geodesics intersect, called a caustic surface, which can be understood in the following way. If the the vector $\si_{,\mu}$ changes by some amount $\de \si_{,\mu}$, it usually means there was a displacement $\de x'^{\nu}$ in the endpoint $x'^{\nu}$ to provoke that change. The two variations are related by
\eq{
\de \si_{,\mu} = {\dd \si_{,\mu} \over \dd x'^{\nu}} \de x'^{\nu} \equiv \si_{,\mu\nu'} \de x'^{\nu}\,,
}
i.e. the required change in $x'^{\nu}$ is
\eq{
\de x'^{\nu} = (\si_{,\nu'\mu})^{-1} \de \si_{,\mu}\,.
}
However, it may happen that there may be a change in $\si_{,\mu}$ without changing the endpoint. Since $\si_{,\mu}$ is tangent to the geodesic, this means that in that case we are in a situation where more than one geodesic is crossing the points $x$ and $x'$. This is equivalent to saying there is a non-zero change $\de \si_{,\mu}$ but at the same time a zero variation in the endpoint $\de x'^{\nu}$. Consequently, the condition
\eq{
(\si_{,\nu'\mu})^{-1} \de \si_{,\mu} = 0
}
gives all points where geodesics intersect, i.e. the caustic surface. The above equation is also equivalent to imposing
\eq{
\textrm{det}(\si_{,\mu\nu'}) = 0\,,
}
from which we realize the importance of the determinant of the matrix $\si_{,\mu\nu'}$. The slightly different definition
\eq{
\Delta = g^{-1/2}(x) \, \textrm{det}(\si_{,\mu\nu'}) \, g^{-1/2}(x')\,,
}
also called the Van Vleck-Morette determinant is the second important quantity in geodesic theory which will make many appearances in this work.

%% file: II.2QFTinCS/main.tex
\chapter{Quantum Field Theory in Curved Spacetime}
\label{cap:chapterII2}
\input{II.2QFTinCS/outline.tex}

\input{II.2QFTinCS/1.section.tex}

\input{II.2QFTinCS/2.section.tex}

\input{II.2QFTinCS/3.section.tex}

\input{II.2QFTinCS/4.section.tex}
\input{II.2QFTinCS/5.section.tex}

%% file: II.2QFTinCS/outline.tex
\section{Outline}

Some approaches to quantum field theory are better when one wishes to consider curved manifolds. The one that most naturally allows the inclusion of gravity is, surprisingly, not the most adopted one in the flat case. One is usually introduced to quantum field theory using the approach of Feynman, where propagators are calculated using diagrams which greatly speed computations. That approach is highly visual and appeals strongly to intuition. Nevertheless, although it is possible to generalise these concepts to curved spacetimes, it becomes very complex to keep track of all diagrams.

Another approach, which is rarely pursued in introductory levels, is the differential formalism used by Schwinger \cite{Schwinger:1951xk,Schwinger:1953tb,Schwinger:1953zza,Schwinger:1953zz,Schwinger:1954zza,Schwinger:2001}, whereby all contact with graphical pictures is foregone and substituted with mathematically rigorous and covariant methods. This approach is undoubtedly more complicated, but it proves to be highly efficient when considering calculations in curved spacetimes.

In this chapter, we will quickly review the main ingredients of the Schwinger approach to quantum field theory, including a very efficient way to renormalize physical quantities in that context.

%% file: II.2QFTinCS/1.section.tex
\section{Quantum Fields in Curved Spacetimes}

A quantum state is specified by values associated to a set of compatible physical properties $\hat{q}_i$ that are related to the number of degrees of freedom of the system. In non-relativistic quantum mechanics, quantum states are characterized by a single parameter $t$ that holds an absolute meaning, in the sense that it represents a global time. Since a state is fully characterized by $\hat{q}_i$ and $t$, it becomes clear that quantum mechanics is not a relativistic theory since it relies on the concept of a single time coordinate for every observer.

To incorporate the principle of relativity in quantum mechanics, we must realize that a state cannot be considered to be fully specified by a single parameter $t$ alone. At any given instant, all points separated by a spatial interval are casually independent since information has a finite propagation speed $c$. Consequently, any measurements performed throughout the points of a spacelike surface cannot have any casual effect on each other; this is the very definition of compatible physical quantites. We must thus come to the realization that, when relativity is in play, there are extra degrees of freedom that must be taken into account in order to obtain fully specified quantum states.

To incorporate these additional degrees of freedom, we start by considering a certain instant $t_0$. Since each small volume element in the three dimensional spacelike surface is physically independent of all other volume elements, we will have a set of dynamical variables $\hat{\chi}_i$ for each of the spatial points $\bs{x}$, which we can use as index. Therefore, all the appropriate degrees of freedom can be accomodated in by using the labels $\hat{\chi}_{i,\bs{x}}(t_0)$, but
\eq{
\hat{\chi}_{i,\bs{x}}(t_0) \equiv \hat{\chi}_i(t=t_0,\bs{x})
}
is the very definition of a field in spacetime, or in this case $n$ fields labeled by the index $i$, for a fixed time coordinate $t=t_0$. This leads to one conclusion: to properly specify a quantum state in a relativistic formulation of quantum mechanics, we must consider dynamical variables $\chi_i(x^{\mu})$ which are operator valued functions of the spacetime coordinates $x^{\mu}$, also called ``quantum fields''. For notational simplicity, we will often use $x$ to denote a spacetime point.

A quantum state compatible with the principle of relativity must thus be parametrized by a spacelike surface $\sigma$, associated to some constant value of the time coordinate $t$. It should also be labeled, like in the previous chapter, by the eigenvalues $\chi_i(x)$ of the dynamical variables $\hat{\chi}_i(x)$ which are quantum fields. However we could as well have considered the states to be characterized by the eigenvalues of functions of $\hat{\chi}_i(x)$. For example, in quantum mechanics the eigenvalues of the angular momentum operator $\bs{L}$ which is the function $\bs{q} \times \bs{p}$, are also frequently used to characterize quantum states. We can thus consider for quantum field theory the most general case where the states are labeled by some eigenvalues $\ze(x)$ (or just $\ze$ for convenience) of functions of the field operators $\hat{\chi}_i(x)$. In summary, a relativistic quantum state in a general curved spacetime will be denoted as
\eq{
\ket{\ze; \sigma}\,,
}
where we use the semicolon to remember that $\sigma$ is a parameter and not a dynamical variable. These states contain all the information we need about the system, so the remaining step is to know what is their dynamical behaviour. It turns out that the single equation
\eq{\label{SAP}
\de \braket{\ze_1; \sigma_1 | \ze_2; \sigma_2} = {i \over \hbar} \bra{\ze_1; \sigma_1} \de \hat{S}_{12} \ket{\ze_2; \sigma_2}
}
governs all of the field's dynamics in a general curved background, where $\ket{\ze_2; \sigma_2}$ is the state on the initial hypersurface $\sigma_2$ and $\bra{\ze_1; \sigma_1}$ is the state in the final hypersurface $\sigma_1$ after the interactions governed by the action $\hat{S}_{12}$ have taken place. Note that, unlike other standard formulations of QFT, the action is an operator, given by
\eq{
\hat{S}_{12} = \int^{\sigma_1}_{\sigma_2} \hat{\ma{L}}[x] \, dv_x\,,
}
where $dv_x$ is the invariant volume element of a $n$-dimensional spacetime
\eq{
dv_x = \sqrt{|g|} \, dx^0 dx^1 dx^2 \cdots dx^n  = \sqrt{|g|} \, d^n x\,,
}
and
\eq{\label{Lft}
\hat{\ma{L}}[x] \equiv \hat{\ma{L}}(\hat{\chi}_i(x),\dd_{\mu}\hat{\chi}_i(x),x)
}
is the Lagrangian density operator, which is a function of the quantum fields of the theory and their derivatives. The $\delta$ is a variation, taken with respect to any variable present in the system. The Eq.~(\ref{SAP}) is not only relativistic, since all the quantities involved are relativistic invariants, but it is also generally covariant, since no use of particular coordinates is present. It is called the Schwinger Action Principle and it is enough to develop all of Quantum Field Theory in any spacetime background. Among other results, it directly implies that
\eq{\label{QFE}
{\de \hat{S}_{12}[\hat{\chi}_i(x)] \over \de \hat{\chi}_j(x')} = 0
}
which is the operational analog of the classical field equations of motion that lead to the Euler-Lagrange equations. A full derivation of the results (\ref{SAP}) and (\ref{QFE}) is present in textbook references \cite{Parker:2009uva,Toms:2012bra} and will be left for the reader. The formalism developed by Schwinger is in some sense a counterpart of the Feynman Path Integral formalism, as it has a differential rather than integral nature.

%% file: II.2QFTinCS/2.section.tex
\section{The effective action}

In this section we will develop the main consequences of the action principle of Eq.~(\ref{SAP}), following the standard literature \cite{Parker:2009uva, Toms:2012bra}. Since we will be interested only in scalar fields in this work, we shall begin by considering $N$ quantum fields denoted by $\hat{\phi}^{i}(x)$ $(i=1,\ldots,N)$ and an equal number of sources $J_i(x)$, such that the action assumes the form
\eq{
\hat{S}_{12} = \int^{\sigma_1}_{\sigma_2} \left\{\hat{\ma{L}}[x]+J_i(x)\hat{\phi}^{i}(x)\right\} \, dv_x\,.
}
The sources are introduced by mathematical convenience and, as we shall see, their job is to be derived with respect to and ultimately taken to zero at the end of the calculations.

Before we proceed any further, it will also prove convenient to use the condensed notation of De Witt, which not only considers repeated indexes to be summed, but also an integration over the corresponding spacetime variables. For example, the quantity $A_{ij}f^{i}h^{i}$ is to be interpreted as
\eq{
A_{ij}f^{i}h^{i} \equiv \sum_{i,j} \int d^{d}x \int d^{d}x' A_{ij}(x,x')f(x)^{i}h(x')^{i}
}
where the index $i$ has the associated coordinate $x$, $j$ has the associated coordinate $x'$, and so on. The extrema of the integrals are taken according to the situation at hand. Using condensed notation, the functional derivative of a functional $F[p(x)]$, which is defined through the relation (with $\e$ small)
\eq{
F[p^i(x) + \e h^i(x)] = F[p^i(x)] + \e \int d^d x' h^j(x') {\de F[p^i(x)] \over \de p^j(x')} + \ma{O}(\e^2)\,,
}
can be compressed into
\eq{
F[p^i + \e h^i] = F[p^i] + \e  {\de F[p^i] \over \de p^j} + \ma{O}(\e^2)\,.
}
We will also use the notation $F_{,i}$ to denote functional derivative. The above result is indistinguishable from the definition of the derivative of a vector function, i.e. as if the fields $p^i(x)$ and functional $F[p^i(x)]$ were swapped for vector components $p^i$ and a function $F[p^i]$ with vector arguments. This means we can conduct all functional calculations in a much simpler way when adopting the condensed notation, as is we were dealing with vector calculus. This fact remains true even for quantum fields, as we only need to be careful with the commutation properties of the fields.

In essence, the point of quantum field theory, and of the Schwinger Action Principle, is to calculate the amplitude $\braket{\ze_1; \sigma_1 | \ze_2; \sigma_2}$, since all objects of interest may be derived from it. We start by diving the action operator in a free field term $\hat{S}_{0}$ plus an interaction term $\hat{S}_1$, i.e.
\eq{
\hat{S}_{12} = \hat{S}_{0} + \hat{S}_1 + J_i\hat{\phi}^i\,.
}
From here, Eq.~(\ref{SAP}) is enough to obtain the result
\eq{\label{AmpS1}
\braket{\ze_1; \sigma_1 | \ze_2; \sigma_2}[J] = \bra{\ze_1; \sigma_1}T\left\{ \exp\left[{i\over \hbar}\left(\hat{S}_1 + J_i\hat{\phi}^i\right)\right] \right\}\ket{\ze_2; \sigma_2}_{0}
}
where the subscript $0$ means that the states are evaluated treating only $\hat{S}_{0}$ as the action, and $T$ is the time ordering operator, defined by
\eq{
\hat{T}\{\hat{A}(x)\hat{B}(x')\} =
\hat{A}(x)\hat{B}(x')~\theta(t-t') + \hat{B}(x')\hat{A}(x)~\theta(t'-t) =
\left\{ \begin{array}{ll}
       \hat{A}(x)\hat{B}(x')\,, \quad \textrm{if} ~ (t-t') > 0 \\
       \\
      \hat{B}(x')\hat{A}(x)\,, \quad \textrm{if} ~ (t-t') < 0
\end{array}
\right.
}
with $\theta$ being the Heaviside function. The derivation of Eq.~(\ref{AmpS1}) is rather lengthy so we refrain from deriving it here. The next step is to define the functional $W[J]$ by
\eq{\label{WJ}
\braket{\ze_1; \sigma_1 | \ze_2; \sigma_2}[J] = e^{i W[J]}
}
so that determining the amplitude $\braket{\ze_1; \sigma_1 | \ze_2; \sigma_2}\equiv \braket{\ze_1; \sigma_1 | \ze_2; \sigma_2}[J=0]$ is equivalent to finding the functional $W[J]$ and taking $J$ to zero. If we differentiate Eq.~(\ref{WJ}) with respect to the $J_i$ and take the source to zero, we obtain
\eq{\label{dJAmp}
{\de \over \de J_i}\braket{\ze_1; \sigma_1 | \ze_2; \sigma_2}[J]\bigg|_{J=0} = {i \over \hbar} {\de W \over \de J_i}e^{i W[0]}
}
and if we apply the same action in Eq.~(\ref{AmpS1}), we get
\eq{\label{dJPhi}
{\de \over \de J_i}\braket{\ze_1; \sigma_1 | \ze_2; \sigma_2}[J]\bigg|_{J=0} = {i \over \hbar} \braket{\ze_1; \sigma_1 | T\{\hat{\phi}^i\} | \ze_2; \sigma_2}_0\,.
}
Defining the time ordered average value of an operator $\hat{A}$, in the states evaluated with $\hat{S}_0$, as
\eq{\label{Aver}
\braket{\hat{A}}_0 = {\braket{\ze_1; \sigma_1 | T\{\hat{A}\} | \ze_2; \sigma_2}_0 \over \braket{\ze_1; \sigma_1 | \ze_2; \sigma_2}_0}\,,
}
we may conclude from Eqs.~(\ref{dJAmp}) and (\ref{dJPhi}) that
\eq{\label{PhiDef}
{\de W \over \de J_i} = \braket{\hat{\phi}^i}_0 \equiv \Phi^i(x)\,,
}
where we use $\Phi^i(x)$ to denote the mean value of the quantum field, commonly known as background field. By induction, one may show the general result
\eq{\label{DeAmpN}
{\de^n \braket{\ze_1; \sigma_1 | \ze_2; \sigma_2}[J]\over \de J_{i_1}\cdots\de J_{i_n}}\bigg|_{J=0} = \left({i \over \hbar}\right)^n \braket{\ze_1; \sigma_1 | T\{\hat{\phi}^{i_1}\cdots\hat{\phi}^{i_n}\} | \ze_2; \sigma_2}_0
}
which demonstrates the pillar importance of the amplitude $\braket{\ze_1; \sigma_1 | \ze_2; \sigma_2}[J]$, since it allows us to calculate all correlation functions, which are the main ingredient of any QFT quantity. We now introduce the so-called effective action $\Gamma[\Phi]$ through the definition
\eq{\label{EA}
\Gamma[\Phi] = W[J] - J_i \Phi^i
}
and differentiate it with respect to the background field, from which we obtain
\eq{
{\de \Gamma \over \de \Phi^j} = {\de W \over \de J_i}{\de J_i \over \de \Phi^j} - {\de J_i \over \de \Phi^j} \Phi^j - J_j\,,
}
where we used the chain rule in $W[J]$, since it can be thought of as a functional of $\Phi$ through an implicit dependence on $J_i$. Inserting Eq.~(\ref{PhiDef}) in the first term on the right hand side gives
\eq{\label{FEJ}
{\de \Gamma \over \de \Phi^j} = - J_j
}
which, after taking $J_j$ at the end of calculations, gives (in all detail and ignoring the index)
\eq{\label{CFE}
{\de \Gamma[\Phi(x)] \over \de \Phi(x')} = 0\,,
}
which is the defining equation for the background field if we know the full form of $\Gamma$. The above equation is the reason why $\Gamma$ is called the effective action, since the equation is in every respect analogous to the principle of least action in classical field mechanics, but with an action functional which is different from the classical action $S$. For this reason, the field $\Phi(x)$ is usually called classical field as well.

The result in Eq.~(\ref{FEJ}) can now be used to eliminate the source variables in Eqs.~(\ref{AmpS1}) and (\ref{EA}), which can be combined to retrieve the relation
\eq{\label{EqEA}
\exp\left({i\over \hbar}\Gamma[\Phi]\right) = \bra{\ze_1; \sigma_1}T\left\{ \exp\left[{i\over \hbar}\hat{S}_1 - {i\over \hbar}{\de \Gamma \over \de \Phi^j} (\hat{\phi}^j-\Phi^j)\right] \right\}\ket{\ze_2; \sigma_2}_{0}\,.
}
Although complicated looking, this equation gives an implicit definition of the effective action, and it is enough to obtain it for any theory.

%% file: II.2QFTinCS/3.section.tex
\section{The leading-order effective action for scalar fields}

The solution of Eq.~(\ref{EqEA}) is often found in a perturbative way. In this section we will find the first order solution for the effective action of any scalar field theory. We will proceed by expressing the field as fluctuating component $\hat{\varphi}$ over the expectation value, i.e.
\eq{\label{RedF}
\hat{\phi}(x) = \hat{\varphi}(x) + \Phi(x)\,,
}
and express the action functional in a Taylor series around the background field
\eq{
\hat{S}[\hat{\phi}] = \hat{S}[\Phi] + \hat{S}_{,i}[\Phi] \hat{\varphi}^i + {1 \over 2} \hat{S}_{,ij}[\Phi] \hat{\varphi}^i \hat{\varphi}^j + \hat{S}_{\textrm{int}}[\Phi,\hat{\varphi}]
}
where we denote all terms of order higher than 2 as $\hat{S}_{\textrm{int}}$. We shall make the choice
\ea{
\hat{S}_0[\hat{\phi}] & = \hat{S}[\Phi] + {1 \over 2} \hat{S}_{,ij}[\Phi] \hat{\varphi}^i \hat{\varphi}^j\,, \label{S0} \\
\hat{S}_1[\hat{\phi}] & = \hat{S}_{,i}[\Phi] \hat{\varphi}^i + \hat{S}_{\textrm{int}}[\Phi,\hat{\varphi}] \label{S1}
}
which enables us to write the argument in the exponential of Eq.~(\ref{EqEA}) as
\eq{
\hat{S}_1 - \Gamma_{,j} (\hat{\phi}^j-\Phi^j) = (\hat{S}_{,j}- \Gamma_{,j}) \hat{\varphi}^j + \hat{S}_{\textrm{int}}[\Phi,\hat{\varphi}]\,.
}
The leading order approximation to the effective action consists in ignoring the above terms altogether which, from Eq.~(\ref{EqEA}), results in
\eq{\label{EALO}
\exp\left({i\over \hbar}\Gamma[\Phi]\right) \approx \braket{\ze_1; \sigma_1 | \ze_2; \sigma_2}_{0}
}
so, in reality, for the leading-order approximation one only needs to be concerned with the $\hat{S}_0$ part of the action. Using the same reasoning as in the last section, we define a functional $W_0[J']$ such that
\eq{\label{Amp0}
\braket{\ze_1; \sigma_1 | \ze_2; \sigma_2}_{0}[J'] = e^{i W_{0}[J']}\,,
}
and introducing a new source term in the action, whereby
\eq{\label{FullS0}
\hat{S}[\hat{\varphi}] = \hat{S}[\Phi] + {1 \over 2} \hat{\varphi}^i \hat{S}_{,ij}[\Phi]  \hat{\varphi}^j + J'_i \hat{\varphi}^i\,,
}
i.e. we take now the field $\hat{\varphi}$ as our dynamical quantity. Deriving the above result with respect to $\hat{\varphi}$ and using the field equation of motion (\ref{QFE}), results in
\eq{\label{Jeq}
J'_i = - D_{ij}  \hat{\varphi}^j
}
where we rellabled $\hat{S}_{,ij}[\Phi] = D_{ij}$ in order to avoid confusion with the functional derivatives, making it clear that $D_{ij}$ does not depend on $\hat{\varphi}$. At this point it becomes extremely useful to define the Green function $G^{ij}(x,x')$ through
\eq{\label{GEdef}
D_{ij}G^{jk} = -\de_i{}^{k}\,,
}
where the condensed notation makes it analogous to the matrix inverse of $D_{ij}$. Contracting both sides of Eq.~(\ref{Jeq}) with $G^{ki}$ gives us
\eq{\label{GJ}
\hat{\varphi}^k = G^{ki}J'_i\,.
}
An analogous version of (\ref{PhiDef}) will arise which, when combined with Eq.~(\ref{GJ}), will give
\eq{
{\de W_0 \over \de J'_i} =  G^{ki}J'_i\,.
}
The solution of the above differential equation is
\eq{\label{PreWJ}
W_0[J'] = {1\over 2} J'_{i}G^{ij}J'_j + W_0[0]
}
which can be straightforwardly checked by taking a variation with respect to $J'$. The only task left to do is to find $W_0[0]$, which is done by taking a variation of the action functional with respect to $D_{ij}$. Defining formal field integration such that
\eq{
A = \int d\Phi^i{\de \over \de \Phi^i} A\,,
}
one readily finds that
\eq{
S[\Phi] = \int \int d\Phi^i d\Phi^j D_{ij}\,.
}
This way, taking a variation in Eq.~(\ref{FullS0}) with respect to $D_{ij}$, we get
\eq{
\de\hat{S} = \int \int d\Phi^i d\Phi^j \de D_{ij}+{1 \over 2} \de D_{ij} \hat{\varphi}^i \hat{\varphi}^j
}
and, when this result is using in conjuction with Eqs.~(\ref{SAP}), (\ref{WJ}) and (\ref{Aver}), one obtains
\eq{\label{DeW0}
\de W_0[0] = \int \int d\Phi^i d\Phi^j \de D_{ij}+{1 \over 2} \de D_{ij} \braket{T\{\hat{\varphi}^i\hat{\varphi}^j\}}[J'=0]\,,
}
after setting $J=0$. On the other hand, using Eqs.~(\ref{WJ}) and (\ref{DeAmpN}) we know that
\eq{
{\de^n W_0[J']\over \de J'_{i_1}\cdots\de J'_{i_n}}\bigg|_{J'=0} = {i \over \hbar} \braket{T\{\hat{\varphi}^{i_1}\cdots\hat{\varphi}^{i_n}\}}[0]\,,
}
so if we apply this result for $n=2$ and use Eq.~(\ref{PreWJ}), we arrive at
\eq{
\braket{T\{\hat{\varphi}^{i}\hat{\varphi}^{j}\}}[0] = -i G^{ij}\,.
}
Inserting the above result in Eq.~(\ref{DeW0}) leads to
\eq{\label{deW0}
\de W_0[0] = \int \int d\Phi^i d\Phi^j \de D_{ij}-{i \over 2} \de D_{ij} G^{ji}\,.
}
We may now note that, due to Eq.~(\ref{GEdef}), we have in a formal sense $G^{ij} = (D^{-1})^{ij}$, i.e. from an operational point of view in the condensed notation, the Green function is the inverse of the matrix $D$. This means one may formally write
\eq{\label{lndet}
\de D_{ij} G^{ji} = \de D_{ij} (D^{-1})^{ji} = \textrm{Tr}[\de D D^{-1}] = \de[\textrm{Tr}\ln(\ell^2 D_{ij})] =\de[\ln \textrm{det}(\ell^2 D_{ij})]
}
where we used the identity $\textrm{Tr}\ln = \ln\textrm{det}$ and introduced an arbitrary constant $\ell$ in order to make the logarithm dimensionless. As usual, one may perform manipulations as if one was dealing with vector calculus, and perform the correct translation in the end to the appropriate funcional form. Inserting Eq.~(\ref{lndet}) in Eq.~(\ref{deW0}) and evidencing out the variation with respect to $D_{ij}$, it becomes clear that
\eq{
W_0[0] = S[\Phi]-{i \over 2} \ln \textrm{det}(\ell^2 D_{ij})\,,
}
implying that, at leading order, the functional $W[J]$ has the form
\eq{\label{FullWJ}
W_0[J'] = S[\Phi]+{1\over 2} J'_{i}G^{ij}J'_j - {i \over 2} \ln \textrm{det}(\ell^2 \hat{S}_{,ij}[\Phi])\,.
}
We now note that, due to Eq.~(\ref{RedF}), we have $\braket{\hat{\varphi}^i} = 0$, so if we take the average of Eq.~(\ref{Jeq}), we readily find that
\eq{
G^{ki}J'_i = 0\,.
}
Using this information in Eq.~(\ref{FullWJ}) and inserting the subsequent result in Eq.~(\ref{Amp0}), we may solve Eq.~(\ref{EALO}) with respect to the effective action, finally obtaining
\eq{\label{LOEA}
\Gamma[\Phi] = S[\Phi] + {i \over 2} \ln \textrm{det}(\ell^2 S_{,ij}[\Phi])
}
which is, of course, valid to leading order. Note that even though we considered only the free field part $\hat{S}_0$ of the action functional, there can still be interaction terms coming from $S_{,ij}[\Phi]$, so this result is valid in all generality. One is thus lead to the conclusion that the effective action is made of the classical action $S[\Phi]$ plus quantum correction terms $\Gamma^{(i)}[\Phi]$, such that
\eq{
\Gamma[\Phi] = S[\Phi] + \sum^{\infty}_{n=1} \hbar^n \Gamma^{(n)}[\Phi]\,,
}
where, in this case, we have just found that
\eq{\label{G1}
\Gamma^{(1)}[\Phi] = {i \over 2} \ln \textrm{det}(\ell^2 S_{,ij}[\Phi])\,.
}
It will also prove useful to write a full functional form for $\Gamma^{(1)}[\Phi]$, which can be done in the following way. First, we note that the quantity $D_{ij}$ will in general have the form $\sqrt{|g|} \ma{F}_{ij}(x,x')$ so, following the spirit of Sec.~(), one may write
\eq{
\de D_{ij} = \bra{x} \de(|\hat{g}|^{1/2}\hat{\ma{F}}_{ij}) \ket{x'}\,.
}
From here, we use Eqs.~(\ref{OpeEqGreen}) and (\ref{InvF}) to derive the slightly different form for the green function
\eq{
G^{ji}(x',x) = \bra{x'} \int^{\infty}_0 e^{-i \tau |\hat{g}|^{1/2} \hat{\ma{F}}_{ji}} d\tau \ket{x}
}
which, when used in conjuction with the completeness relation Eq.~(\ref{CR}) and after some manipulations, gives the result
\eq{\label{NewdDij}
\de D_{ij} G ^{ji} = \de\left(- \int dv_x \int^{\infty}_{0} {d\tau\over \tau} K^{i}{}_{i}(\tau; x,x) \right)
}
where we adopted the heat kernel notation of Eq.~(\ref{HK}). Inserting the relation (\ref{NewdDij}) in Eq.~(\ref{deW0}), will result in the leading-order correction
\eq{\label{G1func}
\Gamma^{(1)}[\Phi] = - {i \over 2} \int dv_x \int^{\infty}_{0} {d\tau\over \tau} \textrm{Tr} K(\tau; x,x)\,,
}
which is the complete functional form of Eq.~(\ref{G1}), where the trace runs through the field indexes of the differential operator. Both versions have their own advantages: (\ref{G1}) is more efficient when dealing with functional derivatives, while (\ref{G1func}) is more useful for renormalization issues.

%% file: II.2QFTinCS/4.section.tex
\section{Renormalization of the effective action}

As is well known in QFT, divergences appear when calculating quantum corrections to a wide variety of quantities, and the process to isolate and remove these divergences is know as renormalization. There is no unique way to renormalize a divergent quantity, but there are some procedures which work best under certain situations. The Schwinger approach to QFT allows us to formulate a rigorous way to do this in curved spacetime, by compacting all leading order corrections into a single quantity $\Gamma^{(1)}$, through the use of the heat kernel $K(\tau;x,x')$. The power of this method lies in the fact that the dynamics of the heat kernel is entirely characterized by Eq.~(\ref{SchroEq}), an equation which has been extensively scrutinized in the literature.

The renormalization of $\Gamma^{(1)}$ will be considered for differential operators of the form (leaving indexes implicit)
\eq{\label{OpeF}
\ma{F} = g^{\mu \nu} \nabla_{\mu} \nabla_{\nu} + Q(x)
}
where $\nabla_{\mu}$ is any covariant derivative (including gauge potentials or connections derived from spacetime curvature) and $Q(x)$ does not contain any derivatives with respect to spacetime indexes. The form considered in Eq.~(\ref{OpeF}) is actually one of the most general ones and all cases considered in this work will fall into that category. When one studies the flat spacetime limit of Eq.~(\ref{SchroEq}), it becomes evident that the divergent behavior comes from the integration in $\tau$. Since the heat kernel will always contain a negative imaginary part from the Feynman boundary conditions, i.e. from the substitution $m^2 \to m^2 - i\e$, it can be shown that the heat kernel is expected to decay exponentially fast for large values of $\tau$. This means the only possible source of divergences lies in the small $\tau$ regime, for which it is well know that the heat kernel posesses an expansion of the form
\eq{\label{smalltau}
K(\tau;x,x') \propto {i \over (4\pi i \tau)^{d/2}} \sum^{\infty}_{k=0} (i\tau)^k E_{k}(x)
}
where the first few coefficients $E_k(x)$ are given by
\ea{
E_0(x) & = I\,, \\
E_1(x) & = {1 \over 6} R I - Q\,, \\
E_2(x) & = \left( -{1\over 30} \square R + {1 \over 72} R^2 -{1\over 180} R^{\mu\nu}R_{\mu\nu} + {1\over 180} R^{\mu\nu\rho\si}R_{\mu\nu\rho\si}\right) I \nonumber \\
       & \hspace{5mm} + {1\over 12} W^{\mu\nu}W_{\mu\nu} + {1 \over 2} Q^2 - {1\over 6} R Q + {1\over 6}\square Q\,,
}
with
\eq{
W_{\mu\nu} = [\nabla_{\mu},\nabla_{\nu}]
}
and where $I$ denotes the identity in whatever indexes appear in the theory. Inserting the expansion (\ref{smalltau}) in Eq.~(\ref{G1func}), it becomes clear that the divergent part of $\Gamma^{(1)}$ is given by
\eq{
\textbf{divp} \, \Gamma^{(1)} = {i\over 2 (4\pi)^{d/2}} \int dv_x \sum^{\infty}_{k=0} \textrm{Tr} E_{k}(x) \, \textbf{divp} \int^{\tau_0}_{0} d\tau (i\tau)^{k-1-d/2}
}
where $\textbf{divp}$ denotes the divergent part and $\tau_0$ is a constant which we may take as small as we like. It remains to evaluate the integral in $\tau$, which gives
\eq{
\int^{\tau_0{\infty}}_{0} d\tau (i\tau)^{k-1-d/2} = (i)^{k-1-d/2} {\tau_0^{k-n/2} \over k-d/2}\,.
}
Evidently, for odd $d$ the integral will vanish when $\tau\to 0$, but it will diverge for $k=d/2$ and even $d$. One may isolate the divergences in many different ways and in this case we will employ the so-called dimensional regularization method, whereby one performs the substitution $d \to d+\e$, where $\e$ is some small quantity assumed to go to zero in the end. By doing this, we conclude that
\eq{
\textbf{divp} \int^{\tau_0{\infty}}_{0} d\tau (i\tau)^{k-1-d/2} = {2i \over \e}
}
which finally leads to the desired result
\eq{\label{divG1}
\textbf{divp} \, \Gamma^{(1)} = -{1\over \e(4\pi)^{d/2}} \int dv_x  \textrm{Tr} E_{d/2}(x)\,.
}
At leading-order, one must then guarantee that there are enough parameters in the action to compensate for the existence of the divergences given by Eq.~(\ref{divG1}). This procedure is not suitable for odd dimensionality, but will only make use of these results for even dimensions in this work.

%% file: II.2QFTinCS/5.section.tex
\section{Renormalization of Vacuum Polarization}

By now it has become clear the most pillar object to compute in QFT is the Green function. One particular quantity that deserves special attention in the area of QFTCS is the vacuum polarization $\braket{\phi^2(x)}$, defined as
\eq{
\braket{\phi^2(x)} \equiv \lim_{x\to x'} G_{\textrm{E}}(x,x')\,.
}
From the results of Sec.~(), we know that $G_{\textrm{E}}(x,x')$ represents the probability amplitude of a particle (associated to the field $\phi$) to propagate from spacetime point $x$ to $x'$. In the limit where $x'$ is taken to $x$, we obtain the probability amplitude for a particle to simply appear at $x$. Therefore, the vacuum polarization is related to rate of particle creation.

In order to understand why this quantity is of particular importance in curved spacetime scenarios, we must begin by noting that the Einstein equations that rule the curvature of spacetime are not applicable in quantum regimes. If one wishes to study the curvature of spacetime taking into account quantum field theory, one must solve instead the semiclassical limit of the Einstein equations, given by
\eq{
G_{\mu\nu} = 8\pi \braket{T_{\mu\nu}}\,,
}
where $\braket{T_{\mu\nu}}$ is the average value of the stress-energy tensor operator. Classically, the stress-energy tensor can be obtain from the action using the relation
\eq{
T_{\mu\nu} = {2 \over \sqrt{|g|}} {\de S \over \de g^{\mu\nu}}\,,
}
so to obtain the quantum field theory average, one must consider the action to be an operator. Using Eqs.~(\ref{SAP}) and (\ref{WJ}), and taking the source to zero, one finds that
\eq{
\braket{T_{\mu\nu}} = {2 \over \sqrt{|g|}} {\de \Gamma \over \de g^{\mu\nu}}\,.
}
If we now look at the leading-order form of $\Gamma$, it becomes clear that $\braket{T_{\mu\nu}}$ will be made of terms involving the Green function and derivatives of it, in the limit where $x\to x'$, i.e. everything becomes centered in calculating $\braket{\phi^2(x)}$.

Like many quantities in QFT, the vacuum polarisation is divergent and must be properly renormalized. In this work, we will calculate various instances of vacuum polarization for the same type of differential operator, given by Eq.~(\ref{OpeF}) with
\eq{\label{speQ}
Q(x) = m^2 + \xi R(x)\,,
}
where $m$ is the scalar field's mass, $R(x)$ the spacetime curvature and $\xi$ the counpling constant of the scalar field to the spacetime curvature. Since the Euclidean Green function is the integral of the heat kernel, the divergences of the vacuum polarization will be the same as for the heat kernel. We have already derived the divergent part of $K(\tau;x,x')$ in the previous section, but here we will concentrate on a method which is also applicable to the odd dimensional case, while at the same time providing a different physical perspective to the problem.

For the specific choice (\ref{speQ}), the differential equation (\ref{SchroEq}) is know to have a solution of the form
\eq{
K(\tau;x,x') = i {\Delta^{1/2}(x,x') \over (4\pi i \tau)^{d/2}} \exp\left[-i\left(m^2 \tau - {\sigma(x,x') \over 2\tau}\right)\right] F(\tau;x,x')\,,
}
where we see the presence of Synge's world function $\si$ and the Van-Vleck Morette determinant.
As for the funcion $F$, it is usually expanded as
\eq{
F(\tau;x,x') = \sum^{\infty}_{k=0} a_k(x,x') (i\tau)^k\,,
}
here the $a_k$ are called heat kernel coefficients, the first few being, in the coincidence limit $x \to x'$,
\ea{
a_0(x) & = 1 \,, \\
a_1(x) & = \left({1\over 6} - \xi R\right) \,,\\
a_2(x) & = {1\over 180} R_{\al\be\g\de}R^{\al\be\g\de} - {1\over 180} R^{\al\be}R_{\al\be} - {1\over 6} \left({1\over 5}-\xi\right)\square R + {1\over 2}\left({1\over 6}-\xi\right)^2 R^2 \,.
}
Under this assumption, we may use Eq.~() to write the Euclidean Green function as
\ea{G(x,x') & =
-  {\Delta^{1/2}(x,x') \over (4\pi i)^{d/2}}  \sum^{\infty}_{k=0} a_k(x,x') \int^{\infty}_{0} {d\tau \over \tau^{d/2}} (i\tau)^k \, \exp\left[-i\left(m^2 \tau - {\sigma(x,x') \over 2\tau}\right)\right] \\
& = -  {\Delta^{1/2}(x,x') \over (4\pi i)^{d/2}}  \sum^{\infty}_{k=0} a_k(x,x') \left(-{\partial \over \partial m^2}\right)^k \left(\int^{\infty}_{0} {d\tau \over \tau^{d/2}} \, \exp\left[-i\left(m^2 \tau - {\sigma(x,x') \over 2\tau}\right)\right]\right)\,. \\
}
Defining now the variables $z^2 = -2 m^2 \sigma$ and $u = -2im^2\tau/z$, one may express the integral in $\tau$ in the form
\eq{
\int^{\infty}_{0} {d\tau \over \tau^{d/2}} \, \exp\left[-i\left(m^2 \tau - {\sigma \over 2\tau}\right)\right] = \left(-{z \over 2im^2}\right)^{1-d/2} \int^{-i\infty}_{0} {du \over u^{d/2}} \exp\left[ {z \over 2} (u-{1/u})\right]
}
and the closed form of this integral is know to be \cite{Gradshteyn:2007}
\eq{\label{intu}
\int^{-i\infty}_{0} {du \over u^{d/2}} \exp\left[ {z \over 2} (u-{1/u})\right] = -i\pi H^{(2)}_{d/2-1}(z)
}
where $H^{(2)}_{n}(z)$ is also called a Henkel function of the second type. Since the variable $\sigma$ is free, we may express the derivatives in $m^2$ with respect to $z$ instead, and together with Eq.~(\ref{intu}) this will result in
\eq{
G(x,x') = -{\pi \Delta^{1/2}(x,x') \over (4\pi)^{d/2}}  \sum^{\infty}_{k=0} a_k(x,x') \sigma^{k+1-d/2}  \left({1 \over z}{\partial \over \partial z}\right)^k \left[z^{d/2-1} H^{(2)}_{d/2-1}(z)\right]\,.
}
Using now the useful formula related to Henkel functions
\eq{
\left({1 \over z}{\partial \over \partial z}\right)^k \left[z^{\mu} H^{(2)}_{\mu}(z)\right] = z^{\mu-k} H^{(2)}_{\mu-k}(z)
}
as well as
\eq{
H^{(2)}_{\nu}(i|z|) = 2 i^{\nu} I_{\nu}(|z|) + {2 \over \pi} i(-i)^{\nu} K_{\nu}(|z|)\,,
}
mutiplying by $i$ and keeping only the real part, one finally obtains
\eq{\label{GEz}
G_{\textrm{E}}(x,x') = {2\Delta^{1/2} \over (4\pi)^{d/2}}  \sum^{\infty}_{k=0} a_k(x,x') (2m^2)^{\nu} |z|^{-\nu} K_{\nu}(|z|)
}
where we have defined
\eq{
\nu = {d \over 2} - 1 -k\,.
}
At first, this seems like a good method to calculate the finite part of the vacuum polarization itself, but the heat kernel coefficients $a_k$ rise in complexity quite fast, having been calculated only up to $k=3$ \cite{Gilkey:1975iq}. In the end, we need the coincidence limit $x\to x'$, which implies $\si(x,x') \to 0$. Thus, to find the divergent part of the Green function in the coincidence limit, one needs to express every quantity in terms of $\si$, including the heat kernel coefficients and the Van Vleck-Morette determinant. This has been done in \cite{Christensen:1978yd}, where it is found that
\ea{
\Delta^{1/2}(x,x') & = 1+ {1\over 12} R_{\al\be} \si^{\al}\si^{\be} -{1\over 24} R_{\al\be;\g} \si^{\al}\si^{\be}\si^{\g} \nonumber \\
& \hspace{5mm} + \left({1 \over 288} R_{\al\be}R_{\g\de} + {1\over 360} R^{\rho\tau}{}_{\al\be} R_{\rho\g\tau\de} + {1\over 80} R_{\al\be;\g\de}\right)\si^{\al}\si^{\be}\si^{\g}\si^{\de} + \cdots \\
a_1(x,x') & = \left({1\over 6}-\xi\right)R - {1\over 2} \left({1\over 6} - \xi\right)R_{;\al} \si^{\al} + \bigg[-{1\over 90}R_{\al\rho}R^{\rho}{}_{\be} + {1\over 180} R^{\rho\tau}R_{\rho\al\tau\be} + {1\over 180} R_{\rho\tau\kappa\al}R^{\rho\tau\kappa}{}_{\be} \nonumber \\
& \hspace{5mm} + {1\over 120} R_{\al\be;\rho}{}^{\rho} + \left({1\over 40} - {1\over 6}\xi\right)R_{;\al\be}\bigg]\si^{\al}\si^{\be} + \cdots \\
a_2(x,x') & = -{1\over 180} R^{\rho\tau}R_{\rho\tau} + {1 \over 180}R^{\al\be\rho\tau}R_{\al\be\rho\tau} + {1\over 6} \left({1\over 5} - \xi\right)R_{;\rho}{}^{\rho} + {1\over 2} \left({1\over 6} - \xi\right)^2 R^2 + \cdots\,.
}
We now rewrite the above expressions in the form
\ea{
\Delta^{1/2} & = \Delta^{1/2}_0 + \Delta^{1/2}_1 + \Delta^{1/2}_2 + \cdots \,, \\
a_k & = a^0_k + a^1_k + a^2_k + \cdots \,,
}
where each term represents an increasing power of $\si$, and express the term $a_k \Delta^{1/2}$ as
\eq{
a_k(x,x') \Delta^{1/2} = [a_k][\Delta^{1/2}] + \sum^{\infty}_{p=1} \sum^p_{j=0} a^j_k \Delta^{1/2}_{p-j}
}
where it is custom to define $(\cdots)^0 \equiv [(\cdots)]$. Inserting this result in Eq.~(\ref{GEz}), noting that $[\Delta^{1/2}]=1$ and keeping only the divergent powers of $z$, one obtains the formula valid for even dimensions (see \cite{Thompson:2008bk} for more details)
\eq{\label{GdivE}
G_{\textrm{div}}(x,x') = {2\over (4\pi)^{d/2}} \sum^{k_d}_{k=0} \left\{ [a_k] (2m^2)^{\nu} |z|^{-\nu} K_{\nu}(|z|) + \sum^{\nu}_{n=1} \sum^{2n}_{p=1} \sum^{p}_{j=0} {2^{2n-1}(-m^2)^{\nu-n} \Gamma(n) \over \Gamma(\nu-n+1)} {a^{j}_{k}\Delta^{1/2}_{p-j} \over (\si^{\rho}\si_{\rho})^n}\right\}\,,
}
and
\eq{\label{GdivO}
G_{\textrm{div}}(x,x') = {2\over (4\pi)^{d/2}} \sum^{k_d}_{k=0} \left\{ [a_k] (2m^2)^{\nu} |z|^{-\nu} K_{\nu}(|z|) + \sum^{\nu+1/2}_{n=1} \sum^{2n}_{p=1} \sum^{p}_{j=0} {2^{2n-2}(-m^2)^{\nu+n+1} \Gamma\left(n-{1\over 2}\right) \over \Gamma\left(\nu-n+{3\over 2}\right)} {a^{j}_{k}\Delta^{1/2}_{p-j} \over (\si^{\rho}\si_{\rho})^{n-1/2}}\right\}
}
for the odd dimensional case, where we define
\eq{
k_d =
\left\{ \begin{array}{ll}
       {d-2 \over 2}\,, \quad \textrm{for} \quad d ~ \textrm{odd}\,, \\
       \\
       {d-3 \over 2}\,, \quad \textrm{for} \quad d ~ \textrm{even}\,.
\end{array}
\right.
}

%% file: II.3BTZ/main.tex
\chapter{Green functions of a massive scalar field in a BTZ spacetime}
\label{cap:chapterII3}
\input{II.3BTZ/outline.tex}
\input{II.3BTZ/1.section.tex}

\input{II.3BTZ/2.section.tex}

\input{II.3BTZ/conclusions.tex}

%% file: II.3BTZ/outline.tex
\section{Introduction}

Here we will develop the first application of QFTCS techniques, where we will consider quantum scalar field in a 3-dimensional curved spacetime, namely the Ba\~{n}ados-Teitelboim-Zanelli (BTZ) solution of the Einstein equations in $1+2$ dimensions \cite{Banados:1992wn}. This case has already been studied for a massless minimally coupled scalar field, which was shown to possess closed form answers \cite{Shiraishi:1993nu}. The massive non-minimally coupled case has not been dealt with yet in the literature, and as such it will be the one to consider in this chapter. We shall see that some analytic results are also possible, although at a much higher computational demand, which will require the derivation of new addition formulas for a ``generalized'' associated Lagrange type functions.

%% file: II.3BTZ/1.section.tex
\section{Quantum scalar field in a BTZ spacetime}

We shall consider a quantum scalar field subject to the action
\eq{\label{SBTZ}
S[\phi] = {1\over 2} \int d^3 x \sqrt{g} \left[
g^{\mu \nu} \partial_\mu \phi \, \partial_\nu \phi + m_{\phi}^2 \phi^2 + \xi R \phi^2
\right]\,,
}
which is the type considered in Chapter 8, through Eq.~(\ref{OpeF}). The field will be taken in the presence of a background geometry characterized by a BTZ metric of a non-rotating black hole. In the $(\tau,r,\phi)$ coordinates with Euclidean time, the latter is given by
\eq{
ds^2 = f(r) d\tau^2 + {1 \over f(r)} dr^2 + r^2 d\theta^2
}
with
\eq{
f(r) = \lambda r^2 - M\,,
}
where $M$ is the mass of the black hole and $\la$ is the cosmological constant. The horizon radius is $r_h = \sqrt{M/\la}$ and the Ricci curvature is given by $R = - 6 \la$. The Euclidean Green function is a solution of
\eq{\label{DifEqBTZ}
(\square_E - m_{\phi}^2 - \xi R) G_E(x,x') = - {\de^{(3)}(x-x') \over \sqrt{g}}
}
and we shall want the solution which is at thermal equilibrium with the black hole, i.e. we want the temperature to be equal to the Hawking temperature of the BTZ black hole, given by
\eq{
T = {\kappa \over 2\pi}
}
where $\kappa = \sqrt{M \la}$ is the surface gravity of the black hole. According to Eqs.~(\ref{GExpN}) and (\ref{GExpL}) in the particular case $N=0$, such an expansion for the Euclidean Green function may be shown to have the form
\eq{
G_\textrm{E}(x,x') = {\kappa \over 4\pi^2} \sum^{\infty}_{n = -\infty} e^{i n \kappa \Delta \tau} \sum^{\infty}_{m = -\infty} e^{i m \Delta \theta} G_{nm}(r,r')\,.
}
Upon inserting the expanded form of $G_{\textrm{E}}(x,x')$ in Eq.~(\ref{DifEqBTZ}), one obtains the equation for the radial Green function
\eq{
\left\{{d^2 \over dr^2}+\left({1 \over r} + {f' \over f}\right){d \over dr} - \left( {n^2 \kappa^2 \over f} + {m^2 \over r^2 f} + {m^2_{\phi}+\xi R \over f}\right)\right\} G_{nm}(r,r') = -{\de(r-r') \over r f}\,.
}
Standard results in differential equations allow us to express the general solution of this equation as
\eq{\label{radGFBTZ}
G_{nm}(r,r') = - {1 \over r f}{p_{nm}(r_<) q_{nm}(r_>) \over \ma{W}_r\{p_{nm},q_{nm}\}}
}
where the $p$ and $q$ are solutions of the homogeneous equation
\eq{\label{homoBTZ}
\left\{{d^2 \over dr^2}+\left({1 \over r} + {f' \over f}\right){d \over dr} - \left( {n^2 \kappa^2 \over f} + {m^2 \over r^2 f} + {m^2_{\phi}+\xi R \over f}\right)\right\} \chi_{nm}(r) = 0\,,
}
and $\ma{W}_r\{p_{nm},q_{nm}\}$ is the Wrosnkian of the two solutions with respect to the variable $r$. The solutions $\chi_{nm}=p_{nm}$ and $\chi_{nm}=q_{nm}$ will denote the ones regular at the horizon and infinity, respectively. We also adopt the notation $r_< \in \textrm{min}\{r,r'\}$ and $r_> \in \textrm{max}\{r,r'\}$. The task now is then to find the for of the homogeneous solutions. This can be done by defining the new variable
\eq{\label{defY}
y = 1- {r^2_h \over r^2}
}
and expressing the homogeneous equation solutions as
\eq{
\chi_{nm}(r) = y^{\mu/2} (1-y)^{{1\over 4} - {\e \over 2}} w(y)\,.
}
This change of variables changes Eq.~(\ref{homoBTZ}) into one of hypergeometric form
\eq{\label{homoY}
y(1-y){d^2 w \over dy^2} + \left[c-(a+b+1)y\right]{dw \over dy} -ab w =0
}
with
\ea{
a & = {\mu \over 2} - {\nu \over 2} - {\e \over 2} \,, \\
b & = {1 \over 2}+{\mu \over 2}+{\nu \over 2} - {\e \over 2}\,, \\
c & = 1+ \mu\,,
}
and
\ea{
\mu & = n\,, \\
\nu & = -{1\over 2} + i{m \over \sqrt{M}} \,, \\
\left({1\over 2} + \e\right)^2 & = 1+{{m^2_{\phi} \over \la}-6\xi}\,. \label{edef}
}
A solution of this equation can always be written as
\eq{
w(y) = c_1 w_1(y) + c_2 w_2(y)\,,
}
where $w_1(y)$ and $w_2(y)$ are two of Kummer's 24 solutions \cite{Gradshteyn:2007}. A general solution of Eq.~(\ref{homoBTZ}) is thus given by
\eq{
\chi_{nm}(r) = y^{\mu/2} (1-y)^{{1\over 4} - {\e \over 2}} (c_1 w_1(y) + c_2 w_2(y))\,.
}
Now, any of Kummer's 24 solutions to the hypergeometric equation can be expressed as a linear combination of two other of those 24 solutions. We will use this fact and take linear combinations $d_1$ and $d_2$ of the coefficients $c_1$ and $c_2$ in such a way that the general solution of Eq.~(\ref{homoY}) is given by
\eq{\label{genSol}
\chi_{nm}(r) = z^{{1\over 2}} (d_1 \, \ma{P}^{-n}_{\nu}(z) + d_2 \, \ma{Q}^{n}_{\nu}(z))\,,
}
where we have defined the variable $z = \sqrt{1-y}$ and the functions
\ea{
\ma{P}^{n}_{\nu}(z) = & \, {2^{n} (z^2-1)^{-{n\over 2}} \over \Gamma(1-n)}F\left[{1 \over 2}-{n \over 2}+{\nu \over 2}-{\e \over 2},-{n\over 2}-{\nu \over 2}-{\e \over 2}, 1-n, 1-z^2\right] \label{GenP} \,,\\
\ma{Q}^{n}_{\nu}(z) = & \, (-1)^{n} 2^{n-1} {\Gamma\left( 1+{n \over 2} + {\nu \over 2} + {\e \over 2}\right) \Gamma\left( {1 \over 2} +{n \over 2} + {\nu \over 2} -{\e \over 2} \right) \over \Gamma\left( {3 \over 2} + \nu\right)} z^{-1-n-\nu+\e} (z^2-1)^{{n\over 2}} \nonumber \\
& \times F\left[ 1+{n\over 2}+{\nu \over 2}+{\e \over 2}, {1\over 2}+{n\over 2}+{\nu\over 2}-{\e \over 2}, {3 \over 2} + \nu, {1 \over z^2} \right]\,, \label{GenQ}
}
which reduce to the associated Legendre functions in the limit $\e = 0$ and thus to the correct result for the conformally coupled case \cite{Shiraishi:1993nu}. In choosing Eq.~(\ref{genSol}), we have used the solutions
\ea{
w_1(y) & = F\left[a,b;c;y\right] \nonumber \\
       & = F\left[{1\over 2} + {\mu \over 2} + {\nu \over 2} - {\e \over 2}, {\mu \over 2} - {\nu \over 2} - {\e \over 2}; 1+\mu; y\right] \label{w1}\,, \\
w_2(y) & = (1-y)^{-b} F\left[b,c-a;1-a+b;{1\over 1-y}\right] \nonumber \\
       & = (y-1)^{-1/2+\e/2-\mu/2-\nu/2} F\left[1 + {\mu \over 2} + {\nu \over 2} + {\e \over 2}, {1\over 2}    + {\mu \over 2} + {\nu \over 2} - {\e \over 2} ; {3 \over 2} + \nu; {1\over 1-y} \right] \label{w2} \,,
}
for the hypergeometric equation. As for the Wronskian of the homogeneous solutions, it is given by
\ea{
\ma{W}\{\ma{P}^{-n}_{\nu}(z),\ma{Q}^{n}_{\nu}(z)\} = (-1)^n {z^{2\e} \over z^2-1}\,, \nonumber \\
}
and using the following formulas derived in Appendix \ref{ap:b}
\ea{
\ma{P}^{n}_{\nu}(z) & = 2^{2n} {\Gamma\left(1 + {n\over 2} + {\nu \over 2} + {\e \over 2}\right) \Gamma\left({1 \over 2} + {n\over 2} + {\nu \over 2} - {\e \over 2}\right)\over \Gamma\left(1 - {n\over 2} + {\nu \over 2} + {\e \over 2}\right) \Gamma\left({1 \over 2} - {n\over 2} + {\nu \over 2} - {\e \over 2}\right)} \, \ma{P}^{-n}_{\nu}(z)\,, \label{parP} \\
\ma{Q}^{n}_{\nu}(z) & = 2^{2n} {\Gamma\left(1 + {n\over 2} + {\nu \over 2} + {\e \over 2}\right) \Gamma\left({1 \over 2} + {n\over 2} + {\nu \over 2} - {\e \over 2}\right)\over \Gamma\left(1 - {n\over 2} + {\nu \over 2} + {\e \over 2}\right) \Gamma\left({1 \over 2} - {n\over 2} + {\nu \over 2} - {\e \over 2}\right)} \, \ma{Q}^{-n}_{\nu}(z)\,, \label{parQ}
}
we may obtain Wronskians associated to different linear combinations. Thus, for any two solutions
\ea{
p_{nm}(r) & = z^{{1\over 2}} (a_1 \, \ma{P}^{-n}_{\nu}(z) + a_2 \, \ma{Q}^{n}_{\nu}(z))\,, \nonumber \\
q_{nm}(r) & = z^{{1\over 2}} (b_1 \, \ma{P}^{-n}_{\nu}(z) + b_2 \, \ma{Q}^{n}_{\nu}(z))\,, \nonumber \\
}
we will have the general Wronskian
\eq{
\ma{W}_{z}\{p_{nm},q_{nm}\} = (a_1 b_2 - a_2 b_1) {(-1)^{n+1} \over r_h}  {z^{2\e+3} \over z^2-1}\,.
}
Since all quantities appearing in Eq.~(\ref{radGFBTZ}) have been obtain, the two-point Green function comes out, in general, as
\eq{
G_E(x,x') = {\kappa \over 4\pi^2} \sum^{\infty}_{n = -\infty} e^{i \kappa n \Delta \tau} \sum^{\infty}_{m = -\infty} e^{i m \Delta \theta} {(-1)^n \over M} (z\,z')^{1/2-\e} {(a_1 \, \ma{P}^{-n}_{\nu}(z) + a_2 \, \ma{Q}^{n}_{\nu}(z)) (b_1 \, \ma{P}^{-n}_{\nu}(z') + b_2 \, \ma{Q}^{n}_{\nu}(z')) \over (a_1 b_2 - a_2 b_1)}\,.
}
Finally, one must take a note of rigor and recall that the solutions (\ref{w1}) and (\ref{w2}) are not properly defined in the interval $0<z<1$ \cite{Erdelyi:1953} and, because of Eq.~(\ref{defY}), this is exactly the interval we are interested in. To resolve this issue, we take the established convention for the Ferrer functions
\ea{
\ma{P}^{n}_{\nu}(z) & = {1 \over 2}\left(e^{i \pi \mu/2}\ma{P}^{n}_{\nu}(z+i 0) + e^{-i \pi \mu/2}\ma{P}^{n}_{\nu}(z-i 0)\right)\,, \\
\ma{Q}^{n}_{\nu}(z) & = {1 \over 2}e^{-i \pi \mu}\left(e^{-i \pi \mu/2}\ma{Q}^{n}_{\nu}(z+i 0) + e^{i \pi \mu/2}\ma{Q}^{n}_{\nu}(z-i 0)\right)\,,
}
which are properly defined for any $z$ satisfying $|z|<1$. Defining $z = \cos\rho$, $z' = \cos\rho'$, with $\rho \in [0,{\pi \over 2}]$, we finally have the general form for two-point Green function
\ea{
G_E(x,x') & = {\kappa \over 4\pi^2} \sum^{\infty}_{n = -\infty} e^{i \kappa n \Delta \tau} \sum^{\infty}_{m = -\infty} e^{i m \Delta \theta} {(-1)^n \over M} \cos\rho^{1/2-\e}{\cos\rho'}^{1/2-\e} \nonumber \\
& \hspace{5mm} \times {(a_1 \, \ma{P}^{-n}_{\nu}(\cos\rho) + a_2 \, \ma{Q}^{n}_{\nu}(\cos\rho)) (b_1 \, \ma{P}^{-n}_{\nu}(\cos\rho') + b_2 \, \ma{Q}^{n}_{\nu}(\cos\rho')) \over (a_1 b_2 - a_2 b_1)}\,.
}
The conformally coupled case corresponds to $a_2 = b_1 = 0$, $a_1 = b_2 = 1$ and $\e = 0$, reducing to previously obtained results \cite{Shiraishi:1993nu}. The full two-point Green function is obtained by performing the two sums in the energy and angular modes, which will be pursued in the next section.

%% file: II.3BTZ/2.section.tex
\section{Two-point Green functions and their coincidence limits}

We now wish to obtain the full for of the Green function by explicitely perform the summations. This has been done in \cite{Shiraishi:1993nu} for the conformally coupled case, where the homogenous solutions where associated Legendre polynomials. For that particular case, there are well known addition formulas that allow the explicit summation straightforwardly. In particular, for $z_1 = \cos \rho_1$, $z_2=\cos\rho_2$ we have
\ea{
\sum^{\infty}_{n=-\infty} (-1)^k P^{-n}_{\nu}(\cos\rho_1) P^{n}_{\nu}(\cos\rho_2)\cos(n \varphi) & = P_{\nu}(\cos\rho_1\cos\rho_2 + \sin\rho_1\sin\rho_2\cos \varphi)\,, \label{AddP}\\
\sum^{\infty}_{k=-\infty} (-1)^k P^{-n}_{\nu}(\cos\rho_1) Q^{n}_{\nu}(\cos\rho_2)\cos(n \varphi) & = Q_{\nu}(\cos\rho_1\cos\rho_2 + \sin\rho_1\sin\rho_2\cos \varphi)\,. \label{AddQ}
}
In this case, however, in the author's best knowledge, there are no general addition formulas for the ``associated Legendre''-like functions defined in Eqs.~(\ref{GenP}) and (\ref{GenQ}). The analog general case with the substitutions $P^{n}_{\nu}\to \ma{P}^{\mu}_{\nu}$ and $Q^{n}_{\nu}\to \ma{Q}^{\mu}_{\nu}$ is beyond reach for now but Appendix \ref{ap:b} provides the derivation of some particular cases which are in fact the most necessary. First of all, we will restrict ourselves to $\e \in \mathbb{Z}$ which induces a restriction of the mass with respect to the coupling constant through Eq.~(\ref{edef}), but still allows for an infinity of cases other than the minimally coupled one. We will also only consider $\varphi \in \{0,\pi\}$. With these restrictions, the resulting addition formulas for the ``generalized associated Legendre functions'' $\ma{P}$ and $\ma{Q}$ may be summarised from Appendix \ref{ap:b} as follows. Defining $z = \cos\rho$ and $z' = \cos(\rho+\delta\rho)$, we have
\ea{
\sum^{\infty}_{n=-\infty} (-1)^n \ma{P}^{-n}_{\nu}(\cos\rho) \ma{P}^{n}_{\nu}(\cos\rho') & = (\cos\rho \cos(\rho + \delta\rho))^{\e} \sum^{2\e}_{k=0}\sum^{k}_{l=0} c_{\nu,kl} \cos(\rho + \delta\rho)^{-l} \cos\rho^{-k+l} \nonumber \\
& \times \left({\sin(\delta\rho) \over 2\sin(\rho + \delta\rho)}\right)^{k} \sum^{\infty}_{p=0} {(k)_p \over p!} \left({\sin\rho \over \sin(\rho + \delta\rho)}\right)^p P_{\nu-l-p}(\cos(\delta\rho))\, \label{nPP}
}
\ea{
\sum^{\infty}_{n=-\infty} \ma{P}^{-n}_{\nu}(\cos\rho) \ma{P}^{n}_{\nu}(\cos\rho') & = (-\cos\rho \cos(\rho + \delta\rho))^{\e} \sum^{\infty}_{k=0}\sum^{\e}_{l=0} d_{\nu,kl} \cos(\rho + \delta\rho)^{-l} \cos\rho^{-k+l+\nu} \nonumber \\
& \times \left({\sin(\delta\rho) \over 2\sin(\rho + \delta\rho)}\right)^{k-\nu} \sum^{\infty}_{p=0} {(k-\nu)_p \over p!} \left({\sin\rho \over \sin(\rho + \delta\rho)}\right)^p P_{\nu-l-p}(\cos(\delta\rho))\, \label{pPP}
}
\ea{
\sum^{\infty}_{n=-\infty} (-1)^n \ma{P}^{-n}_{\nu}(\cos\rho) \ma{Q}^{n}_{\nu}(\cos\rho') & = (\cos\rho \cos(\rho + \delta\rho))^{\e} \sum^{2\e}_{k=0}\sum^{k}_{l=0} c_{\nu,kl} \cos(\rho + \delta\rho)^{-l} \cos\rho^{-k+l} \nonumber \\
& \times \left({\sin(\delta\rho) \over 2\sin(\rho + \delta\rho)}\right)^{k} \sum^{\infty}_{p=0} {(k)_p \over p!} \left({\sin\rho \over \sin(\rho + \delta\rho)}\right)^p Q_{\nu-l-p}(\cos(\delta\rho))\, \label{nPQ}
}
\ea{
\sum^{\infty}_{n=-\infty} \ma{P}^{-n}_{\nu}(\cos\rho) \ma{Q}^{n}_{\nu}(\cos\rho') & = (-\cos\rho \cos(\rho + \delta\rho))^{\e} \sum^{\infty}_{k=0}\sum^{\e}_{l=0} d_{\nu,kl} \cos(\rho + \delta\rho)^{-l} \cos\rho^{-k+l+\nu} \nonumber \\
& \times \left({\sin(\delta\rho) \over 2\sin(\rho + \delta\rho)}\right)^{k-\nu} \sum^{\infty}_{p=0} {(k-\nu)_p \over p!} \left({\sin\rho \over \sin(\rho + \delta\rho)}\right)^p Q_{\nu-l-p}(\cos(\delta\rho))\, \label{pPQ}
}
where the coefficients $c_{\nu,kl}$ and $d_{\nu,kl}$ are given by
\ea{
c_{\nu,kl} & = {(1+\e)_{l}(-\e)_{l}(1+\e)_{k-l}(-\e)_{k-l} \over (1+\nu)_{k-l} (-\nu)_{l} (k-l)! \, l!} \,, \\
d_{\nu,kl} & = {\Gamma(1+\nu) \Gamma(1-\nu+\e) \over \Gamma(1-\nu) \Gamma(1+\nu+\e)}{(1+\e)_{l}(-\e)_{l}(1-\nu+\e)_{k-l}(-\nu-\e)_{k-l} \over (1-\nu)_{k-l} (-\nu)_{l} (k-l)! \, l!} \,.
}
Some comments regarding these addition formulas are in order. First, all of the formulas reduce to the proper limits of Eqs.~(\ref{AddP}) and (\ref{AddQ}) in the limit $\e\to 0$. Secondly, note that $\e$ appears as the upper index in all summations, meaning one needs to include an increasing amount of terms proportionally to $\e$. Finally, even for small $\e$, some of the formulas, namely (\ref{pPP}) and (\ref{pPQ}), are not as usefull since they still involve infinite sums. This can be partially resolved by using integral representations of the Legendre functions, for which at least the summation in the $p$ index can be performed analitically. Despite this drawback, the right hand sides of all formulas still converge much faster than the left hand sides.

Another important point to take into account is the coincidence limit $\de \rho \to 0$ of the addition formulas. These are derived in Appendix \ref{ap:b} and are listed as follows:
\ea{
\sum^{\infty}_{n=-\infty} (-1)^n \ma{P}^{-n}_{\nu}(\cos\rho) \ma{P}^{n}_{\nu}(\cos\rho) & = (\cos\rho)^{2\e}\,, \label{nPPc} \\
\sum^{\infty}_{n=-\infty} \ma{P}^{-n}_{\nu}(\cos\rho) \ma{P}^{n}_{\nu}(\cos\rho) & = (-1)^{\e}(\cos\rho)^{2\e} \sum^{\infty}_{k=0} \left(\sum^{\e}_{l=0} d_{\nu,kl}\right) (2\cos\rho)^{-k+\nu} P_{\nu-k}(\cos\rho)\,, \label{pPPc} \\
\sum^{\infty}_{n=-\infty} (-1)^n \ma{P}^{-n}_{\nu}(\cos\rho) \ma{Q}^{n}_{\nu}(\cos\rho) & = (\cos\rho)^{2\e} \left\{\lim_{\delta\rho \to 0} Q_{\nu}(\cos(\delta\rho)) + \sum^{2\e}_{k=1} \left(\sum^{k}_{l=0} c_{\nu,kl}\right) (2\cos\rho)^{-k}  Q_{k-1}(\cos\rho)\right\} \,, \label{nPQc} \\
\sum^{\infty}_{n=-\infty} \ma{P}^{-n}_{\nu}(\cos\rho) \ma{Q}^{n}_{\nu}(\cos\rho') & = (-1)^{\e}(\cos\rho)^{2\e} \sum^{\infty}_{k=0} \left(\sum^{\e}_{l=0} d_{\nu,kl}\right) (2\cos\rho)^{-k+\nu} Q_{\nu-k}(\cos\rho)\,. \label{pPQc}
}
These formulas are essential if one wishes to calculate vacuum polarisations for any boundary conditions of choice. In the coincidence limit, the only term which automatically originates a divergence at every point in spacetime is the first term in the right hand side of Eq.~(\ref{nPQc}). Its regularisation has already been thoroughly analysed in \cite{Shiraishi:1993nu}.

%% file: II.3BTZ/conclusions.tex
\section{Conclusions}

In this chapter we have obtain the most general analytic form of thermal Green functions in a non-rotating BTZ spacetime, and the respective coincidence limits of every possible term it might contain. The majority of the work is contained in Appendix \ref{ap:b}, where new addition formulas had to be derived in order to express the sum over the energy modes as a series of simpler Legendre functions. This shows clearly the sharp increase in difficulty that is present when one wishes to consider massive fields.

The results obtained can be applied to any boundary conditions of choice, and it entirely depends on the physical context at hand. However, we had to restrict to cases where the mass of the field and its coupling to the curvature were related by integer values of the parameter $\e$ of Eq.~(\ref{edef}). A further generalisation to any values of the free parameters is an interesting problem, although a very complicated one.

%% file: II.4Lifshitz/main.tex
\chapter{Vacuum polarization in a Lifshitz spacetime}
\label{cap:chapterII4}
\input{II.4Lifshitz/outline.tex}
\input{II.4Lifshitz/1.section.tex}

\input{II.4Lifshitz/2.section.tex}

\input{II.4Lifshitz/3.section.tex}

\input{II.4Lifshitz/4.section.tex}

\input{II.4Lifshitz/5.section.tex}

%% file: II.4Lifshitz/outline.tex
\section{Introduction}

In this chapter we will perform the first calculation of a vacuum polarization of this thesis. Unlike in Chapter 9, the final result will not be analytic, as we will have to numerically solve the main differential equations of interest. This is a delicate problem since renormalization requires us to carefully handle the divergences that naturally appear in the calculations.

We will be considering a quantum scalar field in a general black hole solution which asymptotes to a Lifshitz background. The motivation for this choice of spacetime geometry comes from the extension of the AdS/CFT correspondence to nonrelativistic condensed matter systems \cite{Hartnoll:2009sz,Herzog:2009xv}, which essentially encompasses the idea that a strongly coupled theory can be described
in terms of a gravitational weakly coupled dual. We will not pursue applications of this correspondence, however, limiting ourselves instead to the computations from the gravity side.

%% file: II.4Lifshitz/1.section.tex
\section{Quantum scalar field in a Lifshitz spacetime}

Moving on to the problem at hand, let us consider a quantum scalar field described by the action
\eq{\label{SLif}
S[\phi] = {1\over 2} \int d^4x \sqrt{g} \left[
g^{\mu \nu} \partial_\mu \phi \, \partial_\nu \phi + m_{\phi}^2 \phi^2 + \xi R \phi^2
\right]\,,
}
where we included both a mass term $m$ and a coupling $\xi$ to the curvature. This field will be propagating in a Lifshitz black hole geometry
\eq{
ds^2 = \ell^2 \left(r^{2z} {f}(r) d\tau^2 + r^{-2} {u}(r) d^2r + r^2 d^2\Omega_2 \right)\,.
\label{lbh}
}
where $z$ is a parameter related to the scaling of the theory from the condensed matter side of the AdS/CFT correspondence. Explicit solutions of black hole metrics of the type (\ref{lbh}) have been constructed in a variety of models mostly using numerical approximations, with some examples obtained analytically (see, for example, Refs.~\cite{Taylor:2008tg,Pang:2009ad,Danielsson:2009gi,Mann:2009yx,Bertoldi:2009vn,Balasubramanian:2009rx,Brynjolfsson:2009ct,Pang:2009pd}). Black hole solutions with Lifshitz asymptotics were constructed in Ref.~\cite{Taylor:2008tg} and later generalized in Ref.~\cite{Pang:2009ad}. Reference~\cite{Danielsson:2009gi} obtained black hole solutions for $z=2$ and $d=2$ by means of numerical approximation in the same model field theory proposed in \cite{Kachru:2008yh} (Einstein gravity with a cosmological constant plus two- and three-form gauge fields). A class of Lifshitz topological black holes were obtained in \cite{Mann:2009yx} also for $z=2$ and $d=2$. Lifshitz black holes with arbitrary $z$ were obtained in Refs.~\cite{Bertoldi:2009vn}. Other analytical solutions were found in Ref.~\cite{Balasubramanian:2009rx}. Charged solutions with arbitrary $z$ have been obtained in Ref.~\cite{Brynjolfsson:2009ct} and generalizations obtained in Ref.~\cite{Pang:2009pd}. Solutions with $z=3/2$ have been obtained in Ref.~\cite{Cai:2009ac}. All the above cited solutions have the structure of Eq.~(\ref{lbh}).

For the purpose of this work, it will simply be a free parameter, and $\ell$ is a length scale associated to the physical context. In the following, we will limit ourselves to consider a probe scalar, therefore the underlying theory leading to the solution (\ref{lbh}) will be unimportant. The form of the metric functions $f$ and $u$ will not be specified (except for their asymptotic behavior) and the analytical results that we will present are valid in general. Also, the computations will be carried out for a general value of $z$, which we will only specify at the end in the numerical evaluations, used to illustrate the results for specific examples. The method presented here refines and generalizes a similar one developed to compute the same quantity for AdS black holes \cite{Flachi:2008sr} that corresponds to $z=1$ and $f=u^{-1}$ and which the more general results presented here should reproduce.

The Euclidean Green function for an action of the type (\ref{SLif}) will obey a differential equation exactly of the type (\ref{OpeF})
\eq{
\left(\Box - m_{\phi}^2 -\xi R \right) G_E(x,x') = - {\delta^{(4)}(x-x')\over \sqrt{|g|}}\,,
}
and spherical symmetry allows one to express the Green's function through Eq.~(\ref{GExpL}) as
\eq{
G_{\rm E}(x,x') = {\kappa \over 4\pi^2} \sum^{\infty}_{n=-\infty} \sum^{\infty}_{l=0} \left(l+{1\over 2}\right) e^{i\omega_n \Delta\tau} P_l(\cos\gamma)G_{nl}(r,r')\,,
\label{Gansatz}
}
where $\cos\gamma =\cos\theta\cos\theta' + \sin\theta\sin\theta' \cos(\varphi-\varphi')$ and $P_{l}$ are the Legendre polynomials. Using Eqs.~(\ref{deTau}) and (\ref{deO}), it is straightforward to obtain the differential equation for the radial Green's function
\eq{
\left[{d^2\over dr^2}
+ \left({3+z \over r} + {f'\over 2f} - {u'\over 2u}\right) {d\over dr}
-{u \over r^2}\left({\omega_n^2 \over r^{2z}f} +{l(l+1)\over r^2} +m_{\phi}^2 +\xi R \right)
\right] G_{nl}(r,r') =-\sqrt{{u \over f}} {\delta(r'-r) \over r^{z+3}}\,.
\label{eqt}
}
We have set $\ell=1$, thus fixed our units according to this choice. The solutions to the above equation can be expressed in terms of those of the homogeneous equation denoted here by $\chi_{nl}(r)$, where $n$ and $l$ are the radial and angular quantum numbers, respectively. Following the usual notation, the solutions will be indicated as $p_{nl}(r)$ and $q_{nl}(r)$ according to the regularity at the horizon and at infinity

At infinity, the geometry has to recover the Lifshitz solution so $f\rightarrow 1$ and $u \rightarrow 1$. Therefore, in the asymptotically far region, the homogeneous equation becomes
\eq{ \label{homsol}
\left[{d^2\over dr^2}
+ {3+z\over r} {d\over dr} -\left({m_{\phi}^2+\xi R_{\infty} \over r^2} \right)
\right] q_{nl}(r) =0~.
}
where $R_{\infty}= -2(z+1)^2 - 4$ is the Ricci scalar at large distance. The solution asymptotically is then
\eq{\label{homsol2}
q_{nl}(r) \sim r^{\Delta_\pm}\,,
}
where
\eq{
{\Delta_\pm}=
-{z+2\over 2} \pm
\sqrt{ \left({z+2 \over 2}\right)^2+m_{\phi}^2+\xi R_{\infty} }\,.
}
Some care should be paid in selecting the correct solutions, or, in other words, the correct range of parameters. In the present case, one can easily observe that the parameters $m$, $\xi$ and $z$ (we assume $z \geq 1$ and $\xi \geq 0$) must satisfy the relation
\eq{
m_{\phi}^2 \geq -\left(
{z+2\over 2}
\right)^2 + 2 \xi \left((z+1)^2+2\right) \equiv \mu_\star^2,
}
and, in order for the solution asymptoting for $r^{\Delta_-}$ to fall off sufficiently rapidly, the condition
\eq{
m_{\phi}^2 \leq 2 \xi \left((z+1)^2+2\right) = \mu_\star^2 + \left({z+2 \over 2}\right)^2
}
should also be satisfied. Within the region $\mu_\star^2 \leq m_{\phi}^2 \leq \mu_\star^2 + ((z+2)/2)^2$ both solutions are acceptable, while for $m_{\phi}^2 \geq \mu_\star^2 + ((z+2)/2)^2$ only the
$\Delta_-$ solution falls off sufficiently rapidly. Setting $z=2$ and $\xi=0$ recovers known bounds. In the following, we will assume that the parameters satisfy the second inequality and take the solution relative to $\Delta_-$ as the only one acceptable.

In the near horizon limit, we expect that $g_{\tau \tau} = r^{2z} f \to 0$ and $g_{rr} = u/r^2 \to \infty$ while at the same time $g_{\tau\tau}g_{rr} = r^{2z-2}f u \sim 1$. One way to find a solution in this limit is to rescale the coordinates as
\eq{
dr_* = dr/f^*
}
with
\eq{
f^*=r^{z+1}\sqrt{{f \over u}}~,
}
which allows us to rewrite the homogeneous equation in the form
\eq{
\bigg[ {d^2 \over dr_*^2}  - (r^{2z}f)\bigg( {l(l+1) \over r^2} + f^{*'} {r^{z+1} \over \sqrt{f\,u}} +m_{\phi}^2+\xi R \bigg) - \omega^2_n\bigg](r\chi_{nl}(r)) = 0~,
}
where the derivative is with respect to the variable $r$. Thus the horizon limit implies that
\eq{
\bigg[ {d^2 \over dr_*^2} - \omega^2_n\bigg](r p_{nl}(r)) = 0~,
}
leading to
\eq{
p_{nl}(r) \sim {e^{\pm \omega_n r_*} \over r}~.
}
The case with the $+$ sign is the solution regular at the horizon. The radial Green's function can then be shown to have the form
\eq{
G_{nl}(r',r)={1 \over r^{z+3}} \sqrt{u\over f}~{p_{nl}(r_<)q_{nl}(r_>)\over q_{nl}p'_{nl}-p_{nl}q'_{nl}}
}
where the primes denote differentiation with respect to the variable $r$. The Green's function can thus be expressed in the following form:
\eq{
\label{G2}
G_E(x,x') = {\kappa \over 4\pi^2} \sum^{\infty}_{n=-\infty} \sum^{\infty}_{l=0} \left(l+{1\over 2}\right) e^{i\omega_n \Delta \tau} P_l(\cos\gamma) {1 \over r^{z+3}}\sqrt{u\over f} {p_{nl}(r_<)q_{nl}(r_>)\over q_{ln}(r)p'_{ln}(r)-p_{ln}(r)q'_{ln}(r)}~.
}
Due to the diverging behavior, it is not possible to evaluate numerically the above expression before regularization. To by-pass this problem we shall adopt the standard procedure of approximating the solutions by means of a WKB approach and cast (\ref{G2}) in a form suitable for renormalization.
The WKB form of the solutions is
\ea{\label{WKBansatz}
p_{nl}(r) & = r^{a} W^{b} \exp\left({+ \int_{r_s}^{r} W^{c}(u) H(u)} du\right)\,, \\
q_{nl}(r) & = r^{a} W^{b} \exp\left({- \int_{r_s}^{r} W^{c}(u) H(u)} du\right)\,,
}
where $H(u)= u^{d} g^{e}(u) f^{o}(u)$. By inserting this ansatz in the homogeneous equation, we are left with the associated differential equation for the function $W$. We choose the powers of each term in Eq.~(\ref{WKBansatz}) in such a way so as to eliminate the terms with ($\pm$)-signature. Additionally, we require that the term $r^a$ be consistent with the limit of Eq.~(\ref{homsol2}). All of these restrictions amount to setting
\eq{
a=-(z+2)/2~,~~b=-1/2~,~~c=1~, d=-1~,~~e=1/2~,~~o=-1/2~,
}
and the homogeneous equation becomes
\eq{\label{wkbeqt}
W^2 = \Psi + a_1 {W'\over W} +a_2 {W^{'2}\over W^2} +a_3 {W^{''}\over W}\,,
}
where
\ea{
\Psi & = \left[\left(l+{1\over 2}\right)^2-{1\over 4}\right] {f \over r^2}+{\omega_n^2\over r^{2z}} + \si \,, \\
\sigma & = \left(m_{\phi}^2 +\xi R\right)f +\left({z\over 2}+1\right)^2 {f\over u}
+ {r\over 2}\left(1+ {z \over 2}\right) {f'\over u} - {r \over 2}\left(1+{z \over 2}\right) {f u'\over u^2}\,,
}
and
\ea{
a_1 & = {r\over 2} {f\over u} -{r^2\over 4} {f \over u} \left[{u'\over u} - {f'\over f}\right]\,, \\
a_2 & = -{3r^2\over 4} {f\over u}\,, \\
a_3 & = {r^2\over 2}{f\over u}\,.
}
One may now choose to express the solution iteratively as
\ea{\label{SolIt}
W=W^{(0)} + W^{(1)} + \cdots\,,
}
where we will truncate at next-to-leading order. To find the coefficients of each order, we introduce an auxiliary parameter $\la$, multiply each term of Eq.~(\ref{SolIt}) by increasing powers of $\la$, substitute in Eq.~(\ref{wkbeqt}), expand in powers of $\la$ and equate equal powers on both sides. One may then show that the general iterative solution will be of the form
\eq{\label{Wwkb}
{1\over W} = {1\over \sqrt{\Psi}} \left(1 + \de_1 \Psi + \de_2 \Psi + \cdots\right)\,,
}
where $\de_n \Psi$ represents a term corresponding to the $n^{th}$ other approximation. We choose to represent the inverse of $W$ since it is the quantity which will appear in the calculations, as we shall see shortly. For the first two orders, we have
\ea{
\de_1 \Psi & = - {a_1 \over 4} {\Psi'\over \Psi^2} + \left( {a_3-a_2\over 8} \right) {\Psi'^2\over \Psi^3} - {a_3 \over 4} {\Psi^{''}\over \Psi^2} \,, \label{d1psi} \\
\de_2 \Psi & = -\frac{a_1^2 }{8 } {\Psi'' \over \Psi^3} + \frac{11 a_1^2}{32} {\Psi'^2 \over \Psi^4} + \left(\frac{19 a_1 a_3}{16}-\frac{a_1 a_2}{4}\right) \frac{\Psi'' \Psi' }{\Psi^4} + \left(\frac{17 a_1 a_2}{32}-\frac{25 a_1 a_3}{32}\right) \frac{\Psi'^3}{\Psi^5} - \frac{a_1 a_3}{4} {\Psi^{(3)} \over \Psi^3} \nonumber \\
& + \left(-\frac{a_2^2}{8}+\frac{41 a_2 a_3}{32}-\frac{45 a_3^2}{32}\right) \frac{\Psi'' \Psi'^2 }{\Psi^5} + \left(\frac{27 a_2^2}{128}-\frac{51 a_2 a_3}{64}+\frac{75 a_3^2}{128}\right) \frac{\Psi'^4 }{\Psi^6} + \left(\frac{15 a_3^2}{32}-\frac{a_2 a_3}{8}\right) \frac{\Psi''^2 }{\Psi^4} \nonumber \\
& + \left(\frac{a_3^2}{2}-\frac{a_2 a_3}{4}\right) \frac{\Psi^{(3)} \Psi'}{\Psi^4} - \frac{a_3^2 }{8 } {\Psi^{(4)} \over \Psi^3}\,. \label{d2psi}
}
Note that $W$ depends on the modes $n$ and $l$, so we will usually write $W_{nl}$ whenever simplicity allows it. For future convenience, we shall also sometimes adopt the notation $W_n(l)$ and $\Psi(l)$ which makes explicit the argument $l$ while carrying an implicit dependence on $r$. The derivatives are also taken with respect to the variable $r$.

%% file: II.4Lifshitz/2.section.tex
\section{Renormalization}

In this section we will construct the regulated coincidence limit of the Green's function. First of all, we take the spatial coincidence limit $(r',\Omega') \to (r,\Omega)$, define $\kappa = 2\pi / \beta$ and insert the WKB ansatz Eq.~(\ref{WKBansatz}) in Eq.~(\ref{G2}) to obtain
\eq{\label{G}
G_E(x,x') = {\kappa\over 8 \pi^2} \sum^{\infty}_{n=-\infty} e^{i n \kappa \varepsilon} \sum^{\infty}_{l=0}{(l+1/2)\over r^{z+2} W_{n}(l)} ~,
}
where we have redefined $\varepsilon = \Delta \tau$. Both sums, over $l$ and $n$, are divergent, although for different reasons. We will show how to deal with both of them in the following.

\subsection{Regularization in the $l$ modes}

The summation in the angular $l$ modes is an artifact derived from the way we decided to expand the angular part of the Green function modes. It arises in the same sense that a coordinate singularity appears in the ``North pole'' of the spherical coordinates. In the case of the sum, the divergence is clear when one takes the large $l$ limit. Looking at the asymptotic behavior of $W_{n}(l)$ for large $l$,
\eq{
W_n(l) \sim \tilde{W}_n(l) \sim \left(l + {1 \over 2}\right){\sqrt{f} \over r} + \mathcal{O}\left[\left(l + {1 \over 2}\right)^{-1}\right]
}
we see, from Eq.~(\ref{G}), that we are summing the term $1/(r^{z+1}\sqrt{f})$ in an unbound fashion, so there is a divergence that goes as $\sum^{\infty}_{l=1} 1$. We can neatly deal with this issue by noting that
\eq{
\de(\e) \sim \sum^{\infty}_{n=-\infty} e^{i n \kappa \varepsilon}\,,
}
is zero as long as $\varepsilon \neq 0$, or equivalently $x \neq x'$. Thus, as long as $\varepsilon \neq 0$, we are free to add multiples of $\delta(\varepsilon)$ to the Green function in Eq.~(\ref{G}), since these multiples will still be $0$ and the final result will not be altered. In particular, we may choose multiples of the form
\eq{
{\kappa \over 8\pi^2} \sum_n e^{i n \kappa \varepsilon} \sum_l R_l(r)
}
where $R_l(r)$ is arbitrary and independent of $n$. Hence, if we choose $R_l(r) = 1/(r^{z+1}\sqrt{f})$ and subtracting this term in Eq.~(\ref{G}), we will cancel the asymptotic terms for large $l$, thus removing the divergence in the summation over $l$ when the limit $\e \to 0$ is taken. Consequently, a regularized version of Eq.~(\ref{G}), with respect to the angular modes summation, is given by
\eq{\label{Gregl}
G_E(x,x') = {\kappa\over 8 \pi^2} \sum^{\infty}_{n=-\infty} e^{i n \kappa \varepsilon} \sum^{\infty}_{l=0}\left[{l+1/2\over r^{z+2} W_{n}(l)}-{1 \over r^{z+1} \sqrt{f}}\right]\,.
}

\subsection{Regularization in the $n$ modes}

While the divergences due to the $l$ summation can be eliminated without the need for renormalization, the more serious divergences appears in the coincidence limit $\varepsilon \rightarrow 0$ due to the summation over $n$.
As it was seen in Chapter 8, these UV divergences can be traced back to fact that the coincident limit of the Green's function comes from the product of quantum fields which are being evaluated at the same spacetime point. Such divergences can be cured by subtraction of appropriate counterterms. The problem of isolating the divergent terms for a general spherical symmetric metric has been addressed in general in \cite{Christensen:1978yd}, where general formulas have been obtained. We obtain these by direct use of Eq.~(\ref{GdivE}) for $d=4$, together with the results for the world function derivatives \cite{Christensen:1978yd}
\ea{
\si^{\tau} & = -\e + {\e^3 \over 24} {(f')^2 \over f h} + {\e^5 \over 120} \left( {f'^4 \over 8f^2 h^2} + {3\over 16} {(f')^3 h' \over f h^3} - {3\over 8} {f'^2 f'' \over f h^2} \right) + \ma{O}(\e^7) \,, \\
\si^{r} & = {\e^2 f' \over 4 h} - {\e^4 \over 24} \left(-{f'^2 h' \over 8 h^3} + {f' f'' \over 4 h^2}\right) + \ma{O}(\e^6) \,, \\
\si^{\theta_i} & = 0, \quad i=1,\ldots,d-2\,,
}
valid for any spherically symmetric metric of the form
\eq{
ds^2 = f(r) d\tau^2 + h(r) d^2r + r^2 d^2\Omega_{d-2}\,,
}
and separation $\e = \tau - \tau'$ in the timelike coordinate only. Using Eq.~(\ref{RelSi}), one obtains the world function
\eq{\label{si}
\si(x,x') = {f \over 2} \e^2 - {1\over 96 h} f'^2 \e^4 + \ma{O}(\e^6)
}
up to sixth order in the timelike separation. For the particular case of the metric (\ref{GdivE}), we get
\ea{
G_{\textrm{E}\,\rm div.}(x,x') & = {1 \over 16\pi^2}\bigg\{ {4 \over \varepsilon^2 f r^{2z}} + \left(m_{\phi}^2+\left(\xi-{1\over 6}\right)R\right)
\left(\ln\left[{m_{\phi}^2 f \varepsilon^2 r^{2z} \over 4}\right]+2\gamma\right) - m_{\phi}^2 + {r^2 \over 6u}{f'^2 \over f^2} -{r^2 \over 6u}{f'' \over f}
\nonumber \\
& \hspace{15mm} + {z r \over 6u}{u' \over u}+{r^2 \over 12u}{f' \over f}{u' \over u} -{r \over 2u}{f' \over f} - {2z \over 3u}\bigg\}\,,
}
where $\gamma$ is Euler's constant. Finite terms of order $\si^0$ are also included, as is common practice. In order to renormalize the coincidence limit of the Green's function, we must subtract the above expression from Eq.~(\ref{Gregl}) in the limit $\varepsilon=0$. In order to do this, one needs to recast the counterterms in a form more suitable for the subsequent evaluation. This is usually done by using the Abel-Plana sum formula
\eq{\label{APsum}
\sum^{\infty}_{n=m} g(n) = {g(m) \over 2} + \int^{\infty}_{m} g(\tau) d\tau + i \int^{\infty}_{0} {dt \over e^{2\pi t}-1} \left[ g(m+it) - g(m-it) \right]\,.
}
With the particular choices
\eq{
g(n) = \cos(n\kappa\e) (n^2 \kappa^2 + m_{\phi}^2 f r^{2z})^{1/2}
}
and
\eq{
g(n) = \cos(n\kappa\e) (n^2 \kappa^2 + m_{\phi}^2 f r^{2z})^{-1/2}
}
one may choose $m=0$, calculate the second integral in Eq.~(\ref{APsum}) and use the same equation to rewrite the counterterms as
\eq{\label{Gdiv2}
\lim_{x'\to x} G_{E\,\rm div.}(x,x') = {\kappa \over 8 \pi^2}\bigg[\Delta_1+ \Delta_{2}+\Delta_{3}+\Delta_{4}
-\sum^{\infty}_{n=1}\bigg({2\omega_n \over r^{2z}f} + \left[m_{\phi}^2-\left(\xi-{1\over 6}\right)R\right]{1\over \omega_n}\bigg)\bigg]\,,
}
as is done in in \cite{Anderson:1989vg}, where we have defined
\ea{
\Delta_1 & \equiv -\sum^{\infty}_{n=1}\left[ {2\over r^{2z}f}\left(\sqrt{\omega^2_n+m_{\phi}^2r^{2z}f} -\omega_n-{m_{\phi}^2r^{2z}f \over 2\omega_n}\right) +\left(\xi-{1\over 6}\right)R\left({1\over \sqrt{\omega^2_n+m_{\phi}^2r^{2z}f}}-{1\over \omega_n}\right) \right] \\
\Delta_{2} & \equiv {m_{\phi}^2 \over 2\kappa} \ln\left(m_{\phi}^2r^{2z}f\right) - {m_{\phi}^2 \over \kappa} \ln\left(\kappa + (\kappa^2+m_{\phi}^2r^{2z}f)^{1/2}\right) \nonumber \\
& \hspace{10mm}+{2i\over r^{2z}f} \int^{\infty}_{0}{dt \over e^{2\pi t}-1}\left(\left[(1+it)^2\kappa^2+m_{\phi}^2r^{2z}f\right]^{1/2}-\left[(1-it)^2\kappa^2+m_{\phi}^2r^{2z}f\right]^{1/2}\right),\\
\Delta_{3} & \equiv {\left(\xi-{1\over 6}\right)R} \bigg[{\ln\left(m_{\phi}^2 r^{2z}f\right)\over \kappa}-{2\over \kappa}\ln\left(\kappa + (\kappa^2+m_{\phi}^2r^{2z}f)^{1/2}\right)+{1\over \sqrt{\kappa^2+m_{\phi}^2r^{2z}f}} \nonumber\\
& \hspace{10mm} + 2\kappa i \int^{\infty}_{0}{dt \over e^{2\pi t}-1}\left({1\over \left[ (1+it)^2\kappa^2+m_{\phi}^2r^{2z}f\right]^{1/2}}-{1\over \left[(1-it)^2\kappa^2+m_{\phi}^2r^{2z}f\right]^{1/2}}\right)\bigg], \\
\Delta_{4} & \equiv {1\over 2\kappa}\left({r^2 \over 6u}{f'^2 \over f^2} -{r^2 \over 6u}{f'' \over f}+{z r \over 6u}{u' \over u}+{r^2 \over 12u}{f' \over f}{u' \over u} -{r \over 2u}{f' \over f} - {2z \over 3u}-m_{\phi}^2\right).
}

From the above results, the coincidence limit is readily obtained by subtracting the divergent expression (\ref{Gdiv2}) from the unrenormalized result (\ref{Gregl}). After subtraction, the coincidence limit can be safely taken leading to a regular (finite) expression
\ea{\label{phi}
\braket{\phi^2(x)}_{\rm ren} & = {\kappa \over 8 \pi^2}\bigg\{ \sum^{\infty}_{l=0}\left({l+1/2\over r^{z+2} W_{0 }(l)}-{1 \over r^{z+1} \sqrt{f}}\right) + \sum^{\infty}_{n=1}\bigg[2\sum^{\infty}_{l=0}\left({l+1/2\over r^{z+2} W_{n}(l)}-{1 \over r^{z+1} \sqrt{f}}\right)
\nonumber \\
& \hspace{10mm} + {2\omega_n \over r^{2z}f} + \left[m_{\phi}^2-\left(\xi-{1\over 6}\right)R\right]{1\over \omega_n}\bigg]-\Delta_1-\Delta_{2}-\Delta_{3}-\Delta_{4}\bigg\}.
}
In the next section we shall manipulate this result in such a way that it becomes more prone to a numerical evaluation.

%% file: II.4Lifshitz/3.section.tex
\section{Regularity and Summations}

Although Eq.~(\ref{phi}) is finite by construction, the renormalized vacuum polarization, as written in (\ref{phi}) is not yet suitable for straightforward numerical evaluation. First of all, individual pieces are divergent. Thus, it is more instructive to see how these can be combined in order for the divergences to cancel. Secondly, the sum over the angular and energy modes is numerically nontrivial, and it can expedited by appropriately expressing the sums. In this section, we will prove the regularity and outline how the summations are performed.

A useful way to proceed is to add and subtract the WKB approximation of $\braket{\phi^2(x)}$, i.e.
\ea{\label{WKBtrick}
\braket{\phi^2(x)}_{\rm ren} & = \braket{\phi^2(x)}_{WKB} + [\braket{\phi^2(x)}_{\rm ren} - \braket{\phi^2(x)}_{WKB}]\,, \\
& \equiv \braket{\phi^2(x)}_{WKB} + \de \braket{\phi^2(x)}\,,
}
such that the result is divided in a truncated WKB approximation plus some remainder. While this manipulation is only formal, it is clear that the reminder of the WKB approximation, $\de \braket{\phi^2(x)} = \braket{\phi^2(x)} - \braket{\phi^2(x)}_{WKB}$, is regular and can be calculated numerically, modulo a convenient combination of the individual terms as we will describe below. Denoting $\tilde{W}$ as the truncated WKB function, and following previous work \cite{Anderson:1989vg,Flachi:2008sr}, we write
\ea{\label{phiterms}
\braket{\phi^2(x)} = {\kappa \over 8 \pi^2}\{\Upsilon_0+\Sigma_1+\Sigma_2-\Delta\}
}
where we have defined $\Delta = \Delta_1+\Delta_{2}+\Delta_{3}+\Delta_{4}$,
\ea{
\Sigma_1 & \equiv \sum^{\infty}_{l=0}\left(l+{1\over 2}\right)\left({1\over r^{z+2} W_{0}(l)}-{1\over r^{z+2} \tilde{W}_{0}(l)}\right) + 2\sum^{\infty}_{n=1}\sum^{\infty}_{l=0}\left(l+{1\over 2}\right)\left({1\over r^{z+2} W_{n}(l)}-{1\over r^{z+2} \tilde{W}_{n}(l)}\right)\,, \nonumber \\
\Sigma_2 & = \sum^{\infty}_{n=1}\left[2\Upsilon_n+{2\omega_n \over r^{2z}f} + \left[m_{\phi}^2-\left(\xi-{1\over 6}\right)R\right]{1\over \omega_n}\right] \label{Sigma2}
}
and
\eq{
\Upsilon_n \equiv \sum^{\infty}_{l=0}\left({l+1/2\over r^{z+2} \tilde{W}_{n}(l)}-{1 \over r^{z+1} \sqrt{f}}\right)~.
}
Every term except for $\Sigma_2$ is explicitly finite, so we need only to verify that all the divergences in this term cancel. Obvious divergences come from the second and third terms, proportional to $\omega_n$ and $\omega_n^{-1}$, in (\ref{Sigma2}) and should cancel with terms coming from $\Upsilon_n$. In order to isolate these diverging terms, we can apply once more the Abel-Plana formula and recast $\Upsilon_n$ as
\eq{\label{Ups}
\Upsilon_n = {1\over r^{z+2}}\bigg\{{1\over 4\tilde{W}_n(0)}+\left[\int^{\infty}_0\left({\tau+1/2\over \tilde{W}_{n}(\tau)}-{r\over \sqrt{f}}\right)d\tau - {r\over 2\sqrt{f}}\right] + i\int^{\infty}_0{d\tau \over e^{2\pi \tau}-1}\left({i\tau+1/2 \over \tilde{W}_n(i\tau)}-{-i\tau+1/2 \over \tilde{W}_n(-i\tau)}\right)\bigg\}\,.
}
The first term in Eq.~(\ref{Ups}) can be expressed in terms of the following Epstein-Hurwitz $\zeta$
function:
\eq{
\mathcal{Z}_q \equiv \sum^{\infty}_{n=1} \left(\omega^2_n+r^{2z}\sigma(r)\right)^{-q/2}
}
as some straightforward calculations show
\ea{\label{P1}
{1\over 2r^{z+2}}\sum^{\infty}_{n=1}{1\over \tilde{W}_n(0)}
& = {1\over 4r^2} \bigg\{2\mathcal{Z}_1+zr^{2z-2}\left[a_1r-2z\left({a_2-a_3\over 2}\right)-(2z+1)a_3\right]\mathcal{Z}_3 \nonumber \\
&
- r^{4z-2}
\left[
\left({a_1r\over 2}-2\left({a_2-a_3\over 2}\right)z\right)\left(2z\sigma+r\sigma'\right)
-a_3\left(z(2z+1)\sigma-{r^2\sigma''\over 2}\right)
\right]\mathcal{Z}_5 \nonumber \\
&- \left({a_2-a_3\over 2}\right){r^{4z}\over 2}
\left(2z\sigma r^{z-1} +r^z\sigma'\right)^2\mathcal{Z}_7 \bigg\}.
}
Having expressed the result in terms of the above Epstein-Hurwitz $\zeta$-function, isolating the divergences is only a matter of simple power-counting. In Eq.~(\ref{P1}), only the term multiplied by $\mathcal{Z}_1$ is responsible for the divergences, which can be extracted by expanding $\mathcal{Z}_1$ in powers of $\omega_n$ and retaining the terms proportional to $c_1 \omega_n + c_2 \omega_n^{-1}$. Simple steps then lead to the divergent piece of (\ref{P1}), denoted by ${\rm div}_1$
\eq{\label{divP1}
{\rm div}_1
={1 \over 2r^2 \omega_n}~.
}

The next contribution to $\Sigma_2$ is the term in square brackets in Eq.~(\ref{Ups}), namely
\begin{eqnarray}
{2\over r^{z+2}}\sum^{\infty}_{n=1}\left[\int^{\infty}_0\left({\tau+1/2\over \tilde{W}_{n}(\tau)}-{r\over \sqrt{f}}\right)d\tau - {r\over 2\sqrt{f}}\right]
 =  A_1 + A_2 \,, \nonumber \\
A_1  \equiv {2\over r^{z+2}}\sum^{\infty}_{n=1}\left[\int^{\infty}_0\left({\tau+1/2\over \Phi(\tau)^{1/2}}-{r\over \sqrt{f}}\right)d\tau - {r\over 2\sqrt{f}}\right]\,, \\
\hskip -3cm
A_2 \equiv -{1\over 2r^{z+2}}\sum^{\infty}_{n=1}\int^{\infty}_0 {(\tau+1/2) \, \Psi(\tau)\over \Phi(\tau)^{3/2}} \, d\tau\,. \hfill\hskip 1cm
\label {P2}
\end{eqnarray}
The first term $A_1$ can be evaluated by direct integration leading to
\eq{
A_1
= -{2 \over r^{2z} f} \mathcal{Z}_{-1}~,
}
from which the divergent contribution ${\rm div}_2$ can be extracted:
\eq{\label{divP21}
{\rm div}_2
= {2 \over r^{2z}f}\left(\omega_n + {r^{2z} \sigma \over 2\omega_n}\right)
}
For the other term, $A_2$, inserting the explicit expression for $\Psi$, integrating over $\tau$, and summing over the frequencies gives
\ea{
A_2
& = {a_1 \over 6f} \bigg\{\left({4+2z \over r}-2{f' \over f}\right)\mathcal{Z}_1 -r^{2z-1}(2z\sigma +r \sigma')\mathcal{Z}_3\bigg\}
 - {a_2 -a_3\over 60f}\bigg\{\left({4(3z^2+4z+8) \over r^2}-{8(z+4) \over r}{f' \over f} + 8{f'^2 \over f^2}\right)\mathcal{Z}_1
\nonumber \\
& -4r^{2z-1}\left({(3z+2) \over r}-{f' \over f}\right)(2z\sigma +r \sigma')\mathcal{Z}_3 + 3 r^{4z-2}(2z\sigma +r \sigma'^2)\mathcal{Z}_5\bigg\}
\nonumber \\
& - {a_3 \over 6f}\bigg\{\left({12+2z(2z+1) \over r^2}-{8 \over r}{f' \over f}+2{f'' \over f}\right)\mathcal{Z}_1 - r^{2z-2}(2z(2z+1)\sigma-r^2\sigma'')\mathcal{Z}_3\bigg\}\,.
}
In this case too, the only divergent contribution ${\rm div}_3$, which simple steps allow us to isolate, comes from $\mathcal{Z}_1$:
\eq{\label{divP22}
{\rm div}_3
= {1\over 6\omega_n} \bigg\{{z(4-z) \over 2u}-{z\over 2}{r\over u}{f'\over f}-{r^2\over u}{f''\over f} +{r^2\over 2u}{f'^2\over f^2}
+{r\over 2u}{u'\over u}\left(r{f'\over f}-(z+2)\right)\bigg\}~.
}

The last term in Eq.~(\ref{Ups}) to consider is
\eq{\label{P3}
A_3  \equiv
{2i\over r^{z+2}}\sum^{\infty}_{n=1}\int^{\infty}_0{d\tau \over e^{2\pi \tau}-1}\left({i\tau+1/2 \over \tilde{W}_n(i\tau)}-{-i\tau+1/2 \over \tilde{W}_n(-i\tau)}\right)~.
}
Observing that the dominant contribution to the integral comes from the $\tau \sim 0$ region of integration, we make the expansion
\eq{\label{expP3}
\left({i\tau+1/2 \over \tilde{W}_n(i\tau)}-{-i\tau+1/2 \over \tilde{W}_n(-i\tau)}\right) = -i \sum^{\infty}_{j=1} c_{nj} \tau^{2j-1}
}
for small $\tau$ and proceed by direct integration. Some calculations give
\eq{\label{Expan}
\left({i\tau+1/2 \over \tilde{W}_n(i\tau)}-{-i\tau+1/2 \over \tilde{W}_n(-i\tau)}\right) = -i \sum^{\infty}_{j=1} c_{nj}\tau^{2j-1}
}
where
\eq{\label{Cnj}
c_{nj} \equiv {i^{2j} \over (j-1)!}{d^{j-1} \over dx^{j-1}}\left({2 \over \tilde{W}_n(x)} - {1 \over j}{\tilde{W}'_n(x) \over \tilde{W}^2_n(x)}\right)\bigg|_{x=0}\,
}
the exponential in the denominator ensures that values around $\tau=0$ will dominate the integral, so to find the divergent part we need only to worry about such values. Expanding the quantity inside parentheses in Eq.~(\ref{P3}) in a Taylor series around $\tau=0$, we obtain
\eq{\label{Expan}
\left({i\tau+1/2 \over \tilde{W}_n(i\tau)}-{-i\tau+1/2 \over \tilde{W}_n(-i\tau)}\right) = -i \sum^{\infty}_{j=1} c_{nj}\tau^{2j-1}
}
where
\eq{\label{Cnj}
c_{nj} \equiv {i^{2j} \over (j-1)!}{d^{j-1} \over dx^{j-1}}\left({2 \over \tilde{W}_n(x)} - {1 \over j}{\tilde{W}'_n(x) \over \tilde{W}^2_n(x)}\right)\bigg|_{x=0}\,
}
is independent of $\tau$. Inserting Eq.~(\ref{Expan}) in Eq.~(\ref{P3}) and integrating over $\tau$, we are led to
\ea{\label{P3f}
A_3
& = {2\over r^{z+2}}\sum^{\infty}_{n=1}\sum^{\infty}_{j=1} c_{nj} \left(\int^{\infty}_0 {d\tau \over e^{2\pi \tau}-1}\tau^{2j-1}\right) \nonumber \\
& = {2\over r^{z+2}}\sum^{\infty}_{n=1}\sum^{\infty}_{j=1} c_{nj} {\Gamma(2j)\zeta(2j) \over (2\pi)^{2j}} \nonumber \\
& = {1\over r^{z+2}}\sum^{\infty}_{n=1}\sum^{\infty}_{j=1} {(-1)^{j-1} \over 2j} c_{nj}  B_{2j}
}
where $B_{2j}$ are the Bernoulli numbers and the coefficients $c_{nj}$ are those coming from the Taylor expansion of the integrand in Eq.~(\ref{P3}),
\eq{\label{Cnj}
c_{nj} \equiv {i^{2j} \over (j-1)!}{d^{j-1} \over dx^{j-1}}\left({2 \over \tilde{W}_n(x)} - {1 \over j}{\tilde{W}'_n(x) \over \tilde{W}^2_n(x)}\right)\bigg|_{x=0}.
}
Identification of the divergent part in Eq.~(\ref{P3f}) is possible by looking at the asymptotic behavior of the coefficients $c_{nj}$ for large $n$ and checking if it contains some terms which will lead to infinities when summed over $n$. In fact,
it is straightforward to see that for $n \gg 1$ only the $j=1$ term in Eq.~(\ref{Cnj}) leads to a divergence,
\eq{\label{Cn1}
c_{n1} = -{2r^z \over \omega_n}.
}
This means that, for large values of $n$, the value of $\mathcal{P}_{3}$ will be dominated by the sum over large $n$ with coefficients (\ref{Cn1}), which will give rise to an infinity. Thus, using $B_{2} = 1/6$, we reach the conclusion that the general term of the sum for large $n$, which corresponds to the $n$th term of the divergent part, will be
\eq{\label{divP3}
{\rm div}_4
= -{1 \over 6r^2\omega_n}~.
}
From here, summing (\ref{divP1}), (\ref{divP21}), (\ref{divP22}) and (\ref{divP3}) one shows that the final expression is regular at any radius and for any parameter $z$.

%% file: II.4Lifshitz/4.section.tex
\section{Numerical computations}

Having demonstrated the regularity of the results, the remaining task is to compute the sums over $n$, a problem that we can approach exactly. Practically, the problem has been reduced to calculating the zeta functions $\mathcal{Z}_q$. Only $\mathcal{Z}_1$ and $\mathcal{Z}_{-1}$ contain divergences, and, using the proof of the regularity, we can simply regulate these functions by subtracting the corresponding diverging contributions. This translates to the following definition,
\ea{
\tilde{\mathcal{Z}}_{-1} & = \kappa \sum^{\infty}_{n=1}\left(\sqrt{n^2+v^2}-n-{v^2\over 2n}\right),~~~~~ \\
\tilde{\mathcal{Z}}_{1} & = \kappa^{-1} \sum^{\infty}_{n=1}\left({1 \over \sqrt{n^2+v^2}}-{1 \over n}\right), \label{regz}
}
with $v^2 \equiv {r^{2z}\sigma \over \kappa^2}$. Numerical evaluation can then be performed in different ways, depending on the magnitude of $v^2$. For large $v^2$, one may adopt the Chowla-Selberg formula \cite{elizalde,cs}
\eq{\label{CS}
\sum^{\infty}_{n=1}(n^2+\rho^2)^{-s} = -{\rho^{-2s}\over 2} +{\sqrt{\pi}\over 2}{\Gamma(s-1/2)\over \Gamma(s)}\rho^{1-2s}
+{2\pi^{s}\over \Gamma(s)}\rho^{-s+1/2}\sum^{\infty}_{p=1} p^{s-1/2}K_{s-1/2}(2\pi p \rho)\,,
}
to recast the zeta functions as
\eq{
\mathcal{Z}_q = \kappa^{-q}\bigg(-{v^{-q}\over 2}+{\sqrt{\pi}\,\Gamma((q-1)/2)\over 2 \Gamma)(q/2)}v^{1-q} + {2\pi^{q/2}\over \Gamma(q/2)}v^{(1-q)/2}\sum^{\infty}_{p=1}p^{(q-1)/2}K_{(q-1)/2}(2\pi p v)\bigg)\,,\nonumber
}
which is regular for any $q \neq -1, 1$. For either $q=-1, 1$, following the logic we have explained before, one may simply subtract the divergent portion obtaining
\eq{\label{Zm1Reg}
\tilde{\mathcal{Z}}_{-1} = \kappa \bigg({1\over 12} - {v\over 2} +{v^2\over 4}\left[1-2\ln(v/2)+\gamma\right] - {v\over \pi}\sum^{\infty}_{p=1}{K_{-1}(2\pi p v)\over p}\bigg)\,,
}
and
\eq{
\tilde{\mathcal{Z}}_{1} = -{1\over \kappa}\left({1\over 2v}+\ln(v/2)+\gamma-2\sum^{\infty}_{p=1}K_0(2\pi p v)\right)\,.
}
Owing to the presence of the Bessel functions, the evaluation of the sums can be carried out numerically very easily.

When the value of $v^2$ is small, then we may proceed by splitting the summation range in two parts: one up to a value $\bar n \gg v^2$ plus a reminder. We can then expand the reminder for small $v^2$ and complete the infinite sums by adding and subtracting appropriate terms. The procedure is identical to that developed in Ref.~\cite{Flachi:2008sr}. The other regular term, involving  spurious divergent summations over $n$ is $\Delta_1$ that can also be treated along the same lines described above, i.e. expressing it in terms of regularized zeta functions.

The numerical procedure can be implemented straightforwardly and essentially it comes down to calculating every term of (\ref{phiterms}). The contributions $\Delta_1$, $\Delta_{2}$, $\Delta_{3}$, $\Delta_{4}$ and $\Upsilon_0$ do not pose any particular complication and their evaluation can be carried out directly. The term $\Sigma_1$ describes basically the reminder of the WKB approximation and it can be calculated by first solving the homogeneous equation numerically, followed by the subtraction of the approximate result using the WKB expansion up to leading order. The WKB expansion improves for large $l$ and $n$, however, computationally, this term is the most expensive to calculate. For $\Sigma_1$ we expedited this procedure by using a sampling method to compute the sums. Finally, the term $\Sigma_2$ consists of the sum of $A_1$, $A_2$ and $A_3$.  Since we have explicitly shown its regularity, we may substitute the diverging functions $\mathcal{Z}_{q}$ (for $q=-1, 1$) with their regularized counterparts (\ref{regz}), while the functions $\mathcal{Z}_{q}$ (for $q>1$) can be evaluated easily due to the fast convergence of the sums over the Bessel functions. With these preliminaries, the terms $A_1$ and $A_2$ can, then, be calculated directly. The remaining term $A_3$ can be calculated by adding and subtracting the expansion (\ref{expP3}) to (\ref{P3}) and using the WKB approximation for the terms containing $W_n$, which are then expanded in a Taylor series up to some order $j_{max}$. The advantage of using this approach is that the subtraction term quickly drops to zero for relatively small value of $j_{max}$, leading to a faster numerical convergence.

So far the treatment has been independent of the explicit form of the metric functions, so now we must specify them. Some examples are illustrated in Figs.~1 and 2, where we considered
\eq{
f = {1 \over u} = 1- \left({r_h \over r}\right)^{z+3}
}
for which the black hole temperature is given by
\eq{
T_{\textrm{BH}} = {1 \over 4\pi} {g_{00,1} \over \sqrt{g_{11} \, g_{00}}}\bigg|_{r=r_h} = {z+3 \over 4\pi} \, r_h^z\,.
}
The plots for the vacuum polarization are for the specific values $\xi=0$, $m=0.01$, with $z=1$ and $z=2$, for different values of horizon radius $r_0$.

\begin{figure}[h]
\centering
\includegraphics[width=0.5\textwidth]{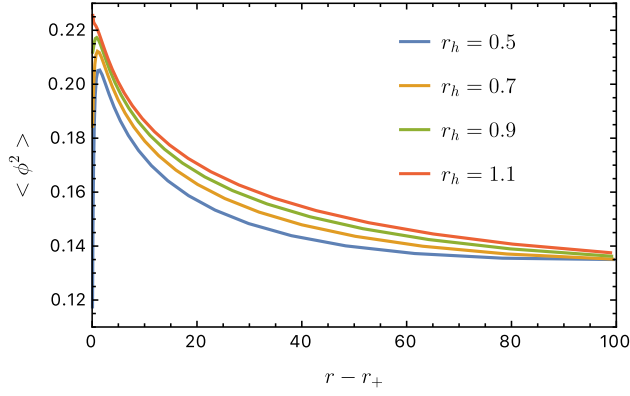}
\caption{Vacuum polarization for $\xi = 0$, $m=0.01$ and $z=1$. The result is finite for all values of $r-r_h$.}
\end{figure}

\begin{figure}[h]
\centering
\includegraphics[width=0.5\textwidth]{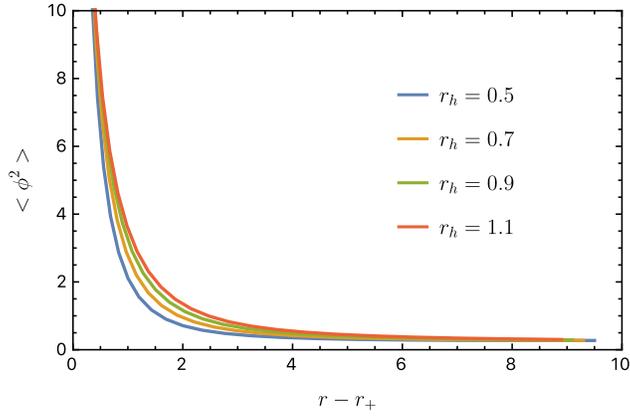}
\caption{Vacuum polarization for $\xi = 0$, $m=0.01$ and $z=2$. The result is finite for all values of $r-r_h$, although it appears to diverge near the horizon due to vertical axis scale. For $r_h = 0.5, 0.7, 0.9, 1.1$, the values at the horizon are about 67, 51, 43, 37, respectively.}
\end{figure}

As a simple consistency check, we may investigate the asymptotic values for large values of the radial coordinate in the case $z=1$. In such a limit, it is readily seen that the metric functions reduce to an AdS type, for which we know that the analytic asymptotic value of the vacuum polarization is given by
\eq{
\braket{\phi^2}_{\textrm{AdS}} \simeq -{1 \over 48 \pi^2}\,.
}
We verify that this is indeed the value obtained for $z=1$. In fact, for large values of $r-r_h$, both plots are well fitted as
\eq{
\braket{\phi^2} = C_1 + {C_2 \over r^{2z} f(r)}\,,
}
where for $z=1$ it is confirmed that $C_1 = -{1 \over 48 \pi^2}$.

%% file: II.4Lifshitz/5.section.tex
\section{Conclusions}

Lifshitz black holes are black hole solutions exhibiting scaling. These solutions are important ingredients in the construction of gravitational dual of Lifshitz field theories, allowing one to investigate finite temperature effects. In this paper we have addressed the problem of calculating the coincidence limit of the Green's function for a massive, nonminimally coupled bulk scalar field, i.e., the vacuum polarization $\braket{\phi^2}$.

It is important to establish
general properties for quantum observables in black hole spacetimes,
in the effort to understand the quantum backreaction problem and
related issues. The calculation of the vacuum polarization is
relevant in two respects. First, it conveys information about the
regularity of quantum observables outside the horizon. Put simply, if
$\braket{\phi^2}$ is regular, then the vacuum expectation value of the quantum energy-momentum
tensor will also be regular. Second, the vacuum polarization
represents the scalar condensate that therefore provides information
about the possibility that symmetry breaking occurs near the black
hole. The computation of quantum vacuum effects is a notoriously difficult task and, here, we have adapted the methodology used in the case of asymptotically AdS black holes to the case of Lifshitz black holes.

The basic approach relies on the use of the WKB approximation and point splitting regularization together, allowing us to express the full solution as a WKB approximated part plus a remainder. This proves to be very effective to explicitly confirm the cancellation of diverging parts, while at the same time providing a regular set up for a numerical calculation. The WKB part is directly computed by using the analytic results expressed in terms of regulated generalized zeta functions, which in turn converge very rapidly due to the fast decay of the modified Bessel functions appearing in them. The remainder part is calculated by numerically solving the mode equation and subtracting the WKB counterparts. The convergence in this case is quite fast as well, since this component is of order $\mathcal{O}(l^{-5},n^{-5})$.

We have dealt with the most general case of spherically symmetric Lifshitz solutions. We then have considered a particular form for the metric functions in order to obtain a numerical result. We chose a function which asymptotes to an AdS case, for which an analytic result had already been calculated, and used it to check with our results which correctly reproduced the expected behavior.

%% file: II.55Dcharged/main.tex
\chapter{Vacuum polarization around a 5-dimensional charged black hole}
\label{cap:chapterII5}
\input{II.55Dcharged/outline.tex}
\input{II.55Dcharged/1.section.tex}

\input{II.55Dcharged/2.section.tex}

\input{II.55Dcharged/3.section.tex}

\input{II.55Dcharged/conclusions.tex}

%% file: II.55Dcharged/outline.tex
\section{Introduction}

Numerous initial studies in vacuum polarization in curved
spacetimes~\cite{Christensen:1978yd,Candelas:1980zt,Candelas:1984pg,Fawcett:1983dk,Anderson:1989vg}
focused in four, i.e., (3+1), dimensions and had the aim of improving
the understanding of particle production in curved spacetimes and
various aspects of black hole evaporation.
However, although calculations in higher dimensional spacetimes have been performed (see \cite{Frolov:1989rv,Shiraishi:1993ti, Decanini:2005eg,Thompson:2008bk}, for example) they have also been scarce, especially regarding numerical calculation.

The goal of this chapter is to explore the standard calculations of vacuum polarization to higher dimensional spacetimes, using the same WKB approximation method adopted in the previous chapters. In this case, we shall focus on a 5-dimensional charged black hole spacetime.

%% file: II.55Dcharged/1.section.tex
\section{Quantum scalar field in a 5-dimensional spherically symmetric metric}

We are interested in the Euclidean Green function of a scalar field in a 5-dimensional spacetime, whose dynamics is governed by the action
\eq{\label{S5C}
S[\phi] = {1\over 2} \int d^5x \sqrt{g} \left[
g^{\mu \nu} \partial_\mu \phi \, \partial_\nu \phi + m_{\phi}^2 \phi^2 + \xi R \phi^2
\right]\,.
}
By Eq.~(\ref{GEdef}), we have that the corresponding Green function satisfies the differential equation
\eq{\label{gf5}
  \left(\square_{\textrm{E}} - m_{\phi}^2 - \xi R\right)G_\textrm{E}(x,x') = - {\de^{(5)}(x-x') \over \sqrt{|g|}}\,,
}
where $\square_\textrm{E}$ is the d'Alembertian operator with Euclidean signature, $m_{\phi}$ is the scalar field mass, $\xi$ is the coupling constant, $R$ is the spacetime curvature. In this work, we will consider the background to be a five-dimensional black hole
described by a five-dimensional metric of the type
\eq{\label{RS5}
ds^2 = -f(r) dt^2 + f^{-1}(r)dr^2 + r^2 d\Omega^2_3\,,
}
where $t$ and $r$ are the time and radial coordinates, respectively, $d\Omega^2_3$ represents the line element of a 3-sphere, and
$f(r)$ is some function of $r$. We assume that at infinity $f(r)$ goes as $1/r^2$ as it should for a five-dimensional spherical asymptotically flat spacetime, and we also assume that $f(r)$ contains an horizon at some radius $r_+$.

Performing a Wick rotation $t = -i \tau$ on the time coordinate, we obtain the Euclidean metric
\eq{
ds_{\rm E}^2 = f(r) d\tau^2 + {dr^2 \over f(r)} + r^2 d\Omega^2_3\,,
}
which is positive definite everywhere outside the horizon. In order to consider a thermalized field, the coordinate $\tau$ must be periodic with period $\be$ equal to
\eq{\label{temp}
\be = 4\pi \left({df \over dr}\right)^{-1}_{r = r_+}\,.
}
The quantity $T_{\rm BH} = \be^{-1}$ will then be the characteristic temperature of the black hole. Working in the Hartle-Hawking vacuum state, we may write the finite temperature Euclidean Green function in the mode-sum representation using Eqs.~(\ref{GExpN}) and (\ref{GExpL})
\eq{\label{ms}
  G_{\rm E}(x,x') = {\kappa \over 4\pi^3} \sum^{\infty}_{n=-\infty} e^{i \omega_n \e} \sum^{\infty}_{l=0} (l+1) C^{(1)}_l(\cos \g) G_{nl}(r,r')
}
where $\kappa \equiv 2\pi / \be$, $\e = \tau - \tau'$, $\omega_n \equiv \kappa n$, $\g$ is the geodesic distance in the 3-sphere, and $C^{(1)}_l(x)$ is a Gengenbauer polynomial. Inserting the mode-sum expansion, Eq.~(\ref{ms}), in Eq.~(\ref{gf5})
leads to the differential equation for the radial Green function
\eq{\label{rgf}
{d^2 G_{nl} \over dr^2} + \left({3\over r} + {f'\over f}\right){d G_{nl} \over dr} - \left({\omega_n^2\over f^2} + {l(l+2) \over f r^2} +{m^2+\xi R \over f}\right)G_{nl} = - {\delta(r-r') \over r^{3}}\,.
}
The solution to Eq.~(\ref{rgf}) can be expressed in terms of solutions of the corresponding homogeneous equation. In particular, if $p_{nl}(r)$ and $q_{nl}(r)$ are solutions of the homogeneous equation regular at the horizon and infinity, respectively, then the radial Green function can be written as
\eq{\label{rgfrr}
G_{nl}(r,r') = C_{nl} \, p_{nl}(r_<) q_{nl}(r_>)\,,
}
where, as usual, $r_<$ and $r_>$ denote the largest and the smallest values of the set $\{r,r'\}$. The quantity $C_{nl}$ is a normalization constant, given by
\eq{
C_{nl} = - {1 \over r^3 f(r)} {1 \over \mathcal{W}(p_{nl}(r),q_{nl}(r))}
}
where ${\mathcal W}(p,q)$ is the Wronskian of the two functions. We now want to find the solution of Eq.~(\ref{rgf}). We will first present the approximate
limiting solutions at infinity and at the horizon and then develop the general solution, since the limiting solutions serve as boundary conditions for the general solution.

The form of $p_{nl}$ and $q_{nl}$ of the Green function in Eq.~(\ref{rgfrr}), solution of Eq.~(\ref{rgf}), can be obtained by expressing the homogeneous equation in two limits, namely, the near-infinity limit and the near-horizon limit.

Starting with the near-infinity limit, i.e., the large $r$ limit, the homogeneous equation of Eq.~(\ref{rgf}) becomes
\eq{
\left\{{d^2 \over dr^2} + {3 \over r} {d \over dr} - (\omega^2_n + \mu^2 + \xi R)\right\}q_{nl}(r) = 0\,,
}
the solution of which, regular at infinity, is of the form
\eq{\label{qsol}
q_{nl}(r) \sim r^{-3/2} e^{- r \sqrt{\omega^2_n + \mu^2 + \xi R}}\,.
}

The near-horizon limit may be obtained by using the tortoise coordinate $r_*$, defined through
\eq{
dr_* = {dr \over f(r)}\,,
}
in terms of which, in the near-horizon limit and for $n \neq 0$, the homogeneous equation of Eq.~(\ref{rgf}) becomes
\eq{\label{nn0}
\left({d^2 \over dr^2_*} - \omega^2_n\right) p_{nl}(r)=0\,.
}
The solution of Eq.~(\ref{nn0}), regular at the horizon, is given by
\eq{\label{psol}
p_{nl}(r) \sim {e^{-\omega_n r_*} \over r}\,.
}
In the case $n = 0$, the homogeneous equation of Eq.~(\ref{rgf}), in the near-horizon limit, becomes
\eq{
{d \over dr}(\ln p_{0l}(r)) = {1 \over f'(r)}\left( {l(l+2) \over r^2} + \mu^2 + \xi R \right)\,,
}
the solution of which
goes as
\eq{\label{psol0}
p_{0l}(r) \sim \exp\left\{ \int^{r}_{r_+} \left( {l(l+2) \over u^2} + \mu^2 + \xi R \right) {du \over f'(u)} \right\}\,.
}
These limiting solutions will be especially important when performing numerical computations, since they will provide the boundary conditions necessary to solve Eq.~(\ref{rgf}) numerically.

We shall now display a general solution of Eq.~(\ref{rgf}) by following the standard procedure developed in \cite{Candelas:1980zt,Candelas:1984pg}, which makes use of a WKB approximation. We begin by using the following ansatz for the solutions of the homogeneous equation for the radial Green function
\ea{
p_{nl}(r) & = {1 \over \sqrt{r^{3} W(r)}} \exp \left\{ + \int^{r}_{r_+} {W(u) \over f(u)} \,  du\right\} \label{psolWKB}\,, \\
q_{nl}(r) & = {1 \over \sqrt{r^{3} W(r)}} \exp \left\{ - \int^{r}_{r_+} {W(u) \over f(u)} \, du\right\} \label{qsolWKB}\,,
}
where $W$ is the WKB function to be determined. As in the previous chapter, the above expressions are chosen specifically to eliminate all sign dependent terms once inserted in the homogeneous equation of Eq.~(\ref{rgf}), while at the same time satisfying both the near-horizon and large $r$ limits which are going to be calculated below. We will omit the $n$ and $l$ indices in the WKB function $W(r)$ whenever necessary for notational convenience. In the end, we are left with the homogeneous equation
\eq{\label{WKBeq}
W^2 = \Phi + a_1 {W' \over W} + a_2 {W'^2 \over W^2} + a_3 {W'' \over W}\,,
}
where
\ea{
\Phi & = \left((l+1)^2-1\right){f \over r^2} + \si(r)\,, \\
\si & = \omega^2_n +(m_{\phi}^2 + \xi R)f + {3f^2 \over 4 r^2} + {3f f' \over 2r}\,,
}
and
\eq{
a_1 = {f f' \over 2}, \quad a_2 = -{3 \over 4} f^2, \quad a_3 = {f^2 \over 2}\,,
}
where a prime in the functions $W$ and $f$ denotes a derivative with respect to the coordinate $r$.
Inserting Eqs.~(\ref{psolWKB}) and (\ref{qsolWKB}) in Eq.~(\ref{rgfrr}), taking the radial coincidence limit and using the fact the Wronskian is given by $\mathcal{W}(p(r),q(r)) = -f/(2W)$, we obtain
\eq{
G_{nl}(r,r) = {1 \over 2 r^3 W_{nl}(r)}
}
The solution to Eq.~(\ref{WKBeq}) can now be expressed iteratively as $W = W_0 + W_1 + \cdots$. Since the form of Eq.~(\ref{WKBeq}) is exactly the same as Eq.~(\ref{wkbeqt}), we obtain the same expansion, i.e.
\eq{
{1 \over W} = {1 \over \sqrt{\Psi}}(1+\de_1 \Psi + \de_2 \Psi + \cdots)\,.
}
For renormalization purposes, we may only be concerned with the first order approximation and consequently with the approximated solution $\tilde{W}$ truncated at first order,
\eq{\label{tildeW}
{1 \over \tilde{W}} = {1+\de_1 \Phi \over \sqrt{\Phi}}\,,
}
or, writing explicitly,
\eq{\label{tildeW1}
{1 \over \tilde{W}} = {1 \over \sqrt{\Phi}} + \al_1 {1 \over \Phi^{5/2}} + \al_2 {(l+1)^2 \over \Phi^{5/2}} + \al_3 {1 \over \Phi^{7/2}} + \al_4 {(l+1)^2 \over \Phi^{7/2}} + \al_5 {(l+1)^4 \over \Phi^{7/2}}\,,
}
with
\ea{
\al_1 & = \frac{r \left((a_1 r-4 a_3) f'-a_1 r^3 \sigma
   '+a_3 r f''-a_3 r^3 \sigma ''\right)}{4 r^4} \nonumber \\
   & \hspace{5mm} +\frac{f(6 a_3-2 a_1 r)}{4 r^4}\,, \\
\al_2 & = \frac{f (2 a_1 r-6 a_3)-r \left((a_1 r-4
   a_3) f'+a_3 r f''\right)}{4 r^4}\,, \\
\al_3 & = -\frac{(a_2-a_3) \left(-r f'+2 f+r^3 \sigma
   '\right)^2}{8 r^6}\,, \\
\al_4 & = -\frac{(a_2-a_3) \left(r f'-2 f\right) \left(-r
   f'+2 f+r^3 \sigma'\right)}{4 r^6}\,, \\
\al_5 & = -\frac{(a_2-a_3) \left(r f'-2 f\right)^2}{8 r^6}\,.
}

%% file: II.55Dcharged/2.section.tex
\section{Renormalization}
\label{Renorm}

We may now take the spatial coincidence limit, for which the Euclidean Green function given in
Eq.~(\ref{ms}) can then be approximated as
\eq{\label{GeDt}
G_{\rm WKB}(x,x') = {\kappa \over 8\pi^3r^3} \sum^{\infty}_{n=-\infty} e^{i \omega_n \e} \sum^{\infty}_{l=0} {(l+1)^2 \over \tilde{W}_{nl}(r)}\,.
}
The Euclidean Green function in Eq.~(\ref{GeDt}) is divergent both in the angular and energy modes, i.e., in the $l$ and $n$ modes, respectively. As in the four dimensional case of Chapter 10, the divergence in the angular $l$ modes is purely mathematical and can be promptly removed. On the other hand, the divergent terms in the energy $n$ modes are physical and must be canceled by some counterterms in order to obtain a fully renormalized result. First we regularize the $l$ modes and afterward the $n$ modes.

\subsection{Regularization in the $l$ modes}

The summation in the angular modes for large $l$ will be divergent so long as terms of $(l+1)$ with powers larger than $-1$ are present. Expanding $(l+1)^2/\tilde{W}$ for large $(l+1)$, we obtain
\eq{
\mathcal{T}_l(r) = {r \over \sqrt{f}}(l+1) + {r \over 32 f^{3/2} (l+1)}(-16r^2 \omega^2_n+16f - 4f^2-16f\si + 4rf f'' + r^2 f'^2 - 4r^2f f'')\,,
}
which diverges in the final sum of Eq.~(\ref{GeDt}). This divergence is not physical, and can be removed by subtracting the quantity

\eq{\label{Tl}
{\kappa \over 8\pi^3r^3} \sum^{\infty}_{n=-\infty} e^{i \omega_n \e{\Delta \tau}} \sum^{\infty}_{l=0} \mathcal{T}_l(r)\,,
}
from Eq.~(\ref{GeDt}). The term involving $\omega^2_n$ is irrelevant, since the summation in $n$ will give $\zeta(-2)$, which is zero. This means the dependence of $\mathcal{T}_l$ is purely on $l$, and so, Eq.~(\ref{Tl}) is a multiple of $\de(\e)$. Therefore, since $\e \neq 0$ at this stage, we are effectively subtracting 0, canceling the divergent large $l$ behavior in the process. After the subtraction we may take the full coincidence limit, for which the Green function becomes
\eq{\label{GeDn}
G_{\rm WKB}(x,x) = {\kappa \over 8\pi^3r^3} \sum^{\infty}_{n=-\infty} \sum^{\infty}_{l=0} \left\{{(l+1)^2 \over \tilde{W}_{nl}} - \mathcal{T}_l\right\}\,.
}

\subsection{Regularization in the $n$ modes}

We now proceed to the regularization of the $n$ modes, physically associated to UV divergences. We will isolate the divergent pieces of Eq.~(\ref{GeDn}) and explicitly see that they cancel with the counterterms provided by the point-splitting method developed in \cite{christensen}.

The Green function (\ref{GeDn}) can be written as
\eq{
G_{\rm WKB}(x,x) = {\kappa \over 8\pi^3r^3} \left(G_0 + 2\sum^{\infty}_{n=1} G_n \right)\,,
}
where we have defined $G_n$ as
\eq{\label{Gn}
  G_n = \sum^{\infty}_{l=0} \left(
  {(l+1)^2 \over \tilde{W}_{nl}} - \mathcal{T}_l\right)
}
and have made use of the fact that
$\sum^{\infty}_{n=-\infty} G_n =G_0+2\sum^{\infty}_{n=1} G_n $.
The term $G_0$ is finite by construction, so all divergences must be contained within $G_n$. In particular, powers of $n$ larger than $-1$ will result in infinity after the summation. To obtain an expression for $G_n$, we shall apply the Abel-Plana sum formula (\ref{APsum}) to Eq.~(\ref{Gn}) and expand for large $n$, arriving at the following divergent part of the Green function
\eq{\label{Gdiv}
G_{\rm div} = {\kappa \over 8 \pi^3 f^{3/2}} \sum^{\infty}_{n=1} \bigg[ \bigg( \mu^2 f - {f \over r^2} + {6\xi f \over r^2} + {f^2 \over r^2} - {6 \xi f^2 \over r^2} + {5f f' \over 4r} - {6\xi f f' \over r} - {f'^2 \over 16} + {f f'' \over 4} - \xi f f''\bigg) \ln \omega_n + \omega^2_n \ln \omega_n\bigg]\,.
}
The divergent terms of the form $1/\omega_n$ cancel out, as expected from Eq.~(\ref{GdivO}), i.e. from spacetimes with odd dimensions (see \cite{Thompson:2008bk}). To obtain a finite renormalized result we must subtract the counterterms given in
Eq.~(\ref{Gdiv}) from Eq.~(\ref{GeDn}), i.e.,
\eq{\label{ren1}
G_{\rm reg} = G_{\rm WKB}-G_{\rm div}\,.
}

In order to check that Eq.~(\ref{Gdiv}) is the correct divergent part we use the generic method already used in Chapter 10. Choosing the point split to lie in the $\tau$ coordinate, the world function, given by Eq.~(\ref{si}), becomes
\eq{
\si = {f \over 2} \e^2 - {f f'^2\over 96} \e^2 + \mathcal{O}(\e^6)\,,
}
and the Schwinger-DeWitt counterterms are then obtained from Eq.~(\ref{GdivO}), giving
\eq{\label{GDS}
G_{\rm SD} = \lim_{\e \to 0}\bigg\{ {1 \over 16 \pi \sqrt{f} \e} \left(\left( {1\over 6}-\xi \right)R-m_{\phi}^2 - {f' \over 4r} + {f'^2 \over 16 f}\right) + {1 \over 8 \pi^2 f^{3/2}\e^3 } \bigg\}\,.
}
Now, we must express the counterterms as a sum in energy modes, and in order to do that, we must convert the inverse powers of $\e$ into sums. Although this case had not been dealt in the literature before, there is a simple way to accomplish this. One starts by applying the Abel-Plana formula (\ref{APsum}) to the function $\left(a^2 \omega_n^2 \right)^s \cos\left(\omega_n \varepsilon\right)$, taking the first and third derivative with respect to $s$ and then setting $s=0$. These steps will lead to the results
\ea{
\lim_{\e \to 0}{1 \over \e} & = - {2 \kappa \over \pi} \sum^{\infty}_{n=1}\ln \omega_n + \mathcal{O}(\e)\,, \label{e1} \\
\lim_{\e \to 0}{1 \over \e^3} & = {\kappa \over \pi} \sum^{\infty}_{n=1}\omega^2_n\ln \omega_n + \mathcal{O}(\e)\,. \label{e2}
}
Inserting Eqs.~(\ref{e1}) and (\ref{e2}) into Eq.~(\ref{GDS}), one immediately arrives at Eq.~(\ref{Gdiv}), thus confirming the existence of a finite result for the vacuum polarisation.

%% file: II.55Dcharged/3.section.tex
\section{Numerical results for the five-dimensional electrically charged
Reissner-Nordstr\"om black hole}
\label{Num}

In obtaining $G_{\rm reg}$ one has made use of the WKB approximation, since $G_{\rm reg} = G_{\rm WKB}-G_{\rm div}$. We want to go a step further and obtain
a more exact result. The remainder $\de G$ between the exact value of the Euclidean Green function $G_{\rm E}$ and the WKB approximated Green function $G_{\rm WKB}$, i.e., $\de G = G_{\rm E} - G_{\rm WKB}$, is usually ignored because it is considered negligible. However, here, in our numerical calculation, we take care of this remainder $\de G$. Thus, instead of writing the approximated vacuum polarization expression as usual, $\braket{\phi^2(x)}_{\textrm{ren.}} = G_{\rm reg}$, we use the exact value for the fully renormalized vacuum polarization as
\eq{
\braket{\phi^2(x)}_{\textrm{ren.}} = G_{\rm reg} + \de G\,.
}
The quantity $G_{\rm reg}$ can be evaluated directly using Eq.~(\ref{ren1}), in its explicitly finite form, which can be written as a combination of Hurwitz-Zeta functions, just like in the four dimensional Lifshitz case. We obtain
\eq{\label{eqsdf}
G_{\rm reg} = {\kappa \over 4 \pi^3 r^3} \sum_{n=1}^{\infty} \left\{Z_{reg}^{(1)}+\sum_{k=2}^{6} d_{2k+1} Z^{(2k+1)} +J_{reg}+P_{reg}\right\}\,,
}
where
\eq{
Z^{(k)} = \sum_{n=1}^\infty \left(\omega_n^2 + \si\right)^{-k/2}, \quad Z^{(1)}_{reg} = {1\over 2} \sum_{n=1}^{\infty} \left[\left(\omega_n^2 + \si\right)^{-1/2} - {1\over \omega_n} \right]\,,
}
and
\ea{
d_{5}  & = -\frac{3 f^2 \left(-f^{'2} x^2+f^{'} f r+3 f^2\right)}{32 r^4} \,, \\
d_{7}  & = \frac{3 f^3 \left(-4 f^{'3} r^3+33 f^{'2} f r^2+30 f' f^2 r-105 f^3\right)}{256 r^6}\,, \\
d_{9}  & = \frac{9 f^4 \left(11 f^{'4} r^4-78 f^{'3} f r^3+13 f^{'2} f^2 r^2+606 f^{'} f^3 r+525 f^4 \right)}{1024 r^8}\,, \\
d_{11} & = -\frac{27 f^6 (f' r+f)^2 \left(-101 f^{'2} r^2+361 f' f r+693 f^2\right)}{4096 r^{10}}\,, \\
d_{13} & = \frac{93555 f^8 (f' r+f)^4}{65536 r^{12}}\,.
}
For numerical purposes, the expression of $Z^{(1)}_{reg}$ is regulated in the same was as Eq.~(\ref{Zm1Reg}). The expression for $J_{reg}$ can be calculated explicitly in terms of hypergeometric functions. However, the result is very long and will not be reported here, although it can be obtained more or less straightforwardly using a symbolic manipulation program.
It should also be noted that the upper bound on the sum over $k$ in (\ref{eqsdf}) reflects the order of the WKB approximation, since higher orders increases the numbers of zeta functions to be added up. The fact that the WKB approximation part of the vacuum polarization can be expressed as a series of such zeta functions was already noted in Ref.~\cite{Flachi:2008sr}. Rather than showing the explicit results for the integrals, it is more instructive to explicitly report the diverging contributions generated by each term, which we subtract in order to regulate each expression:
\eq{
J_{reg} = \int_0^\infty  j_n^{(5)}(x) dx - \left\{ -{1\over 3 \omega_n} + {r^3 \omega_n^2\over 2 f^{3/2}} \left(1+\ln\left({r \omega_n \over 2 \sqrt{f}}\right)\right) + \tilde{a}_1 \left(1+\ln\left({r \omega_n \over 2 \sqrt{f}}\right)\right)\right\}
}
where
\eq{
\tilde a_1 = {r^3\over 8 \sqrt{f}} \left( m^2 - [a_1] +{f'\over r} - {f^{'2} \over 4 f} + {f^{''}\over 3} \right),
}
and
\eq{
P_{reg} = i \int_0^\infty {j^{(5)}_n(i x) -j^{(5)}_n (-i x) \over e^{2\pi x}-1 } dx + {1\over 6\omega_n}.
}

In the numerical results that follow, we have used the WKB approximation and calculated numerically the remainder $\de G$, which is the most computationally demanding term. In the process of numerically calculating the remainder, we used Eqs.~(\ref{psol}) and (\ref{psol0}) for the first point in the numerical range of the solution (near-horizon limit) and Eq.~(\ref{qsol}) for the last point (large radius limit). Of course, if we were to increase the order of the WKB approximation in $G_{\rm reg}$, it would reduce the magnitude of the remainder $\de G$. We have opted to use the WKB approximation up to second order since it in general yields accurate results.

In what follows, we specify that the metric given in Eq.~(\ref{RS5}) is the metric for a five-dimensional electrically charged Reissner-Nordstr\"om black hole, such that $f(r)$ is given by
\eq{\label{rn2}
f(r) = 1 - {2 m \over r^2} + {q^2 \over r^4}\,,
}
where $m$ is the mass parameter and $q$ is the electrically charge parameter. The metric function $f(r)$ given in Eq.~(\ref{rn2}) has an event horizon with radius
\eq{
r_+ = \left( m + \sqrt{m^2 - q^2} \right)^{1/2}\,.
}
It has another horizon, the Cauchy horizon, with radius $r_- = \left( m - \sqrt{m^2 - q^2} \right)^{1/2}$, but it does not enter into our calculations.
In addition, for the function $f(r)$ given in Eq.~(\ref{rn2}), the inverse Hawking temperature defined in Eq.~(\ref{temp})
is
\eq{
\be = {(m + \sqrt{m^2-q^2})^{5/2} \over (m^2 - q^2 + m\sqrt{m^2-q^2})}\pi \,.
}
For completeness we remark that the parameters $m$ and $q$ appearing in Eq.~(\ref{rn2}) are related to the black hole ADM mass $M$ and electrical charge $Q$, through the relations
\ea{
m  & = {4 G_5 M \over 3 \pi}\,, \\
q^2  & = {4 \pi \over 3}G_5 Q^2\,,
}
respectively, where $G_5$ is the gravitational constant for a five-dimensional spacetime.

\begin{figure}[h!]
  \centering
    \includegraphics[width=0.5\textwidth]{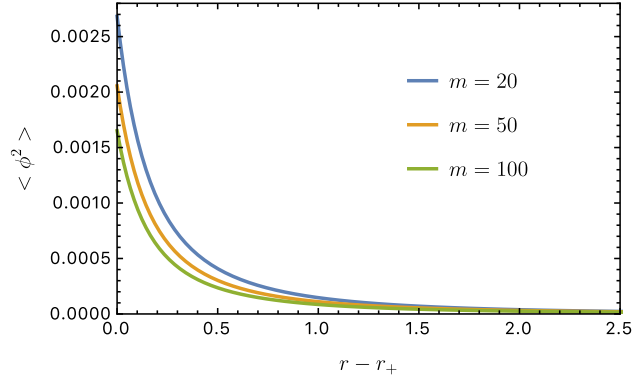}
    \caption{
      Plots of the vacuum polarization
      $\braket{\phi^2}_{\textrm{ren}}-\braket{\phi^2}_{\infty}$ as a
      function of the coordinate distance from the black hole horizon
      radius, i.e., $r-r_+$, for three black hole masses $m$. The charge
      and scalar field mass are fixed as $q = 10$ and $\mu = 0$,
      respectively.}
\end{figure}

\begin{figure}[h!]
  \centering
    \includegraphics[width=0.5\textwidth]{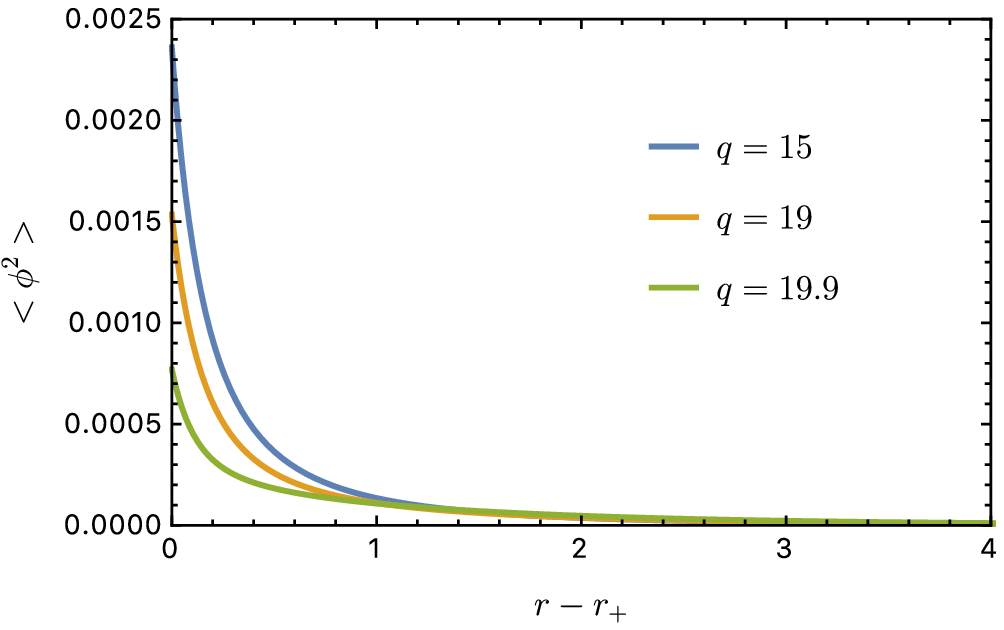}
    \caption{Plots of the vacuum polarization
      $\braket{\phi^2}_{\textrm{ren}}-\braket{\phi^2}_{\infty}$ as a
      function of the coordinate distance from the black hole horizon
      radius, i.e., $r-r_+$, for three black hole charges $q$. The
      back hole and scalar field masses are fixed as $m = 20$ and $\mu
      = 0$, respectively.}
\end{figure}

\begin{figure}[h!]
  \centering
    \includegraphics[width=0.5\textwidth]{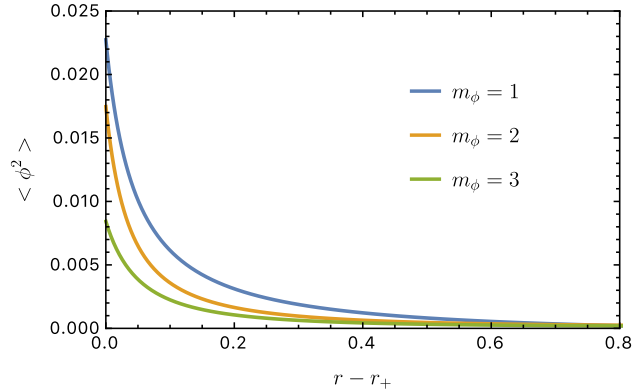}
      \caption{Plots of the vacuum polarization
        $\braket{\phi^2}_{\textrm{ren}}-\braket{\phi^2}_{\infty}$ as a
        function of the coordinate distance from the black hole
        horizon radius, i.e., $r-r_+$, for three scalar field masses
        $\mu$. The mass and charge of the black hole are fixed as $m =
        20$ and $q = 10$, respectively.}
\end{figure}

In Figs.~1-3, we plot $\braket{\phi^2}_{\textrm{ren}}-\braket{\phi^2}_{\infty}$, i.e., the renormalized vacuum polarization normalized to zero at infinity, as a function of the coordinate distance from the five-dimensional Reissner-Nordstr\"{o}m black hole horizon radius, i.e., $r-r_+$, for three different values of the black hole mass, black hole electric charge, and scalar field mass, respectively. For each parameter choice, we find finite values at the horizon with no problems of convergence.  Note that, since we are dealing with a five-dimensional spacetime, the trace of the Maxwell stress-energy tensor does not vanish, as the general formula for $d$ dimensions \cite{Tangherlini:1963bw}
\eq{
T^{\mu}{}_{\mu} = \left({d-4\over 4}\right) {\Gamma[(d-1)/2] \over 2\pi^{(d-1)/2}}F^{\mu\nu}F_{\mu\nu}
}
shows. Consequently, due to the Einstein equations, the Ricci scalar of the Reissner-Nordstr\"{o}m metric is not zero, unlike the four-dimensional case. Despite this, we choose to set $\xi=0$, which is an assumption commonly used, as the mass already introduces a non-trivial factor into the problem. In Fig.~1, we see that the value of the vacuum polarization at the horizon decreases with increasing black hole mass. This is expected, as the black hole temperature decreases and so it is harder to produce excitations in the quantum field. In Fig.~2, the value at the horizon decreases with increasing charge, i.e., as the black hole approaches the extremal limit. This is again expected, as an extremal black hole has zero temperature. In Fig.~3, we see that increasing scalar field mass induces a larger vacuum polarization at the horizon.

%% file: II.55Dcharged/conclusions.tex
\section{Conclusions}

In this work we have extended our previous results \cite{Flachi:2016bwr} and calculated the renormalized vacuum polarization for a massive scalar field around a five-dimensional electrically charged black hole. We have followed the standard approach which makes use of the WKB approximation to extract the
infinities present both in the angular and energy modes of the mode-sum expanded Green function. We have also compared the explicit divergent part with the Schwinger-DeWitt counterterms to get a fully renormalized result for the vacuum polarization. Terms up to second order were used in the approximation, which provided numerical results illustrating the behavior of the vacuum polarization as a function of the various parameters. A simple understanding of the finer features of the vacuum polarization $\braket{\phi^2}_{{\rm ren}}$ in the various cases is difficult due to the complexity of the calculations involved.

%% file: II.65D6D/main.tex
\chapter{Vacuum Polarization in Higher Dimensional Spacetimes}
\label{cap:chapterII6}
\input{II.65D6D/outline.tex}

\input{II.65D6D/1.section.tex}

\input{II.65D6D/2.section.tex}

\input{II.65D6D/3.section.tex}
\input{II.65D6D/4.section.tex}

\input{II.65D6D/conclusions.tex}

%% file: II.65D6D/outline.tex
\section{Introduction}

Developing the formalism adopted in previous chapters for the calculation of $\braket{\hat{\phi}^2}$ to spacetimes of arbitrary dimensions will be the goal of this chapter. The same WKB approach will be considered, and the more sensible shortcomings of the approximation will become evident once one performs the calculations with free dimensionality.

We will continue here the analysis of quantum vacuum polarization around higher dimensional black holes and present the results for the vacuum polarization outside a $D=5$ uncharged black hole as a check. In addition, we also prove the renormalization of the vacuum polarization for $D=6$ in the large mass limit where we can compare our results with those of Ref.~\cite{Thompson:2008bk}.

%% file: II.65D6D/1.section.tex
\section{Quantum scalar field in a Schwarzschild-Tangherlini spacetime}

A static neutral black hole in $d$ spacetime dimensions is described by the Schwarschild-Tangherlini metric (\ref{LE2}), whose Euclideanized version reads
\eq{
ds^2 = f(r) d\tau^2 + f^{-1}(r) {dr^2} + r^{2} d\Omega_{d-2}^2
}
with
\eq{
f(r) = 1 - \left({r_h \over r}\right)^{d-3}\,.
}
A quantum scalar field in the presence of such a background is described by the action
\eq{
S = {1\over 2} \int d^d x \sqrt{g} \left[
g^{\mu \nu} \partial_\mu \phi \partial_\nu \phi + m_{\phi}^2 \phi^2 + \xi R \phi^2
\right]\,.
}
Although the Ricci scalar is identically zero in this case, we still keep the curvature coupling term in order to derive more general results. The Euclidean Green function $G_E(x,x')$ satisfies the differential equation
\eq{\label{GEN}
\left(\Box - m_{\phi}^2 -\xi R \right) G_E(x,x') = - {\delta^{(d)} (x-x') \over \sqrt{g}}\,.
}
Since the background spacetime is spherically symmetric, one may use the expasion of Eqs.~(\ref{GExpN}) and (\ref{GExpL}) in order to express the Green function in thermal equilibrium with the black hole as
\eq{\label{GreenExp}
G_{\textrm{E}}(x,x') = {\kappa\over 4\pi^2} {\Gamma\left({N\over 2}\right) \over \pi^{(N/ 2)}} \sum^{\infty}_{n=-\infty} e^{i \omega_n \Delta\tau} \sum_{l=0}^{\infty} \left(l+{N \over 2}\right) C^{(N/2)}_l\left(\cos \g\right) G_{nl}(r,r')\,.
}
We recall that $N=d-3$ and that the Gegenbauer polynomials $C^{(N/2)}_l\left(\cos \g\right)$ generalize to higher dimensions the Legendre functions and result from the summation over the azimuthal quantum numbers of the hyper-spherical harmonics in $d$ dimensions (relevant formulae can be found in Ref.~\cite{Gradshteyn:2007}). Inserting the expanded form of the Green function in Eq.~(\ref{GEN}) and representing the Dirac delta on the right hand side using Eqs.~(\ref{deTau}) and (\ref{deO}), we obtain the differential equation for the radial Green function
\eq{
{d^2 G_{nl} \over dr^2} + \left({N+1\over r} + {f'\over f}\right){d G_{nl} \over dr} - \left({\omega_n^2\over f^2} + {l(l+N) \over f r^2} +{m_{\phi}^2+\xi R \over f}\right)G_{nl} = - {\delta(r-r') \over r^{N+1}}
}
As in the previous chapters, the full solution for the radial Green function can be written in terms of the two independent solutions $p_{nl}$ and $q_{nl}$ of the homogeneous radial wave equation
\eq{\label{HomoEqN}
{d^2 \chi_{nl}(r) \over dr^2} + \left({N+1\over r} + {f'\over f}\right){d \chi_{nl}(r) \over dr} - \left({\omega_n^2\over f^2} + {l(l+N) \over f r^2} +{m_{\phi}^2+\xi R \over f}\right){\chi_{nl}}(r) = 0\,,
}
assuming the form
\eq{\label{GFN}
G_{nl}\left(r,r'\right) = {1\over r^{N+1} f} {p_{nl}(r_{<}) q_{nl}(r_{>}) \over \ma{W}_{r}\{p_{nl},q_{nl}\}}\,.
}
As we have seen in previous chapters, the direct numerical evaluation of (\ref{GreenExp}) is impeded by the diverging nature of the coincidence limit. To bypass the problem we will again use the point-splitting method and take the coincidence limit along all directions but the timelike one. We then use the WKB approximation to explicitly extract the divergences. With the divergences in hands, we compute the counter-terms using the Schwinger - De Witt expansion and subtract them off. This way of proceeding is nothing but a generalization to higher dimensions of the method developed by Candelas \cite{Candelas:1980zt} and later refined by Anderson \cite{Anderson:1994hg}.

Proceeding the same way as in previous chapters, we must find the asymptotic form of the homogeneous equation solutions in the near horizon and infinity limits. Beginning with the behavior near infinity, we get from the homogeneous equation (\ref{HomoEqN}) the limiting form
\eq{
\left\{{d^2 \over dr^2} + {(N+1) \over r} {d \over dr} - (\omega^2_n + m_{\phi}^2 + \xi R)\right\}q_{nl}(r) = 0\,,
}
the solution of which, regular at infinity, is of the form
\eq{\label{qsolN}
q_{nl}(r) \sim r^{-(N+1)/2} e^{- r \sqrt{\omega^2_n + m_{\phi}^2 + \xi R}}\,.
}
As for the near-horizon limit, we again make use of a tortoise coordinate $r_*$ defined through
\eq{
dr_* = {dr \over f(r)}\,,
}
in terms of which, in the near-horizon limit and for $n \neq 0$, the homogeneous equation of Eq.~(\ref{HomoEqN}) becomes
\eq{\label{nn0N}
\left({d^2 \over dr^2_*} - \omega^2_n\right) p_{nl}(r)=0\,.
}
The solution of Eq.~(\ref{nn0N}), regular at the horizon, is given by
\eq{\label{psolN}
p_{nl}(r) \sim {e^{-\omega_n r_*} \over r}\,.
}
In the case $n = 0$, the homogeneous equation of Eq.~(\ref{HomoEqN}), in the near-horizon limit, becomes
\eq{
{d \over dr}(\ln p_{0l}(r)) = {1 \over f'(r)}\left( {l(l+N) \over r^2} + m_{\phi}^2 + \xi R \right)\,,
}
the solution of which goes as
\eq{\label{psol0N}
p_{0l}(r) \sim \exp\left\{ \int^{r}_{r_+} \left( {l(l+N) \over u^2} + m_{\phi}^2 + \xi R \right) {du \over f'(u)} \right\}\,.
}

With the asymptotic forms of $p_{nl}$ and $q_{nl}$ at our disposal, we now take the general WKB ansatz of Eq.~(\ref{WKBansatz}) and insert it in the homogeneous equation (\ref{HomoEqN}). The different powers of the ansatz are then fixed by the choice which eliminates the sign changing terms in the resulting homogeneous equation as well as keeping the powers consistent with the know limit solutions of Eqs.~(\ref{qsolN}) and (\ref{psolN}). We are left with the WKB form for the homogeneous solutions
\ea{\label{AnsatzN}
p_{nl}(r) & = {1 \over \sqrt{r^{1+N} W}}\exp\left({+ \int_{r_h}^{r} W(u) {du \over f(u)}}\right) \,, \\
q_{nl}(r) & = {1 \over \sqrt{r^{1+N} W}}\exp\left({- \int_{r_h}^{r} W(u) {du \over f(u)}}\right) \,.
}
The homogeneous equation then assumes the exact same form as Eq.~(\ref{wkbeqt}), i.e.
\eq{\label{Wwkb56}
W^2 = \Psi + a_1 {W'\over W} +a_2 {W^{'2}\over W^2} +a_3 {W^{''}\over W}\,,
}
with the analogous quantities
\ea{
\Psi & = \left[\left(l+{N\over 2}\right)^2-{N^2\over 4}\right] {f \over r^2}+ \omega_n^2 + \si \,, \\
\sigma & = \left(m_{\phi}^2 +\xi R\right)f + \left({N^2\over 4}-{1\over 4}\right){f^2 \over r^2} + \left({N+1 \over 2}\right){f f' \over r}\,,
}
and
\eq{
a_1  = {f f' \over 2}\,, \quad a_2  = -{3 f^2 \over 4}\,, \quad a_3  = {f^2 \over 2}\,.
}
Since Eq.~(\ref{Wwkb56}) has the exact same form as Eq.~(\ref{Wwkb}), we write the WKB function using the same iterative approximation, such that we again have
\eq{
{1\over W} = {1\over \sqrt{\Psi}} \left(1 + \de_1 \Psi + \de_2 \Psi + \cdots\right)\,, \nonumber
}
where the first two orders are still given by Eqs.~(\ref{d1psi}) and (\ref{d2psi}). Depending on the dimensionality at hand, we will need to retain increasingly more orders $\de_n \Psi$ to account for all diverging pieces. For $d=4,5$ we have already seen that $\de_1 \Psi$ is enough, while we shall see that for $d=6$ there are still some leftover diverging terms in $\de_2 \Psi$.

%% file: II.65D6D/2.section.tex
\section{Renormalization}
\label{Renorm}

Following the usual protocol, we now take the spatial coincidence limit, obtaining for the radial Green function
\eq{
G_{nl}(r,r) = {1 \over 2r^{N+1} W}\,,
}
and thus, the complete Green function
\eq{\label{GeDtN}
G_{\rm E}(x,x) = {\kappa \over 8\pi^2 r} {\Gamma\left({N \over 2}\right) \over (\pi r^2)^{{N \over 2}} \Gamma(N)} \sum_{n=-\infty}^{\infty} e^{i \kappa n \e} \sum_{l=0}^{\infty} \left(l+{N \over 2}\right)  {(l+1)_{N-1} \over W}\,,
}
where we define the Pochhammer symbol
\eq{
(x)_a = {\Gamma(x+a) \over \Gamma(x)}\,.
}
As it happened for four and five spacetime dimensions, the summation both in the energy and angular modes will be divergent, and we will have to deal with them separatly.

\subsection{Regularization in the $l$ modes}

In this case, terms of $(l+N/2)$ with powers larger than $-1$ will exist, so one must remove them using the same trick applied in previous chapters. For general $N$, we must expand
\eq{
\left(l+{N \over 2}\right)  {(l+1)_{N-1} \over W}
}
for large $(l+N/2)$, giving some expression $\mathcal{T}_l(r)$. This expression will contain powers of $n$ in even combinations, which vanish when summed over $n$ in a $\zeta$ regularized way, so the latter quantity can be constructed as a function of powers of $l$ only.
The divergent contribution coming from the angular modes summation can then be safely removed by subtracting the term
\eq{\label{TlN}
{\kappa \over 8\pi^2 r} {\Gamma\left({N \over 2}\right) \over (\pi r^2)^{{N \over 2}} \Gamma(N)} \sum^{\infty}_{n=-\infty} e^{i \omega_n \e} \sum^{\infty}_{l=0} \mathcal{T}_l(r)\,,
}
from Eq.~(\ref{GeDtN}), since Eq.~(\ref{TlN}) is a multiple of $\de(\e)$. After the subtraction we may take the full coincidence limit, in which the Green function becomes
\eq{\label{GeDnN}
G_{\rm E}(x,x) = {\kappa \over 8\pi^2 r} {\Gamma\left({N \over 2}\right) \over (\pi r^2)^{{N \over 2}} \Gamma(N)} \sum_{n=-\infty}^{\infty}\sum_{l=0}^{\infty} \left[ \left(l+{N \over 2}\right)  {(l+1)_{N-1} \over W} - \mathcal{T}_l(r)\right]\,.
}

\subsection{Regularization in the $n$ modes}

To regularize the summation in the $n$ modes, we must extract the physical divergences contained in it. Using the same procedure as in Chapter 11, we rewrite the coincidence limit of the Green function as
\eq{
G_{\rm E}(x,x) = {\kappa \over 8\pi^2 r} {\Gamma\left({N \over 2}\right) \over (\pi r^2)^{{N \over 2}} \Gamma(N)} \left(G_0 + 2\sum_{n=1}^{\infty} G_n \right)
}
where we defined
\eq{
G_n = \sum_{l=0}^{\infty} \left[ \left(l+{N \over 2}\right)  {(l+1)_{N-1} \over W_{nl}} - \mathcal{T}_l(r)\right]\,.
}
As before, $G_0$ is finite by construction, so all divergences must be contained within $G_n$. Direct application of the Able-Plana formula (\ref{APsum}) will allows us to perform the summation in the $l$ modes, thus leaving expressions in term of the modes $n$. An expansion for large $n$ will reveal the powers that will contribute to the overall divergent piece of the Green function.

To explicitly confirm the cancellation of the divergences, we express the counterterms given by Eqs.~(\ref{GdivE}) or (\ref{GdivO}) in terms in summations in energy modes, by using higher power cases of Eqs.~(\ref{e1}) and (\ref{e2}). This process becomes increasingly complex for higher dimensions due to essentially two reasons: as the number of dimensions grows, we need more heat kernel coefficients $a_k$, which themselves are very complicated as $k$ grows; as a consequence, increasingly more (lengthy) corrections to the WKB approximation become necessary, turning the countertems cancellation very heavy, even when using a symbolic manipulation software.

After the cancellation of the divergent terms, we are left with a finite expression written in terms of the finite leftover parts of the WKB approximation, with the vacuum polarization written in the form of a renormalized WKB part plus a remainder to be calculated numerically, akin to Eq.~(\ref{WKBtrick}), i.e.,
\eq{
\braket{\phi^2(x)}_{\rm ren} = \braket{\phi^2(x)}^{\rm ren}_{WKB} + \de \braket{\phi^2(x)}\,.
}

%% file: II.65D6D/3.section.tex
\section{Five dimensional case - Neutral static black hole}

\begin{figure}[h!]
\begin{center}
\unitlength=1mm
\unitlength=1mm
\begin{picture}(180,60)
   \includegraphics[height=5cm]{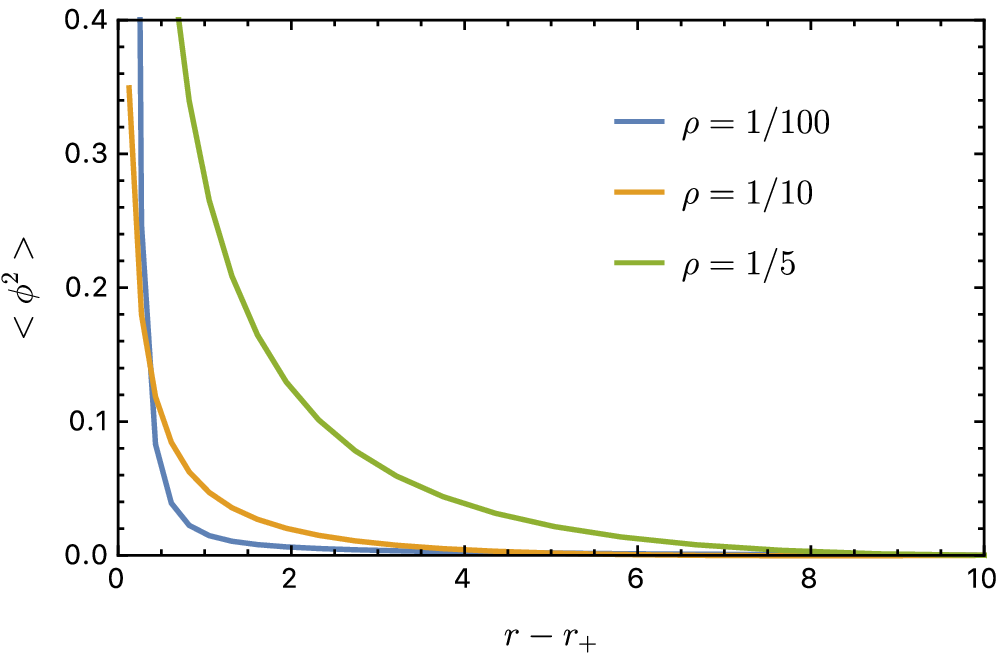}
   \includegraphics[height=5cm]{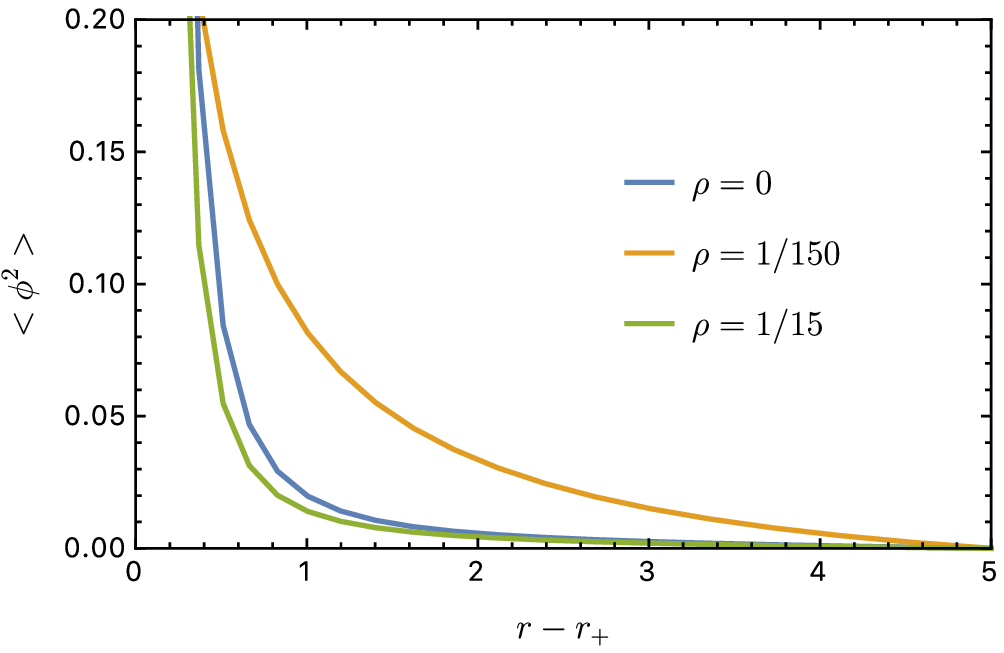}
\end{picture}
\end{center}
\caption{Profile of the renormalized vacuum polarization for $M_{BH}=5$ (left-panel) and $M_{BH}=15/2$ (right-panel), for various values of the parameter $\rho={m\over 2M_{BH}}$.}
\label{fig1}
\end{figure}

Although we have already given a treatment for the calculation of the vacuum polarization in a five dimensional spacetime, we revisit here this case by giving explicit results for a neutral non-rotating black hole. The regulated Green function will have the same form as Eq.~(\ref{eqsdf}), which we recall again for convenience
\eq{
G_{\rm reg} = {\kappa \over 4 \pi^3 r^3} \sum_{n=1}^{\infty} \left\{Z_{reg}^{(1)}+\sum_{k=2}^{6} d_{2k+1} Z^{(2k+1)} +J_{reg}+P_{reg}\right\}\,.
}
Using the metric function of a neutral non-rotating black hole
\eq{
f(r) = 1 - \left({2M_{\rm BH} \over r}\right)^{2}\,,
}
where $M_{\rm BH}$ is the black hole mass, we obtain the plots of Fig.~(\ref{fig1}), for different values of the parameter
\eq{
\rho = {m_{\phi} \over 2M_{\rm BH}}\,.
}

%% file: II.65D6D/4.section.tex
\section{Six dimensional case - Large $m_{\phi}$ limit renormalization}

The six-dimensional case is computationally more demanding and extracting the divergences is not trivial. The reason for the increased complexity comes from the fact that higher order WKB terms give rise to additional ultraviolet divergent contributions and this leads to algebraically very cumbersome combinations of hypergeometric functions. Computing the counter-terms also becomes more difficult as the dimensionality increases. Some simplification can be achieved in the specific limit of large mass and here, as a check on the method, we have limited our analysis to this case. The procedure to extract the divergences is basically the same as in five dimensions and consists in operating on the Abel-Plana rearrangement of $G_n$. Keeping the first two leading terms in a large mass expansions we find the following diverging behavior for the Green function:
\eq{
G_{\rm div} = {\kappa\over 128 \pi^4} \sum^{\infty}_{n=1}{1 \over \omega_n }\left\{m_{\phi}^4 + m_{\phi}^2 \left[\frac{4 \omega_n^2}{f}+2\left(\frac{1}{6} - \xi \right)\left(f''+{8f'\over r} +{12f\over r^2} -{12\over r^2} \right)\right]\right\}\,,
}
where we are considering only the first two terms in a large mass expansion. In extracting the divergences above all the terms above are generated by the second integral in the Abel-Plana rearrangement (\ref{APsum}) in six dimensions. The form of the divergences can be quickly understood from the general form of the heat-kernel coefficients and dimensional analysis. It can be noted that the last term multiplying $m_{\phi}^2$ is the scalar curvature in six dimensions. The counter-terms can be extracted similarly to what we did in five dimensions, giving, in the large mass limit,
\eq{
G_{\rm div} = \left({m_{\phi}^4 \over 64 \pi^3} - {[a_1] m_{\phi}^2 \over 32\pi^3}\right)\log\varepsilon  -  {m_{\phi}^2 \over 16 \pi^3 f}
\frac{1}{\varepsilon^2}\,.
}
Using the expressions from Refs.~(\cite{Candelas:1984pg,Anderson:1989vg}) that relate $\log \varepsilon$ and $1/\varepsilon^2$ with sums of powers of $\omega_n$, given by
\ea{
\ln\e & = -{1 \over \kappa} \sum^{\infty}_{n=1} {1 \over \omega_n} + \ma{O}(\e)\,, \\
{1 \over \e^2} & = -\kappa \sum^{\infty}_{n=1} \omega_n \cos(\omega_n \e) + {\kappa^2 \over 12} + \ma{O}(\e)\,,
}
our result exactly compensates the divergences. In fact, it coincides with the result of Ref.~\cite{Thompson:2008bk} in the same limit.

%% file: II.65D6D/conclusions.tex
\section{Conclusions}

In this paper we have studied the renormalized vacuum polarization for a higher dimensional Schwarzschild-Tangherlini black hole. We have presented a general approach for computing the vacuum polarization and fully analyzed the problem in five dimensions. We have also extracted the divergences using the WKB approximation and explicitly calculated the counter-terms, proving the regularity of the result. Finally, we have then evaluated the renormalized expression numerically. In six dimensions, we have limited ourselves to proving the regularity of the vacuum polarization in the large mass limit due to the rapid growing of the problem's complexity. In the latter case, we extracted the divergences using the WKB approximation, calculated the counter-terms and explicitly verified the regularity. Our results for the counter-terms in the large mass limit coincide with those of Ref.~\cite{Thompson:2008bk}. Straightforward generalizations of the calculations of this paper include other black hole geometries. However, the most interesting and non-trivial generalization is the increase in number of dimensions. It should also be possible to relax the high mass approximation and work out in full detail along similar lines the six dimensional case with some effort.

%% file: II.7SymRest/main.tex
\chapter{Symmetry restoration of a scalar field through a charged black hole}
\label{cap:chapter7}
\input{II.7SymRest/outline.tex}
\input{II.7SymRest/1.section.tex}
\input{II.7SymRest/2.section.tex}

\input{II.7SymRest/3.section.tex}

\input{II.7SymRest/4.section.tex}

\input{II.7SymRest/5.section.tex}

\input{II.7SymRest/conclusions.tex}

%% file: II.7SymRest/outline.tex
\section{Introduction}

Excluding interactions from the action simplifies the calculations, and is an interesting problem by itself. However, the inclusion of interacting terms is more realistic since new effects come into play, enriching the amount of physical results. In particular, black holes provide a perfect trigger of symmetry breaking and phase transitions of various kinds (see \cite{Gregory:2013hja,Burda:2015yfa} for example). The physics is in fact remarkably simple and clearly explained in Refs.~\cite{Hawking:1980ng,Fawcett:1981fw,Moss:1984zf}. Due to gravitational redshift, the radiation emitted by a black hole looses energy as it propagates through spacetime, decreasing the overall local energy density as the distance from the horizon increases. Inversely, an observer at infinity will feel a temperature increase as he approaches the black hole. This makes it possible for a system in a broken phase at large distances to have its symmetry restored sufficiently close to the black hole. Although this problem has been considered before for neutral black holes (see for example \cite{Hawking:1980ng}), explicit numerical solutions have never been worked out. This chapter will be dedicated to this purpose.

%% file: II.7SymRest/1.section.tex
\section{Physical setup}

We shall be interested in a massive self-interacting quantum scalar field $\hat{\phi}$, following a theory with an action
\eq{\label{AIF}
S[\hat{\phi}] = - {1\over 2} \int dv_x \left\{ g^{\mu\nu}\hat{\phi}_{,\mu}\hat{\phi}_{,\nu} + (m_{\phi}^2+\xi R) \hat{\phi}^2 - {\la \over 2} \hat{\phi}^4 \right\}
}
where $\la$ is the self-interaction coupling constant. We shall refrain from specifying the background geometry of spacetime until it becomes necessary, so as to maintain as much generality as possible. In order to study any aspect of this theory, we must have access to the effective action which, by using Eq.~(\ref{LOEA}), we know to be of the form
\eq{
\Gamma[\Phi] = S[\Phi] + {i \over 2} \ln \textrm{det}(\ell^2 S_{,\Phi\Phi}[\Phi])
}
up to leading order. Since we are dealing with a single component scalar field, the second derivative in the field components is simply $S_{,ij} \equiv S_{,\Phi\Phi}$. The goal now is to use the effective action to obtain the background field $\Phi(x)$, which can be done by developing Eq.~(\ref{CFE}). Reintroducing field indexes for the moment, one uses Eq.~(\ref{LOEA}) to obtain
\ea{
{\de \Gamma[\Phi^k] \over \de \Phi^l} & = {\de S[\Phi^k] \over \de \Phi^l} + {i \over 2} S^{-1}_{ij} {\de S^{ji}[\Phi^k] \over \de \Phi^l} \\
& = \sqrt{|g|} \left(\square \Phi^l- \la \de_{ij} \Phi^i \Phi^j \Phi^l +(m_{\phi}^2+\xi R)\Phi^l - 3 \la \braket{\hat{\phi}^2} \Phi^l \right)\,.
}
In deriving the above result we have used the fact that
\eq{
{\de S^{ji}[\Phi^k] \over \de \Phi^l} = 6 \la \Phi^l \de^{ji} \sqrt{|g|}
}
and $\de^{ji} = - S_{,}{}^{j}{}_{p} G^{p}{}_{i}$, which can be read off of Eq.~(\ref{GEdef}). Using the result $G^{i}{}_{i} \equiv \braket{\hat{\phi}^2}$ finally introduces the vacuum polarization into the equation. The field equation of motion Eq.~(\ref{CFE}) for the background field of one component is thus
\eq{\label{BFE}
\square \Phi - \Phi (\la \Phi^2 - (m_{\phi}^2+\xi R) + 3 \la \braket{\hat{\phi}^2}) = 0\,.
}
Before we proceed any further, however, one must check that the theory at hand is renormalizable.

%% file: II.7SymRest/2.section.tex
\section{Renormalization of the effective action}
\label{renormEA}

In order to check the renormalizability of the theory described by the action of Eq.~(\ref{AIF}), we will have to consider each quantity appearing in the action as a bare quantity prior to renormalization, i.e. we must consider the matter action
\eq{\label{BA}
S_M[\Phi] = - {1\over 2} \int dv_x \left\{ (\dd^{\mu}\Phi_{B})(\dd_{\mu}\Phi_{B}) + (m_{\phi B}^2+\xi_B R) \Phi_{B}^2 - {\la_B \over 2} \Phi_B^4 \right\}
}
where the subscript $B$ denotes a bare quantity that must be renormalized. It happens that, as we shall see shortly, in order to renormalize the theory in a curved spacetime, we must also consider the gravitational action with curvature square terms, of the form
\eq{
S_G = {1\over 2} \int dv_x \left\{ \Lambda_{B} + \kappa_B R + \al_{1B} R^{\mu\nu\rho\si} R_{\mu\nu\rho\si} + \al_{2B} R^{\mu\nu}R_{\mu\nu} + \al_{3B} R^2\right\}\,.
}
We will adopt dimensional regularization as in Chapter (QFTCS), whereby we consider the number of spacetime dimensions to be $4+\e$, with $\e$ some positive quantity. In order to obtain properly adimensional quantities in the action, we must multiply each of them by some power of a length scale $\ell$, chosen so that the volume of extra dimensions is $\ell^{\e}$. We will also assume no dependence of the background field on the coordinates of the extra dimensions.

Standard renormalization procedure now dictates that each bare quantity must be decomposable into some finite quantity plus an infinite counterterm, and that these counterterms are in precisely the same number as the infinities appearing in the effective action. Putting the latter reasoning into practice, we define
\ea{
\kappa_B & = \ell^{-\e} (\kappa + \de\kappa)\,, \label{kB} \\
\al_{iB} & = \ell^{-\e}(\al_i + \de \al_i), \,\, (i=1,2,3)\,, \label{aB} \\
\Phi_{B} & = \ell^{-\e/2} (1+\de Z) \Phi_{B}\,, \label{PhiB} \\
m_{\phi B}^2 & = m_{\phi}^2 + \de m_{\phi B}^2\,, \label{mB} \\
\xi_{B} & = \xi + \de \xi\,, \label{xiB} \\
\la_{B} & = \ell^{\e} (\la + \de \la)\,, \label{lB}
}
where the counterterms are denoted with a $\de$ and where we have multiplied by a proper power of $\ell^{\e}$ in each term in order to assure correct dimensions. If the theory is renormalizable, we should be able to attribute a counterterm to each singular term of the effective action. We must then calculate all the singular contributions of the effective action, a task which has already been performed in Chapter 2 in all generality. In fact, we only need to note that from the action Eq.~(\ref{AIF}) one may extract that (with field indexes)
\eq{\label{SMij}
S_M{}_{,ij} = \left[-\de_{ij}\square_x - \de_{ij} m^2_{\phi}-\de_{ij} \xi R + 3 \la \Phi_i \Phi_j \right] \de(x,x')\,,
}
i.e., we are considering the particular case of Eq.~(\ref{OpeF}) with
\eq{\label{QIF}
Q(x) = m^2_{\phi}+\xi R - 3 \la \Phi^2(x)\,.
}
All divergent pieces are thereby contained within Eq.~(\ref{divG1}), which for $d=4$ and a one-component scalar field is
\eq{\label{divGIF}
\textbf{divp} \, \Gamma^{(1)} = -{1\over 16\pi^2 \e} \int dv_x E_{2}(x)
}
with
\ea{\label{E2}
E_{2}(x) & = \left({1\over 72} + {1 \over 2}\xi^2 - {1\over 6}\xi\right) R^2 + {1\over 180} \left(R^{\mu\nu\rho\si}R_{\mu\nu\rho\si} - R^{\mu\nu}R_{\mu\nu}\right) + \left(\xi  - {1 \over 6}\right) m^2_{\phi} R \nonumber \\
& \hspace{5mm} + {1\over 2} m^4_{\phi} + {1\over 2} m^2_{\phi} \la \Phi^2 + {1\over 2}\left( \xi - {1\over 6}\right)\la R \Phi^2 + {1\over 8} \la^2 \Phi^4\,.
}
The Laplacian terms are not present, since one has
\eq{
\int d^4x \sqrt{|g|} \nabla_{\mu}\nabla^{\mu}\Phi = \int d^4x \, \dd_{\mu}(\sqrt{|g|} \dd^{\mu}\Phi) = \sqrt{|g|} \dd^{\mu}\Phi \bigg|^{\infty}_{-\infty} = 0
}
and both $\Phi$ and $\dd_{\mu}\Phi$ are expected to go to 0 at spatial infinity, where we have also used the relation
\eq{
\sqrt{|g|} \nabla_{\mu} v^{\mu} = \dd_{\mu}(\sqrt{|g|} v^{\mu})
}
valid for any vector field $v^{\mu}$. Since the derivative of the Ricci curvature $\dd_{\mu}R$ is also expected to go to zero at infinity, we conclude that all Laplacian terms in Eq.~(\ref{divG1}) for the choice (\ref{QIF}) go to zero after spacetime integration.

On the other side, inserting the couplings defined through Eqs.~(\ref{kB})-(\ref{lB}) and focusing on the counterterm part of the action, we obtain
\ea{\label{divPG1}
\textbf{divp} \, \Gamma^{(1)} & = \int dv_x \bigg\{ \de\Lambda + \de\kappa R + \de\al_{1} R^{\mu\nu\rho\si} R_{\mu\nu\rho\si} + \de\al_{2} R^{\mu\nu}R_{\mu\nu} + \de\al_{3} R^2 + \de Z (\dd^{\mu}\Phi_{B})(\dd_{\mu}\Phi_{B}) \nonumber \\
& \hspace{5mm} - \left( {1\over 2} \de m^2_{\phi} + m^2_{\phi} \de Z \right) \Phi^2 - \left({1\over 2}\de\xi + \xi \de Z\right) R\Phi^2 - \left({1\over 4!} \de\la + {1\over 3!} \la \de Z\right)\Phi^4 \bigg\}\,.
}
Inserting Eq.~(\ref{E2}) in Eq.~(\ref{divGIF}) and comparing with Eq.~(\ref{divPG1}), we see that the counterterms are exactly of the same type as the divergent terms of the effective action, with the exception of $\de Z (\dd^{\mu}\Phi_{B})(\dd_{\mu}\Phi_{B})$, which has no associated singular part in Eq.~(\ref{E2}). This implies that one does not need to renormalize the background field at leading order through Eq.~(\ref{PhiB}), and so we may set
\eq{
\de Z = 0
}
without any consequences. As a consequence, the effective action is renormalized at leading order, i.e. it remains finite for $\e \to 0$ if one performs the choices
\ea{
\de \Lambda & = {1 \over 32\pi^2 \e} m^4_{\phi}\,, \\
\de \kappa & = {1\over 16\pi^2 \e} \left(\xi - {1\over 6}\right) m^2_{\phi}\,, \\
\de \al_1 & = {1\over 16\pi^2 \e} {1\over 180} \,, \\
\de \al_2 & = -{1\over 16\pi^2 \e} {1\over 180} \,, \\
\de \al_3 & = {1\over 16\pi^2 \e} \left({1\over 72} + {1 \over 2}\xi^2 - {1\over 6}\xi\right) \,, \\
\de m^2_{\phi} & = -{1\over 16\pi^2 \e} m^2_{\phi} \la \,, \\
\de \xi & = {1\over 16\pi^2 \e} \left( {1\over 6} - \xi \right) \la \,, \\
\de \la & = - {1\over 8\pi^2 \e} \la^2 \,.
}
Hence, it is proved that the theory of a self-interacting scalar field in curved spacetime described by the action (\ref{AIF}) is renormalizable.

%% file: II.7SymRest/3.section.tex
\section{Calculation of the vacuum polarization}

In order to extract the background field from Eq.~(\ref{BFE}), one needs to first calculate the vacuum polarization $\braket{\hat{\phi}^2}$, which is the coincidence limit of the Green function associated to Eq.~(\ref{SMij}). One therefore needs to solve
\eq{\label{phi2IF}
\left[ \square_x - Q(x) \right]G_{\rm E}(x,x') = - {\de^{(4)}(x-x') \over \sqrt{|g|}}
}
the obtain the vacuum polarization, where $Q(x)$ is given by Eq.~(\ref{QIF}). Unsurprisingly, analytically solving the systems of Eqs.~(\ref{BFE}) and (\ref{phi2IF}) is an impossible task, so we must resort to some approximation method. As we shall detail in the next section, we will make use of a so-called self-consistent approximation, which will require us to evaluate multiple instances of the vacuum polarization. Despite having used the standard WKB approach in all previous chapters, this approach is very sensitive at the horizon, and very time consuming if one wishes to increase the numerical accuracy of the results. For this reason, we will employ a recent alternative method to calculate the vacuum polarization which, although having its own drawbacks, has a much faster convergence under certain conditions.{}

The approach we will use to calculate the vacuum polarization was developed in \cite{Taylor:2017sux}. In essence, it consists in choosing the point-splitting of the world function $\si(x,x')$ in such a way that one may express the divergent Green mode functions in a closed form, which are subtracted mode by mode from the original mode functions to give a finite result, with the original Green function expressed in a spherical harmonic form of Eqs.~(\ref{GExpN}) and (\ref{GExpL})
\eq{\label{GEexp}
G_{\rm E}(x,x') = {\kappa \over 4\pi^2} \sum^{\infty}_{n=-\infty} e^{i n \kappa \Delta\tau} \sum^{\infty}_{l=0} \left(l+{1\over 2}\right)  P_l(\cos\gamma)G_{nl}(r,r')	\,.
}
We will highlight here this process for the particular case $d=4$. The divergent part of the Green function, given in Eq.~(\ref{GdivE}), can be put in the so-called Hadamard parametrix form
\eq{\label{GdivT}
G_{\rm div}\left( x , x ^ { \prime } \right) = \frac { 1 } { 8\pi^2 } \left\{ \frac { \Delta^{1/2}(x,x') } { \sigma\left( x , x ^ { \prime } \right) } + V \left( x , x ^ { \prime } \right) \log \left( 2 \sigma(x,x') / \ell ^ { 2 } \right) \right\}
}
where the parameter $\ell$ is introduced to ensure a dimensionless logarithm argument, and where the function $V(x,x')$ can be expanded in powers of $\si$ as
\ea{
V \left( x , x ^ { \prime } \right) & = \sum _ { p = 0 } ^ { \infty } V _ { p } \left( x , x ^ { \prime } \right) \sigma ^ { p }\,.
}
Inserting the Hadamard parametrization in Eq.~(\ref{phi2IF}), one finds the differential equation for $V_p(x,x')$
\eq{\label{EqVp}
2 ( p + 1 ) ( p + 2 ) V _ { p + 1 } + 2 ( p + 1 ) \sigma ^ { a } \nabla _ { a } V _ { p + 1 }  - 2 ( p + 1 ) V _ { p + 1 } \Delta ^ { - \frac { 1 } { 2 } } \sigma ^ { a } \nabla _ { a } \Delta ^ { \frac { 1 } { 2 } }  + \left( \square - m ^ { 2 } - \xi R \right) V _ { p } = 0
}
for $p\neq 0$ and
\eq{\label{EqV0}
2 V _ { 0 } + 2 \sigma ^ { a } \nabla _ { a } V _ { 0 } - 2 V _ { 0 } \Delta ^ { - \frac { 1 } { 2 } } \sigma ^ { a } \nabla _ { a } \Delta ^ { \frac { 1 } { 2 } }  + \left( \square - m ^ { 2 } - \xi R \right) \Delta^{1/2} = 0
}
for $p=0$. The departing point from standard approaches is to now assume an expansion of the world function in the form
\eq{
\si(x,x') = \sum_{ijk} \si_{ijk}(r) \e^{i+j+k} w^i \Delta r^j s^k
}
where $\Delta r = r-r'$ and
\eq{
w^2 = {2\over \kappa^2}(1-\cos(\kappa \Delta \tau))\,, \quad s^2 = f(r) w^2 + 2r^2 (1-\cos\g)
}
are referred to as ``extended coordinates''. We treat each extended coordinate as quantities of order $\e$, where $\e$ is some small quantity. The coefficients $\si_{ijk}(r)$ are then found by inserting the expansion in Eq.~(\ref{RelSi}) and equating each order individually.  The functions $V _ { p } \left( x , x ^ { \prime } \right)$ can also be expanded in powers of extended coordinates, i.e.
\eq{
V _ { p } \left( x , x ^ { \prime } \right) = \sum _ { i j k } v _ { i j k } ^ { ( p ) } ( r ) \e ^ { i + j + k } w ^ { i } \Delta r ^ { j } s ^ { k }
}
where the coefficients $v _ { i j k } ^ { ( p ) } ( r )$ can be obtained by inserting the above expansion in Eqs.~(\ref{EqVp}) and Eq.~(\ref{EqV0}). We will proceed by considering a metric of the form
\eq{\label{ChargedBH}
ds^2 = f(r) d\tau^2 + f^{-1}(r) dr^2 + r^2 d\theta^2 + r^2\sin^2\theta d\varphi^2\,,
}
where it proves very useful to use the following formulas in the calculation of Laplacians
\eq{
{1 \over \sin^2\theta} \left({\dd \cos\g \over \dd \varphi}\right)^2 + \left( {\dd \cos\g \over \dd\theta} \right)^2 = 1-\cos^2\g\,,
}
\eq{
\left[{1 \over \sin\theta} {\dd \over \dd\theta} \left( \sin\theta {\dd \over \dd\theta} \right) + {1\over \sin^2\theta} {\dd \over \dd \varphi^2}\right](\cos^n\g) = n(n-1)\cos^{n-2}\g - (n+1)n \cos^n\g\,.
}
Once the desired expansion coefficients are calculated, one may insert them in Eq.~(\ref{GdivT}), obtaining the expansion for the divergent part of the Green function
\ea{
\frac { \Delta^{1/2} } { \sigma } + V \log \left( 2 \sigma / \ell ^ { 2 } \right) & = \sum _ { i = 0 } ^ { 2 } \sum _ { j = 0 } ^ { i } \mathcal { D } _ { i j } ^ { ( + ) } ( r ) \e ^ { 2 i - 2 } \frac { w ^ { 2 i + 2 j } } { s ^ { 2 j + 2 } } + \log \left( \e ^ { 2 } s ^ { 2 } / \ell ^ { 2 } \right) \sum _ { i = 0 } ^ { 1 } \sum _ { j = 0 } ^ { i } \mathcal { T } _ { i j } ^ { ( 1 ) } ( r ) \e ^ { 2 i } s ^ { 2 i - 2 j } w ^ { 2 j } \nonumber \\
& + \mathcal { T } _ { 10 } ^ { ( r ) } ( r ) \e ^ { 2 } s ^ { - 2 } w ^ { 4 } + \mathcal { D } _ { 1 , 1 } ^ { ( - ) } ( r ) + \mathcal { O } \left( \e ^ { 4 } \log \e \right)
}
along with some finite terms which go to zero in the coincidence limit. These terms are not strictly necessary, but they speed the overall converge of numerical calculations. The coefficients of the Hadamard parametrix are found to be
\ea{
\mathcal { D } _ { 00 } ^ { ( + ) } ( r ) & = 2\,, \\
\mathcal { D } _ { 10 } ^ { ( + ) } ( r ) & = \frac{f' f}{6 r}-\frac{f'f}{12} -\frac{f^2}{6 r^2}+\frac{f}{6 r^2} \,, \\
\mathcal { D } _ { 11 } ^ { ( + ) } ( r ) & = -\frac{f' f^2}{6 r}+\frac{f'^2 f}{24} +\frac{f^3}{6 r^2}-\frac{f^2}{6 r^2}-\frac{\kappa ^2 f}{6}  \,, \\
\mathcal { D } _ { 20 } ^ { ( + ) } ( r ) & = -\frac{7 f' f^{(2)} f^2}{360 r}+\frac{f'^2 f^{(2)} f}{576} +\frac{f' f^{(3)} f^2}{240} -\frac{19 f' f^3}{360 r^3}+\frac{f' f^2}{36 r^3}+\frac{43 f'^2 f^2}{1440 r^2}-\frac{f'^2 f}{288 r^2}+\frac{\kappa^2 f' f}{72 r} \nonumber \\
& -\frac{f'^3 f}{288 r}+\frac{7 f^{(2)} f^3}{360 r^2}-\frac{f^{(2)} f^2}{144 r^2}-\frac{\kappa ^2 f^{(2)} f}{144} +\frac{(f^{(2)})^2 f^2}{320} -\frac{f^{(3)} f^3}{120 r}+\frac{19 f^4}{720 r^4} -\frac{f^3}{36 r^4}+\frac{f^2}{720 r^4} \nonumber \\
& -\frac{\kappa ^2 f^2}{72 r^2}+\frac{\kappa ^2 f}{72 r^2} \,, \\
\mathcal { D } _ { 21 } ^ { ( + ) } ( r ) & = \frac{11 f' f'' f^3}{720 r}-\frac{11 f'^2 f'' f^2}{2880}+\frac{13 f' f^4}{180 r^3}-\frac{f' f^3}{18 r^3}-\frac{67 f'^2 f^3}{1440 r^2}+\frac{f'^2 f^2}{96 r^2}-\frac{\kappa ^2 f' f^2}{24 r}+\frac{\kappa^2 f'^2 f}{144} +\frac{f'^3 f^2}{96r} \nonumber \\
& -\frac{f'^4 f}{2880}-\frac{11 f'' f^4}{720 r^2}+\frac{f'' f^3}{144 r^2}+\frac{\kappa ^2 f'' f^2}{144} -\frac{13 f^5}{360 r^4}+\frac{f^4}{18 r^4}-\frac{7 f^3}{360 r^4}+\frac{\kappa ^2 f^3}{24 r^2}-\frac{\kappa ^2 f^2}{24 r^2}-\frac{\kappa^4 f}{45}  \,, \\
\mathcal { D } _ { 22 } ^ { ( + ) } ( r ) & = -\frac{f' f^5}{36 r^3}+\frac{f' f^4}{36 r^3}+\frac{f'^2 f^4}{48 r^2}-\frac{f'^2 f^3}{144 r^2}+\frac{\kappa ^2 f' f^3}{36 r}-\frac{\kappa ^2 f'^2 f^2}{144} -\frac{f'^3 f^3}{144 r}+\frac{f'^4 f^2}{1152}+\frac{f^6}{72 r^4}-\frac{f^5}{36 r^4} \nonumber \\
& +\frac{f^4}{72 r^4}-\frac{\kappa ^2 f^4}{36 r^2}+\frac{\kappa ^2 f^3}{36 r^2}+\frac{\kappa ^4 f^2}{72}  \,, \\
\mathcal { D } _ { 11 } ^ { ( - ) } ( r ) & = -{f' \over 6r}\,, \\
\ma{T}^{(r)}_{10}(r) & = \frac{f' f'' f^2}{144r}-\frac{f'^2 f''f}{576} +\frac{Q f' f^2}{24 r}-\frac{Q f'^2 f}{96} -\frac{f' f^3}{72 r^3}+\frac{f' f^2}{72 r^3}+\frac{7 f'^2 f^2}{288 r^2}+\frac{f'^2 f}{288 r^2}+\frac{\kappa ^2 f' f}{36r} \nonumber \\
& -\frac{f'^3 f}{144r}-\frac{f'' f^3}{144r^2}+\frac{f'' f^2}{144r^2}+\frac{\kappa ^2 f'' f}{144} -\frac{Qf^3}{24 r^2}+\frac{Q f^2}{24 r^2}+\frac{\kappa ^2 Q f}{24}-\frac{f^4}{72r^4}+\frac{f^3}{36 r^4}-\frac{f^2}{72r^4} \nonumber \\
& +\frac{\kappa ^2 f^2}{72 r^2}-\frac{\kappa ^2f}{72 r^2}\,, \\
\ma{T}^{(l)}_{00}(r) & = \frac{f'}{3r}+\frac{f''}{12}+\frac{f}{6r^2}+\frac{Q}{2}-\frac{1}{6 r^2} \,, \\
\ma{T}^{(l)}_{10}(r) & = -\frac{Q' f'}{48} -\frac{f'f''}{120 r}-\frac{f'f^{(3)}}{240} +\frac{f' f}{40r^3}+\frac{Q f'}{24r}+\frac{f'^2}{120 r^2}-\frac{Q'f}{12 r}-\frac{f'' f}{30r^2}+\frac{Qf''}{48} +\frac{f''^2}{480}-\frac{Q''f}{48} \nonumber \\
& -\frac{7 f^{(3)} f}{240 r}-\frac{f^{(4)} f}{240}+\frac{f^2}{120r^4}+\frac{Q^2}{16}-\frac{1}{120 r^4}\,,
}
\ea{
\ma{T}^{(l)}_{10}(r) & = -\frac{Q' f' f}{48} -\frac{f'f'' f}{40 r}-\frac{f'f^{(3)} f}{480} +\frac{f' f^2}{30r^3}-\frac{f' f}{24r^3}+\frac{f'^2 f}{240r^2}+\frac{Q' f^2}{24 r}-\frac{Qf'' f}{48} +\frac{f'' f^2}{40r^2}-\frac{f''^2f}{240} \nonumber \\
& +\frac{f^{(3)} f^2}{80 r}+\frac{f^{(4)} f^2}{480} +\frac{Q f^2}{24 r^2}-\frac{Qf}{24 r^2}-\frac{f^3}{60r^4}+\frac{f}{60 r^4} \,.
}
The idea now is to express the terms $ w ^ { 2 i + 2 j } / s ^ { 2 j + 2 } $ and $\log \left(s ^ { 2 } / \ell ^ { 2 } \right) s ^ { 2 i - 2 j } w ^ { 2 j }$ in the same spherical harmonic decomposition of the Euclidean Green function, i.e.
\ea{
{w ^ { 2 i \pm 2 j } \over s ^ { 2 \pm 2 j }} & = \sum^{\infty}_{n=-\infty} e^{i n \kappa \Delta \tau} \sum^{\infty}_{l=0} \left(l+{1\over 2}\right) P_l(\cos\gamma) \Psi^{(\pm)}_{nl}(i,j|r) \,, \\
\log \left(s ^ { 2 } / \ell ^ { 2 } \right) s ^ { 2 i - 2 j } w ^ { 2 j } & = \sum^{\infty}_{n=-\infty} e^{i n \kappa \Delta \tau} \sum^{\infty}_{l=0} \left(l+{1\over 2}\right) P_l(\cos\gamma) \chi_{nl}(i,j|r) \,,
}
where $\Psi_{nl}(i,j|r)$ and $\chi_{nl}(i,j|r)$ are the mode functions of the decomposition. The surprising fact is that both mode functions can be written in closed form. Bypassing a good deal of intermediate calculations, we cite the results of \cite{Taylor:2017sux}:
\ea{
\eta(r) & = \sqrt { 1 + \frac { f ( r ) } { \kappa ^ { 2 } r ^ { 2 } } } \,, \\
{ \Psi }^{(+)} _ { n l }( i , j | r ) & = \frac { 2 ^ { 2 i - j - 2 } ( - 1 ) ^ { n } i ! \Gamma \left( i + \frac { 1 } { 2 } \right) } { \sqrt{\pi} \kappa ^ { 2 i + 2 j } r ^ {2 j + 2 } \Gamma \left( j + 1 \right) } \left( \frac { 1 } { \eta } \frac { d } { d \eta } \right) ^ { j } \bigg\{ \frac { P _ { l  } ( \eta ) Q _ { l } ( \eta ) } { ( i - n ) ! ( i + n ) ! } \nonumber \\
& + \sum _ { p = \max \{ 1 , n - i \} } ^ { i + n } \frac { P _ { l } ^ { - p } ( \eta ) Q _ { l } ^ { p } ( \eta ) } { ( i + p - n ) ! ( i - p + n ) ! } + \sum^{i-n} _ { p = \max \{ 1 , - n - i \} } \frac { P _ { l } ^ { - p } ( \eta ) Q _ { l } ^ { p } ( \eta ) } { ( i + p + n ) ! ( i - p - n ) ! } \bigg\}\,, \\
{ \Psi^ { ( - ) }   _ { n l } ( i , j | r ) } & = \frac { 2 ^ { 2 i - 2 j - 1 } ( - 1 ) ^ { n + j } ( i - j ) ! \Gamma \left( i - j + \frac { 1 } { 2 } \right) } { \sqrt{\pi} \kappa ^ { 2 i - 2 j } r ^ { 2 - 2 j } \Gamma \left( 1 - j \right) } \sum _ { k = 0 } ^ { j } ( - 1 ) ^ { k } \left( \begin{array} { l } { j } \\ { k } \end{array} \right) \frac { ( l + {1\over 2} + j - 2 k ) } { ( l + {1\over 2} - k ) _ { j + 1 } } \nonumber \\
& \times \bigg\{ \frac { P _ { l + j - 2 k } ( \eta ) Q _ { l + j - 2 k } ( \eta ) } { ( i - j - n ) ! ( i - j + n ) ! } + \sum^{i-j+n} _ { p = \max \{ 1 , n - i + j \} } \frac { P _ { l + j - 2 k } ^ { - p } ( \eta ) Q _ { l + j - 2 k } ^ { p } ( \eta ) } { ( i - j + p - n ) ! ( i - j - p + n ) ! } \nonumber \\
& + \sum _ { p = \max \{ 1 , - n - i + j \} } ^ { i - j - n } \frac { P _ { l + j - 2 k } ^ { - p } ( \eta ) Q _ { l  + j - 2 k } ^ { p } ( \eta ) } { ( i - j + p + n ) ! ( i - j - p - n ) ! } \bigg\}\,, \\
{ \chi } _ { n l } ( i , j | r ) & = ( - 1 ) ^ { n } \frac { 2 ^ { 2 j - 2 } r ^ { 2 i - 2 j } ( i - j ) ! j ! \Gamma \left( j + \frac { 1 } { 2 } \right) } { \sqrt{\pi} \kappa ^ { 2 j } } \sum _ { k = 0 } ^ { 1 + i - j } ( - 1 ) ^ { k } \left( \begin{array} { c } { 1 + i - j } \\ { k } \end{array} \right) \frac { \left( l + i - j - 2 k + \frac { 3 } { 2 } \right) } { ( l + {1\over 2} - k ) _ {1 + i - j } } \nonumber \\
& \times \bigg\{ \frac { P _ { l + 1 + i - j - 2 k } ( \eta ) Q _ { l + 1 + i - j - 2 k } ( \eta ) } { ( j - n ) ! ( j + n ) ! } + \sum _ { p = \max \{ 1 , n - j \} } \frac { P _ { l + 1 + i - j - 2 k } ^ { - p } ( \eta ) Q _ { l + 1 + i - j - 2 k } ^ { p } ( \eta ) } { ( j + p - n ) ! ( j - p + n ) ! } \nonumber \\
& + \sum _ { p = \max \{ 1 , - n - j \} } \frac { P _ { l + 1 + i - j - 2 k } ^ { - p } ( \eta ) Q _ { l + 1 + i - j - 2 k } ^ { p } ( \eta ) } { ( j + p + n ) ! ( j - p - n ) ! } \bigg\}\,, \quad \textrm{for} \quad l>i-j\,, \\
{ \chi } _ { n l } ( i , j | r ) & = \frac { \kappa } { 2 \pi }  \left( 2 r ^ { 2 } \right) ^ { i - j -1} ( - 1 ) ^ { l } \left( \frac { 2 } { \kappa ^ { 2 } } \right) ^ { j } \bigg[ \frac { d } { d \lambda } ( \lambda + 1 - l ) _ { l } \left( \frac { 2 r ^ { 2 } } { \ell ^ { 2 } } \right) ^ { \lambda - i + j } \int _ { 0 } ^ { 2 \pi / \kappa } ( 1 - \cos \kappa t ) ^ { j } e ^ { - i n \kappa t } \nonumber \\
& \times \left( z ^ { 2 } - 1 \right) ^ { \frac { 1 } { 2 } \left( 1 + \lambda \right) } \mathcal { Q } _ { l } ^ { - 1 - \lambda } ( z ) d t \bigg] _ { \lambda = i - j }\, \quad \textrm{for} \quad  l \leq i - j\,.
}
Taking the coincidence limit $\Delta \tau\to 0$ and $\g \to 0$, we conclude that the divergent part of the Euclidean Green function can be expressed as
\eq{
G_{\rm E \, div} = {\kappa \over 4\pi^2} \sum^{\infty}_{n=-\infty} \sum^{\infty}_{l=0} \left(l+{1\over 2}\right)  G^{S}_{nl}(r) - {f'\over 48\pi^2}
}
where
\eq{
G^{S}_{nl}(r) = {1\over \kappa} \bigg\{\sum _ { i = 0 } ^ { 2 } \sum _ { j = 0 } ^ { i } \mathcal { D } _ { i j } ^ { ( + ) } ( r )  \Psi^{(+)} _ { n l }( i , j | r ) + \mathcal { T } _ { 10 } ^ { ( \mathrm { r } ) } ( r ) { \Psi }^ { ( - ) }  _ { n l } \left( 2, 0 | r \right) + \sum _ { i = 0 } ^ { 1 } \sum _ { j = 0 } ^ { i } \mathcal { T } _ { i j } ^ { ( 1 ) } ( r ){ \chi } _ { n l } ( i , j | r ) \bigg\}\,.
}
Subtracting the above quantity from the unrenormalized Green function of Eq.~(\ref{GEexp}) and rearranging the terms, we obtain the renormalized vacuum polarization in the form
\eq{\label{phi2RenT}
\braket{\hat{\phi}^2} _ { \rm ren } = \frac { \kappa } { 4 \pi^2 } \sum _ { l = 0 } ^ { \infty }  \bigg( l + {1\over 2} \bigg) \bigg\{ G _ { 0 l } ( r ) - G _ { 0 l } ^ { S } ( r ) + 2 \sum _ { n = 1 } ^ { \infty } \left( G _ { n l } ( r ) - G _ { n l } ^ {S} ( r ) \right) \bigg\} + {f'\over 48\pi^2} \,.
}
This is the expression we will use to numerically calculate the regularized vacuum polarization. From the way it is computed, it is clear that one cannot construct an explicit finite quantity, having instead to subtract two diverging quantities to obtain a finite one. On the other hand, a lot of computational power is gained from this, since we only need to perform the calculations for a few tens of modes in order to obtain good convergence at the horizon. In the next section we discuss in more detail the pros and cons of this method.

%% file: II.7SymRest/4.section.tex
\section{Symmetry restoration outside a charged black hole}
\label{SymR}

The goal is to compute the background field $\Phi$ for given values of mass $m_{\phi}$ and coupling constant $\la$, on the outside of a charged black hole described by the metric (\ref{ChargedBH}) with
\eq{
f(r) = {(r-r_+)(r-r_-) \over r^2}
}
where $r_+$ is the event horizon and $r_-$ the Cauchy horizon, given by
\eq{
r_{\pm} = M_{\rm BH} \pm \sqrt{M_{\rm BH}^2 - Q_{\rm BH}^2}\,,
}
with $M_{\rm BH}$ the mass of the black hole and $Q_{\rm BH}$ its charge. A black hole of this type has a surface gravity of
\eq{
\kappa = {r_+ - r_- \over 2r^2_+}\,.
}
Assuming a solitonic type configuration for the background field, we will have $\Phi \equiv \Phi(r)$, so the differential equation Eq.~(\ref{BFE}) obeyed by the same field will reduce to
\eq{\label{BFEr}
\left\{{d^2 \over dr^2} + \left({2 \over r}+{f' \over f}\right){d \over dr} - \left({m_{\phi}^2-3\la \braket{\phi^2}_{\rm ren} \over f}\right)\right\} \Phi + {\la \over f}\Phi^3=0
}
where the Ricci curvature is eliminated since $R=0$ identically for a Reissner-Nordstr\"om black hole in four dimensions. Note that we have taken only the vacuum polarization to be renormalized, since we have seen in Sect.~\ref{renormEA} that no renormalization is needed for $\Phi$ at leading order.
{}
To solve Eq.~(\ref{BFEr}), we will employ a self-consistent approximation in the following way. First, we compute $\braket{\hat{\phi}^2}_{\rm ren}$ for the case where no background field is present, i.e. for $Q = m^2_{\phi}$. Then we insert the result in Eq.~(\ref{BFEr}) and compute the resulting $\Phi$. After that, we compute again the vacuum polarization but now with $\Phi$ inserted, i.e. such that we have the effective mass squared of Eq.~(\ref{QIF}), given by $Q(r) = m^2_{\phi} - 3 \la \Phi^2(r)$. We take the resulting $\braket{\hat{\phi}^2}_{\rm ren}$ and put it back into Eq.~(\ref{BFEr}), giving a new function for $\Phi$. These steps are repeated until the results for the vacuum polarization and background field stop changing appreciably.

We will first give the details involved in the calculation of the vacuum polarization and then follow with the procedure used to calculate the background field.

\subsection{Calculation of $\braket{\hat{\phi}^2}$}

Each iteration of the self-consistent approximation we use, will involve a calculation of $\braket{\hat{\phi}^2}_{\rm ren}$ for a given function $Q(r)$. Although the quantity to calculate is directly given by Eq.~(\ref{phi2RenT}), there are some careful points to consider in calculating it. One particularly important aspect to consider is the calculation of the Green function modes $G_{nl}(r)$ which, using Eq.~(\ref{phi2IF}), are seen to satisfy the equation
\eq{\label{GEeqV}
{d^2 G_{nl} \over dr^2} + \left({2 \over r} + {f' \over f}\right){d G_{nl} \over dr} - \left({\kappa^2 n^2 \over f^2}+{l(l+1) \over f r^2}+{m_{\phi}^2 - 3 \la \Phi^2(r) \over f}\right) G_{nl} = - {\de^{(4)}(r-r') \over r^2}\,.
}
One then employs a similar reasoning of previous chapters to conclude that
\eq{
G_{nl}(r,r') = C_{nl} \, p_{nl}(r_<) q_{nl}(r_>)\,,
}
where
\eq{
C_{nl} \, \ma{W}\{p_{nl}(r),q_{nl}(r)\} = -{1 \over r^2 f(r)}\,.
}
By integrating the above relation between $r$ and $\infty$, one is able to extract the relation
\eq{
C_{nl} \, q_{nl}(r) = p_{nl}(r) \int^{\infty}_{r} {dr' \over r'^{2} f(r') (p_{nl}(r'))^2}\,,
}
which essentially asserts that we only need to compute $p_{nl}(r)$ in order to find $q_{nl}(r)$, and consequently $G_{nl}(r,r')$. The mode function $p_{nl}(r)$ obeys the homogeneous version of Eq.~(\ref{GEeqV}), i.e.
\eq{\label{DifEqPnl}
{d^2 p_{nl} \over dr^2} + \left({2 \over r} + {f' \over f}\right){d p_{nl} \over dr} - \left({\kappa^2 n^2 \over f^2}+{l(l+1) \over f r^2}+{Q(r) \over f}\right) p_{nl} = 0\,.
}
For $n=0$ and $Q(r)=0$, we have the analytic solution
\eq{
p_{nl}(r) = P_{0l}\left({2r \over r_+ + r_-} - {r_+ + r_- \over r_+ + r_-}\right)\,,
}
which motivates the definition
\eq{
\xi = {2r \over r_+ + r_-} - {r_+ + r_- \over r_+ + r_-}\,, \quad 1 \leq \xi < \infty\,,
}
and the subsequent change of variables
\eq{
\zeta = \xi-1 = {1\over \kappa r_+} \left({r\over r_+}-1\right)\,, \quad 0 \leq \zeta < \infty
}
or, inversely,
\eq{
r = r_+ (1+\kappa r_+ \zeta)\,.
}
The variable $\zeta$ is exactly of the type which facilitates a Frobenius analysis of a function at the origin, or at the horizon in the variable $r$. This is ideal, since $p_{nl}(r)$ can only be evaluated numerically in this case, and it will require input boundary conditions, which we consequently choose to impose at some value very close to the horizon.

We shall then employ the Frobenius method to derive the boundary conditions at the horizon, i.e. near $\zeta = 0$. The goal is to obtain an expansion of the form
\eq{\label{pnlZeta}
p_{nl}(\zeta) = \zeta^{\rho} \sum^{\infty}_{k=0} a_k \zeta^k\,,
}
which amounts to finding the coefficients $a_k$. Using
\eq{
{d \over dr} = {1\over \kappa r^2_+}{d \over d\zeta}\,, \quad {d^2 \over dr^2} = {1\over \kappa^2 r^4_+}{d^2 \over d\zeta^2}
}
we rewrite Eq.~(\ref{DifEqPnl}) in the canonical form
\eq{\label{ZetaEq}
\zeta^2 {d^2 p_{nl} \over d\zeta^2} + \zeta h(\zeta) {d p_{nl} \over d\zeta} + w(\zeta) p_{nl} = 0
}
with
\ea{
h(\zeta) & = {r_+ - r_- +2\kappa r^2_+ \zeta \over r_+ - r_- + \kappa r^2_+ \zeta} = 2\left({1+\zeta \over 2+\zeta}\right) = \sum^{\infty}_{k=0} \left({\zeta \over 2} \right)^k \equiv \sum^{\infty}_{k=0} h_k \zeta^k\,, \\
w(\zeta) & = {-n^2 (r_- \zeta - r_+ (2+\zeta))^4 - r^4_+ 4\zeta (4l(l+1)+(r_- \zeta-r_+ (2+\zeta))^2 Q) \over 16 \, r^4_+ (2+\zeta)^2} \equiv \sum^{\infty}_{k=0} w_k \zeta^k\,.
}
The procedure is to insert the expansion (\ref{pnlZeta}) in Eq.~(\ref{ZetaEq}) and solve for the coefficients $a_k$ order by order, obtaining in general an equation which expresses each $a_k$ in terms of coefficients with higher and/or lower indexes $k$. For example, for $k=0$, we obtain
\eq{
\rho(\rho+1) + h_0 \rho + w_0 = 0\,,
}
also called the fiducial equation. Since $h_0=1$ and $w_0 = -n^4/4$, we obtain the two possible roots $\rho_1 = n/2$ and $\rho_2=-n/2$. We will consider only the first root $\rho_1$, since it is the one which gives a regular result at the horizon, i.e. it is the one associated to $p_{nl}$. Using this information, and after rearranging Eq.~(\ref{ZetaEq}) in powers of $\zeta^{\rho_1 + k}$, one is lead to the recurrence relation
\ea{
& -a_k \left(r_+^4 \left(k^2+k-l (l+1)+Q r_+ (2 r_--3
   r_+)\right)+\frac{1}{2} (2 k+1) n r_+^4-\frac{1}{4} n^2 r_+^2
   \left(6 r_-^2-12 r_- r_++5 r_+^2\right)\right) \nonumber \\
   & +r_+^3
   a_{k+1} \left(2 r_+ (k (2 k+5)-l (l+1)+3)+n r_+ (4 k-n+5)+2 n^2
   r_--2 Q r_+^3\right) \nonumber  \\
   &-\frac{1}{16} a_{k-2} (r_--r_+)^2
   \left(n^2 (r_- -r_+)^2+4 Q r_+^4\right)+\frac{1}{2} r_+ a_{k-1}
   (r_--r_+) \left(n^2 (r_--r_+)^2+Q r_+^3 (3
   r_+-r_-)\right) \nonumber \\
   & - a_{k+2} 4 (k+2) r_+^4 (k+n+2) = 0\,.
}
Using $a_{-3} = a_{-2} = a_{-1} = 0$ and $a_0 = 1$, one finally obtains the desired expansion Eq.~(\ref{pnlZeta}) and with it the value of $G_{nl}$ at the horizon, as well as its derivate. The values of $p_{nl}$ outside the horizon are then obtained by numerically solving Eq.~(\ref{DifEqPnl}) subject to the calculated boundary conditions. For this purpose, we used the function \textbf{NDSolve} of the \textit{Mathematica} software, which gives the option to regulate the number of digits for working precision, precision goals and accuracy goals.

\subsection{Calculation of $\Phi$}

We now move on to the calculation of the background field. The differential equation (\ref{BFEr}) has some interesting features which make it fairly challenging to directly obtain a numeric solution (\cite{Hawking:1980ng}). The main feature is the fact that the system behaves as having an effective mass given by Eq.~(\ref{QIF}), i.e. it acts as if under the influence of an effective potential
\eq{\label{Veff}
V_{\rm eff}(\Phi) = -{1\over 2}(m^2+\xi R-3\la \braket{\phi^2})\Phi^2 + {1\over 4} \la \Phi^4\,.
}
For each radial distance, this potential affects the solution in a way which is strongly dependent on $\braket{\phi^2}$. At large distances from the horizon the metric is essentially flat, so we have the general result for the vacuum polarization at any temperature and any mass \cite{Hewitt:2015}
\eq{
\braket{\phi(\infty)^2}_{T} = {1\over 2\pi^2} \int^{\infty}_{m} {\sqrt{\omega^2-m^2} \over e^{\omega / T} - 1} \, d\omega\,.
}
At large radii, the derivative terms of Eq.~(\ref{BFEr}) are negligible, so we can solve for the field, obtaining
\eq{\label{PhiInf}
\Phi(\infty) = \sqrt{{m^2 \over \la} - 3 \braket{\phi(\infty)^2}_{T}} \quad \textrm{for} \quad T <T_c\,,
}
where $T_c$ is the critical temperature, which is defined as the temperature for which the square root of Eq.~(\ref{PhiInf}) becomes negative. In the case where $T_{\rm bh} >T_c$, the temperature is too high to allow a non-zero minima for the potential, which means we have
\eq{
\Phi(\infty) = 0  \quad \textrm{for} \quad T >T_c\,,
}
so the symmetry of the potential is never broken. Evidently, we will be interested in the case where symmetry restoration occurs only at some finite distance from the horizon, meaning we will only consider temperatures at infinity where Eq.~(\ref{PhiInf}) is valid. The result (\ref{PhiInf}) will thus be used as the boundary condition at infinity.

As one proceeds from a positive value for $\Phi$ at infinity in the direction of the horizon, the first derivative starts out from a small negative value, and the second derivative from a small positive value, where the sign of these variations is largely dictated by the effective mass value. As the background field starts decreasing considerably, the effective mass will start to increase, and at some point the first and second derivative will switch signs, preventing the background field from reaching negative values, instead saturating in some value very close to zero. This is interpreted as a restoration of the symmetry of the effective potential, and the point where the derivatives switch sign will be called the bubble radius. In the next section we will explicitly find numerical representations of this physical picture.

%% file: II.7SymRest/5.section.tex
\section{Numerical results}
\label{NumRes}

In order to solve Eq.~(\ref{BFEr}) numerically, for each iteration of the self-consistent approximation, we divide the problem into three stages. First, from the effective potential of Eq.~(\ref{Veff}) at various radii we obtain an approximate value for the size of the bubble. This size can also be calculated by equating the Tolman temperature to the asymptotic critical temperature. The corresponding value of the field at the bubble radius will serve as a boundary condition, for the next step of the procedure, which consists in solving Eq.~(\ref{BFEr}) inside and outside the bubble separately.

Inside the bubble, we choose a point as close as possible to the horizon and evaluate the minima of the effective potential at that radius, for which we can find the corresponding value of $\Phi$. Using the latter value together with the determined value of the background field at the bubble radius, we may find the solution inside the bubble. For the region outside the bubble, it remains only to find the value of the field at infinity, i.e. some large value of the radius which saturates the value of the field. Since the field will be considered to be at thermal equilibrium with the black hole, its value at infinity will be given by Eq.~(\ref{PhiInf}) at the black hole temperature
\eq{\label{BHtemp}
T_{\rm bh} = {r_+-r_- \over 4\pi r^2_+}\,.
}
Employing the self-consistent approximation described in Sec.~(\ref{SymR}), we obtain the results for the background field in Fig.~1 and for the vacuum polarization in Fig.~2. The results have been checked using a slightly different approach which fixes the value of the field asymptotically and uses the value of the derivative as a shooting parameter. We have verified in a number of cases that the solutions obtained in the two ways coincide to the numerical accuracy we have used.
\begin{figure}[h]
\centering
\includegraphics[width=0.5\textwidth]{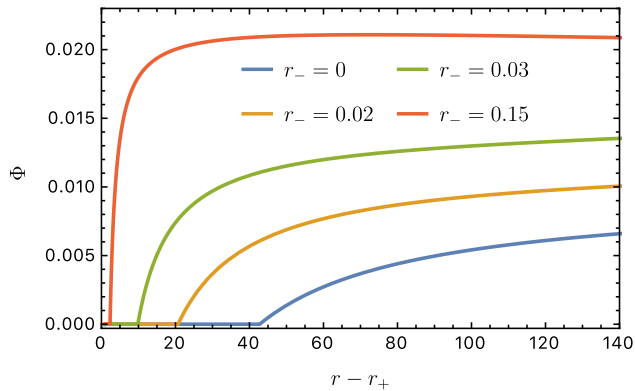}
\caption{Profile of the background field for $\la = 7.1\times 10^{-3}$, $m = 0.01$, $r_+=1$ and varying $r_-$.}
\end{figure}
\begin{figure}[h]
\centering
\includegraphics[width=0.5\textwidth]{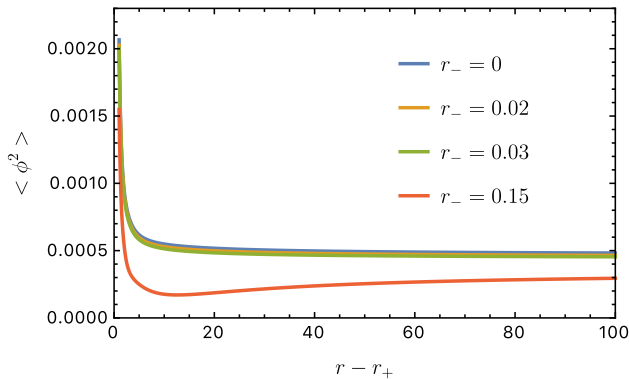}
\caption{Profile of the vacuum polarization for $\la = 7.1\times 10^{-3}$, $m = 0.01$, $r_+=1$ and varying $r_-$.}
\end{figure}

Some comments of the numerical results are in order. As the Cauchy radius $r_-$ is increased, i.e. as the black hole charge increases, the black hole becomes colder, as one may confirm through Eq.~(\ref{BHtemp}). As the temperature decreases, it becomes harder to excite the quantum field modes, so the vacuum polarization everywhere, which is evident from Fig.~2. From the point of view of the background field, symmetry restoration is more probable with increasing temperature, so colder black holes will have a smaller bubble of restored symmetry, something that is also clear from Fig.~1. Another interesting fact is that the overall form of the vacuum polarization is more affected by smaller bubbles. This is because $\braket{\phi^2}$ stabilizes quickly in a distance relatively small from the horizon, so the background field can only alter the form of the vacuum polarization curve when its region of larger variations (i.e. the bubble) is situated near the horizon. For larger bubbles, we see that the effect of the background field on the vacuum polarization are not so distinct. At the horizon, the field is negligible, so the value of $\braket{\phi(r_+)^2}$ is mostly unaltered. At infinity, the field $\Phi$ gives the largest finite contribution which translates in a smaller effective mass, which in turn increases the vacuum polarization.

With respect to the numerical calculations, we observe that the solutions stabilize relatively fast, at the order of three or four iterations of the self-consistent approximation. Regarding the vacuum polarization, the method employed here converges quickly on the horizon, where only some tens of modes are necessary to obtain a good result. However, at large radii (at the order of some hundreds of $r_+$), the convergence becomes slower, requiring the sum of some hundreds of modes. Since the order of magnitude of each Green mode function becomes very small for large distances, we are faced with the task of calculating very accurately hundreds of differences between very small numbers in Eq.~(\ref{phi2RenT}). As a consequence, we must find the numerical solutions of the homogeneous version of Eq.~(\ref{DifEqPnl}) for each mode with a very high precision, which revealed to be a considerable heavy and fine-tuned task for the symbolic manipulation software Mathematica used for the purpose. These shortcomings increased the overall computational time, which was reasonably lessened by parallelizing the code and using it in a computer cluster. In this regard, the numeric efficiency may be improved by adopting a different method to find the numerical solutions for the mode functions.

%% file: II.7SymRest/conclusions.tex
\section{Conclusions}

In this work we have constructed soliton-like bubble solutions for a self-interacting quantum scalar field around a charged four dimensional black hole. The method we have adopted includes self-consistently one-loop quantum effects encoded in the scalar vacuum polarization. The latter was calculated generalizing to the present case a new method developed in \cite{Taylor:2017sux}. The results we have obtained clearly support the picture where a broken symmetry is restored sufficiently near the black hole horizon, due to the increase of the local temperature associated to gravitational redshift. We confirmed this intuitive picture by extending the results of Refs.~\cite{Fawcett:1981fw,Moss:1984zf} and by explicitly constructing the solutions for the solitonic configuration. In particular, we observe that as the black hole becomes hotter, there is an increasingly bigger bubble-like region around the black hole where the temperature is high enough to induce a phase change in the background field. In contrast, as charge increases the bubble gets smaller. Interestingly, the vacuum polarization is seen to be considerably influenced only after the bubble, where the background field is strong enough.

%% file: 98.Conclusions/conclusions.tex
\chapter{Conclusions}
\label{cap:conclusions}

In this thesis we studied a number of physical systems at finite temperature under the influence of gravity. In Part I, we sought out to study the thermodynamic properties of black holes by using thin matter shells and the junction conditions. The same procedure was applied to a variety of different spacetimes, consisting in imposing the junction conditions on the thin shell such that the interior and exterior spacetimes to the shell formed together a single solution of the Einstein equations. This led to the specific mass and pressure necessary for the shell to remain static. By inserting those in the first law of thermodynamics, we were able to obtain the entropy differential in each situation, were the thermal equation of state remained an arbitrary function of the gravitational radius of the system. An ansatz was then given for this undetermined function, thus allowing the calculation of a specific entropy as well as an analysis of the intrinsic thermodynamic stability of the shell. The shells were then taken to their gravitational radius, leading to the Bekenstein-Hawking entropy. This result is by no means trivial, since there is no reason a priori for a system to have the Bekenstein-Hawking entropy once it was assumed to have a constant Hawking temperature throughout its distribution. In fact, since the shell is exactly at the event horizon of the black hole when it is taken to its gravitational radius and in that limit the usual black hole entropy is recovered, then we are strongly inclined to believe that this is evidence that the degrees of freedom of a black hole are situated at its event horizon.

A case which revealed to be particularly interesting was the extremally charged case, where it was seen that there were a number of choices one could take to attack the problem. An a priori extremal shell could be consired, or a charged shell could be made extremal only after it was taken to its gravitational radius. Different results where obtained depending on the approach, resulting in the final conclusion that an extremal black hole may have any entropy between zero and the Bekenstein-Hawking entropy, thus belonging to a entire different class of object compared to other black holes.

Part II of the thesis was dedicated to quantum systems at finite temperature in curved spacetimes. Vacuum polarisation effects where chosen as the quantity to calculate, for being the most simple ones to study while being physically relevant at the same time. The approach taken in the first instances was the standard one, which involved expressing the Green function as a sum in energy and angular modes, followed by the numerical calculation of the homogeneous equation solution. The numerical computations revolved around using a WKB approximation to the approximate result as an explicit divergent piece plus some finite remainder. The divergent pieces were always made of two components: one due to the angular modes summation and another due to the energy modes. The latter could always be removed with clever mathematical machinery, but the former could only be cancelled by introducing diverging counterterms to renormalize the result. After proper subtraction, the result was an explicitely finite quantity, which represented the renormalized vacuum polarisation. This method was shown to work for higher dimensional spacetimes, although it was clear that the degree of complexity rose very rapidly.

In the last chapter, a more realistic scenario was considered, where the quantum field was interacting with itself, thereby opening the door to phase transitions. After obtaining the differential equations for the vacuum polarisation and the background field, we applied a self-consistent approximation to numerically solve the system of equations. It was seen that, if the black hole was hot enough, a bubble would form around the black hole where the field was hot enough to exist in a symmetric phase.

%% file: AppendiceA/appendixA.tex
\chapter{Equations of thermodynamic stability for an electrically charged system}
\label{ap:a}

Reproducing the approach followed in \cite{Callen:1985}, in this appendix we shall show the derivation of the equations of thermodynamic stability for an electrically charged system, given by Eqs.~(\ref{Stab1}), (\ref{Stab2}) and (\ref{Stab3}).

We start by considering two identical subsystems, each with an entropy $S = S(M,A,Q)$, where $M$ is the internal energy of the system (equivalent to the rest mass), $A$ is its area and $Q$ its electric charge. The usual state variables of a thermodynamic system are the internal energy $U$, volume $V$ and other conserved quantities $N$, like the number of particles, for example. However, the system we wish to study is a thin shell, and thus it is more natural to use the variables $(M,A,Q)$. Thermodynamic stability is guaranteed if $dS = 0$ and $d^2 S < 0$ are both satisfied, or in other words, if the entropy is an extremum and a maximum respectively.

Now suppose we keep $A$ and $Q$ constant and remove a positive amount of internal energy $\Delta M$ from one subsystem to the other. The total entropy of the two subsystems goes from the value $2 S(M,A,Q)$ to $S(M+\Delta M, A, Q) + S(M-\Delta M,A,Q)$. If the initial entropy $S(M,A,Q)$ is a maximum, then the sum of initial entropies must be greater or equal to the sum of final entropies, i.e.
\begin{equation}\label{Beq1}
S(M+\Delta M, A, Q) + S(M-\Delta M,A,Q) \leq 2 S(M,A,Q).
\end{equation}
Expanding $S(M+\Delta M, A, Q)$ and $S(M-\Delta M, A, Q)$ in a Taylor series to second order in $\Delta M$, we see that Eq.~(\ref{Beq1}) becomes
\begin{equation}\label{Beq2}
\left(\frac{\dd^2 S}{\dd M^2}\right)_{A,Q} \leq 0
\end{equation}
in the limit $\Delta M \to 0$. The same reasoning applies if we fix $M$ and $Q$ instead and apply a positive change of area $\Delta A$, so we must have
\begin{equation}\label{Beq3}
S(M, A+\Delta A, Q) + S(M,A-\Delta A,Q) \leq 2 S(M,A,Q).
\end{equation}
which in the limit $\Delta A \to 0$ gives
\begin{equation}\label{Beq4}
\left(\frac{\dd^2 S}{\dd A^2}\right)_{M,Q} \leq 0.
\end{equation}
If we fix $M$ and $A$ and make a positive change $\Delta Q$ on the charge, we have
\begin{equation}\label{Beq5}
S(M, A, Q+\Delta Q) + S(M,A,Q-\Delta Q) \leq 2 S(M,A,Q).
\end{equation}
and so it follows that
\begin{equation}\label{Beq6}
\left(\frac{\dd^2 S}{\dd Q^2}\right)_{M,A} \leq 0.
\end{equation}
However, if we keep only one quantity fixed, like $Q$ for example, we must also have a final sum of entropies smaller than the initial sum if we apply a simultaneous change of area and internal energy rather than separately, i.e.
\begin{equation}\label{Beq7}
S(M+\Delta M, A+\Delta A, Q) + S(M-\Delta M,A-\Delta A,Q) \leq 2 S(M,A,Q).
\end{equation}
This inequality is satisfied by Eq.~(\ref{Beq2}) and Eq.~(\ref{Beq4}), but it also implies a new requirement. If we expand the left side in a Taylor series to second order in $\Delta M$ and $\Delta A$, and use the abbreviated notation $S_{ij} = \dd^2 S/\dd i \dd j$, we get
\begin{equation}
S_{MM} (\Delta M)^2 + 2 S_{MA} \Delta M \Delta A + S_{AA} (\Delta A)^2 \leq 0.
\end{equation}
Multiplying both sides by $S_{MM}$ and adding and subtracting $S_{MA}^2 (\Delta A)^2$ to the left side, allows the last inequality to be written in the form
\begin{equation}
(S_{MM} \Delta M + S_{MA} \Delta A)^2 + (S_{MM}S_{AA} - S_{MA}^2)(\Delta A)^2 \geq 0.
\end{equation}
Since the first term in the left side is always greater than zero, we see that it is sufficient to have
\begin{equation}\label{Beq8}
\left(\frac{\dd^2 S}{\dd M^2}\right) \left(\frac{\dd^2 S}{\dd A^2}\right) - \left(\frac{\dd^2 S}{\dd M \dd A}\right)^2 \geq 0.
\end{equation}
This concludes the derivation of Eqs.~(\ref{C1})-(\ref{C3}). However, we can repeat the same calculations but fixing $M$ and $A$ in turns. It is now straightforward to see that, when fixing $M$, we must have
\begin{equation}
S_{AA} (\Delta A)^2 + 2 S_{AQ} \Delta A \Delta Q + S_{QQ} (\Delta Q)^2 \leq 0,
\end{equation}
which is satisfied by
\begin{equation}\label{Beq9}
\left(\frac{\dd^2 S}{\dd A^2}\right) \left(\frac{\dd^2 S}{\dd Q^2}\right) - \left(\frac{\dd^2 S}{\dd A \dd Q}\right)^2 \geq 0.
\end{equation}
Finally, by fixing $A$ follows the inequality
\begin{equation}
S_{MM} (\Delta M)^2 + 2 S_{MQ} \Delta M \Delta Q + S_{QQ} (\Delta Q)^2 \leq 0
\end{equation}
which implies the sufficient condition
\begin{equation}\label{Beq10}
\left(\frac{\dd^2 S}{\dd M^2}\right) \left(\frac{\dd^2 S}{\dd Q^2}\right) - \left(\frac{\dd^2 S}{\dd M \dd Q}\right)^2 \geq 0.
\end{equation}

The last case left consists of doing a simultaneous change in all the state variables of the system, i.e.,
\begin{equation}\label{Beq11}
S(M+\Delta M, A+\Delta A, Q+\Delta Q) + S(M-\Delta M,A-\Delta A,Q-\Delta Q) \leq 2 S(M,A,Q).
\end{equation}
To investigate the sufficient differential condition that this inequality implies, one must first expand $S(M+\Delta M, A+\Delta A, Q+\Delta Q)$ and $S(M-\Delta M,A-\Delta A,Q-\Delta Q)$ in a Taylor series to second order in $\Delta M$, $\Delta A$ and $\Delta Q$, which can be shown to lead to
\begin{equation}\label{Beq12}
S_{MM} (\Delta M)^2 + S_{AA} (\Delta A)^2 + S_{QQ} (\Delta Q)^2 + 2 S_{MA} \Delta M \Delta A + 2 S_{MQ} \Delta M \Delta Q + 2 S_{QA} \Delta A \Delta Q \leq 0.
\end{equation}
Multiplying the above relation by $S_{MM}$, noting that
\begin{align}
(S_{MM} \Delta M + & S_{MA} \Delta A + S_{MQ} \Delta Q)^2 = S_{MM}^2 (\Delta M)^2 + S_{MA}^2 (\Delta A)^2 + S_{MQ}^2 (\Delta Q)^2 + \nonumber \\
& + 2 S_{MM} S_{MA} \Delta M \Delta A + 2 S_{MM} S_{MQ} \Delta M \Delta Q + 2 S_{MA} S_{MQ} \Delta A \Delta Q
\end{align}
and inserting this on Eq.~(\ref{Beq12}), gives
\begin{align}
(S_{MM} \Delta M + & S_{MA} \Delta A + S_{MQ} \Delta Q)^2 + (S_{MM}S_{AA} - S_{MA}^2)(\Delta A)^2 + (S_{MM}S_{QQ} - S_{MQ}^2)(\Delta Q)^2 +\nonumber \\
& + 2 (S_{MM}S_{QA}-S_{MA}S_{MQ}) \Delta A \Delta Q \geq 0.
\end{align}
Recalling Eq.~(\ref{Beq8}) and Eq.~(\ref{Beq10}), and noting that the first term in the above inequality is always positive, we conclude that the condition
\begin{equation}
\left(\frac{\dd^2 S}{\dd M^2}\right) \left(\frac{\dd^2 S}{\dd Q \dd A}\right) - \left(\frac{\dd^2 S}{\dd M \dd A}\right) \left(\frac{\dd^2 S}{\dd M \dd Q}\right) \geq 0
\end{equation}
is sufficient to satisfy Eq.~(\ref{Beq11}).

\cleardoublepage

%% file: AppendiceB/appendixB.tex
\chapter{Addition theorems for the functions $\ma{P}^{\mu}_{\nu}$ and $\ma{Q}^{\mu}_{\nu}$}
\label{ap:b}

\section{General result}

Using the approach followed in \cite{Erdelyi:1953}, we shall derive special cases of addition theorems for the functions $\ma{P}^{\mu}_{\nu}$ and $\ma{Q}^{\mu}_{\nu}$, introduced in Chapter 9. In order to generalize the addition theorem for the associated Legendre polynomials, we must first find some key relations between $\ma{P}$ and $\ma{Q}$, as defined by Eqs.~(\ref{parP}) and (\ref{parQ}). The first relation is a generalization of Whipple's formula, which is obtained by performing the substitutions $\nu \to -n -1/2$, $n \to -\nu-1/2$ and $z \to z(z^2-1)^{-1/2}$ on $\ma{Q}^{n}_{\nu}(z)$ and comparing it to $\ma{P}^{n}_{\nu}(z)$. We find that
\eq{\label{pmn}
\ma{P}^{n}_{\nu}(z) = { 2^{n+\nu+3/2} i (-1)^{\nu} (z^2-1)^{-1/4-\e/2} z^{2\e} \over \Gamma\left( {1\over 2}  - {n \over 2} - {\nu \over 2} + {\e \over 2} \right) \Gamma\left(   - {n \over 2} - {\nu \over 2} - {\e \over 2} \right)} \ma{Q}^{-\nu-{1\over 2}}_{-n-{1\over 2}}\left[z(z^2-1)^{-1/2}\right]\,,
}
which, in the limit $\e \to 0$, reduces to 3.3 (14) of \cite{Erdelyi:1953}. The next formula we're going to need is a generalization of 3.7 (10) of \cite{Erdelyi:1953}. Using 7.7 (29) with the substitution $a \to 1$, $b \to z$, $v \to -1/2-\e$, $\rho \to -n$ and $\mu \to 1/2 + \nu$, we obtain
\eq{
\ma{Q}^{n}_{\nu}(z) = {(-1)^n (z^2-1)^{n/2} z^{1/2-\e} \over 2} \Gamma\left( 1 + {n \over 2} + {\nu \over 2} + {\e \over 2} \right) \Gamma\left(- {n \over 2} - {\nu \over 2} - {\e \over 2} \right) \int^{\infty}_{0} J_{\nu+{1\over 2}}(t) \, J_{-{1\over 2} - \e}(z t) \, t^n \, dt
}
It is also possible to show that this can be written as
\eq{\label{qmn}
\ma{Q}^{n}_{\nu}(z) = (-1)^n (z^2-1)^{n/2} z^{1/2-\e} {1 \over 2} (-1)^{{1\over 4} - {\e \over 2}} \int^{\infty}_{0} \pi (-1)^{-{\nu \over 2} - {1\over 4}} J_{\nu+{1\over 2}}(it) \, H^{(1)}_{-{1\over 2} - \e}(i z t) \, t^n \, dt\,.
}
where we used the result (straightforwardly proven in a symbolic manipulation software like Mathematica)
\eq{
\int^{\infty}_{0} J_{\nu+{1\over 2}}(t) \, J_{-{1\over 2} - \e}(z t) \, t^n \, dt = \sin\left[\pi \left( -{n \over 2} -{\nu \over 2} -{\e \over 2} \right) \right] (-1)^{{1\over 4} - {\e \over 2}} \int^{\infty}_{0} I_{\nu + {1\over 2}}(t) H^{(1)}_{-{1\over 2} - \e}(i z t) t^n \, dt\,.
}
Using now the integral expansion of the Bessel function (derived from 7.7 (31) of \cite{Erdelyi:1953})
\eq{
\pi (-1)^{-{\nu \over 2} - {1\over 4}} J_{\nu+{1\over 2}}(it) = \left[\int^{\pi}_{0} e^{t \cos x} \cos\left(\nu x + {x\over 2}\right) dx - \cos\left(\nu \pi\right) \int^{\infty}_{0} e^{-\nu x - {t\over 2} - t \cosh x} dx \right]
}
in Eq.~(\ref{qmn}), we obtain
\ea{
\ma{Q}^{n}_{\nu}(z) = & {1 \over 2} (-1)^{n} (z^2-1)^{n/2} z^{1/2-\e}\,, \nonumber \\
& \times \left\{ \int^{\pi}_{0} dx \cos\left(\nu x + {x\over 2}\right) A - \cos\left(\nu \pi\right)\int^{\infty}_{0} dx \, e^{-\nu x}  B \right\}\,,
}
\ea{
A = & \int^{\infty}_0  e^{t \cos x}  \, (-1)^{{1\over 4} - {\e \over 2}} H^{(1)}_{-{1\over 2} - \e}(i z t) \, t^n \,dt = {\Gamma\left({3 \over 2}+n+\e\right)\Gamma\left({1 \over 2}+n-\e\right) \over \Gamma\left({3 \over 2}+n\right) 2^n z^{n+1}\sqrt{\pi}} \nonumber \\
& \times F\left[{1\over 2} + n - \e,{3\over 2}+n+\e; {3\over 2} + n; {z + \cos(x) \over 2z}\right]\,, \nonumber \\
B = & \int^{\infty}_{0} e^{- {t\over 2} - t \cosh x} \, (-1)^{{1\over 4} - {\e \over 2}} H^{(1)}_{-{1\over 2} - \e}(i z t) \, t^n \, dt \,.
}
Inserting this representation in Eq.~(\ref{pmn}) and noting that $n$ is an integer, we obtain
\ea{
\ma{P}^{n}_{\nu}(z) = & {\Gamma\left( - \nu -\e \right) 2^{1+n+\nu} (-1)^n 2^{\e}\over \sqrt{\pi} \Gamma\left({1\over 2}-{n \over 2}-{\nu \over 2}+{\e \over 2}\right) \Gamma\left(-{n \over 2}-{\nu \over 2}-{\e \over 2}\right)} \int^{\pi}_{0} g(x) \cos\left(n x\right) dx
}
where
\ea{
g(x) = & 2^{\nu-\e} z^{\nu + \e} {\Gamma\left(1-\nu+\e\right) \over \Gamma\left(1-\nu\right)} F\left[- \nu - \e,1-\nu+\e; 1-\nu; {z-\sqrt{z^2-1} \cos(x) \over 2z}\right]\,.
}
We can also write
\eq{
\ma{P}^{n}_{\nu}(z) = {\Gamma\left(1+n+\nu+\e\right) \over 2 \pi \Gamma\left(1+\nu+\e\right)} { \Gamma\left({1\over 2}-{n \over 2}-{\nu \over 2}-{\e \over 2}\right) \over \Gamma\left({1\over 2}-{n \over 2}-{\nu \over 2}+{\e \over 2}\right)} \int^{\pi}_{-\pi} g(x) e^{i n x} dx
}
hence, substituting $x=\Phi-\psi$, we may observe that
\eq{\label{finalPmn}
\ma{P}^{n}_{\nu}(z) \cos(n \psi) = {\Gamma\left(1+n+\nu+\e\right) \over 2 \pi \Gamma\left(1+\nu+\e\right)} { \Gamma\left({1\over 2}-{n \over 2}-{\nu \over 2}-{\e \over 2}\right) \over \Gamma\left({1\over 2}-{n \over 2}-{\nu \over 2}+{\e \over 2}\right)} \int^{2\pi}_{0} g(\Phi-\psi) \cos(n \Phi) d\Phi\,.
}
Now, consider the sum
\eq{
S \equiv  {\Gamma\left({1\over 2}-{\nu \over 2}+{\e \over 2}\right) \over \Gamma\left({1\over 2}-{\nu \over 2}-{\e \over 2}\right)} \ma{P}^{0}_{\nu}(z) + 2 \sum^{\infty}_{n=1} {\Gamma\left(1+\nu+\e\right) \over \Gamma\left(1+n+\nu+\e\right)} {\Gamma\left({1\over 2}-{n \over 2}-{\nu \over 2}+{\e \over 2}\right)  \over \Gamma\left({1\over 2}-{n \over 2}-{\nu \over 2}-{\e \over 2}\right)} \ma{P}^{n}_{\nu}(z) \cos[n (v-\psi)]\,.
}
Using Eq.~(\ref{finalPmn}), we obtain
\eq{
S = {b_0 \over 2} + \sum^{\infty}_{n=1} b_n \cos(n \, 0)
}
where
\eq{
b_n \equiv {1\over \pi}\int^{2\pi}_{0} g[\Phi-(v-\psi)] \cos(n \Phi) d\Phi
}
are the coefficients of the Fourier expansion of the function $g[\Phi-(v-\psi)]$. Thus, we conclude that
\eq{
S = g[-(v-\psi)]
}
and so we prove the following expansion:
\ea{
S = & \sum^{\infty}_{n=-\infty} {\Gamma\left(1+\nu+\e\right) \over \Gamma\left(1+n+\nu+\e\right)} {\Gamma\left({1\over 2}-{n \over 2}-{\nu \over 2}+{\e \over 2}\right)  \over \Gamma\left({1\over 2}-{n \over 2}-{\nu \over 2}-{\e \over 2}\right)} \ma{P}^{n}_{\nu}(z) e^{-i n \psi} e^{i n v} \nonumber \\
= & \sum^{\infty}_{n=-\infty} b_n e^{i m v} \nonumber \\
= & 2^{\nu-\e} z^{\nu + \e} {\Gamma\left(1-\nu+\e\right) \over \Gamma\left(1-\nu\right)} F\left[- \nu - \e,1-\nu+\e; 1-\nu; {z-\sqrt{z^2-1} \cos(v-\psi) \over 2z}\right] \nonumber \\
\equiv & f_1(v) \,.
}
Setting $w=0$, changing $\nu \to -\nu-1$ and noting, by direct inspection, that $\ma{P}^{n}_{\nu}(z) = \ma{P}^{n}_{-\nu-1}(z)$, we obtain the version
\ea{
& \sum^{\infty}_{n=-\infty} {\Gamma\left(-\nu+\e\right) \over \Gamma\left(n-\nu+\e\right)} {\Gamma\left(1-{n \over 2}+{\nu \over 2}+{\e \over 2}\right)  \over \Gamma\left(1-{n \over 2}+{\nu \over 2}-{\e \over 2}\right)} \ma{P}^{n}_{\nu}(z) e^{i n v}  \nonumber \\
= & \sum^{\infty}_{n=-\infty} a_n e^{i m v} \nonumber \\
= & 2^{-1-\nu-\e} z^{-1-\nu + \e}  {\Gamma\left(2+\nu+\e\right) \over \Gamma\left(2+\nu\right)} F\left[1+ \nu - \e,2+\nu+\e; 2+\nu; {z-\sqrt{z^2-1} \cos(v) \over 2z}\right] \nonumber \\
\equiv & f_2(v) \,.
}
Now we note that
\ea{
\sum^{\infty}_{n=-\infty} a_n \overline{b}_n  & \sum^{\infty}_{n=-\infty} {\Gamma\left(1+\nu+\e\right) \over \Gamma\left(1+n+\nu+\e\right)} {\Gamma\left({1\over 2}-{n \over 2}-{\nu \over 2}+{\e \over 2}\right)  \over \Gamma\left({1\over 2}-{n \over 2}-{\nu \over 2}-{\e \over 2}\right)} {\Gamma\left(-\nu+\e\right) \over \Gamma\left(n-\nu+\e\right)} {\Gamma\left(1-{n \over 2}+{\nu \over 2}+{\e \over 2}\right)  \over \Gamma\left(1-{n \over 2}+{\nu \over 2}-{\e \over 2}\right)} \ma{P}^{n}_{\nu}(z) \ma{P}^{n}_{\nu}(z') e^{i n \psi}  \nonumber \\
= & \sum^{\infty}_{n=-\infty} {\Gamma\left(1+\nu+\e\right) \over \Gamma\left(1+\nu-\e\right)} 4^{-\e} {\Gamma\left({1\over 2}-{n \over 2}+{\nu \over 2}-{\e \over 2}\right) \Gamma\left({1\over 2}+{n \over 2}-{\nu \over 2}+{\e \over 2}\right)  \over \Gamma\left({1\over 2}+{n \over 2}+{\nu \over 2}+{\e \over 2}\right) \Gamma\left({1\over 2}-{n \over 2}-{\nu \over 2}-{\e \over 2}\right)} \nonumber \\
& \times \left( 4^{-n} {\Gamma\left(1-{n \over 2}+{\nu \over 2}+{\e \over 2}\right) \Gamma\left({1\over 2}-{n \over 2}+{\nu \over 2}-{\e \over 2}\right)  \over \Gamma\left(1+{n \over 2}+{\nu \over 2}+{\e \over 2}\right) \Gamma\left({1\over 2}+{n \over 2}+{\nu \over 2}-{\e \over 2}\right)} \right) \ma{P}^{n}_{\nu}(z) \ma{P}^{n}_{\nu}(z') e^{i n \psi} \nonumber \\
= & {\Gamma\left(1+\nu+\e\right) \over \Gamma\left(1+\nu-\e\right)} 4^{-\e} \sum^{\infty}_{n=-\infty}  {\cos\left[{\pi \over 2}\left(n+\nu+\e\right)\right] \over \cos\left[{\pi \over 2}\left(n-\nu+\e\right)\right]}  \ma{P}^{-n}_{\nu}(z) \ma{P}^{n}_{\nu}(z') e^{i n \psi}
}
and that, using Parseval's theorem, this is equal to
\ea{
\sum^{\infty}_{n=-\infty} a_n \overline{b}_n & = {1\over 2\pi} \int^{\pi}_{-\pi} \left(\sum^{\infty}_{p = - \infty} a_p e^{i p \Phi} \right) \left(\sum^{\infty}_{k = - \infty} \overline{b}_k e^{- i k \Phi} \right) \, d\Phi \nonumber \\
= & {1\over 2\pi} \int^{\pi}_{-\pi} f_1(\Phi) f_2(\Phi) \, d\Phi \,.
}
Thus, we obtain the addition theorem
\ea{
& {1\over 2\pi} \int^{\pi}_{-\pi} 2^{-1} z_1^{\nu+\e}z_2^{-1 -\nu + \e} {\Gamma\left(1+\nu-\e\right) \over \Gamma\left(1+\nu+\e\right)} {\Gamma\left(1-\nu+\e\right) \over \Gamma\left(1-\nu\right)}  {\Gamma\left(2+\nu+\e\right) \over \Gamma\left(2+\nu\right)} \times  \nonumber \\
& \times F\left[- \nu - \e,1-\nu+\e; 1-\nu; {z_1-\sqrt{z_1^2-1} \cos(\Phi-\psi) \over 2z_1}\right] \nonumber \\
& F\left[1+ \nu - \e,2+\nu+\e; 2+\nu; {z_2-\sqrt{z_2^2-1} \cos(\Phi) \over 2z_2}\right] d\Phi= \sum^{\infty}_{n=-\infty}  {\cos\left[{\pi \over 2}\left(n+\nu+\e\right)\right] \over \cos\left[{\pi \over 2}\left(n-\nu+\e\right)\right]}  \ma{P}^{-n}_{\nu}(z_1) \ma{P}^{n}_{\nu}(z_2) e^{i n \psi}\,.
}
or, in a more useful form,
\ea{
& {1\over 2\pi} \int^{\pi}_{-\pi} (z_1 z_2)^{\e} {\Gamma\left(1+\nu-\e\right) \over \Gamma\left(1+\nu+\e\right)} {\Gamma\left(1-\nu+\e\right) \over \Gamma\left(1-\nu\right)}  {\Gamma\left(2+\nu+\e\right) \over \Gamma\left(2+\nu\right)} {\left[z_1+\sqrt{z_1^2-1} \cos(\Phi-\psi)\right]^{\nu} \over \left[z_2+\sqrt{z_2^2-1} \cos(\Phi)\right]^{\nu+1}}  \nonumber \\
& \times F\left[1+ \e,-\e; 1-\nu; 1-{z_1+\sqrt{z_1^2-1} \cos(\Phi-\psi) \over 2z_1}\right] \, F\left[1+\e,-\e; 2+\nu; 1-{z_2+\sqrt{z_2^2-1} \cos(\Phi) \over 2z_2}\right] d\Phi \nonumber \\
& = \sum^{\infty}_{n=-\infty}  {\cos\left[{\pi \over 2}\left(n+\nu+\e\right)\right] \over \cos\left[{\pi \over 2}\left(n-\nu+\e\right)\right]}  \ma{P}^{-n}_{\nu}(z_1) \ma{P}^{n}_{\nu}(z_2) e^{i n \psi}\,.
}
which reduces to the Legendre addition theorem in the limit $\e \to 0$. To find a closed form for the left hand side, one must express the integral as a complex counter integral, as is done in \cite{Whittaker:1902}. This case however, is more complicated.

Our job now is to transform the above integral in a complex integral and make use of the formula
\eq{\label{Theorem}
P_{\nu}(z) = {1 \over 2\pi} \int_{C} 2^{-\nu} {(t^2-1)^{\nu} \over (t-z)^{\nu+1}} \, dt
}
where $C$ is a path containing the point $z$. For that, following \cite{Whittaker:1902}, we define the complex variable
\eq{
t = {e^{i \Phi } \left(e^{-i\psi} \sqrt{z_1-1}\sqrt{z_2^2+1}-z1\sqrt{z_2-1}\right) + z_1 \sqrt{z_2+1}-e^{i \psi} \sqrt{z_1^2-1} \sqrt{z_2-1} \over e^{i\Phi} \sqrt{z_2-1} + \sqrt{z_2+1}}\,,
}
which, for $\Phi \in [-\pi,\pi]$, can be shown to draw a circle $C$ containing the points $t=1$ and $t=z_*$, with
\eq{
z_* = z_1 z_2 - \sqrt{z_1^2-1}\sqrt{z_2^2-1}\cos\psi\,.
}
Now, defining
\ea{
y_1 & = z_1 + \sqrt{z_1^2-1} \cos(\Phi - \psi)\,, \\
y_2 & = z_2 + \sqrt{z_2^2-1} \cos(\Phi)\,,
}
it can be shown that
\eq{
{d\Phi \over y_2} = {dt\over (t-z)}
}
and that
\eq{\label{R}
{y_1 \over y_2} = 2^{-1} \left({t^2-1 \over t-z}\right) \equiv R\,,
}
so the integral can be written as
\ea{\label{AlmostInt}
& {1\over 2\pi} \int_{C} (z_1 z_2)^{\e} {\Gamma\left(1+\nu-\e\right) \over \Gamma\left(1+\nu+\e\right)} {\Gamma\left(1-\nu+\e\right) \over \Gamma\left(1-\nu\right)}  {\Gamma\left(2+\nu+\e\right) \over \Gamma\left(2+\nu\right)}  2^{-\nu} {(t^2-1)^{\nu} \over (t-z)^{\nu+1}}\,  \nonumber \\
& \times F\left[1+ \e,-\e; 1-\nu; 1-{y_1 \over 2z_1}\right] \, F\left[1+\e,-\e; 2+\nu; 1-{y_2 \over 2z_2}\right] dt\,.
}

The hypergeometrics can be dealt with by using formula (8.3.6) of \cite{Beals:2010}, where we get
\ea{
F\left[1+ \e,-\e; 1-\nu; 1-{y_1 \over 2z_1}\right] = &
{\Gamma\left(1-\nu\right)\Gamma\left(-\nu\right) \over \Gamma\left(1-\nu+\e\right)\Gamma\left(-\nu-\e\right)} F\left[1+ \e,-\e; 1+\nu; {y_1 \over 2z_1}\right] + \nonumber \\
& {\Gamma\left(1-\nu\right)\Gamma\left(\nu\right) \over \Gamma\left(1+\e\right)\Gamma\left(-\e\right)} \left({y_1 \over 2z_1}\right)^{-\nu} F\left[1+ \e-\nu,-\e-\nu; 1-\nu; {y_1 \over 2z_1}\right] \nonumber \\
& \equiv F_1 + F_2 \label{HG1} \,, \\
F\left[1+ \e,-\e; 2+\nu; 1-{y_2 \over 2z_2}\right] = & {\Gamma\left(1+\nu\right)\Gamma\left(2+\nu\right) \over \Gamma\left(1+\nu-\e\right)\Gamma\left(2+\nu+\e\right)} F\left[1+ \e,-\e; -\nu; {y_2 \over 2z_2}\right] + \nonumber \\
& {\Gamma\left(2+\nu\right)\Gamma\left(-1-\nu\right) \over \Gamma\left(1+\e\right)\Gamma\left(-\e\right)} \left({y_2 \over 2z_2}\right)^{1+\nu} F\left[2+ \e+\nu,1-\e+\nu; 2+\nu; {y_2 \over 2z_2}\right] \nonumber \\
&\equiv F_3 + F_4 \label{HG2} \,.
}

\section{Results for $\psi = 0$ and integer $\e$}

We now simplify the calculations by considering the coincidence limit in the time coordinate, i.e. $\psi = 0$. Using then the definition of $y_1$ and $y_2$, solving for $\cos(\Phi)$ and equating, we obtain
\eq{
{y_1 - z_1 \over \sqrt{z_1^2-1}} = {y_2 - z_2 \over \sqrt{z_2^2-1}}\,,
}
so, using Eq.~(\ref{R}), we get
\ea{
y_1 & = {z_1-\al z_2 \over R - \al}R \,, \\
y_2 & = {z_1-\al z_2 \over R - \al} \,,
}
with
\eq{
\al = {\sqrt{z_1^2-1} \over \sqrt{z_2^2-1}}\,.
}
Using Eqs.~(\ref{HG1}) and (\ref{HG2}), the integral (\ref{AlmostInt}) becomes divided into four parts. However, for integer $\e$, only one term is non-zero. We focus on this term and use the series representation of the hypergeometric functions, together with the Cauchy product formula for infinite series and Eq.~(\ref{Theorem}), obtain the result
\ea{
(z_1 z_2)^{\e}{\sin[\pi(\e + \nu)] \over \sin(\pi \nu)} \sum^{\infty}_{k=0}\sum^{k}_{l=0} {(1+\e)_{l}(-\e)_{l}(1+\e)_{k-l}(-\e)_{k-l} \over (1+\nu)_{k-l} (-\nu)_{l} (k-l)! \, l!} z_2^{-l} z_1^{-k+l} \left({z_1 - \al z_2 \over 2}\right)^{k} \sum^{\infty}_{p=0} {(k)_p \over p!} \al^p P_{\nu-l-p}(z_*)
}
for the integral (\ref{AlmostInt}). The addition theorem for integer $\e$ then becomes
\ea{\label{ATPP}
\sum^{\infty}_{n=-\infty} (-1)^n \ma{P}^{-n}_{\nu}(z_1) \ma{P}^{n}_{\nu}(z_2) = & (z_1 z_2)^{\e} \sum^{\infty}_{k=0}\sum^{k}_{l=0} {(1+\e)_{l}(-\e)_{l}(1+\e)_{k-l}(-\e)_{k-l} \over (1+\nu)_{k-l} (-\nu)_{l} (k-l)! \, l!} z_2^{-l} z_1^{-k+l} \left({z_1 - \al z_2 \over 2}\right)^{k} \nonumber \\
& \times \sum^{\infty}_{p=0} {(k)_p \over p!} \al^p P_{\nu-l-p}(z_*)\,.
}
We are now interested in the limit $z_1 = z_2$, which must be taken with care. Expansion of the summand and subsequent summation is not enough, since the summation in $k$ goes up to infinity, so all orders are relevant. One way to do it is to recall the integral representation
\eq{
P_{\la}(z) = {1 \over \pi} \int^{\pi}_{0} {d\phi \over (z + \sqrt{z^2-1}\cos \phi)^{\la+1}}
}
so that we get
\ea{
\left({z_1 - \al z_2}\right)^{k} \sum^{\infty}_{p=0} {(k)_p \over p!} \al^p P_{\nu-l-p}(z_*) & = \left({z_1 - \al z_2}\right)^{k} \sum^{\infty}_{p=0} {(k)_p \over p!} \al^p {1 \over \pi} \int^{\pi}_{0} {d\phi \over (z_* + \sqrt{z_*^2-1}\cos \phi)^{\nu-l-p+1}} \nonumber \\
& = {1 \over \pi} \int^{\pi}_{0} {1 \over (z_* + \sqrt{z_*^2-1} \cos \phi)^{\nu-l+1}} \left( {z_1-\al z_2 \over 1-z_* \al - \al \sqrt{z_*^2-1} \cos \phi} \right)^k d\phi
}
which, for $z_1$ very close to $z_2$, gives
\eq{\label{Pk1}
{1 \over \pi} \int^{\pi}_{0} \left\{{1 \over (z_1 - \sqrt{z_1^2-1}\cos \phi)^{k}}+\mathcal{O}(z_1-z_2) \right\} d\phi = P_{k-1}(z_1) + \mathcal{O}(z_1-z_2)
}
so the addition theorem in the limit $z_1 \to z_2$ becomes
\eq{
\sum^{\infty}_{n=-\infty} (-1)^n \ma{P}^{-n}_{\nu}(z_1) \ma{P}^{n}_{\nu}(z_1 ) = z_1^{2\e} \sum^{\infty}_{k=0} c_{k m} (2 z_1)^{-k} P_{k-1}(z_1) = z_1^{2\e}
}
with
\eq{
c_{k m} = \sum^{k}_{l=0} {(1+\e)_{l}(-\e)_{l}(1+\e)_{k-l}(-\e)_{k-l} \over (1+\nu)_{k-l} (-\nu)_{l} (k-l)! \, l!} = \sum^{k}_{l=0} (-1)^l {(1+\e-l)_{k} (-\e-l)_{k} \over (1+\nu-l)_{k} (k-l)! \, l!}\,,
}
where we used the result
\eq{
\sum^{\infty}_{k=0} c_{k m} (2 z_1)^{-k} P_{k-1}(z_1) = 1.
}
To get the addition theorem in terms of $\ma{P}$ and $\ma{Q}$, we must find the expression for $\ma{P}_{\nu}^n(-z)$ in terms of those functions. Decomposing the function $\ma{P}^{n}_{\nu}(-z)$ in terms of hypergeometrics with argument $z^2$ and comparing with the same decompositions for $\ma{P}^n_{\nu}$ and $\ma{Q}^n_{\nu}$, we obtain the result (for integer $\e$ only)
\eq{
\ma{P}^n_{\nu}(-z) = (-1)^{\pm \nu}(-1)^{\e+1} \ma{P}^{n}_{\nu}(z) - {2 \over \pi} z^{2\e} (-1)^{\e+1} \sin(\pi\nu) \ma{Q}^{n}_{\nu}(z)\,.
}
The usual trick now consists in taking $z_2 \to -z_2$ and $\psi \to \psi + \pi$ in the addition theorem (\ref{ATPP}). What we get in the end is
\ea{
\sum^{\infty}_{n=-\infty} (-1)^n \ma{P}^{-n}_{\nu}(z_1) \ma{Q}^{n}_{\nu}(z_2) = & \left({z_1 \over z_2}\right)^{\e} \sum^{2\e}_{k=0}\sum^{k}_{l=0} {(1+\e)_{l}(-\e)_{l}(1+\e)_{k-l}(-\e)_{k-l} \over (1+\nu)_{k-l} (-\nu)_{l} (k-l)! \, l!} z_2^{-l} z_1^{-k+l} \left({z_1 - \al z_2 \over 2}\right)^{k} \nonumber \\
& \times \sum^{\infty}_{p=0} {(k)_p \over p!} \al^p Q_{\nu-l-p}(z_*)
}
which has been confirmed numerically to be correct. Note that, in the limit $\e \to 0$, only the term with $k=l=0$ survives, leaving $Q_{\nu}(z_*)$, the expected result from \cite{Shiraishi:1993nu}. We will keep that term separated in the full sum. To take the limit $z_1 = z_2$ and thus obtaining the final desired result, one may note the following relations
\eq{
Q_{\la}(\cos(\theta)) = Q_{-\la-1}(\cos(\theta)) + {\pi \over \tan(\pi \lambda)} P_{\lambda}(\cos(\theta))
}
and
\eq{
Q_{\la}(\cos(\theta)) = {1 \over 2} \int^{\infty}_{0} {dt \over (\cos(\theta)+ i\sin(\theta) \cosh(t))^{\la +1}} + {1 \over 2} \int^{\infty}_{0} {dt \over (\cos(\theta)- i\sin(\theta) \cosh(t))^{\la +1}}
}
where $0 < \theta < \pi$. Now we define $z_1=\cos(\theta_1)$ and $z_2=\cos(\theta_2)$ with $\theta_2 = \theta_1 + \de \theta$. In that case, $z_* = \cos(\de\theta)$, $\al = \sin(\theta_1)/\sin(\theta_1+\de\theta)$ and we have
\eq{
\left({z_1 - \al z_2}\right)^{k} \sum^{\infty}_{p=0} {(k)_p \over p!} \al^p Q_{\nu-l-p}(z_*) = \ma{Q}_1 + \ma{Q}_2
}
where
\ea{
\ma{Q}_1 = & {1 \over \tan(\pi \nu)} \left({\sin(\theta_1 + \de \theta) \over \sin(\theta_1 +\de\theta)-\sin(\theta_1)}\right)^k \times \nonumber \\
& \int^{\pi}_{0} {1 \over (\cos(\de\theta)+i \sin(\de\theta)\cos(t))^{\nu-l+1}} \left({1-\al \over 1- \cos(\de\theta) \al - \al i \sin(\de\theta)\cos(t)}\right)^k dt \\
\ma{Q}_2 = &  \left({\sin(\theta_1 + \de \theta) \over \sin(\theta_1 +\de\theta)-\sin(\theta_1)}\right)^k \times \nonumber \\
& \bigg\{{1 \over 2} \int^{\infty}_{0} {dt \over (\cos(\de\theta)+ i\sin(\de\theta) \cosh(t))^{l-\nu-k}} \left({1-\al \over \cos(\de\theta) + i \sin(\de\theta)\cosh(t)-\al}\right)^k \\
& + {1 \over 2} \int^{\infty}_{0} {dt \over (\cos(\de\theta)- i\sin(\de\theta) \cosh(t))^{l-\nu-k}} \left({1-\al \over \cos(\de\theta) - i \sin(\de\theta)\cosh(t)-\al}\right)^k \bigg\} \nonumber \,.
}
From the above results we can quickly take the limit $\de\theta \to 0$, giving
\eq{
\left({z_1 - \al z_2}\right)^{k} \sum^{\infty}_{p=0} {(k)_p \over p!} \al^p Q_{\nu-l-p}(z_*) \to {\pi \over \tan(\pi \nu)} P_{k-1}(\cos(\theta_1)) + Q_{k-1}(\cos(\theta_1))\,,
}
leading us to the final form of the addition formula in the coincidence limit
\eq{
\sum^{\infty}_{n=-\infty} (-1)^n \ma{P}^{-n}_{\nu}(z_1) \ma{Q}^{n}_{\nu}(z_1) = \lim_{z_2 \to z_1} Q_{\nu}(z_*) + \sum^{2\e}_{k=1} c_{k m} (2z_1)^{-k}  Q_{k-1}(z_1)\,.
}